\documentclass[12pt]{article}
\usepackage{amsmath}
\usepackage{graphicx}
\graphicspath{{png_figs/}}

\usepackage{enumerate}
\usepackage{natbib}
\usepackage{url} % not crucial - just used below for the URL 

\newcommand{\blind}{1}

\usepackage{bookmark}

\newenvironment{proof}{\paragraph{Proof:}}{\hfill$\square$}
\usepackage{subcaption}
\usepackage{float}
\usepackage{enumerate}% http://ctan.org/pkg/enumerate
\usepackage{soul}
\usepackage{amsmath, amssymb} 
\usepackage{mathrsfs}
\usepackage{subcaption}
\usepackage{xcolor}

\usepackage[framemethod=tikz]{mdframed}

\usepackage{mhequ}
\usepackage{xcolor}
\usepackage{hyperref}

\newtheorem{theorem}{Theorem}

\newtheorem{remark}{Remark}
\newtheorem{theorem*}{Theorem S}

\makeatletter

\newcounter{example}[section]

\newcommand{\pr}{\text{Pr}}

\newcommand{\rulesep}{\vrule height -1ex\ }
\renewcommand{\t}[1]{\text{#1}}

\usepackage{xcolor}

\newcommand{\be}{\begin{equation*}
        \begin{aligned}}

\newcommand{\ee}{ \end{aligned}
        \end{equation*}
}

\newcommand{\bel}{\begin{equation}
        \begin{aligned}}

\newcommand{\eel}{ \end{aligned}
        \end{equation}
}
\usepackage{overpic}

\usepackage{float}

\usepackage{algorithm}
\usepackage[noend]{algpseudocode}

\newcommand{\env}[2]{\begin{#1}#2\end{#1}}

\usepackage{subfiles}

\begin{document}

\def\spacingset#1{\renewcommand{\baselinestretch}%
{#1}\small\normalsize} \spacingset{1}

%%%%%%%%%%%%%%%%%%%%%%%%%%%%%%%%%%%%%%%%%%%%%%%%%%%%%%%%%%%%%%%%%%%%%%%%%%%%%%

\if1\blind
{
        \title{Spectral Clustering, Bayesian Spanning Forest, and 
        Forest Process}
\author{Leo L. Duan\footnote{Department of Statistics, University of Florida, {li.duan@ufl.edu}}
\qquad Arkaprava Roy \footnote{Department of Biostatistics, University of Florida, {ark007@ufl.edu}}}
  \maketitle
} \fi

\if0\blind
{
  \bigskip
  \bigskip
  \bigskip
  \begin{center}
    {\LARGE\bf Spectral Clustering, Bayesian Spanning Forest, and 
        Forest Process}
\end{center}
  \medskip
} \fi

\bigskip
\begin{abstract}
Spectral clustering views the similarity matrix as a weighted graph, and partitions the data by minimizing a graph-cut loss. Since it minimizes the across-cluster similarity, there is no need to model the distribution within each cluster. As a result, one reduces the chance of model misspecification, which is often a risk in mixture model-based clustering. Nevertheless, compared to the latter, spectral clustering has no direct ways of quantifying the clustering uncertainty (such as the assignment probability), or allowing easy model extensions for complicated data applications. To fill this gap, we propose the Bayesian forest model as a generative graphical model for spectral clustering. This is motivated by our discovery that the posterior connecting matrix in a forest model has almost the same leading eigenvectors, as the ones used by normalized spectral clustering. To induce a distribution for the forest, we develop a ``forest process'' as a graph extension to the urn process, while we carefully characterize the differences in the partition probability. We derive a simple Markov chain Monte Carlo algorithm for posterior estimation, and demonstrate superior  performance compared to existing algorithms. We illustrate several  model-based extensions useful for data applications, including  high-dimensional and multi-view clustering for images.
\end{abstract}

\noindent%
\vfill

\newpage
\spacingset{1.9} % DON'T change the spacing!
%% Here are the title, author names and addresses

%%%%%%%%%%%%%%%%%%%%%%%%%%%%%%%%%%%%%%%%%%%%%%%%%%%%%%%%%%%%%%%

%%%%%%%%%%%%%%%%%%%%%%%%%%%%%%%%%%%%%%%%%%%%%%%%%%%%%%%%%%%%%%%

\section{Introduction}
Clustering aims to partition data $y_1,\ldots, y_n$ into disjoint groups. There is a large literature ranging from various algorithms such as K-means and DBSCAN \citep{macqueen1967classification,10.5555/3001460.3001507,frey2007clustering} to mixture model-based approaches [reviewed by \cite{fraley2002model}]. In the Bayesian community, model-based approaches are especially popular. To roughly summarize the idea, we view each $y_i$ as generated from a distribution $\mathcal K(\cdot\mid \theta_i)$, where  $(\theta_1,\ldots,\theta_n)$ are drawn from a discrete distribution $ \sum_{k=1}^K w_k \delta_{\theta^*_k}(\cdot)$, with $w_k$ as the probability weight, and $\delta_{\theta^*_k}$ as a point mass at $\theta^*_k$. With prior distributions, we could estimate all the unknown parameters  ($\theta^*_k$'s, $w_k$'s, and $K$) from the posterior.

The model-based clustering has two important advantages. First, it allows important uncertainty quantification such as the probability for cluster assignment $c_i$, $\text{Pr}(c_i=k\mid y_i)$, as a probabilistic estimate that $y_i$ comes from the $k$th cluster ($c_i=k \Leftrightarrow \theta_i =\theta^*_k$). Different from commonly seen asymptotic results in statistical estimation, the clustering uncertainty does not always vanish even as $n\to \infty$. For example, in a two-component Gaussian mixture model with equal covariance, for a point $y_i$ at nearly equal distances to two cluster centers, we would have both $\text{Pr}(c_i=1\mid y_i)$ and $\text{Pr}(c_i=2\mid y_i)$ close to $50\%$ even as $n\to \infty$. For a recent discussion on this topic as well as how to quantify the partition uncertainty, see \cite{wade2018bayesian} and the references within. Second, the model-based clustering can be easily extended to handle more complicated modeling tasks. Specifically, since there is a probabilistic process associated with the clustering, it is straightforward to modify it to include useful dependency structures. We list a few examples from a rich literature: \cite{ng2006mixture} used a mixture model with random effects to cluster correlated gene-expression data,  \cite{muller2010random,park2010bayesian,ren2011logistic} allowed the partition to vary according to some covariates, \cite{guha2016nonparametric} simultaneously clustered the predictors and use them in high-dimensional regression.

On the other hand, model-based clustering has its limitations. Primarily, one needs to carefully specify the density/mass function $\mathcal K$, otherwise, it will lead to unwanted results and difficult interpretation. For example, \cite{coretto2016robust} demonstrated the sensitivity of the Gaussian mixture model to non-Gaussian contaminants, \cite{miller2018robust} and \cite{cai2021finite} showed that when the distribution family of $\mathcal K$ is misspecified, the number of clusters would be severely overestimated. It is natural to think of using more flexible parameterization for $\mathcal K$, in order to mitigate the risk of model misspecification. This has motivated many interesting works, such as modeling $\mathcal K$ via skewed distribution \citep{fruhwirth2010bayesian,lee2016finite}, unimodal distribution \citep{rodriguez2014univariate}, copula \citep{kosmidis2016model}, mixture of mixtures \citep{malsiner2017identifying}, among others. Nevertheless, as the flexibility of $\mathcal K$ increases, the modeling and computational burdens also increase dramatically.

In parallel to the above advancements in model-based clustering, spectral clustering has become very popular in machine learning and statistics.  \cite{von2007tutorial} provided a useful tutorial on the algorithms and a review of recent works. On clustering point estimation, spectral clustering has shown good empirical performance for separating non-Gaussian and/or manifold data, without the need to directly specify the distribution for each cluster. Instead, one calculates a matrix of similarity scores between each pair of data, then uses a simple algorithm to find a partition that approximately minimizes the total loss of similarity scores across clusters (adjusted with respect to cluster sizes). This point estimate is found to be not very sensitive to the choice of similarity score, and empirical solutions have been proposed for tuning the similarity and choosing the number of clusters \citep{zelnik2005self,shi2009data}. There is a rapidly growing literature of frequentist methods on further improving the point estimate [\cite{chi2007evolutionary,rohe2011spectral,NIPS2011_31839b03,lei2015consistency,han2021eigen,lei2022bias}; among others], although, in this article, we focus on the Bayesian perspective and aim to characterize the probability distribution.

Due to the algorithmic nature, spectral clustering cannot be directly used in model-based extension, or produce uncertainty quantification. This has motivated a large Bayesian literature. There have been several works trying to quantify the uncertainty around the spectral clustering point estimate. For example, since the spectral clustering algorithm can be used to estimate the community memberships in a stochastic block model, one could transform the data into a similarity matrix, then treat it as if generated from a Bayesian stochastic block model \citep{snijders1997estimation,nowicki2001estimation,mcdaid2013improved,geng2019probabilistic}. Similarly, one could take the Laplacian matrix (a transform of the similarity used in spectral clustering) or its spectral decomposition, and model it in a probabilistic framework \citep{socher2011spectral,duan2019spiked}.

Broadly speaking, we can view these works as following the recent trend of robust Bayesian methodology, in conditioning the parameter of interest (clustering) on an insufficient statistic (pairwise summary statistics) of the data. See \cite{lewis2021bayesian} for recent discussions.
Pertaining to Bayesian robust clustering, one gains model robustness by avoiding putting any parametric assumption on within-cluster distribution $\mathcal K(\cdot \mid \theta^*_k)$; instead, one models the pairwise information that often has an arguably simple distribution. Recent works include the distance-based P\'olya urn process
\citep{blei2011distance,socher2011spectral},
Dirichlet process mixture model on Laplacian eigenmaps \citep{banerjee2015bayesian},
Bayesian distance clustering \citep{duan2021dist},
generalized Bayes extension of product partition model
\citep{rigon2020generalized}.

This article follows this trend. Instead of modeling $y_i$'s as conditionally independent (or jointly dependent) from a certain within-cluster distribution $\mathcal K(\cdot \mid \theta_k^*)$, we choose to model $y_i$ as dependent on another point $y_j$ that is close by, provided $y_i$ and $y_j$ are from the same cluster. This leads to a Markov graphical model based on a spanning forest, a graph consisting of multiple disjoint spanning trees (each tree as a connected subgraph without cycles). The spanning forest itself is not new to statistics. There has been a large literature on using spanning trees and forests for graph estimation, such as \cite{meila2000learning,meilua2006tractable,edwards2010selecting,byrne2015structural,duan2021bayesian,luo2021bayesian}. Nevertheless, a key difference between graph estimation and graph-based clustering is that --- the former aims to recover both the node partition and the edges characterizing dependencies, while the latter only focuses on estimating the node partition alone (equivalent to clustering). Therefore, a distinction of our study is that we will treat the edges as a nuisance parameter/latent variable, while we will characterize the node partition in the marginal distribution.

Importantly, we formally show that by marginalizing the randomness of edges, the point estimate on the node partition is provably close to the one from the normalized spectral clustering algorithm. As the result, the spanning forest model can serve as the probabilistic model for the spectral clustering algorithm --- this relationship is analogous to the one between the Gaussian mixture model and the K-means algorithm \citep{macqueen1967classification}. Further, we show that treating the spanning forest as random, as opposed to a  fixed parameter (that is unknown), leads to much less sensitivity in clustering performance, compared to cutting the minimum spanning tree algorithm \citep{gower1969minimum}.

On the distribution specification on the node and edges, we take a Bayesian non-parametric approach by considering the forest model as realized from a ``forest process'' --- each cluster is initiated with a  point from a root distribution, then gradually grown with new points from a leaf distribution. We characterize the key differences in the partition distribution between the forest and classic P\'olya urn processes. This difference also reveals that extra care should be exerted during model specification when using graphical models for clustering.

Lastly, by establishing the probabilistic model counterpart for spectral clustering, we show how such models can be easily extended to incorporate other dependency structures. We demonstrate several extensions, including a multi-subject clustering of the brain networks, and a high-dimensional clustering of photo images.

\vspace*{-1cm}
\section{Method}
\vspace*{-1cm}
\subsection{Background on Spectral Clustering Algorithms}
\vspace*{-0.5cm}
We first provide a brief review of spectral clustering algorithms. For data $y_1,\ldots, y_n$, let $A_{i,j}\ge 0$ be a similarity score between $y_i$ and $y_j$, and denote the degree $D_{i,i}= \sum_{j\neq i} A_{i,j}$. To partition the data index $(1,\ldots, n)$ into $K$ sets, $\mathcal V= (V_1,\ldots, V_K)$, we want to solve the following problem:
\bel\label{eq:normalized_graph_cut}
\min_{\mathcal V} \sum_{k=1}^K \frac{\sum_{i\in V_k, j \not \in V_k} A_{i,j}}{
\sum_{i\in V_k} D_{i,i} }.
\eel
This is known as the minimum normalized cut loss. The numerator above represents the across-cluster similarity due to cutting $V_k$ off from the others; and the denominator prevents trivial solutions of forming tiny clusters with small $ \sum_{i\in V_k} D_{i,i}$.

This optimization problem is a combinatorial problem, hence has motivated approximate solutions such as spectral clustering. To start, using the Laplacian matrix $L=D-A$ with $D$ the diagonal matrix of $D_{i,i}$'s, and the normalized Laplacian $N=D^{-1/2} L D^{-1/2}$, we can equivalently solve the above problem via:
\be
\min_{\mathcal V}  \text{tr} (Z'_{\mathcal V} N Z_{\mathcal V}),
\ee
where $Z_{{\mathcal V}: i,k} =   1(i\in V_k) \sqrt{D_{i,i}} / \sqrt{\sum_{i\in V_k}D_{i,i}}$. It is not hard to verify that $Z'_{\mathcal V}Z_{\mathcal V}=I_K$. We can obtain a relaxed minimizer of $Z:Z'Z=I_K$, by simply taking $\hat Z$ as the bottom $K$ eigenvectors of $N$ (with the minimum loss equal to the sum of the smallest $K$ eigenvalues). Afterward, we cluster the rows of $\hat Z$ into $K$ groups (using algorithms such as the K-means), hence producing an approximate solution to \eqref{eq:normalized_graph_cut}.

To clarify, there is more than one version of the spectral clustering algorithms. An alternative version to \eqref{eq:normalized_graph_cut} is called ``minimum ratio cut'', which replaces the denominator $\sum_{i\in V_k} D_{i,i} $ by the size of cluster $|V_k|$. Similarly, continuous relaxation approximation can be obtained by following the same procedures above, except for clustering the eigenvectors of the unnormalized $L$. Details on comparing those two versions can be found in \cite{von2007tutorial}. In this article, we focus on the one based on \eqref{eq:normalized_graph_cut} and the normalized Laplacian matrix $N$. This version is also commonly referred to as ``normalized spectral clustering''.

\vspace*{-0.5cm}
\subsection{Probabilistic Model via Bayesian Spanning Forest}
\vspace*{-0.5cm}
%The spectral clustering algorithms operate on the pairwise similarity matrix $A$, which is some {\em transform} of the original data $y$.
The next question is if there is some partition-based generative model for $y$, that has the maximum likelihood estimate (or, the posterior mode in the Bayesian framework) almost the same as the point estimate from the normalized spectral clustering.

We found an almost equivalence in the spanning forest model. A spanning forest model is a special Bayesian network that describes the conditional dependencies among $y_1,\ldots, y_n$. Given a partition $\mathcal V=(V_1,\ldots, V_K)$ of the data index $(1,\ldots,n)$, consider a forest graph $\mathcal F_{\mathcal V}= (T_1,\ldots, T_k)$, with each $T_k=(V_k, E_k)$ a component tree (a connected subgraph without cycles), $V_k$ the set of nodes and $E_k$ the set of edges among $V_k$. Using $\mathcal F_{\mathcal V}$ and a set of root nodes $\mathcal R_{\mathcal V}=(1^*,\ldots, K^*)$ with $k^*\in V_k$, we can form a graphical model with a conditional likelihood given the forest:
\bel\label{eq:dag_likelihood}
\mathcal L(y ; \mathcal V, \mathcal F_{\mathcal V}, \mathcal R_{\mathcal V}, \theta) =  \prod_{k=1}^{K}
\bigg [ r(y_{k^*} ; \theta) \prod_{(i,j) \in T_k }f(y_i \mid y_j; \theta)
\bigg ],
\eel
where we refer to $r( \cdot ; \theta)$ as a ``root'' distribution, and $f(\cdot\mid y_j;\theta)$ as a ``leaf'' distribution; and we use $\theta$ to denote the other parameter; and we use simplified notation $(i,j)\in G$ to mean that $(i,j)$ is  an edge of the graph $G$.

% Consider the data generated based on a likelihood $\mathcal L(y ; \mathcal F, \theta)$, which depends on a graph $\mathcal F$ known as ``spanning forest'' (or simply, ``forest'') and some other parameter $\theta$. To introduce the necessary background, a spanning forest is the union of $K$ disjoint spanning trees, and each tree is a connected graph. Importantly, (i) each tree has no cycles hence it is the smallest graph connecting $|V_k|$ nodes, and (ii) those trees give a partition of the data index $(1,\ldots,n)=(V_1 \cup\ldots \cup V_K)$. Further, in each tree $T_k$, there is one node. Therefore, we have the graph $\mathcal F= (T_1,\ldots, T_k, )$.

 \begin{figure}[H]
        \begin{subfigure}[t]{.3\textwidth}
        \centering
           \includegraphics[width=.9\linewidth]{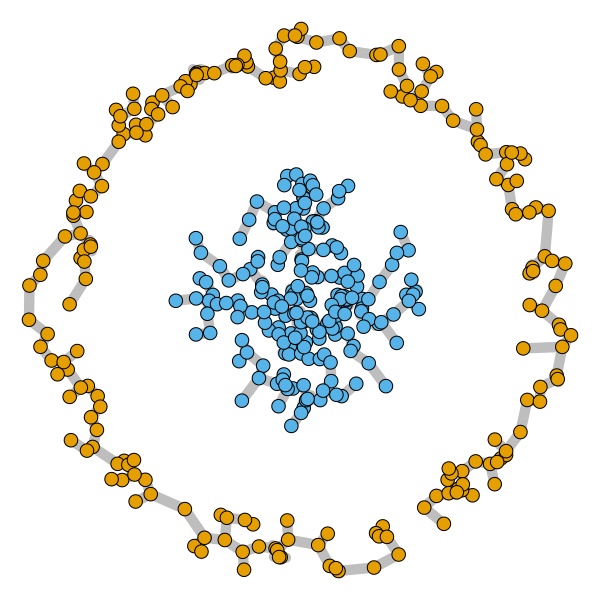}
       \end{subfigure}
               \begin{subfigure}[t]{.3\textwidth}
                       \centering
           \includegraphics[width=.9\linewidth]{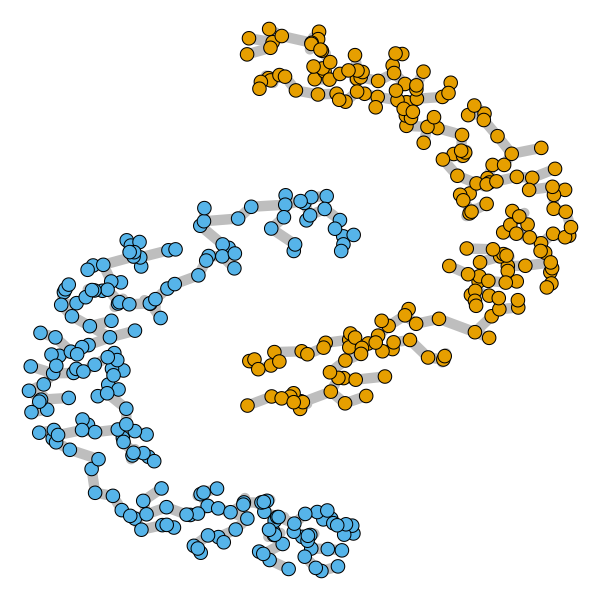}
       \end{subfigure}
               \begin{subfigure}[t]{.3\textwidth}
                       \centering
           \includegraphics[width=.9\linewidth]{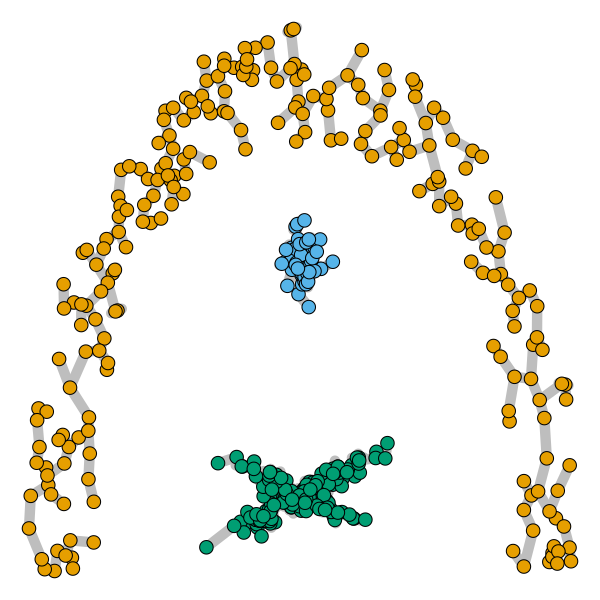}
       \end{subfigure}
        \caption{Three examples of clusters that can be represented by a spanning forest.\label{fig:illustration}}
\end{figure}
\vspace*{-1cm}
\noindent
 Figure \ref{fig:illustration} illustrates the high flexibility of a spanning forest in representing clusters. It shows the sampled $\mathcal F$ based on three clustering benchmark datasets. Note that some clusters are not elliptical or convex in shape. Rather, each cluster can be imagined as if it were formed by connecting a point to another nearby.
 \vspace*{-0.2cm}
\begin{remark}
To clarify, the point estimation on a spanning forest (as some fixed and unknown graph) has been studied \citep{gower1969minimum}. However, a distinction here is that we consider $\mathcal V$ as the parameter of interest, but the edges and roots ($\mathcal F_{\mathcal V},\mathcal R_{\mathcal V}$) as latent variables. The performance differences are shown in the Supplementary Materials S4.6.
\end{remark}
The stochastic view of $(\mathcal F_\mathcal V,\mathcal R_\mathcal V)$ is important, as it allows us to incorporate the uncertainty of edges and avoids the sensitivity issue in the point graph estimate. Equivalently, our clustering model is based on the marginal likelihood that varies with the node partition $\mathcal V$:
\vspace*{-0.5cm}
\bel\label{eq:marginal_lik}
\mathcal L(y ; \mathcal V, \theta)
=\sum_{\mathcal F_{\mathcal V}, \mathcal R_{\mathcal V}}\mathcal L(y ; \mathcal V, \mathcal F_{\mathcal V}, \mathcal R_{\mathcal V}, \theta)\Pi(\mathcal F_\mathcal V, \mathcal R_{\mathcal V}\mid \mathcal V).
\eel
\vspace*{-0.3cm}
\noindent where $\Pi(\mathcal F_\mathcal V, \mathcal R_{\mathcal V}\mid \mathcal V)$ is the latent variable distribution that we will specify in the next section. We can quantify the marginal connecting probability for each potential edge $(i,j)$:
 \vspace*{-0.5cm}
\bel\label{eq:marginal_connecting}
M_{i,j}:=\text{Pr}[F_\mathcal V \ni (i,j)]
\propto \sum_{\mathcal V}\sum_{\mathcal F_{\mathcal V}, \mathcal R_{\mathcal V}}
1[(i,j)\in  F_\mathcal V]
\mathcal L(y ; \mathcal V, \mathcal F_{\mathcal V}, \mathcal R_{\mathcal V}, \theta)\Pi(\mathcal F_\mathcal V, \mathcal R_{\mathcal V}\mid \mathcal V).
\eel
Similar to the normalized graph cut, there is no closed-form solution for directly maximizing \eqref{eq:marginal_lik}. However, closed-form does exist for  \eqref{eq:marginal_connecting} (see Section 4). Therefore, an approximate maximizer of \eqref{eq:marginal_lik}, $\hat {\mathcal V}$, can be obtained via computing the matrix $M$ and searching for $K$ diagonal blocks (after row and column index permutation) that contain the highest total values of $M_{i,j}$'s. Specifically, we can extract the top leading eigenvectors of $M$ and cluster the rows into $K$ groups.

\begin{figure}[H]
        \centering
           \includegraphics[width=.8\linewidth]{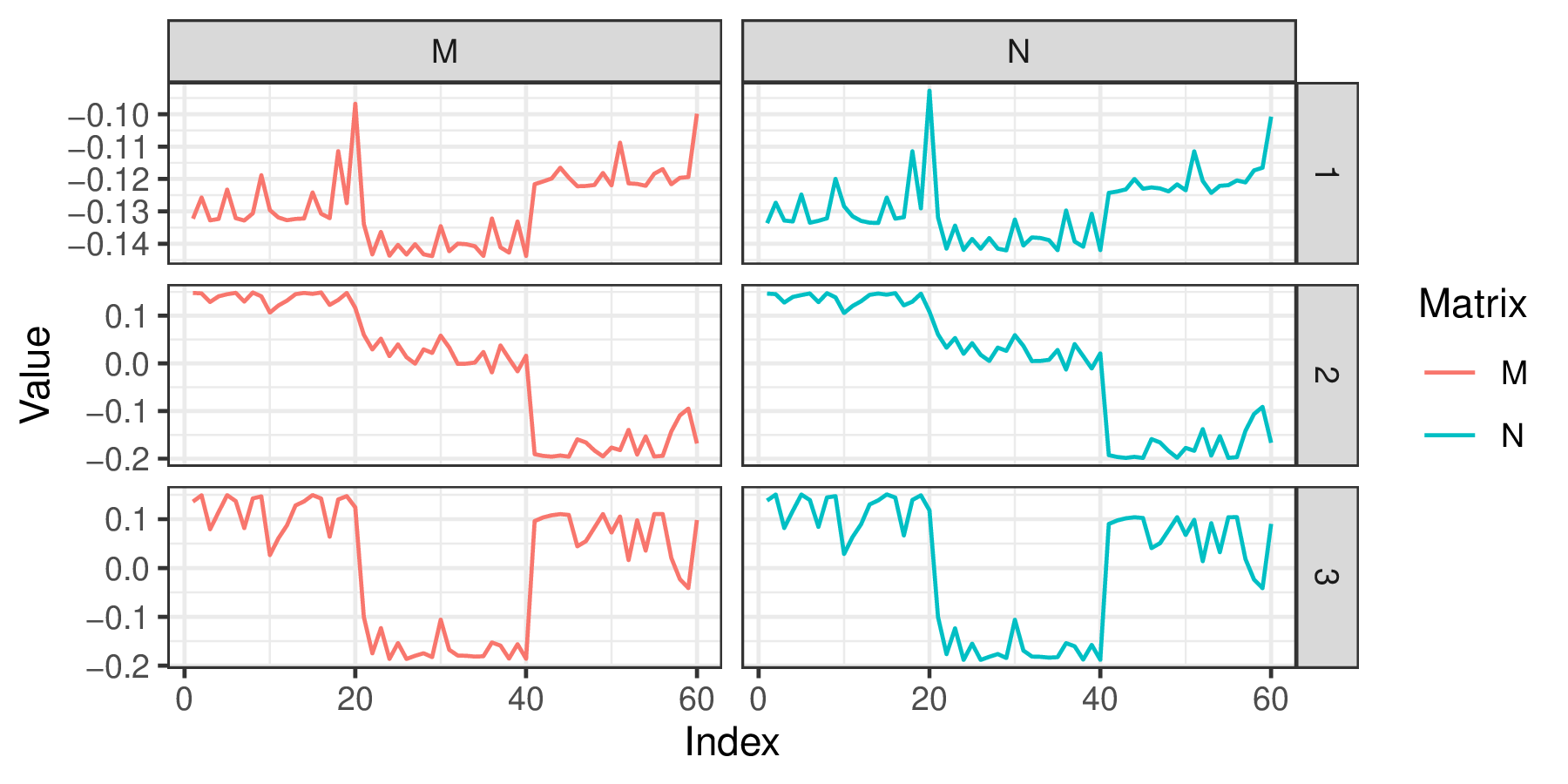}
               \caption{Comparing the eigenvectors of a marginal connecting probability matrix $M$ and the ones of normalized Laplacian $N$.\label{fig:spectral_equiv}}
\end{figure}
\vspace*{-0.5cm}

This approximate marginal likelihood maximizer produces almost the same estimate as the normalized spectral clustering does. This is because the two sets of eigenvectors are almost the same.  Further, it is important to clarify that such closeness does not depend on how the data are really generated. Therefore, to provide some numerical evidence, for simplicity, we generate $y_i$ from a simple three-component Gaussian mixture in $\mathbb{R}^2$ with means in $(0,0),(2,2),(4,4)$ and all variances equal to $I_2$. Figure \ref{fig:spectral_equiv}  compares the eigenvectors of the matrix $M$ and the normalized Laplacian $N$ (that uses $f$ and $r$ to specify $A$, with details provided in Section 4). Clearly, these two are almost identical in values. Due to this connection, the clustering estimates from spectral clustering can be viewed as an approximate estimate for $\hat{\mathcal V}$ in \eqref{eq:marginal_lik}.% And the spanning forest graph provides a probabilistic model for the spectral clustering algorithm.

We now fully specify the Bayesian forest model.
For simplicity, we now focus on continuous $y_i \in \mathbb{R}^p$. For ease of computation, we recommend choosing $f$ as a symmetric function $f(y_i\mid y_j; \theta) = f(y_j\mid y_i;\theta)$, so that the likelihood is invariant to the direction of each edge; and choose $r$ as a diffuse density, so that the likelihood is less sensitive to the choice of a node as root.
In this article, we choose a Gaussian density for $f$ and Cauchy for $r$:
\bel\label{eq:choice_density}
%& f(y_i \mid y_j;\theta) =  {(2\pi\sigma_{i,j})^{-1/2}} \exp \left \{ - \frac {\|y_i-y_j\|_2^2}{2\sigma_{i,j}} \right\},\\
%&r(y_i; \theta) = (\pi \gamma)^{-1}  \frac{1}{ 1+ \|y_{i}-\mu\|_2^2/\gamma^2}.
& f(y_i \mid y_j;\theta) =  {(2\pi\sigma_{i,j})^{-p/2}} \exp \left \{ - \frac {\|y_i-y_j\|_2^2}{2\sigma_{i,j}} \right\},\\
&r(y_i; \theta) =  \frac{\Gamma[(1+p)/2]}{\gamma^p \pi^{(1+p)/2}} \frac{1}{ (1+ \|y_{i}-\mu\|_2^2/\gamma^2)^{(1+p)/2}}.
\eel
where  $\sigma_{ij} >0$ and $\gamma>0$ are scale parameters.
As the magnitudes of distances between neighboring points may differ significantly from cluster to cluster, we use a local parameterization $\sigma_{i,j} =\tilde \sigma_i \tilde \sigma_j$, and will regularize $(\tilde \sigma_1,\ldots, \tilde \sigma_n)$ via a hyper-prior.

\begin{remark}
        In \eqref{eq:choice_density}, we effectively use Euclidean distances $\|y_i-y_j\|_2$. We focus on Euclidean distance in the main text, for the simplicity of presentation and to allow a complete specification of priors. One can replace Euclidean distance with some others, such as Mahalanobis distance and geodesic distance. We present a case of high-dimensional clustering based on geodesic distance on the unit-sphere in the Supplementary Materials S1.1.
 \end{remark}
\vspace*{-1cm}
\subsection{Forest Process  and Product Partition Prior}
\vspace*{-0.5cm}
To simplify notations as well as to facilitate computation, we now introduce an auxiliary node $0$ that connects to all roots $(1^*,\ldots, K^*)$. As the result, the model can be equivalently represented by a spanning tree rooted at $0$:
\be
& \mathcal T=(V_{\mathcal T}, E_{\mathcal T}), \\
& V_{\mathcal T}=\{0\} \cup V_1 \cup \ldots \cup V_K, \; E_{\mathcal T}= \{ (0,1^*),\ldots, (0,K^*)\} \cup E_1\cup\ldots \cup E_K.
\ee
In this section, we focus on the distribution specification for $\mathcal T$. The distribution, denoted by $\Pi(\mathcal T)$, $\Pi(\mathcal T)$ can be factorized according to the following hierarchies: picking the number of partitions $K$, partitioning the nodes into $(V_1,\ldots, V_K)$, forming edges $E_k$ and picking one root $k^*$ for each $V_k$. To be clear on the nomenclature, we call $\Pi(\mathcal F_{\mathcal V}, \mathcal R_{\mathcal V} \mid \mathcal V)$ as the ``latent variable distribution'', $\Pi_0(\mathcal V)$ as the ``partition prior''.

\bel\label{eq:hierarchical}
\Pi(\mathcal T)= \underbrace{
\bigg\{
\Pi_0(K)\Pi_0(V_1,\ldots, V_K \mid K)}_{\Pi_0(\mathcal V)
}
\bigg\}
\underbrace{\prod_{k=1}^K
 \bigg \{\Pi(E_k\mid V_k) \Pi( k^* \mid E_k, V_k) \bigg\}}_{
\Pi(\mathcal F_{\mathcal V}, \mathcal R_{\mathcal V} \mid \mathcal V)}.
\eel

\begin{remark}
In Bayesian non-parametric literature, $\Pi_0(K)\Pi_0(V_1,\ldots, V_K \mid K)$ is known as the partition probability function, which plays the key role in controlling cluster sizes and cluster number in model-based clustering. However, when it comes to graphical model-based clustering (such as our forest model), it is important to note the difference  --- for each partition $V_k$, there is an additional probability $\Pi(E_k, k^* \mid V_k)$ due to the multiplicity of all possible subgraphs formed between the nodes in $V_k$.
\end{remark}

For simplicity, we will use discrete uniform distribution for $\Pi(E_k, k^* \mid V_k)$. Since there are $n_k^{(n_k-2)_+}$  possible spanning trees for $n_k$ nodes [$(x)_+=x$ if $x>0$, otherwise $0$], and  $n_k$ possible choice of roots. We have $\Pi(E_k, k^* \mid V_k) = n_k^{-(n_k-1)}$.

We now discuss two different ways to complete the distribution specification. We first take a ``ground-up'' approach by viewing  $\mathcal T$ as from a stochastic process where the node number $n$ could grow indefinitely. Starting from the first edge $e_1=(0,1)$, we sequentially draw new edges and add to $\mathcal T$, from
\bel\label{eq:fp}
& e_i \mid e_1,\ldots e_{i-1} \sim  \sum_{j=1}^{i-1} \pi^{[i]}_j \delta_{(j,i)}(\cdot) +  \pi^{[i]}_i \delta_{(0,i)}(\cdot),\\
& y_i \mid (j,i) \sim 1(j \ge 1) f(\cdot \mid y_j)+ 1(j=0) r(\cdot),
\eel
with some  probability vector $(\pi^{[i]}_1,\ldots, \pi^{[i]}_i)$  that adds up to one.
 We refer to \eqref{eq:fp} as a forest process. The forest process is a generalization of the P\'olya urn process \citep{blackwell1973ferguson}. For the latter, $e_i=(j,i)$ would make node $i$ take the same value as node $j$, $y_i =y_j$ [although in model-based clustering, one would use notation $\theta_i =\theta_j$ , and $y_i\sim \mathcal K(\cdot \mid \theta_i)$]; $e_i=(0,i)$ would make node $i$ draw a new value for $y_i$ from the base distribution.  Due to this relationship, we can borrow popular parameterization for $\pi^{[i]}_j$ from the urn process literature. For example, we can use the Chinese restaurant process parameterization $\pi^{[i]}_j= 1/(i-1+\alpha)$ for $j=1,\ldots, (i-1)$, and $\pi^{[i]}_i= \alpha/(i-1+\alpha)$ with some chosen $\alpha>0$. After marginalizing over the order of $i$ and partition index [see \cite{miller2019elementary} for a simplified proof of the partition function], we obtain:
\bel\label{eq:crp}
\Pi(\mathcal T)=\frac{\alpha^{K} \Gamma(\alpha)}{\Gamma(\alpha+n)} \prod_{k=1} ^K\Gamma(n_k) n_k^{-(n_k-1)}.
\eel
Compared to the partition probability prior in the Chinese restaurant process, we have an additional $n_k^{-(n_k-1)}$ term that corresponds to the conditional prior weight of for each possible $(k^*,E_k)$ given a partition $V_k$.

To help understand the effect of this additional term on the posterior, we can imagine two extreme possibilities in the conditional likelihood given a $V_k$. If the conditional  $\mathcal L(y_i: i\in V_k \mid k^*,E_k)$ is skewed toward one particular choice of tree $(\hat k^*,\hat E_k)$ [that is, $\mathcal L(y_i: i\in V_k \mid k^*,E_k)$ is large when $(k^*,E_k)= (\hat k^*,\hat E_k)$, but is close to zero for other values of $(k^*,E_k)$], then $n_k^{-(n_k-1)}$ acts as a penalty for a lack of diversity in trees.
 On the other hand, if $\mathcal L(y_i: i\in V_k \mid k^*,E_k)$ is  equal for all possible $(k^*,E_k)$'s, then we can simply marginalize over $(k^*, E_k)$ and be not be subject to this penalty [since  $\sum_{(k^*, E_k)} n_k^{-(n_k-1)}=1$].

 Therefore, we can form an intuition by interpolating those two extremes: if a set of data points (of size $n_k$) are ``well-knit'' such that they can be connected via many possible spanning trees (each with a high conditional likelihood), then it would have a higher posterior probability of being clustered together, compared to some other points (of the same size $n_k$) that have only a few trees with high conditional likelihood.

With the ``ground-up'' construction useful for understanding the difference from the classic urn process, the distribution \eqref{eq:crp} itself is not very convenient for posterior computation. Therefore, we also explore the alternative of a ``top-down" approach. This is based on directly assigning a product
partition probability \citep{hartigan1990partition,barry1993bayesian,crowley1997product,quintana2003bayesian} as
\bel\label{eq:simple_prior}
& \Pi_0(V_1,\ldots, V_K \mid K) =  \frac{\prod_{k=1}^K  n_k^{(n_k-1)}}{
\sum_{\text{all }(V^*_1,\ldots, V^*_K) }\prod_{k=1}^K  |V^*_k|^{(|V^*_k|-1)}
},
\eel
where the cohesion function $n_k^{(n_k-1)}$ effectively cancels out the  probability for each $(k^*,E_k)$. To assign a prior for $K$, we assign a probability
$$\Pi_0(K) \propto \lambda^K \sum_{\text{all }(V^*_1,\ldots, V^*_K) }\prod_{k=1}^K  |V^*_k|^{(|V^*_k|-1)},$$supported on $K\in \{1,\ldots,n\}$ with $\lambda>0$, with $\Pi(E_k, k^* \mid V_k) = n_k^{-(n_k-1)}$, multiplying the terms according to \eqref{eq:hierarchical} leads to
\bel\label{eq:trunc_poisson}
\Pi(\mathcal T) \propto  \lambda^K,
\eel
which is similar to a truncated geometric distribution
 and easy to handle in posterior computation, and we will use this from now on.  In this article, we set $\lambda=0.5$.

\begin{remark}
We now discuss the exchangeability of the sequence of random variables generated from the above forest process. The exchangeability is defined as the the invariance of distribution $\Pi( X_1=x_1,\ldots X_n=x_n ) = \Pi( X_1= x_{\tilde\pi_1},\ldots X_n=x_{\tilde\pi_n})$ under any permutation $(\tilde\pi_1,\ldots,\tilde\pi_n)$ \citep{diaconis1977finite}. For simplicity, we focus on the joint distribution with $\theta$ marginalized out. There are three categories of random variables associated with each node index $i$: the first drawn edge $(j,i)$ that points to a new node $i$ (whose sequence forms $\mathcal T=(\mathcal V, \{E_k, k^*\}_{k=1}^K)$), the cluster assignment of a node $c_i$  (whose sequence forms $\mathcal V$), and the data point $y_i$. It is not hard to see that, since each component tree encodes an order among $\{i: c_i=k\}$, the joint distribution of the data and the forest $\Pi(y_1,\ldots, y_n, \mathcal  T)$ is not exchangeable. Nevertheless, as we marginalize out each $(E_k,k^*)$  to form the clustering likelihood $\mathcal L(y ; \mathcal V)$ as in \eqref{eq:marginal_lik}, and all priors $\Pi_0(\mathcal V)$ presented in this section only depend on the number and sizes of clusters, the joint distribution of the data and cluster labels $\Pi \{ (y_1,c_1),\ldots, (y_n,c_n)\} = \mathcal L( y_i; \mathcal V \} \Pi_0( \mathcal V)$ is exchangeable. Further, we could marginalize over $\mathcal V$, and see that $\Pi(y_1,\ldots, y_n)$ is exchangeable.
\end{remark}
\vspace*{-1cm}
\subsection{Hyper-priors for the Other Parameters}
\vspace*{-0.5cm}
We now specify the hyper-priors for the parameters in the root and leaf densities. To avoid model sensitivities to scaling and shifting of the data, we assume that the data have been appropriately scaled and centered (for example, via standardization), so that the marginally $\mathbb E y \approx 0$ and $\mathbb E \|y_{.,j}- \mathbb E y_{.,j}\|_2^2 \approx 1$ for $j=1,\ldots,p$. To make the root density $r(\cdot)$ close to a small constant in the support of the data, we set $\mu=0$ and $\gamma^2 \sim \text{Inverse-Gamma}(2, 1)$.

For $\sigma_{i,j}$ in the leaf density $f(y_i \mid y_j; \sigma_{i,j})$, in order to likely pick an edge $(i,j)$ with $j$ as a close neighbors of $i$ (that is, $(i,j)$ with small $\|y_i-y_j\|_2$), we want most of $\sigma_{i,j}=\tilde\sigma_i\tilde\sigma_j$ to be small.  We use the following hierarchical inverse-gamma prior that shrinks each $\tilde\sigma_i$, while using a common scale hyper-parameter $\beta_\sigma$ to borrow strengths among $\tilde\sigma_i$'s,
\[
      \begin{aligned}
      &         \beta_\sigma \sim \text{Exp}(\eta_\sigma), \qquad \eta_\sigma \sim \text{Inverse-Gamma}(a_\sigma, \xi_\sigma),\\
      & \tilde\sigma_i \stackrel{iid}\sim  \text{Inverse-Gamma}(b_\sigma, \beta_\sigma) \text{ for } i=1,\ldots,n,
      \end{aligned}
\]
where $\eta_\sigma$ is the scale parameter for the exponential. % One can marginalize over $\eta_\sigma$, and obtain a generalized Pareto prior $\Pi_0(\beta_\sigma)\propto (1+ \beta_\sigma/\xi_\sigma)^{-(1+a_\sigma)}$.
To induce a shrinkage effect {\em a priori}, we use $a_\sigma=100$ and $\xi_\sigma=1$ for a likely small $\eta_\sigma$ hence a small $\beta_\sigma$. Further, we note that the coefficient of variation  $\sqrt{ \text{Var}(\tilde \sigma_i \mid \beta_\sigma)}/{\mathbb{E}(\tilde \sigma_i \mid \beta_\sigma)} = 1/\sqrt{b_\sigma-2}$; therefore, we set $b_\sigma=10$ to have most of $\tilde\sigma_i$ near  $\mathbb{E}(\tilde \sigma_i \mid \beta_\sigma)= \beta_\sigma/(b_\sigma-1)$ in the prior. We use these hyper-prior settings in all the examples presented in this article. %An alternative is to follow \cite{ascolani2022clustering} by putting a hyper-prior such as generalized Gamma  $\Pi_0(\lambda)\propto \lambda^{d_\lambda-1} \exp[- (\lambda/ a_\lambda)^{p_\lambda}]$.

In addition, \cite{zelnik2005self} show good empirical performance in spectral clustering, based on a heuristic of setting $\tilde\sigma_{i}$ to a low order statistic of the distances to $y_i$. We develop a model-based formalization that achieves similar effects. Since the model is more involved than a simple Bayesian spanning forest model, we defer the details to the Supplementary Materials S5.

\vspace*{-0.5cm}
\subsection{Model-based Extensions}
\vspace*{-0.5cm}
Compared to algorithms, a major advantage of probabilistic models is the ease of building useful model-based extensions. We demonstrate three directions for extending the Bayesian forest model.  Due to the page constraint, we defer the details and numeric results of the first two extensions in the Supplementary Materials S1.1 and S1.2.

\noindent\textbf{Latent Forest Model:} First, one could use the realization of the forest process as latent variables in another model $\mathcal M$ for data $(y_1,\ldots,y_n)$,
\be
 z_1,\ldots, z_n \sim \t{Forest Model} (\mathcal T;\theta_z),\qquad y_1,\ldots, y_n \sim \mathcal M(z_1,\ldots, z_n;\theta_y),
\ee
where $\theta_z$ and $\theta_y$ denote the other needed parameters.
For example, for clustering high-dimensional data such as images, it is often necessary to represent each high-dimensional observation $y_i$ by a low-dimensional coordinate   $z_i$ \citep{wu2014spectral,chandra2020escaping}. In the Supplementary Materials, we present a high-dimensional clustering model, using an autoregressive matrix Gaussian for $\mathcal M$ and a sparse von Mises-Fisher for the forest model.

\noindent\textbf{Informative Prior--Latent Variable Distribution:} Second, in applications it is sometimes desirable to have the clustering dependent on some external information $x$, such as covariates  \citep{muller2011product} or an existing partition \citep{paganin2021centered}. From a Bayesian view, this can be achieved via taking an $x$-informative distribution:
\vspace*{-0.3cm}
\be
\mathcal T \sim \Pi(\cdot\mid x), \qquad
 y_1,\ldots, y_n \sim \t{Forest Model} (\mathcal T;\theta).
\ee
In the Supplementary Materials, we illustrate an extension with a covariate-dependent product partition model [PPMx, \cite{muller2011product}] into the distribution of $\mathcal T$.

\noindent\textbf{Hierarchical Multi-view Clustering:}
Third, for multi-subject data $(y^{(s)}_1,\ldots, y_n^{(s)})$ for $s=1,\ldots, S$, we want to find a clustering for every $s$. At the same time, we can borrow strength among subjects, by letting subjects share some similar partition structure on a subset of nodes (while differing on the other nodes). This is known as multi-view clustering. On the other hand, a challenge is that a forest is a discrete object subject to combinatorial constraints, hence it would be difficult to partition the nodes freely while accommodating the tree structure. To circumvent this issue, we propose a latent coordinate-based distribution that gives a continuous representation for $\mathcal T^{(s)}$. Consider a latent $z^{(s)}_i\in \mathbb{R}^d$ for each node $i=1,\ldots,n$, we assign a joint prior--latent variable distribution for $z^{(s)}$ and $\mathcal T^{(s)}$:
\bel\label{eq:np_forest_model}
 &\Pi [z^{(s)}, \mathcal{T}^{(s)}]  \propto \\
 &\lambda^{K[\mathcal{T}^{(s)}]}
 \bigg[\prod_{(i,j)\in \mathcal{T}^{(s)}:i\ge 1,j\ge 1} \exp(- \frac{\|z_i^{(s)}- z_j^{(s)}\|_2^2}{2\rho })  \bigg] \bigg[\prod_{i=1}^n \bigg\{\sum_{k=1}^{\tilde\kappa} v_{i,k}
 \exp ( - \frac{\|z_i^{(s)}-  \eta^*_k\|_2^2}{2 \sigma_z^2})\bigg\} \bigg], \\
& (v_{i,1},\ldots, v_{i,\tilde\kappa}) \sim \t{Dir}(1/\tilde\kappa,\ldots, 1/\tilde\kappa) \qquad  \text{ for }i=1,\ldots n,\\
& \{ y_{1}^{(s)}, \ldots, y_{n}^{(s)}\}  \sim \t{Forest Model} (\mathcal T^{{(s)}})
\qquad  \text{ for }s=1,\ldots S,
\eel
where $v_{i,1},\ldots, v_{i,\tilde\kappa}$  are the weights that vary with $i$ and $\sum_{k=1}^{\tilde \kappa} v_{i,k}=1$, $\rho>0$, and $z^{(s)}\in \mathbb{R}^{n\times d}$ is the matrix form. Equivalently, the above assigns each node a location parameter $\eta_i^{(s)}$, drawn from a hierarchical Dirichlet distribution with shared atoms $\{\eta_1^*,\ldots,\eta_{\tilde\kappa}^*\}$ and probability $(v_{.,1},\ldots,,v_{.,\tilde\kappa})$  \citep{doi:10.1198/016214506000000302}. Further, one could let $\eta_k^*$  vary over node according to some functional using a hybrid Dirichlet distribution \citep{petrone2009hybrid}.

Using a Gaussian mixture kernel on $z_i^{(s)}$, we can now separate $z^{(s)}_i$'s into several groups that are far apart. To make the parameters identifiable and have large separations between groups, we fix $\tilde\eta^{*}_k$'s on the $d$-dimensional integer lattice $\{0,1,2\}^d$ with $d=2$ (hence $\tilde\kappa=9$); and we use $\sigma^2_z=0.01$ and $\rho=0.001$ in this article.

\begin{remark}
   To clarify, our goal is to induce between-subject similarity in the \underline {node partition}, not the tree structure. For example,  for two subjects $s$ and $s'$, when $z_i^{(s)}$ and $z_i^{(s')}$ are both near $\eta^*_k$ for all $i\in C$, then both the spanning forest $\mathcal T^{(s)}$ and $\mathcal T^{(s')}$ will likely cluster the nodes in $C$ together, even though $ T^{(s)}_k$ and $ T^{(s')}_k$ associated with $V_k\supset C$ may be different.
\end{remark}

The posterior can be sampled efficiently using the Gibbs sampling algorithm. We provide the posterior sampling algorithm in the Supplementary Materials S1.3, and illustrate this model in Section 6 of modeling brain regions for multiple subjects.

\vspace*{-0.5cm}
\section{Posterior Computation}
\vspace*{-0.5cm}
\subsection{Gibbs Sampling Algorithm}

We now describe the Markov chain Monte Carlo (MCMC) algorithm. For ease of notation, we use an $(n+1)\times (n+1)$ matrix $S$, with $S_{i,j}= \log f (y_i \mid y_j ; \theta)$, $S_{0,i}=S_{i,0}= \log r (y_i ; \theta)  +\log \lambda$ (for convenience, we use $0$ to index the last row/column), $S_{i,i}=0$, and $A_{\mathcal T}$ to represent the adjacency matrix of $\mathcal T$. We have the posterior distribution
\bel\label{eq:comp_posterior}
\Pi( \mathcal T, \theta \mid y) \propto  \exp \big\{ \t{tr}[S(\theta) A_{\mathcal T}]/2\big\}\Pi_0(\theta).
\eel
Note the above form conveniently include the prior term for the number of clusters, $\lambda^K$, via the number of edges adjacent to node $0$.

Our MCMC algorithm alternates in updating $\mathcal T$ and $\theta$, hence is a Gibbs sampling algorithm. To sample  $\mathcal T$ given $\theta$, we take the random-walk covering algorithm for weighted spanning tree \citep{mosbah1999non}, as an extension of the Andrei--Broder algorithm for sampling uniform spanning tree \citep{broder1989generating,aldous1990random}. For this article to be self-contained, we describe the algorithm below. The above algorithm produces a random sample $\mathcal T$ following the full conditional $\Pi(\mathcal T \mid \theta,y)$ proportional to \eqref{eq:comp_posterior}. It has an expected finish time of $O(n\log n)$. Although some faster algorithms have been developed  \citep{schild2018almost}, we choose to present the random-walk covering algorithm for its simplicity.

\begin{algorithm}
\begin{algorithmic}
\State Start with $V_{\mathcal T}=\{0\}$ and $E_{\mathcal T}=\varnothing$, and set $i\leftarrow0$:
\While{$|V_{\mathcal T}|\neq n+1$}
    \State Take a random walk from $i$ to $j$ with probability
    $\text{Pr}(j \mid i) = \frac{\exp[S_{i,j} (\theta)]}{\sum_{j:j\neq i} \exp[S_{i,j} (\theta)]}.$
\If {$j\not \in V_{\mathcal T}$ } \State Add $j$ to $V_{\mathcal T}$. Add $(i,j)$ to $E_{\mathcal T}$.
\EndIf
\State Update $i \leftarrow j$.
\EndWhile
\end{algorithmic}
\caption{Random-walk covering algorithm for sampling the augmented tree $\mathcal T$.}
\end{algorithm}

We sample $\tilde \sigma_i$ using the following steps,
\be
& (\eta_\sigma \mid .) \sim  \t{Inverse-Gamma}\big(1+ a_\sigma,   \beta_\sigma + \xi_\sigma \big )\\
& (\beta_\sigma \mid .) \sim  \t{Gamma}\bigg\{ 1+ n b_\sigma,   (\sum_{i=1}^n \frac{1}{\tilde\sigma_i}+\frac{1}{\eta_\sigma})^{-1} \bigg\}\\
& (\tilde \sigma_i \mid .) \sim \t{Inverse-Gamma} \bigg [
\frac{p\sum_{j} 1\{(i,j)\in \mathcal T\} }{2} + b_\sigma
 ,
\sum_{j:(i,j)\in \mathcal T} \frac{\|y_i-y_j\|^2_2}{ 2\tilde\sigma_j } + {\beta_\sigma }
\bigg]
\ee

To update %$\mu$ and
 $\gamma$, we use the form of the multivariate Cauchy as a scale mixture of $\text{N}(\mu, \gamma^2 u_{\gamma,i} I_p )$ over $u_{\gamma,i} \sim \text{Inverse-Gamma}(1/2,1/2)$.   We can update  via
\be
& u_{\gamma,i} \sim \text{Inverse-Gamma}( \frac{1+p}{2},  \frac{1}{2} +\frac{\|y_i-\mu\|_2^2}{2\gamma^2}),\\
& \gamma^2 \sim  \text{Inverse-Gamma}(2 + \frac{Kp}{2},  \hat\sigma^2_y + \sum_{i:(0,i)\in \mathcal T}\frac{\|y_i-\mu\|_2^2}{2 u_{\gamma,i}}).\\
%& \mu \sim \text{N} \bigg\{
%[\sum_{i:(0,i)\in \mathcal T} \frac{1}{\gamma^2 u_{\gamma,i}} + 1]^{-1}
%  [ \sum_{i:(0,i)\in \mathcal T} \frac{y_i}{\gamma^2 u_{\gamma,i}}  + \bar y] ,
%   [\sum_{i:(0,i)\in \mathcal T} \frac{1}{\gamma^2 u_{\gamma,i}} + 1]^{-1}  I_p
%   \bigg\}.
\ee
We run the MCMC algorithm iteratively for many iterations. And we discard the first half of iterations as burn-in.

\begin{remark}
         We want to emphasize that the Andrei--Broder random-walk covering algorithm  \citep{broder1989generating,aldous1990random,mosbah1999non}  is an exact algorithm for sampling a spanning tree $\mathcal T$. That is, if $\theta$ were fixed, each run of this algorithm would produce an {\em independent} Monte Carlo sample $\mathcal T \sim \Pi(\mathcal T \mid \theta,y)$.
         Removing the auxiliary node 0 from $\mathcal T$ will produce $K$ disjoint spanning trees. This augmented graph technique is inspired by \cite{boykov2001fast}.

              In our algorithm, since the scale parameters in $\theta$ are unknown, we use Markov chain Monte Carlo that updates two sets of parameters, (i) $(\theta_{[t+1]} \mid \mathcal T_{[t]})$ and (ii) $(  \mathcal T_{[t+1]} \mid \theta_{[t+1]})$ from iteration $[t]$ to $[t+1]$. Therefore, rigorously speaking, there is a Markov chain dependency between $\mathcal T_{[t]}$ and $\mathcal T_{[t+1]}$ induced by $\theta_{[t+1]}$. Nevertheless, since we draw $\mathcal T$ in a block via the random-walk covering algorithm, we empirically find that $\mathcal T_{[t+1]}$ and $\mathcal T_{[t]}$ are substantially different. In the Supplementary Materials S4.4, we quantify the iteration-to-iteration graph changes, and provide diagnostics with multiple start points of $(\mathcal T_{[0]},\theta_{[0]})$.
      \end{remark}

\vspace*{-1cm}
\subsection{Posterior Point Estimate on Clustering}
\vspace*{-0.5cm}
%To produce a point estimate on clustering, $(\hat c_1,\ldots, \hat c_n)$, one could take advantage of our discovered relationship between the Bayesian forest model and the spectral clustering algorithm. This involves first using a posterior point estimate on $\hat\theta$ (such as the posterior mean), then applying a spectral clustering algorithm on a similarity matrix formed by $A_{i,j}=\exp(S_{i,j})$ to produce $K$ clusters.
In the field of Bayesian clustering, for producing point estimate on the partition, it had been a long-time practice to simply track $\text{pr}(c_i=k \mid y)$, then take the element-wise posterior mode over $k$ as the point estimate for $\hat c_i$. Nevertheless, this was shown to be sub-optimal due to that: (i) label switching issue causes unreliable estimates on $\text{pr}(c_i=k \mid y)$; (ii) the element-wise mode can be unrepresentative of the center of distribution for $(c_1,\ldots, c_n)$ \citep{wade2018bayesian}. These weaknesses have motivated new methods of obtaining point estimate of clustering, that transform an $n\times n$ pairwise co-assignment matrix $\{\text{pr}(c_i=c_j\mid y)\}_{\text{all }(i,j)}$ into an $n\times K$ assignment matrix \citep{medvedovic2002bayesian,rasmussen2008modeling,molitor2010bayesian,wade2018bayesian}. More broadly speaking, minimizing a loss function based on the posterior sample (via some estimator or algorithm) is common for producing a point estimate under some decision theory criterion. For example, the posterior mean comes as the minimizer of the squared error loss; in Bayesian factor modeling, an orthogonal Procrustes-based loss function is used for producing the posterior summary of the loading matrix from the generated MCMC samples \citep{assmann2016bayesian}.

We follow this strategy. There have been many algorithms that one could use. For a recent survey, see \cite{dahl2022search}. In this article, we use a simple solution of first finding the mode of $K$ from the posterior sample, then doing a $\hat K$-rank symmetric matrix factorization on $\{\text{pr}(c_i=c_j\mid y)\}_{\text{all }(i,j)}$ and clustering into $\hat K$ groups, provided by \texttt{RcppML} package \citep{debruine2021fast}.

\vspace*{-1cm}

 \section{Theoretical Properties}
\vspace*{-0.5cm}
\subsection{ Convergence of Eigenvectors \label{sec:spectral}}
\vspace*{-0.5cm}
We now formalize the closeness of the eigenvectors of matrices $N$ and $M$ (shown in Section 2.2), by establishing the convergence of the two sets of eigenvectors as $n$ increases.

To be specific, we focus on the normalized spectral clustering algorithm using the similarity  $A_{i,j}=\exp(S_{i,j})$, with $S_{i,j}= \log f (y_i \mid y_j ; \theta)$, $S_{0,i}=S_{i,0}= \log r (y_i ; \theta)  +\log \lambda$. On the other hand, for the specific form, $f(y_i \mid y_j)$ can be any density satisfying $f(y_i\mid y_j, \theta) = f(y_j\mid y_i, \theta)$, $r(y_i; \theta)$ can be any density satisfying $r(y_i;  \theta)>0$. For the associated normalized Laplacian $N$, we denote the first $K$ bottom eigenvectors by  $\phi_1,\ldots, \phi_K$, which correspond to the smallest $K$ eigenvalues.

Let $M$ be the matrix with $M_{i,j}=\t{pr}[\mathcal T \ni  (i,j)  \mid y, \theta ]$ for $i\neq j$ and $M_{i,i}=0$. The Kirchhoff's tree theorem \citep{chaiken1978matrix} gives an enumeration of all $\mathcal T\in \mathbb{T}$,
\bel\label{eq:matrix_tree}
\sum_{\mathcal T\in \mathbb{T}} \prod_{(i,j)\in \mathcal T} \exp(S_{i,j}) = (n+1)^{-1}\prod_{h=2}^{n+1}\lambda_{(h)} (L)
\eel
where $L$ is the Laplacian matrix transform of the similarity matrix $A$; $\lambda_{(h)}$ denotes the $h$th smallest eigenvalue. Differentiating its logarithmic transform with respect to $S_{i,j}$,
\be
M_{i,j} & =  \t{Pr}[ \mathcal T \ni  (i,j)\mid y ]
  =
  \frac{\sum_{\mathcal T\in \mathbb{T},(i,j)\in \mathcal T} \prod_{(i',j')\in \mathcal T} \exp(S_{i',j'})}{\sum_{\mathcal T\in \mathbb{T}} \prod_{(i',j')\in \mathcal T} \exp(S_{i',j'})}
  =
   \frac{\partial \sum_{i=2}^{n+1} \log\lambda_{(i)} (L)}{\partial S_{i,j}}.
\ee
Let $\Psi_1, \ldots, \Psi_K$ be the top $K$ eigenvectors of $M$, associated with eigenvalues $\xi_1\ge \xi_2 \ge \ldots \ge \xi_K$, and $\xi_K > \xi_{K+1}\ge \xi_{K+2}\ge \ldots \ge \xi_{n+1}$. And we can compare with the $K$ leading eigenvectors of $(-N)\in \mathbb{R}^{n\times n}$,  $\phi_1,\ldots, \phi_K$. Using $\Psi_{1:K}$ and $\phi_{1:K}$ to denote two $(n+1)\times K$ matrices, we now show they are close to each other.

\begin{theorem}
There exists an orthonormal matrix $R\in\mathbb{R}^{K\times K}$ and a finite constant $\epsilon>0$,
\be
\|\Psi_{1:K}-\phi_{1:K} R \|_F \le \frac{
  40 \sqrt{K (n+1)}
  }{\xi_{K} -\xi_{K+1}}  \max_{i,j} \left \{ (1+\epsilon)(D^{-1/2}_i-D^{-1/2}_j)^2
A_{i,j} \right\},
\ee
        with probability at least $1- \exp(-n).$
\end{theorem}
\begin{remark}
To make the right-hand side go to zero, a sufficient condition is to have all $ A_{i,j}/D_{i,i}= O(n^{-\kappa})$ with $\kappa>1/2$.
We provide a detailed definition of the bound constant $\epsilon$ in the Supplementary Materials S2.

To explain the intuition behind this theorem,  
our starting point is the close relationship between Laplacian and spanning tree models --- multiplying both sides of Equation \eqref{eq:matrix_tree} by $(n+1)^{-(n-1)}$ shows that the non-zero eigenvalue product of the graph Laplacian $L$ is proportional to the marginal probability of $n$ data points from a spanning forest-mixture model.
Starting from this equality, we can write the marginal inclusion probability matrix of $\mathcal T$ as a mildly perturbed form of the normalized Laplacian matrix. Intuitively, when two matrices are close, their eigenvectors will be close as well \citep{yu2015useful}.
\end{remark}
Therefore, under mild conditions, as $n\to \infty$, the two sets of leading eigenvectors converge. In the Supplementary Materials S4.7, we show that the convergence is very fast, with the two sets of leading eigenvectors becoming almost indistinguishable starting around $n\ge 50$.

Besides the eigenvector convergence, we can examine the marginal posterior
\[
\Pi(\mathcal V\mid \theta, y) \propto \Pi_0(K,V_1,\ldots,V_K) \bigg\{ \prod_{k=1}^K [\sum_{i\in V_k}r(y_i)] \bigg\}\prod_{k=1}^K \bigg\{ n_k^{-1}\prod_{h=2}^{n_k}\lambda_{(h)} (L_k) \bigg\},
\]
where $L_k$ is the unnormalized Laplacian matrix associated with matrix $\{A_{i,j}\}_{i\in V_k,j\in V_k}$. 
Imagine that if we put all indices in one partition $V_1=(1,\ldots,n)$, then $\Pi(\mathcal V\mid \theta, y)$ would be very small due to those close-to-zero eigenvalues. Applying this deduction recursively on subsets of data, it is not hard to see that a high-valued $\Pi(\mathcal V\mid \theta,y)$ would correspond to a partition, wherein each $V_k$ has $\lambda_{(h)} (L_k)$ away from $0$ for any $h\ge 2$.
\vspace*{-0.5cm}
\subsection{Consistent Clustering of Separable Sets}
\label{sec:clustcon}
\vspace*{-0.5cm}
We show that clustering consistency is possible, under some separability assumptions when the data-generating distribution follows a forest process. Specifically,
we establish posterior ratio consistency, as the ratio between the maximum posterior probability assigned to other possible clustering assignments to the posterior probability assigned to the true clustering assignments converges to zero almost surely under the true model \citep{cao2019posterior}.

To formalize the above, we denote the true cluster label for generating $y_i$ by $c^0_i$ (subject to label permutation among clusters), and we define the enclosing region for all possible $y_i:c^0_i=k$ as $R_k^0$ for $k=1,\ldots, K_0$ for some true finite $K_0$. And we refer to $R^0=(R_1^0,\ldots, R_{K_0}^0)$ as the ``null partition''.  By separability, we mean the scenario that $(R_1^0,\ldots, R_{K_0}^0)$ are disjoint and there is a lower-bounded distance between each pair of sets. As alternatives, regions $R=(R_1,\ldots, R_K)$  could be induced by $\{c_1,\ldots,c_n\}$ from the posterior estimate of $\mathcal T$. For simplicity, we assume the scale parameter in $f$ is known and all equal $\sigma_{i,j}=\sigma^{0,n}$.

\vspace{5mm}

{\noindent \underline{Number of clusters is known.}}
We first start with a simple case when we have fixed $K=K_0$. For regularities, we consider  data as supported in a compact region $\mathcal{X}$, and satisfying the following assumptions:
\begin{itemize}
   \item (A1, diminishing scale) $\sigma^{0,n}=C'(1/\log n)^{1+\iota}$ for some $\iota>0$ and $C'>0$. %\leo{do you mean $o(1/n)$?}
    \item (A2, minimum separation) $\inf_{x\in R_k^0,y\in R_{k'}^0}\|x-y\|_2>M_n$, for all $k\neq k'$ with some positive constant $M_n>0$ such that $M_n^2/\sigma^{0,n}=8\tilde m_0\log(n)$  for all $(i,j)$ and is known for some constant $\tilde  m_{0}>p/2+2$.
\item (A3, near-flatness of root density) For any $n$, $\epsilon_1<r(y)<\epsilon_2$ for all $y\in\mathcal{X}$. % The bounds $\epsilon_1$ and $\epsilon_2$ will depend on $\gamma_0$.% which is assumed to be known for simplicity.
\end{itemize}
%To explain those three assumptions, the first one ensures that the partitions are well separated.
%The scale parameter is controlled by our second assumption.
%The flatness of the base distribution is quantified in our last assumption.

% \begin{remark}
% Our assumption (A1) is based on our empirical uniform prior for $\sigma^{0,n}$ using the minimum distances. Assumption (A2) coupled with (A1) suggests that the minimum separation is allowed to decay with sample size but at a rate slower than $\sigma^{0,n}$.
          % In the generative model (1), we may have a constant $\sigma^0$ used in $f^*$ that does not change with $n$ and have the data generated from some $\mathcal T_0$; however, as $n\to \infty$, the data within each $R_k^0$ become very dense, hence we can have an alternative $\mathcal T$ with all distances over edges $(i,j):i\ge 1,j\ge 1$ diminish to zero, while still corresponding to true $c^0_i$'s --- that is, we have a limiting $\mathcal T_{\infty}$ (the graph at the point mass where the posterior converges to) that is different from $\mathcal T_0$ on edges, but equal to $\mathcal T_{0}$ on node partition. To put it another way, under this scenario, the posterior is inconsistent in graph estimation, but still consistent in clustering. %\leo{However, this condition can be slightly relaxed into being $\sigma_{i,j}=m_{i,j}/\sqrt{\log(n)}$ with $L>m_{i,j}>p/2+2$, where $0<L<\infty$. leo: Likely to get attacked here. If we don't have proof of this, it's best that we don't say it.}
% \end{remark}

Under the null partition, $\Pi(\mathcal{T}|y)$ is maximized at $\mathcal T=\mathcal{T}_{\textrm{MST},R^0}$, which contains $K_0$ trees with each $T_k$ being the minimum spanning tree (denoted by subscript ``MST'') within region $R_k^0$.
Similarly, for any alternative $R$, $\Pi(\mathcal{T}|y)$ is maximized at the $\mathcal{T}=\mathcal{T}_{\textrm{MST},R}$.

%The null density is assumed to have $\mathcal{T}$ set to the corresponding MST under the null partitioning.
%Let $d_n=\max_{k}\max_{i,j\in R_{k}^0\cap\mathcal{T}}\|y_i-y_j\|_2$. Here, $R_{k}^0\cap\mathcal{T}$ stands for the set of pairs that belong to the same region $R_{k}^0$ and are connected by an edge in $\mathcal{T}$. We must $d_n$ decrease with, $n$. Otherwise, the total distance of the MST will be divergent, which automatically contradicts the boundedness assumption on $\mathcal{X}$. Hence, there exists $N$ such that for all $n>N$, we have $d_n<M/2$.

%To ensure that the data-space is covered $R_i$'s, we need $\cup_{i=1}^KR_{i}^0\subset\cup_{i=1}^KR_{i}$.

\begin{theorem}
Under  (A1,A2,A3), we have ${\Pi(\mathcal{T}_{\textrm{MST},R}|y)}/{\Pi(\mathcal{T}_{\textrm{MST},R^0}|y)}\rightarrow 0$ almost surely, unless $R_{i}^0\subseteq R_{\xi(i)}$ for some permutation map $\xi(\cdot)$.
\label{knownclust}
\end{theorem}

%Here, the convergence is shown in a probability sense.
%This could be replaced by almost sure sense with some added requirement on $\sigma_{0n}$.
%To show posterior ratio consistency, we need to show,
%, in the spirit of the likelihood ratio test.

{\noindent\underline{Number of clusters is unknown:}} Next, we relax the condition by having a $K$ not necessarily equal to $K_0$.
We show the consistency in two parts for 1)$K<K_0$, and 2) $K>K_0$ separately. In order to show posterior ratio consistency in the second part, we need some finer control on $r(y)$:
\begin{itemize}
    \item (A3') The root density satisfies $\tilde m_1e^{-M/2\sigma^{0,n}}\leq r(y)\leq \tilde  m_2 e^{-M/2\sigma^{0,n}}$ for some $\tilde  m_1<\tilde  m_2$.
\end{itemize}
In this assumption, we essentially assume the root distribution to be flatter with a larger $n$. Then we have the following results.
\begin{theorem}
1) If $K<K_0$, under the assumptions (A1,A2,A3), we have \\${\Pi(\mathcal{T}_{\textrm{MST},R}|y)}/{\Pi(\mathcal{T}_{\textrm{MST},R^0}|y)}\rightarrow 0$ almost surely.

2) If $K>K_0$, under the assumptions (A1,A2,A3'), we have ${\Pi(\mathcal{T}_{\textrm{MST},R}|y)}/{\Pi(\mathcal{T}_{\textrm{MST},R^0}|y)}\rightarrow 0$ almost surely.
\label{unknownclust}
\end{theorem}

{The above results show posterior ratio consistency.
Furthermore, when the true of clusters is known, the ratio consistency result can be further extended to show clustering consistency, which is proved in the Supplementary Materials S3.}

%\begin{remark}
%        The minimum separation condition of Assumption (1) sets the minimum separation among the clusters to be at least $M$, independent of $n$. This is primarily for computational convenience. However, it can be relaxed further to $M_n\rightarrow 0$ with $\sigma_{0,n}^2M_n\rightarrow \infty$ and $M_n/\sigma_{0,n}\rightarrow \infty$. One such sequence is that $M_n=O(1/\log^{\delta} n)$ for $1/2>\delta>0$ and $\sigma_{0,n}=O(1/\sqrt{\log n})$. All the theoretical assertions would still hold with minor modifications in the calculation. This suggests that with increasing sample size, the method will be able to identify clusters having even smaller separations between them.
%\end{remark}

\vspace*{-1cm}
\section{Numerical Experiments}
\vspace*{-0.5cm}
\subsection{Clustering Near-Manifold Data}
\vspace*{-0.5cm}
To illustrate the capability of uncertainty quantification, we carry out clustering tasks on those near-manifold data commonly used for benchmarking clustering algorithms.

\begin{figure}[H]
     \begin{subfigure}[t]{.24\textwidth}
        \includegraphics[height=4cm,width=1\linewidth]{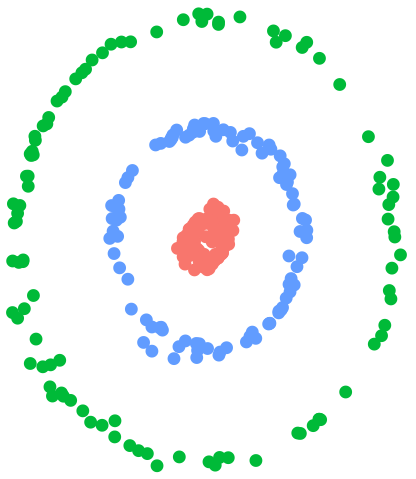}
            \caption{Posterior point estimate.}
    \end{subfigure}
     \begin{subfigure}[t]{.24\textwidth}
        \includegraphics[width=.8\linewidth]{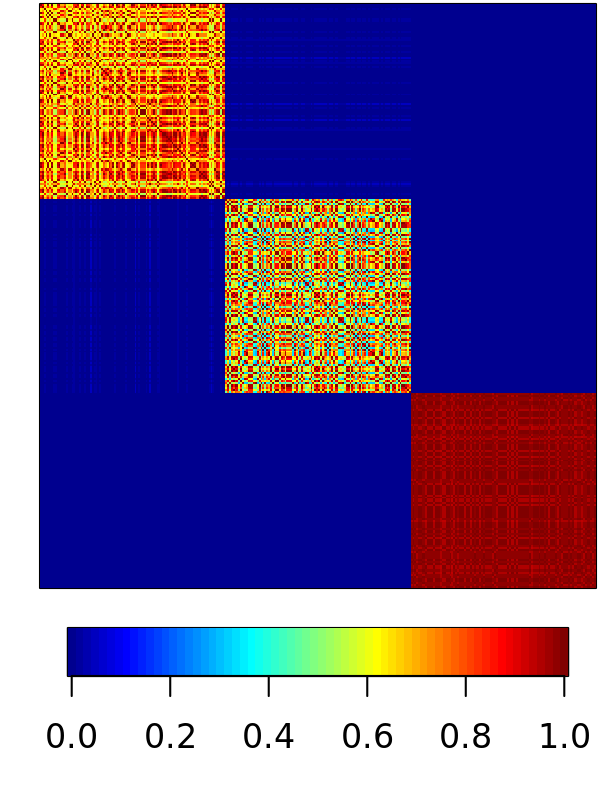}
            \caption{$\text{Pr}(c_i=c_j \mid y)$.}
    \end{subfigure}
    \begin{subfigure}[t]{.24\textwidth}
        \includegraphics[height=4cm,width=1\linewidth]{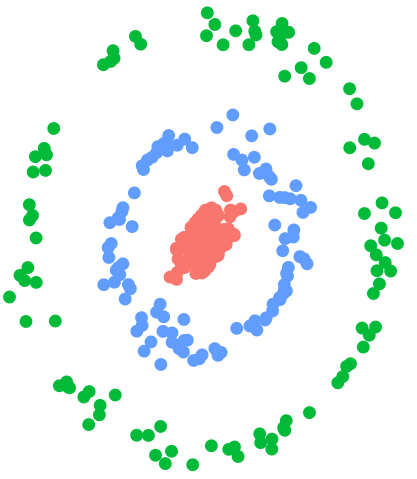}
            \caption{Posterior point estimate.}
    \end{subfigure}
     \begin{subfigure}[t]{.24\textwidth}
        \includegraphics[width=.8\linewidth]{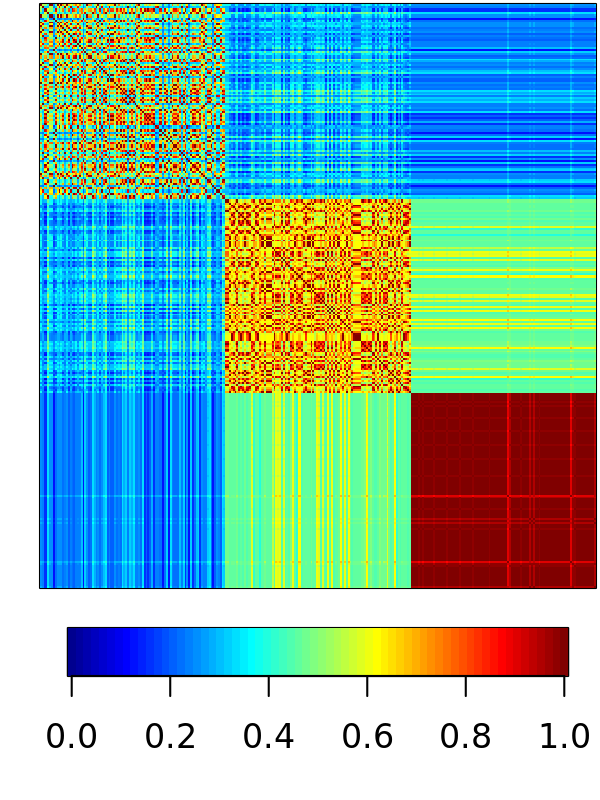}
            \caption{$\text{Pr}(c_i=c_j \mid y)$.}
    \end{subfigure}
     \begin{subfigure}[t]{.24\textwidth}
        \includegraphics[height=4cm,width=1\linewidth]{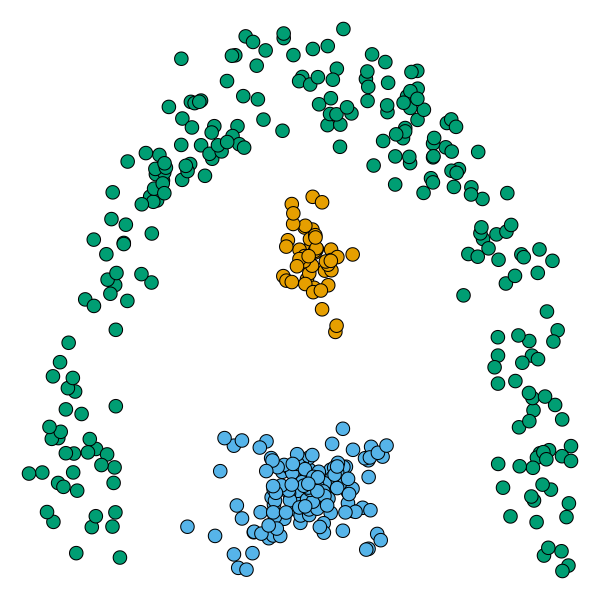}
            \caption{Posterior point estimate.}
    \end{subfigure}
     \begin{subfigure}[t]{.24\textwidth}
        \includegraphics[width=.8\linewidth]{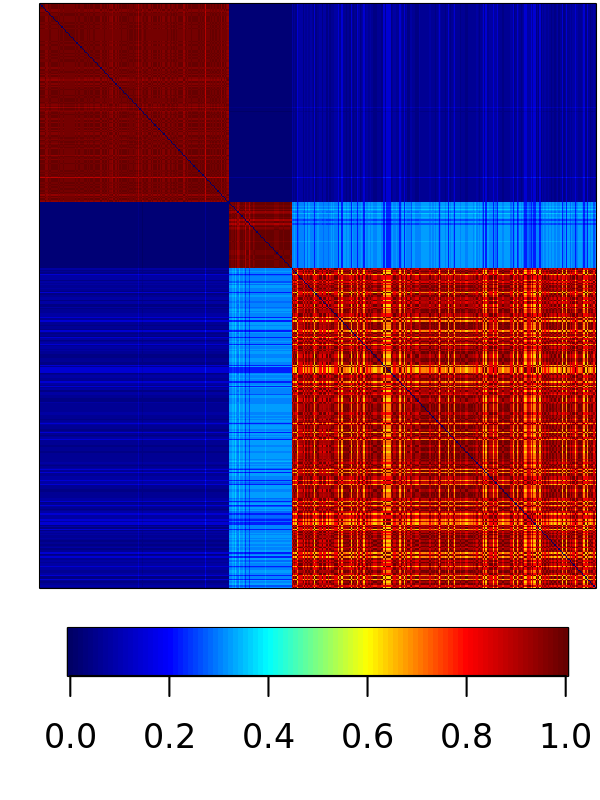}
            \caption{$\text{Pr}(c_i=c_j \mid y)$.}
    \end{subfigure}
    \begin{subfigure}[t]{.24\textwidth}
        \includegraphics[height=4cm,width=1\linewidth]{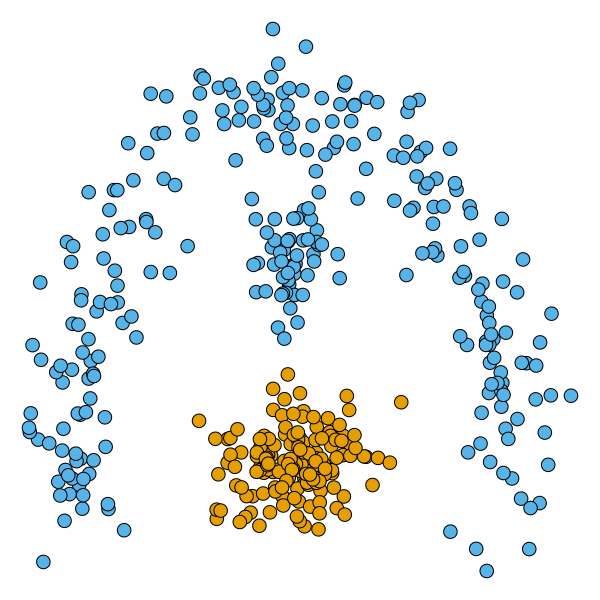}
            \caption{One posterior sample.}
    \end{subfigure}
     \begin{subfigure}[t]{.24\textwidth}
        \includegraphics[width=.8\linewidth]{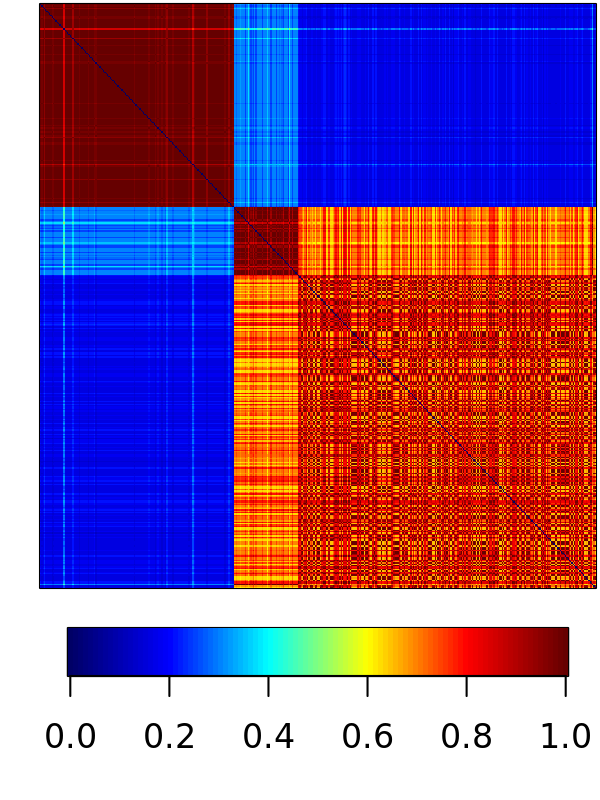}
            \caption{$\text{Pr}(c_i=c_j \mid y)$.}
    \end{subfigure}
        \caption{Uncertainty quantification in clustering data generated near three manifolds. When data are close to the manifolds (Panels a,e), there is very little uncertainty on clustering in low $\pr(c_i=c_j\mid j)$ between points from different clusters (Panels b,f). As data deviate more from the manifolds (Panel c,g), the uncertainty increases (Panels d,h). And in Panel g, the point estimate shows a two-cluster partitioning, while there is about $20\%$ of probability for three-cluster partitioning.
         \label{fig:uq_3rings}
        }
        \end{figure}
\vspace*{-0.5cm}
In the first simulation, we start with $300$ points drawn from three rings of radii $0.2$, $1$ and $2$, with $100$ points from each ring. Then we add some Gaussian noise to each point to create a coordinate near a ring manifold. We present two experiments, one with noises from $\t{N}(0,  0.05^2 I_2)$, and one with noises $\t{N}(0,  0.1^2 I_2)$. As shown in Figure \ref{fig:uq_3rings}, when these data are well separated (Panel a, showing Posterior point estimate), there is very little uncertainty on the clustering (Panel b), with the posterior co-assignment $\pr(c_i=c_j\mid y)$ close to zero for any two data points near different rings. As noises increase, these data become more difficult to separate. There is a considerable amount of uncertainty for those red and blue points: these two sets of points are assigned into one cluster with a probability close to $40\%$ (Panel d). We conduct another simulation based on an arc manifold and two point clouds (Panels e-h), and find similar results. Additional experiments are described in the Supplementary Materials S4.2.

\vspace*{-0.5cm}
\subsection{Uncertainty Quantification for Data from Mixture Model}
In the Supplementary Materials  S4.1 and S4.3, we present some uncertainty quantification results, for clustering data that are from mixture models. We compare the estimates with the ones from Gaussian mixture models, which could correspond to correctly/erroneously specified component distribution. Empirically, we find that the uncertainty estimates on $\text{Pr}(c_i=c_j\mid y)$ and $\text{Pr}(K\mid y)$ from the forest model are close to the ones based on the true data-generating distribution; whereas the Gaussian mixture models suffer from sensitivity in model specification, especially when $K$ is not known.
\vspace*{-0.5cm}
\section{Application: Clustering in Multi-subject Functional Magnetic Resonance Imaging Data}
\vspace*{-0.5cm}
In this application, we conduct a neuroscience study for finding connected brain regions under a varying degree of impact from Alzheimer's disease. The source dataset is resting-state functional magnetic resonance imaging (rs-fMRI) scan data, collected from $S=166$ subjects at different stages of Alzheimer's disease. Each subject has scans over $n=116$ regions of interest using the Automated Anatomical Labeling (AAL) atlas \citep{rolls2020automated,shi2021application} and over $p=120$ time points. We denote the observation for the $s$th subject in the $i$th region by $y_{i}^{(s)}\in \mathbb{R}^{p}$.

The rs-fMRI data are known for their high variability, often characterized by a low intraclass correlation coefficient (ICC), $(1-\hat\sigma^{2}_{\text{within--group}}/\hat\sigma^{2}_{\text{total}})$, as the estimate for the proportion of total variance that can be attributed to variability between groups \citep{noble2021guide}. Therefore, our goal is to use the multi-view clustering to divide the regions of interest for each subject, while improving our understanding of the source of high variability.

    \begin{figure}[H]
              \begin{subfigure}[t]{.23\textwidth}
              \begin{overpic}[height=3.8cm]
         {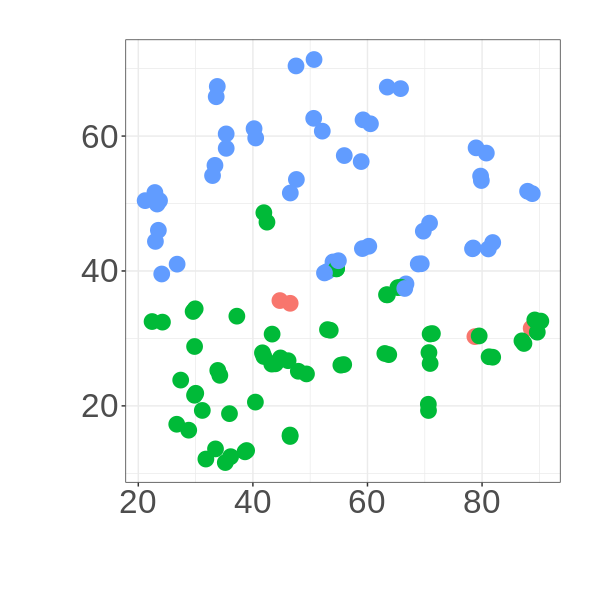}
         \put(20,0){\scriptsize Atlas coordinate 1}
         \put(5,20){\rotatebox{90}{\scriptsize Atlas coordinate 2}}
         \end{overpic}
                         \caption{\scriptsize Clustering of the nodes for one healthy subject.}
    \end{subfigure} \;
              \begin{subfigure}[t]{.23\textwidth}
              \begin{overpic}[height=3.8cm]
         {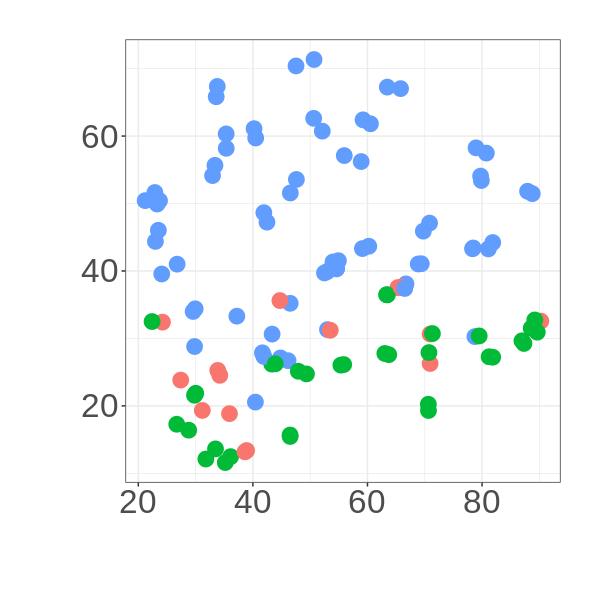}
         \put(20,0){\scriptsize Atlas coordinate 1}
         \put(5,20){\rotatebox{90}{\scriptsize Atlas coordinate 2}}
         \end{overpic}
                         \caption{\scriptsize  Clustering of the nodes for another healthy subject.}
    \end{subfigure} \;
              \begin{subfigure}[t]{.23\textwidth}
              \begin{overpic}[height=3.8cm]
         {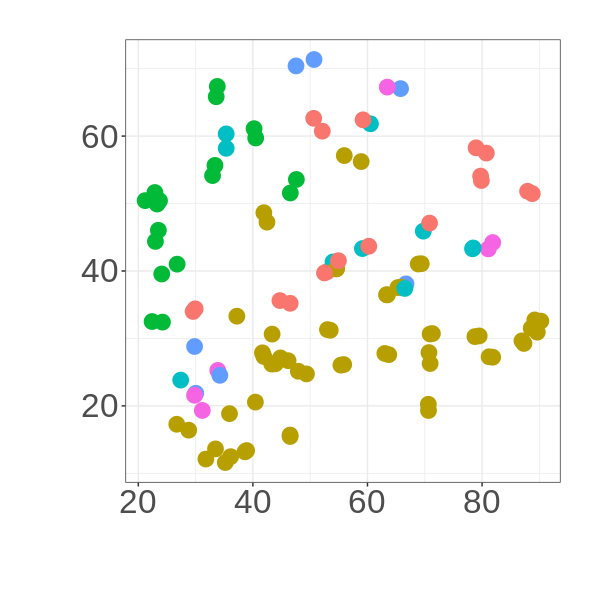}\textbf{}
         \put(20,0){\scriptsize Atlas coordinate 1}
         \put(5,20){\rotatebox{90}{\scriptsize Atlas coordinate 2}}
                  \end{overpic}
                         \caption{\scriptsize Clustering of the nodes for one diseased subject.}
    \end{subfigure} \;
              \begin{subfigure}[t]{.23\textwidth}
                      \begin{overpic}[height=3.8cm]
         {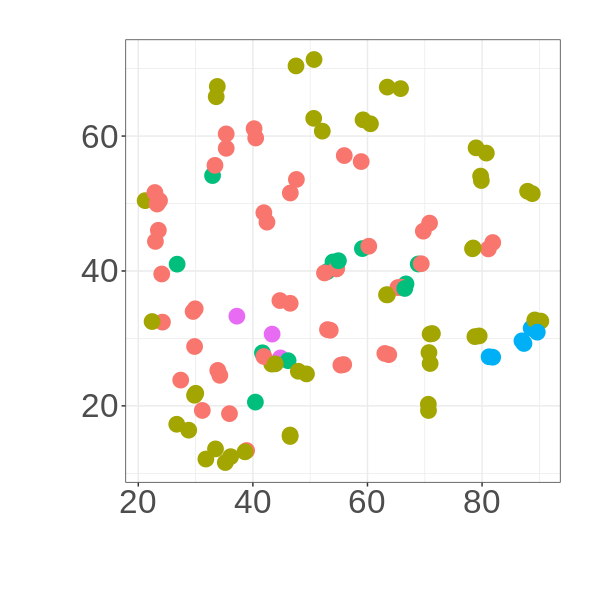}
         \put(20,0){\scriptsize Atlas coordinate 1}
         \put(5,20){\rotatebox{90}{\scriptsize Atlas coordinate 2}}
                  \end{overpic}
                         \caption{\scriptsize Clustering of the nodes for another diseased subject.}
    \end{subfigure}
        \caption{
        Results of brain region clustering (lateral view) for four subjects taken from the healthy and diseased groups. The multi-view clustering model allows subjects to have similar partition structures on a subset of nodes, while having subtle differences on the others (Panels a and b, Panels c and d). At the same time, the healthy subjects show less degree of variability in the brain clustering than the diseased subjects.
        \label{fig:application2}
        }
        \end{figure}
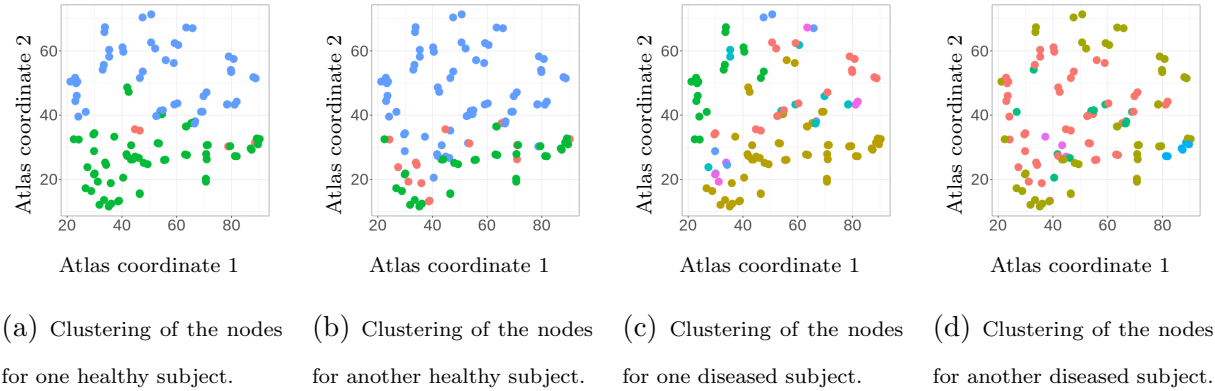
\begin{figure}[H]
              \begin{subfigure}[t]{.3\textwidth}
              \begin{overpic}[height=4.1cm]
         {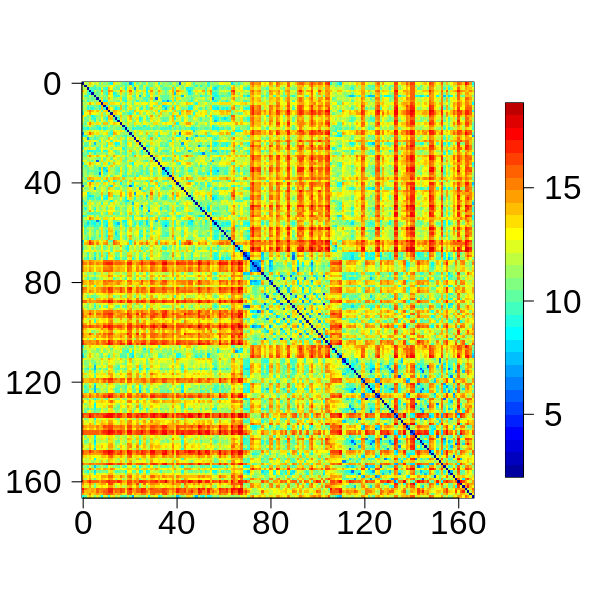}
         \put(30,0){\scriptsize Subject Index}
         \put(-7,30){\rotatebox{90}{\scriptsize Subject Index}}
         \end{overpic}
                         \caption{\scriptsize Pairwise distance of the latent coordinates $\|z^{(s)}-z^{(s')}\|_F$ between subjects, the first 64 subjects are healthy and the latter 102 are diseased.}
    \end{subfigure} \;
              \begin{subfigure}[t]{.3\textwidth}
              \begin{overpic}[height=3.8cm]
         {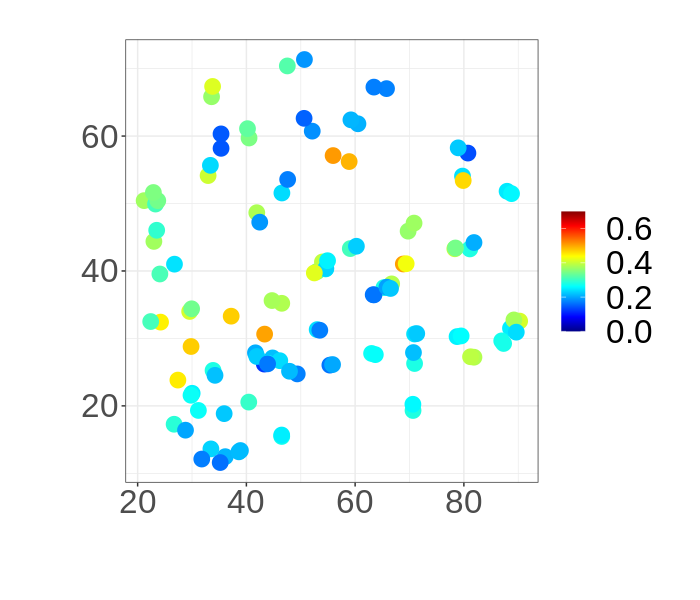}
         \put(20,0){\scriptsize Atlas coordinate 1}
         \put(5,20){\rotatebox{90}{\scriptsize Atlas coordinate 2}}
         \end{overpic}
                         \caption{\scriptsize Within-group variances of $z^{(.)}_i$ for subjects in the healthy group, plotted over the automated anatomical labeling atlas.}
    \end{subfigure} \;
              \begin{subfigure}[t]{.3\textwidth}
              \begin{overpic}[height=3.8cm]
         {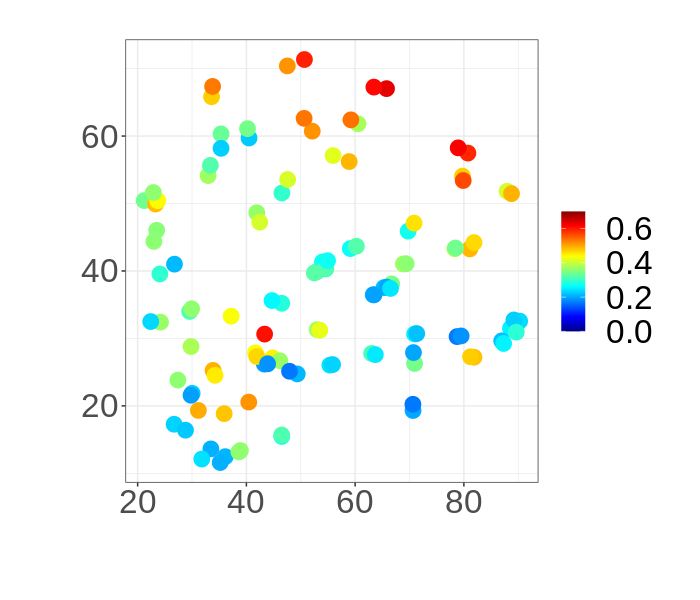}
         \put(20,0){\scriptsize Atlas coordinate 1}
         \put(5,20){\rotatebox{90}{\scriptsize Atlas coordinate 2}}
                  \end{overpic}
                         \caption{\scriptsize Within-group variances of $z^{(.)}_i$ for subjects in the diseased group, plotted over the automated anatomical labeling atlas.}
    \end{subfigure} \;
\\
              \begin{subfigure}[t]{.5\textwidth}\centering
         \includegraphics[height=5cm]{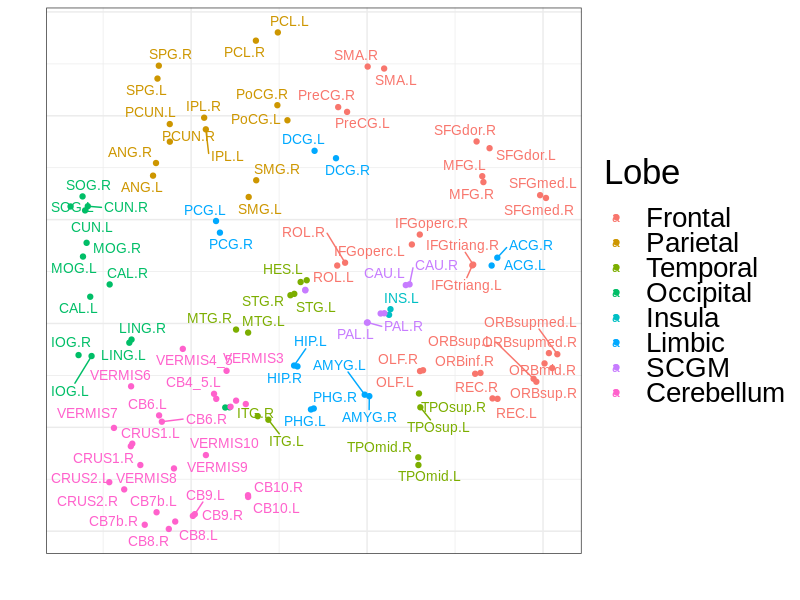}
                         \caption{\scriptsize The regions of interest (ROIs) colored by the associated lobe names, under the automated anatomical labeling atlas.}
    \end{subfigure}
                  \begin{subfigure}[t]{.45\textwidth}
                      \begin{overpic}[height=5.2cm]
         {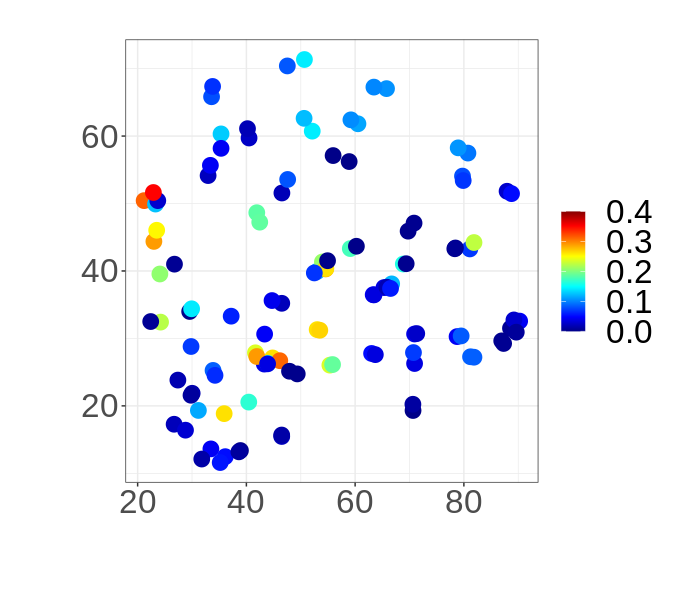}
         \put(20,0){\scriptsize Atlas coordinate 1}
         \put(5,20){\rotatebox{90}{\scriptsize Atlas coordinate 2}}
                  \end{overpic}
                         \caption{\scriptsize Intraclass correlation coefficients for the regions of interest $(1-\hat\sigma^{2}_{\text{within--group},i}/\hat\sigma^{2}_{\text{total},i})$ .}
    \end{subfigure}
              \begin{subfigure}[t]{.45\textwidth}\centering
         \includegraphics[width=1\linewidth]{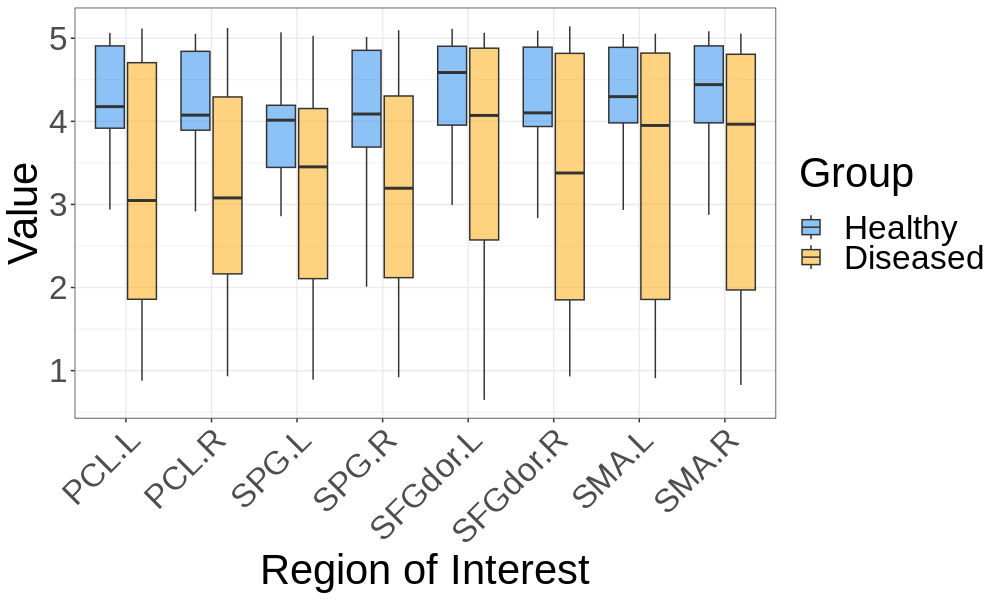}
                         \caption{\scriptsize Boxplot visualization of the latent coordinates for the regions with high variability in the diseased group.}
    \end{subfigure}\;
              \begin{subfigure}[t]{.45\textwidth}\centering
         \includegraphics[width=1\linewidth]{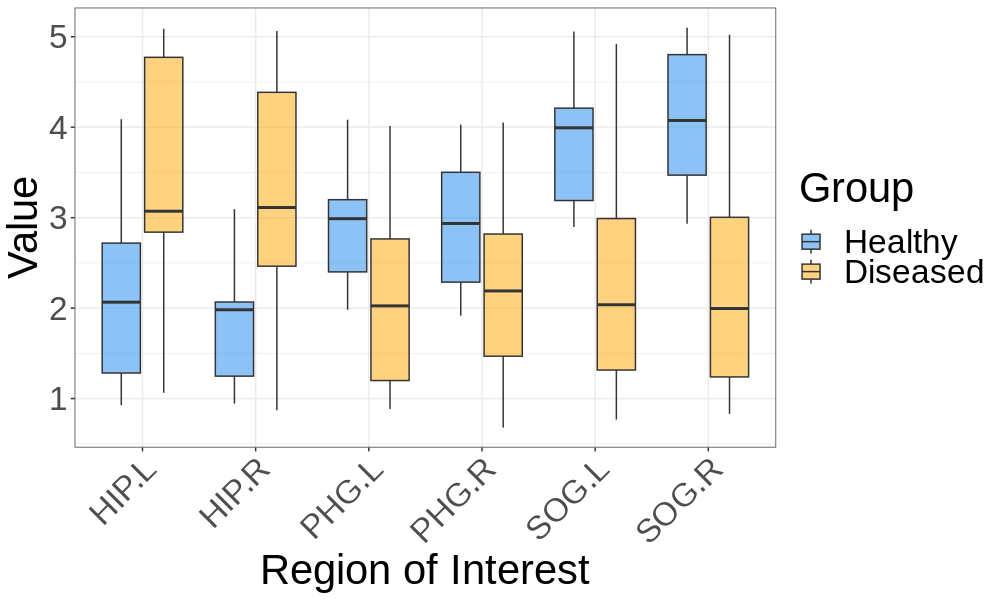}
                         \caption{\scriptsize Boxplot visualization of the latent coordinates for the regions with low variability in the diseased group.}
    \end{subfigure}
        \caption{Using the latent coordinates to characterize the heterogeneity within the subjects.
        \label{fig:application}
        }
        \end{figure}

        \vspace*{-0.5cm}
We fit the multi-view clustering model to the data, by running
 MCMC for $5,000$ iterations and discarding the first $2,500$ as burn-in.
 As shown in Figure \ref{fig:application2}, the hierarchical Dirichlet distribution on the latent coordinates induces similarity between the clustering of brain regions among subjects on a subset of nodes, while showing subtle differences on the other nodes.
On the other hand, some major differences can be seen in the clusterings between the healthy and diseased subjects. Using the latent coordinates (at the posterior mean), we quantify the distances between $z^{(s)}$ and $z^{(s')}$ for each pair of subjects $s\neq s'$.  As shown in Figure \ref{fig:application}(a), there is a clear two-group structure in the pairwise distance matrix formed by $\|z^{(s)}-z^{(s')}\|_F$, and the separation corresponds to the first 64 subjects being healthy (denoted by $s\in g_1$) and the latter 102 being diseased (denoted by $s\in g_2$).

Next, we compute the within--group variances for these two groups, using $\sum_{s\in g_l}\|  z_i^{(s)}- (\sum_{s\in g_l} z^{(s)}_i / |g_l|)\|_F^2/|g_l|$ for $l=1$ and $2$, and plot the variance over each region of interest $i$ on the spatial coordinate of the atlas. Figure \ref{fig:application}(b) and (c) show that, although both groups show some degree of variability, the diseased group shows clearly higher variances in some regions of the brain. Specifically, the paracentral lobule
 (PCL) and superior parietal gyrus (SPG), dorsolateral superior frontal gyrus (SFGdor), and supplementary motor area (SMA) in the frontal lobe show the highest amount of variability. Indeed, those regions are also associated with very low ICC scores [Figure \ref{fig:application}(e)] calculated based on the variance of $z^{(s)}_i$, with pooled estimates $\hat\sigma^{2}_{\text{total},i}= \sum_{s}\|  z_i^{(s)}- (\sum_{s} z^{(s)}_i /S)\|_F^2/S$ and $\hat\sigma^{2}_{\text{within--group},i} =   \sum_{l=1}^2\sum_{s\in g_l}\|  z_i^{(s)}- (\sum_{s\in g_l} z^{(s)}_i /|g_l|)\|_F^2/S$. On the other hand, some regions such as the hippocampus (HIP), parahippocampal gyrus (PHG), and superior occipital gyrus (SOG) show relatively lower variances within each group, hence higher ICC scores.
 
 To show more details on the heterogeneity, we plot the latent coordinates associated with those ROIs using boxplots. Since each $z_i^{(s)}$ is in two-dimensional space, we plot the linear transform $\tilde z_i^{(s)} = z_{i,1}^{(s)} +z_{i,2}^{(s)}$. Interestingly, those 8 ROIs with high variability still seem quite informative for distinguishing the two groups (Figure \ref{fig:application}(f)). To verify, we concatenate those latent coordinates and form an $S\times 16$ matrix, and fit them in a logistic regression model for classifying the healthy versus diseased states. The Area Under the Curve (AUC) of the Receiver Operating Characteristic is 86.6\%. On the other hand, when we fit the 6 ROIs with low variability in logistic regression, the AUC increases to 96.1\%.

 An explanation for the above results is that Alzheimer's disease does different degrees of damage in the frontal and parietal lobes (see the two distinct clusterings in Figure \ref{fig:application2} (c) and (d)), and the severity of the damage can vary from person to person. On the other hand, the hippocampus region (HIP and PHG), important for memory consolidation, is known to be commonly affected by Alzheimer's disease \citep{braak1991neuropathological,klimova2015alzheimer}, which explains the low heterogeneity in the diseased group. Further, to our best knowledge, the high discriminability of the superior occipital gyrus (SOG) is a new quantitative finding, that could be meaningful for a further clinical study.

For validation, without using any group information, we concatenate those $z_i^{(s)}$'s over all $i=1,\ldots,116$ and form an $S\times 232$ matrix and use lasso logistic regression to classify the two groups. When $12$ predictors are selected (as a similar-size model to the one above using 6 ROIs), the AUC is 96.4\%.  Since $z_i^{(s)}$'s are obtained in an unsupervised way, this validation result shows that the multi-view clustering model produces meaningful representation for the nodes in this Alzheimer's disease data. We provide further details on the clusterings, including the number of clusters, and the posterior co-assignment probability matrices in the Supplementary Materials S4.5.

\vspace*{-1cm}
        \section{Discussion}
        \vspace*{-0.5cm}
        In this article, we present our discovery of a probabilistic model for popular spectral clustering algorithms. This enables straightforward uncertainty quantification and model-based extensions through the Bayesian framework. There are several directions worth exploring. First, our consistency theory is conducted under the condition of separable sets, similar to \cite{ascolani2022clustering}. For general cases with non-separable sets, clustering consistency (especially on estimating $K$) is challenging to achieve; to our best knowledge, existing consistency theory only applies to data generated independently from a mixture model \citep{miller2018mixture,zeng2020quasi}. For data generated dependently via a graph, this is still an unsolved problem. Second, in all of our forest models, we have been careful in choosing densities with tractable normalizing constants. One could relax this constraint by using densities $f(y_i\mid y_j,\theta)= \alpha_f g_f(y_i\mid y_j;\theta)$ and $r(y_i;\theta)= \alpha_r g_r(y_i;\theta)$, with $g$ some similarity function, and $(\alpha_f,\alpha_r)$ potentially intractable. In these cases, the forest posterior becomes
        $\Pi(\mathcal T\mid .)\propto  (\lambda\alpha_r/\alpha_f)^K  \prod_{(0,i)\in \mathcal T} g_r(y_i;\theta)
        \prod_{(i,j)\in \mathcal T} g_r(y_i\mid y_j;\theta)$. Therefore, one could choose an appropriate $\tilde \lambda=\lambda\alpha_r/\alpha_f$ (equivalent to choosing some value of $\lambda$), without knowing the value of $\alpha_f$ or $\alpha_r$; nevertheless, how to calibrate $\tilde \lambda$ still requires further study. Third, a related idea is the Dirichlet Diffusion Tree \citep{neal2003density}, which considers a particle starting at the origin, following the path of previous particles, and diverging at a random time. The data are collected as the locations of particles at the end of a time period. Compared to the forest process, the diffusion tree process has the conditional likelihood given the tree invariant to the ordering of the data index, which is a stronger property compared to the marginal exchangeability of the data points. Therefore, it is interesting to further explore the relationship between those two processes.

\spacingset{1} % DON'T change the spacing!

\bibliographystyle{chicago}

\bibliography{ref}

\appendix
\renewcommand{\thesection}{S\arabic{section}}

\section{Model-based Extensions to Forest Model}
\subsection{Extension to High-dimensional Clustering Model}
      For clustering high dimensional data, good performances have been demonstrated through finding a low-dimensional sparse representation $z_i$ for each $y_i$ \citep{vidal2011subspace,wu2014spectral}, and then clustering $z_i$ instead of $y_i$. To briefly review the idea, for high-dimensional data, a useful assumption is that $y_i\in \mathbb{R}^p$ can be ``reconstructed'' using a linear combination of a few other $y_j$'s, that is, $y_i\approx \sum_{j} w_{i,j} y_j$, with $w_{i,i}=0$ and $w_i=(w_{i,1},\ldots, w_{i,n})$ contains only a few non-zeros. 
%     In addition, one often imposes some unit norm constraint on $z_i$, such as $\|z_i\|_1=1$ or $\|z_i\|_2=1$. 

      Although $w_i$ is obtained as a vector of coefficients, it can be viewed as a low-dimensional {\em relative} coordinate,  that can be used instead of the absolute coordinate $y_i\in \mathbb{R}^p$. The key idea is that if $w_i$ and $w_j$ are in different subspaces ($w_i'w_j=0$), then $y_i$ and $y_j$ are likely to be in different clusters. Using a similarity function defined on each pair $(w_i,w_j)$, one could obtain a similarity matrix and then apply the spectral clustering algorithm.

      We now propose a generative distribution.
      We use $W=[w_1', \ldots, w'_n]$ as the $n\times n$ matrix with the $i$th row equal to $w_i$, and $Y$ the $n\times p$ data matrix. We include the reconstruction loss $\|Y-WY\|_F^2$ (with $\|.\|_F$ the Frobenius norm) via a matrix Gaussian distribution:
      \be
        &    Y \sim \text{Matrix-Gaussian} \big\{O,  \sigma^2_y [(I_n-W)'(I_n-W)]^{-1}, I_p\big\}.      \ee
      We note a link between this model and the spatial autoregressive (SAR) model \citep{ord1975estimation}, except that the neighborhood information $W$ is not known. We view each $w_i$ as a transform of another unit-norm vector $z_i$ that satisfies $\|z_i\|_2=1$ and $\|z_i\|_0=d$ (the number of non-zeros is $d$) via
       \be 
        w_{i,k}= \alpha_i z_{i,k}, \; \text{for } k\neq i, \quad
       w_{i,i}=0, \quad z_{i,i}\in \mathbb{R}, 
       \ee
         with $\alpha_i>0$ some scale parameter, and $z_{i,i}$ not necessarily zero.
    And we model $(z_1, \ldots, z_n)$ as from a forest model based on sparse von Mises--Fisher densities:
      \be
      & z_1,\ldots, z_n \sim \text{Forest Model}(\mathcal T), \\
%      & \Pi(z_1,\ldots, z_n, \mathcal T) \propto \lambda^K  \prod_{(i,j)\in \mathcal T:i>0,j>0} (z_i'z_j)_+   \prod_{i=1}^n 1(\|z_i\|_2=1,z_{i,i}=0, \|z_i\|_0 \le d),\\
     & f(z_i \mid z_j; \kappa)\propto  \exp( \kappa z_i'z_j )1( z_i' z_j \neq  0)
     1(\|z_i\|_2=1,\|z_i\|_0 = d),
      \\
     &r(z_i)  %=   \frac{\Gamma(d/2)}{2\pi^{d/2} {n-1 \choose d}}   
     \propto 1(\|z_i\|_2=1,\|z_i\|_0 = d).
      \ee
      The leaf $f(z_i \mid z_j;\kappa)$ is supported in those $(d-1)$-dimensional unit spheres $\mathcal S^{(d-1)}\subset \mathcal S^{(n-1)}$, such that $z_i$ and $z_j$ are not in completely disjoint subspaces.     
          The von Mises--Fisher density in a given $\mathcal S^{(d-1)}$ has a tractable normalizing constant that depends on $\kappa$ only. Further, with $\|z_j\|_0=d$, we can easily tell the number of those $\mathcal S^{(d-1)}$ with $z_i:  z_i'z_j\neq 0$ is equal to ${n\choose d}- {n-d \choose d}$. Similarly, $r$ is a uniform density on all $\mathcal S^{(d-1)}\subset \mathcal S^{(n-1)}$. 
          Therefore, both of the normalizing constants in $f$ and $r$ are available. 
                We refer to the model for $Y$ as a latent forest model.

         One could further assign priors on $\kappa$, $\sigma^2_y$ and $\alpha_i$'s, and develop a Gibbs sampling algorithm 
        for posterior estimation.
In this section, since our main focus is to demonstrate a high-dimensional model extension and compare the point estimates against a few other algorithms,  we use a fast posterior approximation algorithm for the above model.
Specifically, we first use the lasso algorithm to solve for a sparse $\hat W= {\arg\min}_{W:  w_{i,i}=0 \;\forall i} (1/2)\|Y-WY\|_F^2+ \lambda \|W\|_1$ with $\lambda=1$; then for each $\hat w_i$, we take the top $(d-1)$ elements in magnitude, and set the other elements to zero. Then we replace $w_{i,i}$ by $1$ and normalize the vector to produce a unit 2-norm vector $z_i$. Conditioning on the transformed matrix $\hat Z$ and $\kappa$ fixed to $10$, we sample the forest $\mathcal T$ using the random-walk covering algorithm.

To assess the clustering performance, we use the image data from the Yale face database B \citep{georghiades2001few}. This dataset contains single light source images of 10 subjects. We take the ones corresponding to the forward-facing poses under 64 different illumination conditions (shown in Figure S.1). We resize each image to have 48$\times$42 pixels. We label those images by subject id from 1 to 10. Therefore, we have a clustering task with $n=640$ and $p=2,016$.

\begin{figure}[H]
     \begin{subfigure}[t]{.32\textwidth}
        \includegraphics[height=5cm,width=1\linewidth, trim=1cm 2cm 1cm 2cm,clip]{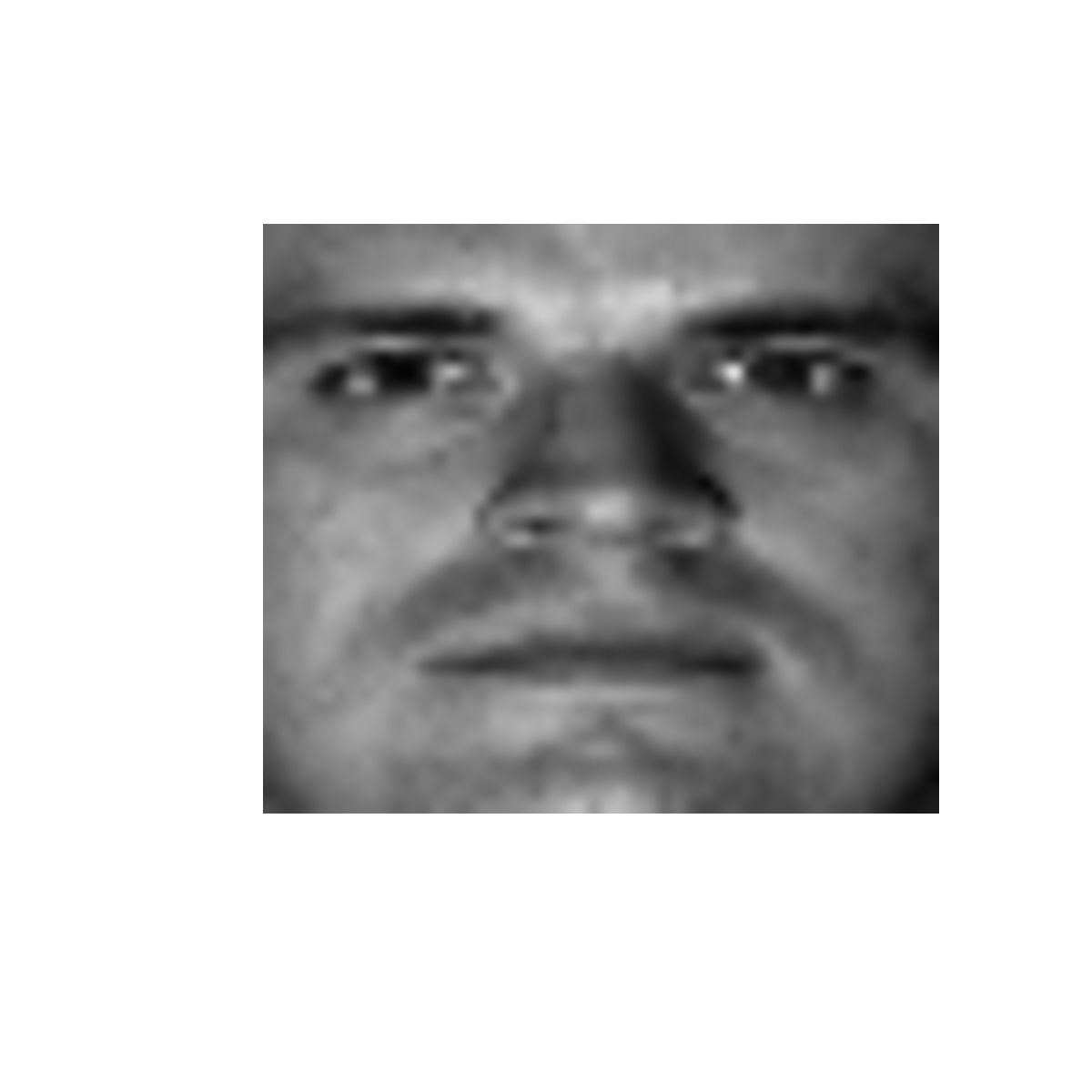}
            \caption{One subject under illumination condition 1.}\;
    \end{subfigure} 
             \begin{subfigure}[t]{.32\textwidth}
        \includegraphics[height=5cm,width=1\linewidth, trim=1cm 2cm 1cm 2cm,clip]{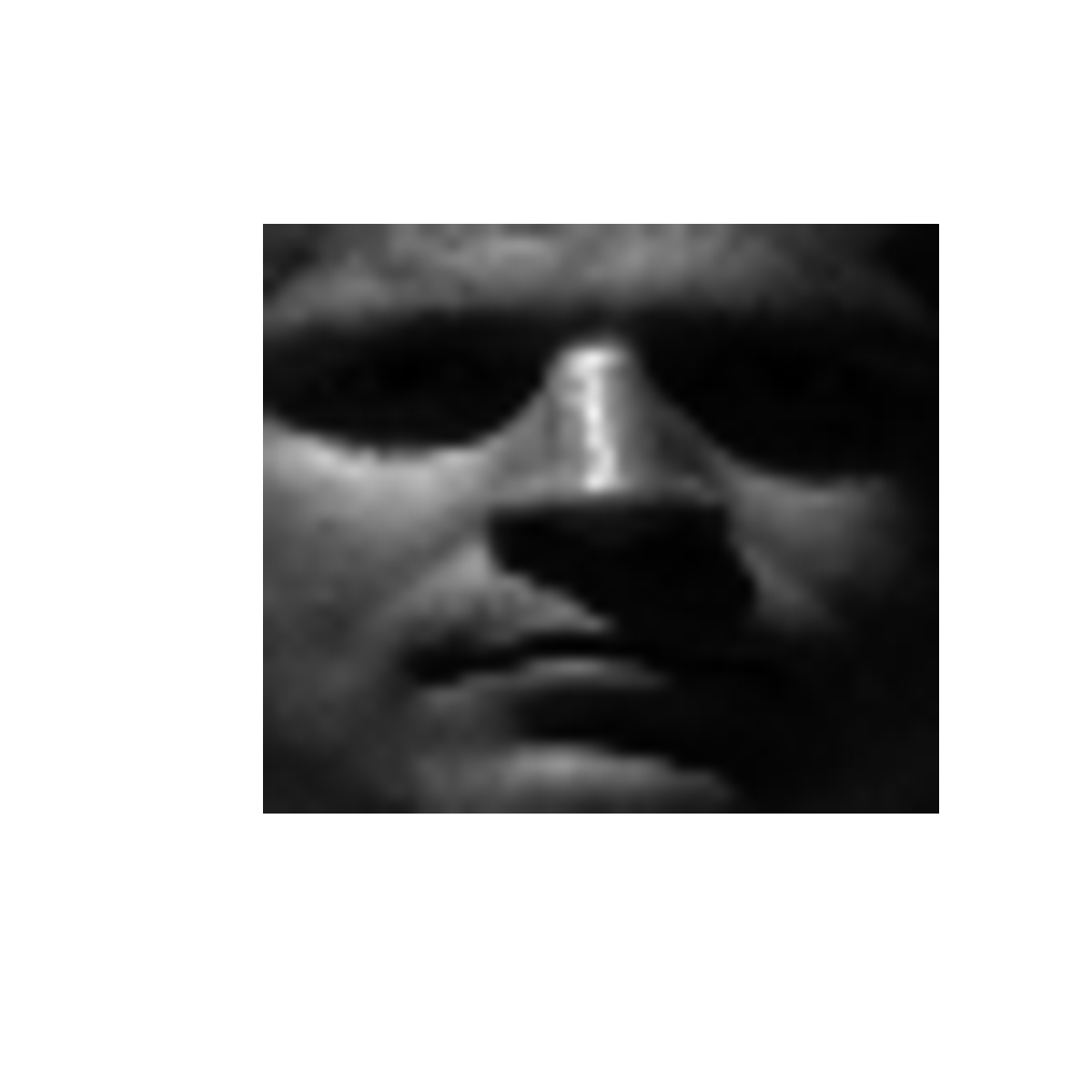}
            \caption{One subject under  illumination condition 2.}
    \end{subfigure} \;
         \begin{subfigure}[t]{.32\textwidth}
        \includegraphics[height=5cm,width=1\linewidth, trim=1cm 2cm 1cm 2cm,clip]{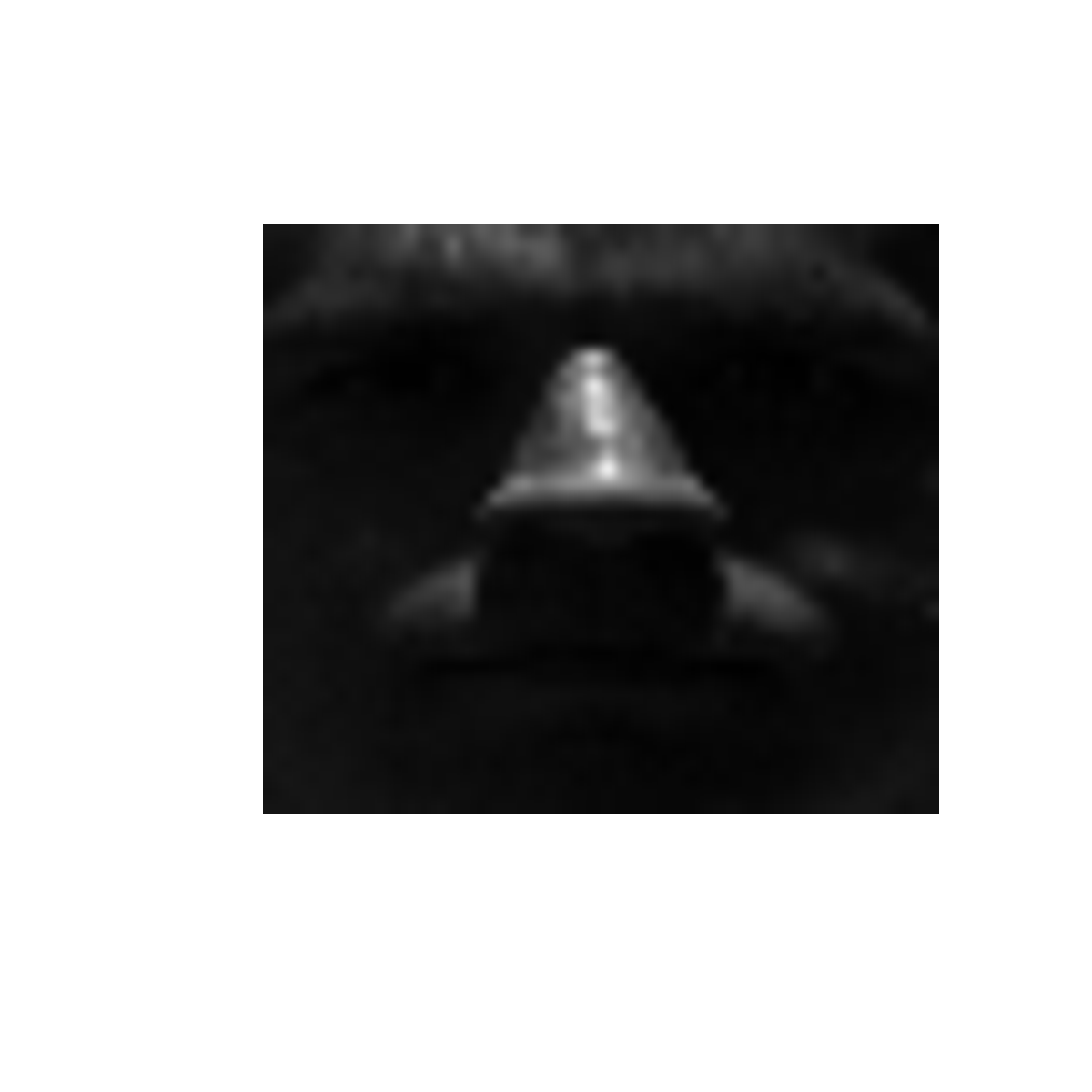}
            \caption{One subject under  illumination condition 3.}
    \end{subfigure} \\
         \begin{subfigure}[t]{.32\textwidth}
        \includegraphics[height=5cm,width=1\linewidth, trim=1cm 2cm 1cm 2cm,clip]{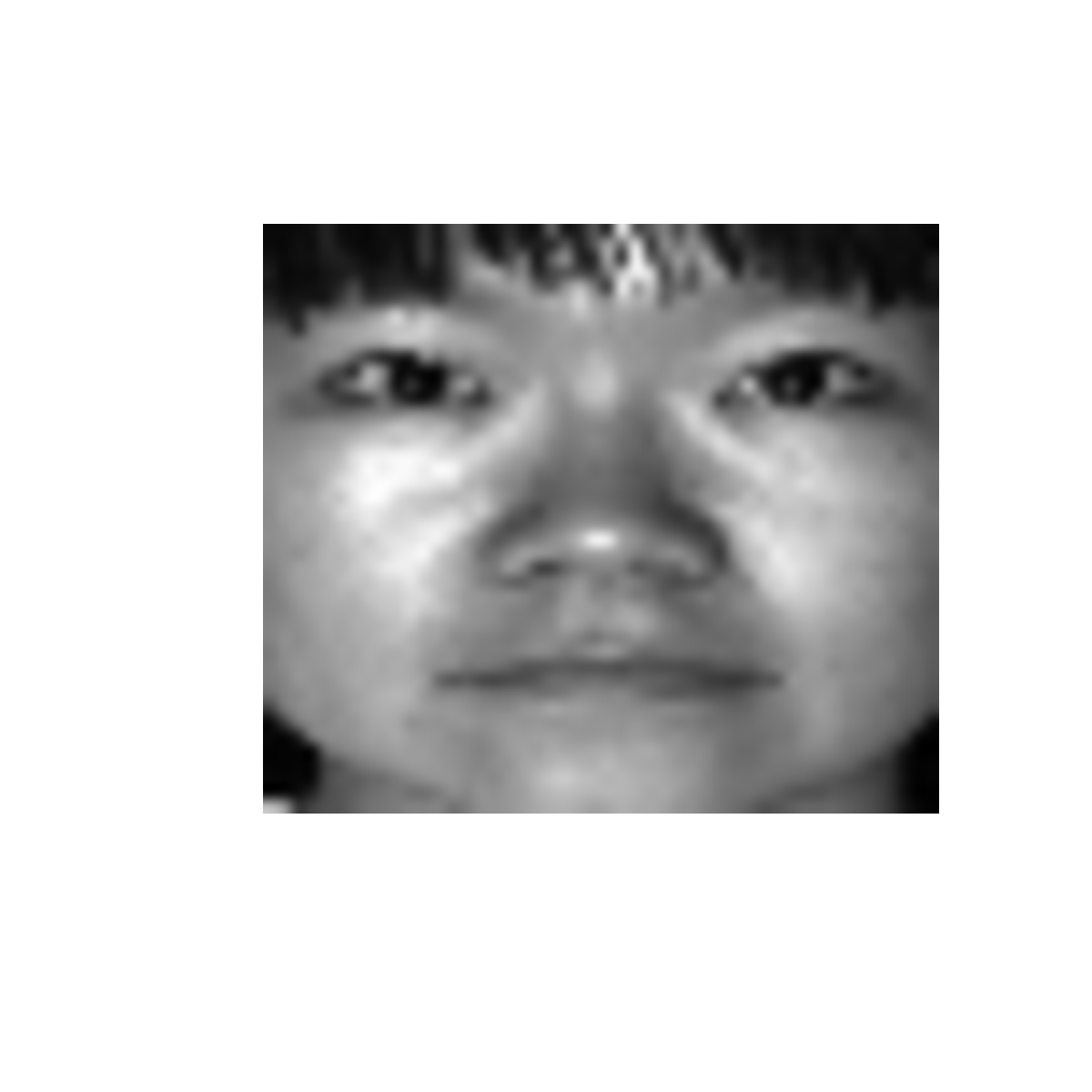}
            \caption{Another subject under  illumination condition 1.}
    \end{subfigure} 
             \begin{subfigure}[t]{.32\textwidth}
        \includegraphics[height=5cm,width=1\linewidth, trim=1cm 2cm 1cm 2cm,clip]{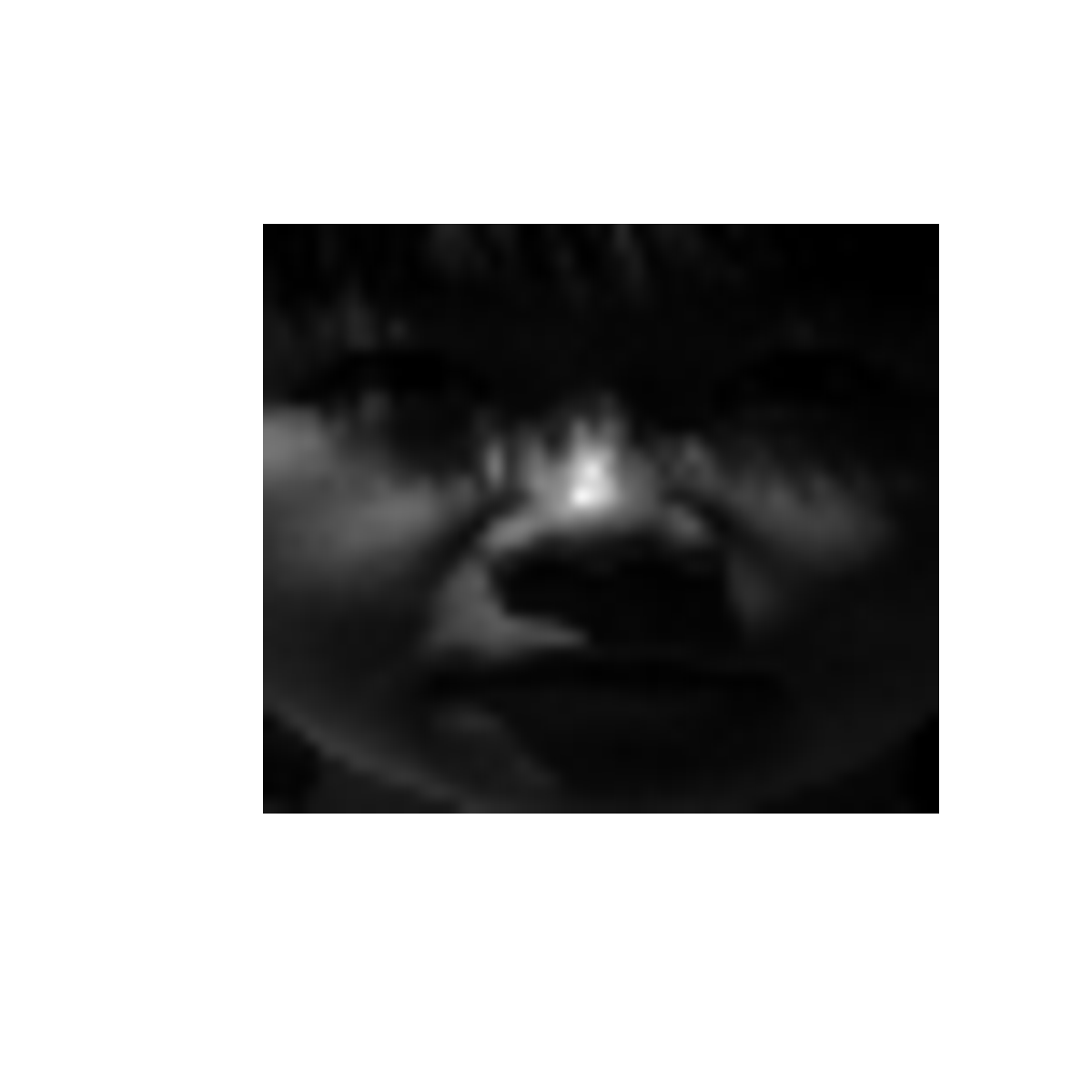}
            \caption{Another subject under  illumination condition 2.}
    \end{subfigure} \;
         \begin{subfigure}[t]{.32\textwidth}
        \includegraphics[height=5cm,width=1\linewidth, trim=1cm 2cm 1cm 2cm,clip]{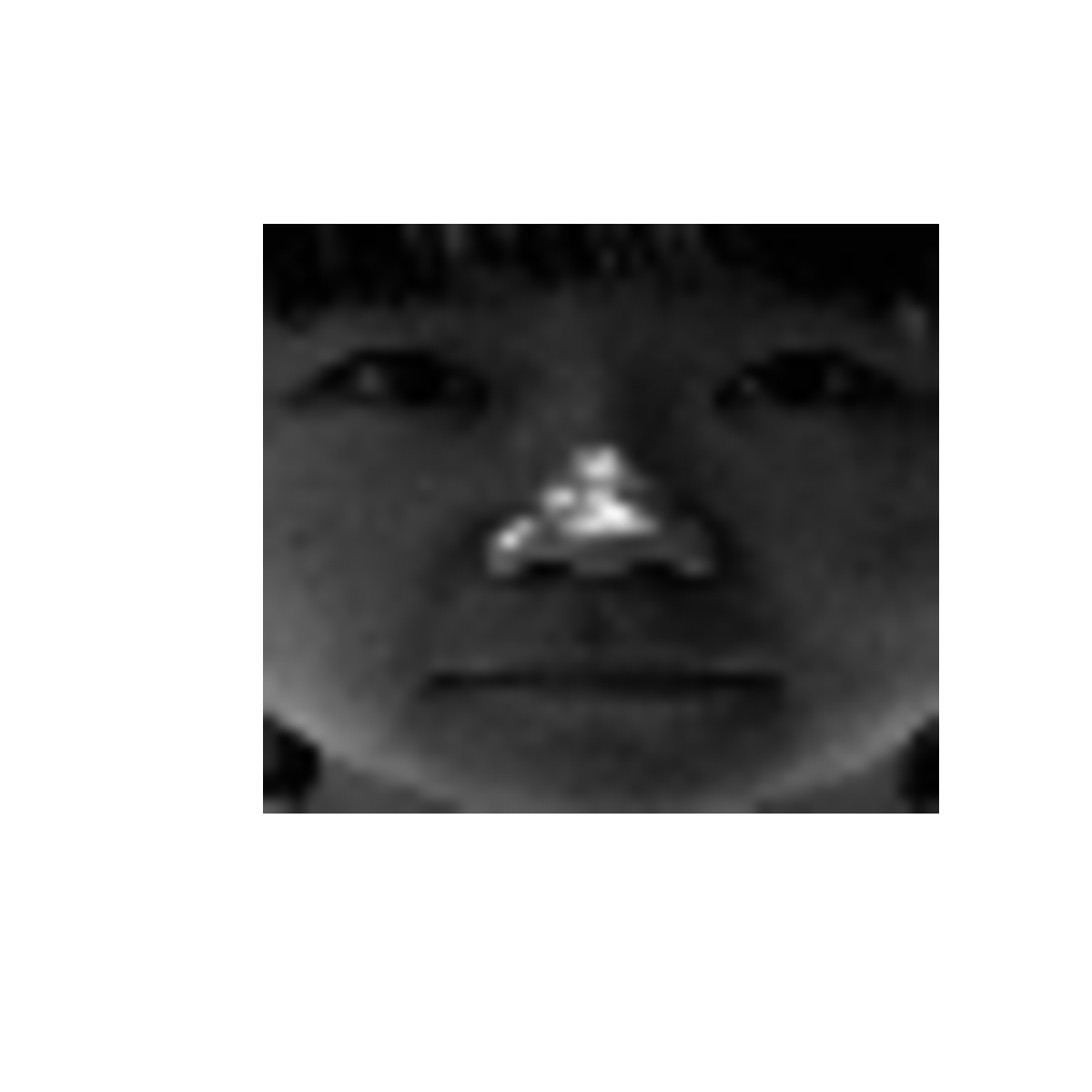}
            \caption{Another subject under  illumination condition 3.}
    \end{subfigure} 
        \caption*{Figure S.1: a few sample photos from the Yale face database B \citep{georghiades2001few}.
        }
\end{figure}

We compare the performance against several popular clustering methods. To produce a point estimate, for the forest model on $z_i$'s, we apply spectral clustering with $K=10$ on the posterior co-assignment probability matrix (as described in the main text); for each of the other methods, we use $K=10$ as the specified parameter.
To evaluate the clustering accuracy, we relabel the point estimate $(c_1,\ldots, c_n)$ using the Hungarian matching algorithm \citep{kuhn1955hungarian}, so that the Hamming distance $\text{dist}_{\text {h}}$ between $(c_1,\ldots, c_n)$ and the subject id's is minimized. Then the clustering accuracy is calculated as $(n-\text{dist}_{\text {h}})/n$. As the accuracy can be sensitive to the initialization of each algorithm, for a fair comparison, we repeat running each algorithm 20 times, and report the mean and the 95\% confidence interval.

\begin{table}[H]
\footnotesize
\centering
    \begin{tabular}{|l |l | l | l | l|}
  \hline            
 Method   & K-means                 & Mclust (VII)                             & Mclust (VEI)                              & Mclust (EII) \\
   \hline           
 Accuracy & 0.18 (0.16, 0.21)                   & 0.24  (0.24, 0.24)                                   & 0.26 (0.26, 0.26)                                    & 0.23 (0.23, 0.23) \\
   \hline          
      \hline            
 Method   & HDDC (AkjBkQkDk)        & HDDC (AkjBQkDk)                          & SpecC on $e^{- \lambda_s\|y_i-y_j\|_2^2}$
          & SpecC on $y_i'y_j$ \\
             \hline           
 Accuracy & 0.324                   & 0.296                                    & 0.35 (0.25, 0.43)                                     & 0.30 (0.28, 0.34) \\
 \hline
 \hline
 Method & K-means on $w_i$   & SpecC on $({w_i}' w_j)_+$ &  SpecC on $|w_{i,j}| + |w_{j,i}|$
        & Forest on $z_i$ \\
             \hline           
 Accuracy & 0.25 (0.18, 0.39)                   & 0.64 (0.52, 0.69)                                    & 0.59 (0.46, 0.68)                                     & 0.82 (0.71, 0.93) \\
    \hline  
\end{tabular}
\caption*{Table S.1: Clustering 640 face photos collected from 10 subjects. \label{tb:yale_b}}
\end{table}

Tabel S.1 shows the results. We use the K-means function from the native R library, the Mclust function in the MCLUST package \citep{scrucca2016mclust} for various Gaussian mixture models, the hddc function in HDclassif \citep{berge2012hdclassif} package for Gaussian mixture models with near low-rank covariance matrices, and the specc function from the kernlab package \citep{karatzoglou2004kernlab} for the spectral clustering algorithm. For those spectral clustering algorithms, we use $w_i$'s as the sparse representation estimated from the lasso regression, without imposing a low-cardinality constraint. For the latent forest model, we use $z_i$'s with cardinality constraint at $d=4$.

Clearly, for this high-dimensional dataset, clustering the sparse representation $z_i$'s (or $w_i$'s) instead of $y_i$'s  has a significantly improved accuracy. Interestingly, we found that K-means on those $w_i$'s produce much worse results than all the spectral clustering algorithms. This suggests that forest models (as a generative model for spectral clustering) give a better fit to those  $w_i$'s, compared to the Gaussian mixture models (as a generative model for K-means). Lastly, compared to the existing spectral clustering algorithms using similarity 
$|w_{i,j}| + |w_{j,i}|$ \citep{vidal2011subspace} or
$({w_i}' w_j)_+$ \citep{wu2014spectral}, imposing a cardinality constraint seemed to further improve the signal that is helpful for clustering. To verify this, we also conducted spectral clustering using similarity $({z_i}' z_j)_+$ and obtained almost the same clustering accuracy as the one from the latent forest model.

As shown in Figure S.2, for this high-dimensional dataset, the pairwise Euclidean distance $\|y_i-y_j\|_2$'s are too noisy to be used for clustering, the inner product on the sparse $z_i$ has much less noise, and the pairwise co-assignment probability matrix produces a very clear partition of 10 clusters.

\begin{figure}[H]
     \begin{subfigure}[t]{.32\textwidth}
        \includegraphics[height=6cm,width=1\linewidth]{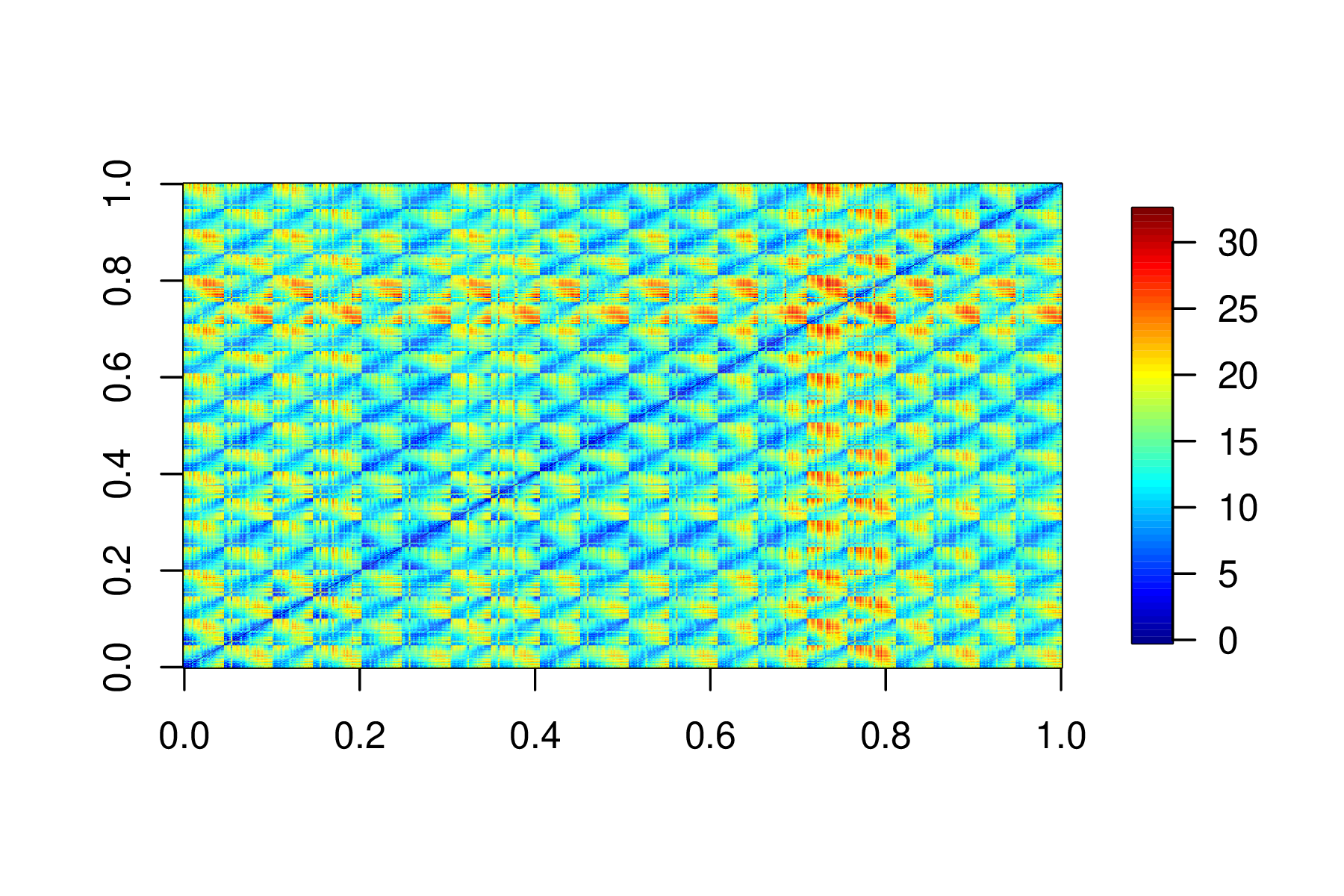}
            \caption{Euclidean distances between $y_i$'s.}\;
    \end{subfigure} 
             \begin{subfigure}[t]{.32\textwidth}
        \includegraphics[height=6cm,width=1\linewidth]{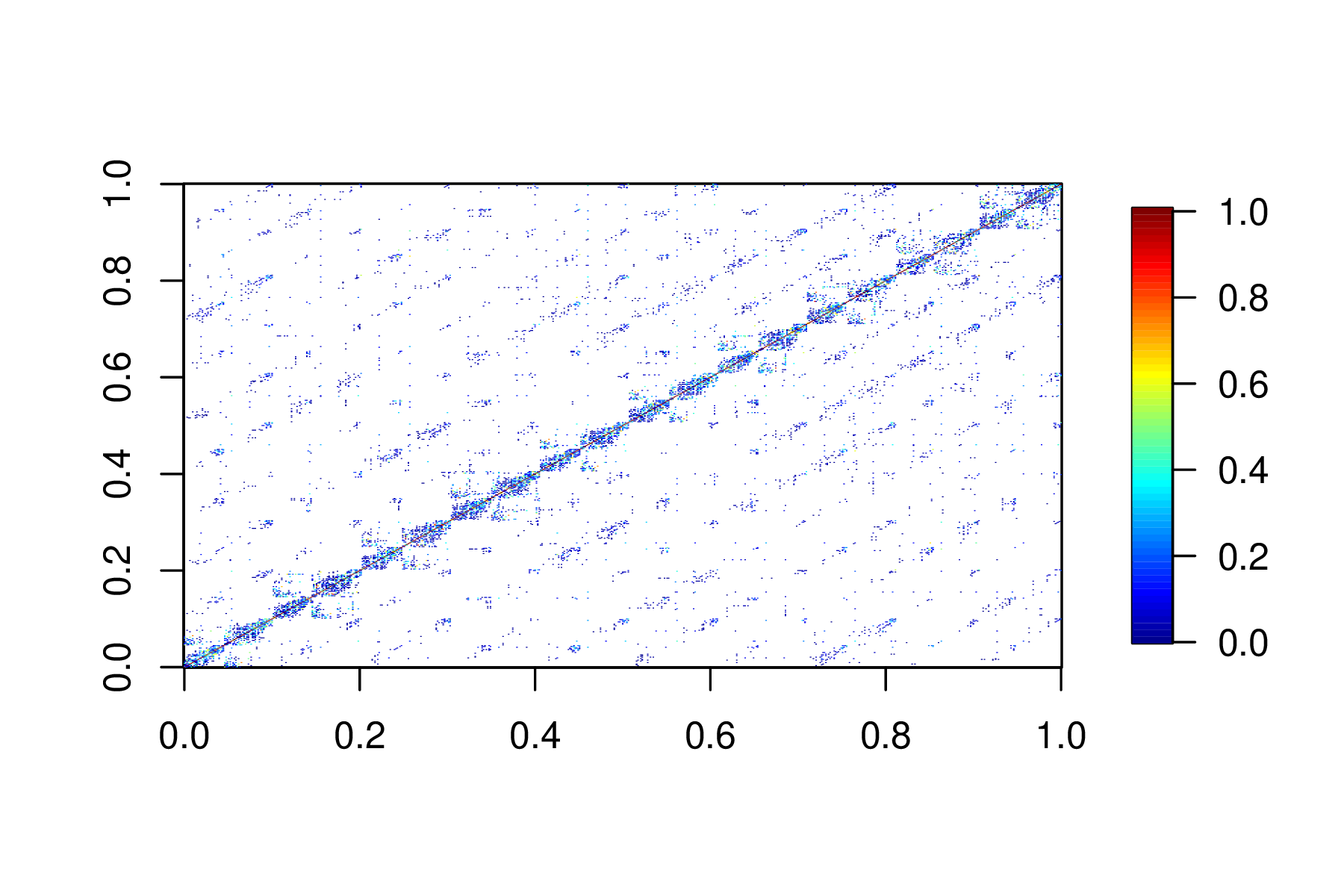}
            \caption{$({z_i}' z_j)$ between the latent $z_i$'s.}
    \end{subfigure} \;
         \begin{subfigure}[t]{.32\textwidth}
        \includegraphics[height=6cm,width=1\linewidth]{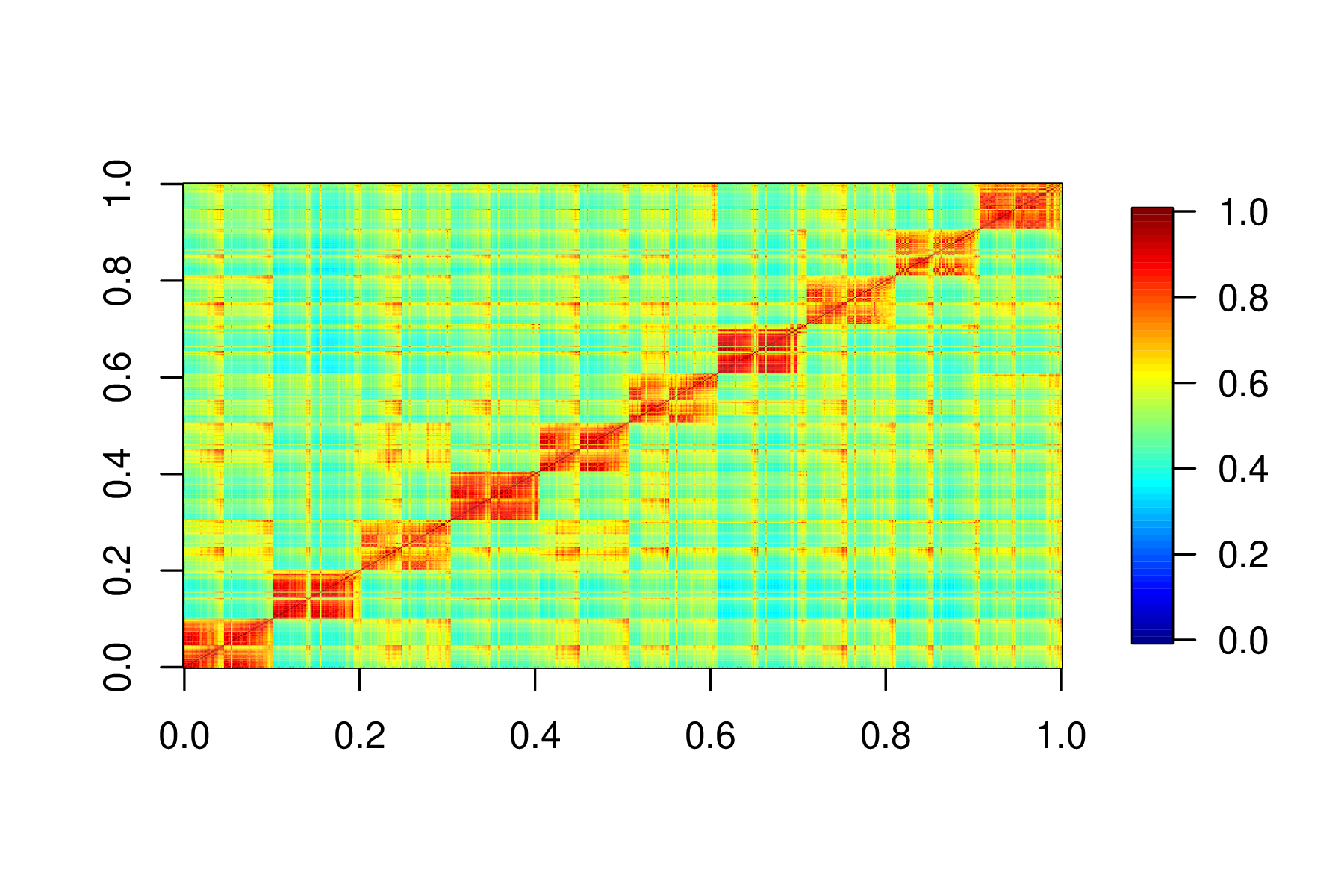}
            \caption{$\text{Pr}(c_i=c_j \mid y)$ from the latent forest model.}
    \end{subfigure} 
        \caption*{Figure S.2: Pairwise information between the observed $y_i$'s, and sparse latent $z_i$'s, and posterior co-assignment probability matrix in the latent forest model.
        }
        \end{figure}

\subsection{Extension to Covariate-dependent Forest Clustering}
 Following \cite{muller2011product}, we now illustrate an extension where the clustering is dependent on external covariates $x_i$'s (each $x_i$ is an $m$-dimension vector). \cite{muller2011product} proposed the following covariate-dependent product partition model (PPMx):
\be
& \Pi_0(V_1,\ldots, V_K \mid K, x) \propto  \prod_{k=1}^K  C(V_k) G(\{x_i\}_{i\in V_k}),
\ee
where $C$ and $G$ together form a modified cohesion function, with $G$
positive-valued and quantifying the overall similarity among those $x_i:i\in V_k$. To specify $G$, \cite{muller2011product} proposed to use 
\be 
G(\{x_i\}_{i\in V_k})=\int [\prod_{i\in V_k} \tilde g_1(x_i; \xi_k)] \tilde  g_2(\xi_k) d \xi_k
\ee
with $\tilde  g_1$ and $\tilde  g_2$ some probability density/mass functions with conjugacy, such as $\tilde g_1$ as multivariate Gaussian $\text{N}(\cdot \mid \mu_k,\Sigma_1)$ and $\tilde g_2$ as Gaussian for $\text{N}(\mu_k \mid 0,\Sigma_2)$, with $\Sigma_1$ and $\Sigma_2$ some fixed parameters. Importantly, the purpose of $G$ is to form a density-based cohesion function as a priori, hence $G$ is not interpreted as the generative distribution for $x_i$'s.

We note that the above $G(\{x_i\}_{i\in V_k})$ effectively treats $x_i:i\in V_k$ as conditionally independent. Now suppose there is a tree $T_k$, we can equivalently form a joint distribution by starting from a $x{_k*} \sim   \tilde g_1(\cdot; \xi_k)$, and then for any $(i,j)\in T_k$, $(x_j-x_i) \sim   \tilde g^*_1(\cdot; \xi_k)$, with $\tilde g^*_1$ the transformed distribution on the difference. Therefore, we have a tree-based similarity function:
\be
G(\{x_i\}_{i\in V_k}; T_k)= \int  \bigg [\tilde g_1(x_{k^*}; \xi_k) \prod_{(i,j)\in T_k} \tilde g^*_1 (x_j-x_i; \xi_k) \bigg]  \tilde  g_2(\xi_k)  d \xi_k.
\ee
In this section, we use Gaussian $\tilde g_1$ and $\tilde g_2$ as mentioned above. We have $\tilde g^*_1$ as $\text{N}(\cdot \mid 0, 2 \Sigma_1)$. After integration, we have 
\be
& G  (\{x_i\}_{i\in V_k}; T_k) = \prod_{(i,j)\in T_k} 
\underbrace{|2\pi (2\Sigma_1)|^{-1/2}\exp \bigg [ - (x_i-x_j)'(4\Sigma_1)^{-1}(x_i-x_j)\bigg]}_{f_0(x_i;x_j)}\times \\
& 
\underbrace{|2\pi \Sigma_1 \Sigma_2|^{-1/2} 
|\Sigma_1^{-1}+\Sigma_2^{-1}|^{-1/2}
\exp \bigg[
-  \frac{1}{2} x_{k^*} '\Sigma^{-1}_1x_{k^*} +
 \frac{1}{2} x_{k^*} ' \Sigma_1^{-1}(\Sigma_1^{-1}+\Sigma_2^{-1})^{-1}\Sigma_1^{-1}x_{k^*}
\bigg]
}_{r_0(x_k^*)},
\ee
where we use $f_0$ and $r_0$ to simplify notation.
Therefore, we can achieve similar effects of PPMx, using an $x$-informative tree distribution: 
\be
&\Pi(E_k\mid V_k) \Pi( {k^*} \mid E_k, V_k)=
\frac{
r_0(x_{k^*}) \prod_{(i,j)\in T_k} f_0(x_i; x_j)
}{
[\sum_{k\in V_k} r_0(x_{k})]
    [\sum_{T'_k}\prod_{(i,j)\in T'_k}f_0(x_i; x_j)]
},\\
& \Pi_0(V_1,\ldots, V_K , K) \propto  \lambda^{K}
\bigg \{
 \prod_{k=1}^K  
[\sum_{k\in V_k} r_0(x_{k})]
    [\sum_{T'_k}\prod_{(i,j)\in T'_k}f_0(x_i; x_j)]\bigg \}.
\ee
Note that if $f_0(x_i; x_j)\propto 1$ for any $(x_i,x_j)$, and $r_0(x_i)\propto 1$ for any $x_i$, then the above would be
$\Pi_0(V_1,\ldots, V_K , K)= \lambda^K n_k^{n_k-1}$,
the same as the distribution we describe in the main text. 

Compared to directly clustering $(y_i,x_i)$ as the joint observation together, a strength of the above approach (and PPMx methods in general) is that as a priori,
we can directly control the influence from $x_i$ to clustering, by adjusting the parameters in $G$. For example, we use $\Sigma_1=\Sigma_2 = \eta S_n$ with $S_n$ the empirical covariance of $x_i$'s and $\eta>0$ an adjustable hyper-parameter. This leads to 
 ${f_0(x_i;x_j)} = {|2\pi (2\Sigma_1)|^{-1/2}\exp  [ - (x_i-x_j)'(4\Sigma_1)^{-1}(x_i-x_j)]}$
and 
${r_0(x_{k^*})}= |2\pi (2\Sigma_1) |^{-1/2} \exp [-  x_{k^*} ' (4\Sigma_1)^{-1}x_{k^*}]$. As $\eta$ increases, the influence of $x_i$ becomes weaker. Note that if we were to use $x_i$ in a likelihood, we would not have such flexibility.

To illustrate this model, we use the Palmer Penguins dataset provided in the ``palmerpenguins'' package \citep{horst2020palmerpenguins}.  To clarify, for such a clean dataset, existing approaches such as the Gaussian mixture model can also produce a similarly good accuracy; our goal here is to illustrate the high extensibility of the forest model via distribution specification.

 We remove the duplicated data entries and have a sample of size $n=334$. The dataset has observations about three species of Antarctic penguins, containing the length and depth measurements of each penguin's bill (in mm). These two variables contain strong signals for distinguishing between species, and we denote each record of (length, depth) by $y_i$. In addition, the dataset also has measurements of flipper length (in mm) and body mass (in grams), and we denote each record by $x_i$.

As shown in Figure S.3, the forest model without using covariates correctly estimates most species labels. On the other hand, including the external information from $x_i$ further increases the accuracy.

\begin{figure}[H]
     \begin{subfigure}[t]{.22\textwidth}
        \includegraphics[height=5cm,width=1\linewidth]{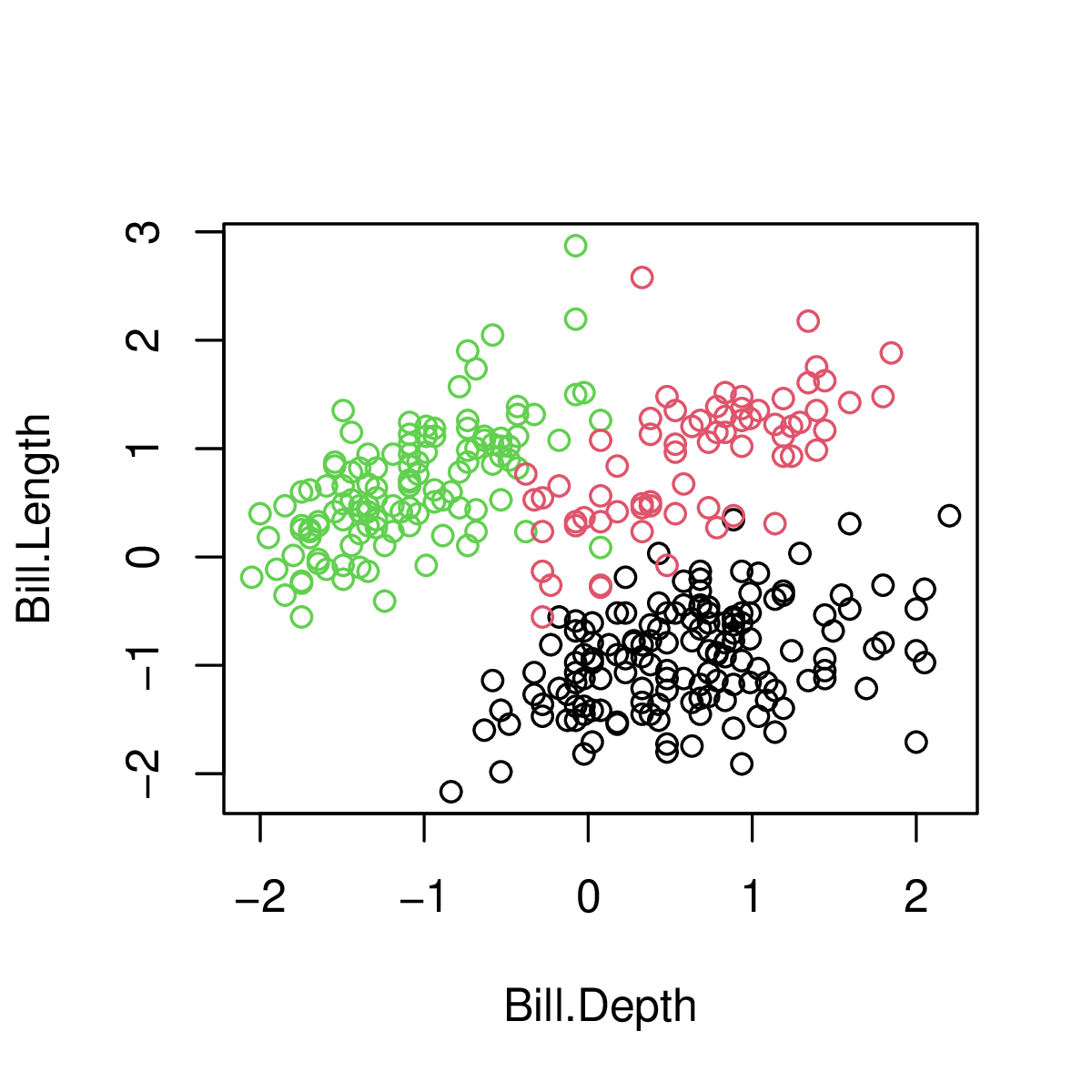}
            \caption{Penguin data, colored by species.}\;
    \end{subfigure} 
%             \begin{subfigure}[t]{.22\textwidth}
%        \includegraphics[height=4cm,width=1\linewidth]{penguin_mclust}
%            \caption{Gaussian mixture estimate.}
%    \end{subfigure} \;
         \begin{subfigure}[t]{.22\textwidth}
        \includegraphics[height=5cm,width=1\linewidth]{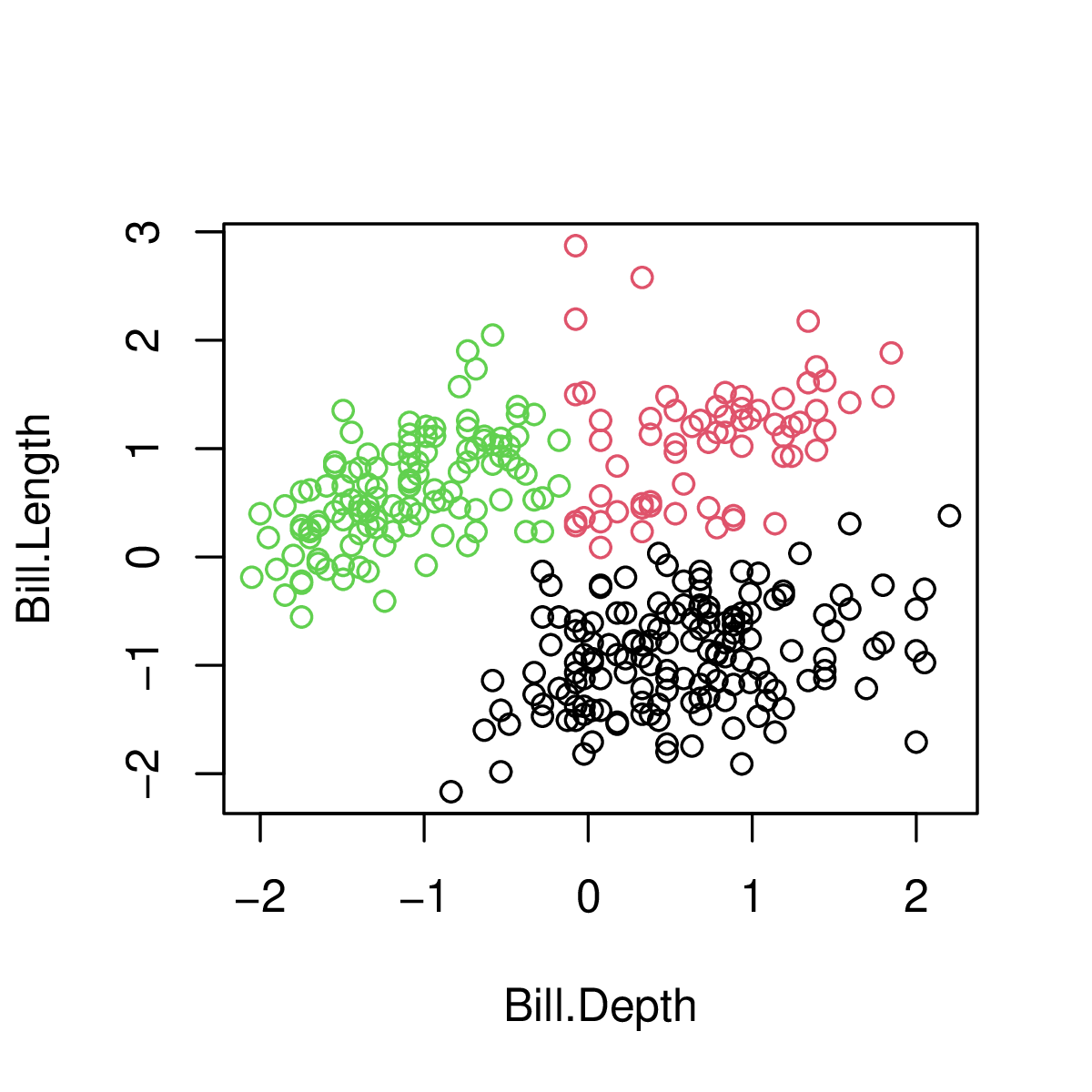}
            \caption{Forest model estimate without using covariates.}
    \end{subfigure} 
             \begin{subfigure}[t]{.22\textwidth}
        \includegraphics[height=5cm,width=1\linewidth]{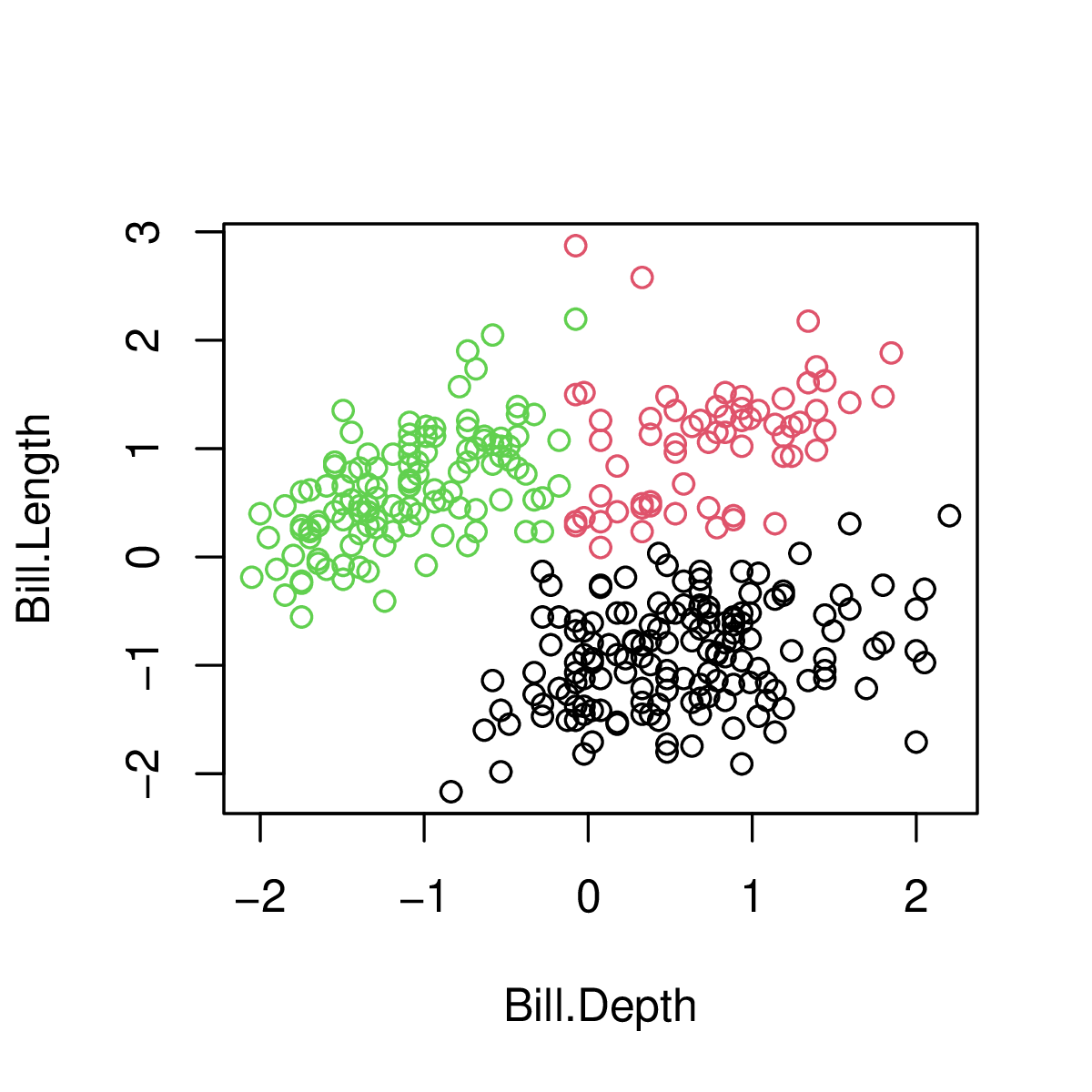}
            \caption{Forest model estimate using $x_i$ as external covariates with $\eta=2$.}
    \end{subfigure} 
                 \begin{subfigure}[t]{.22\textwidth}
        \includegraphics[height=5cm,width=1\linewidth]{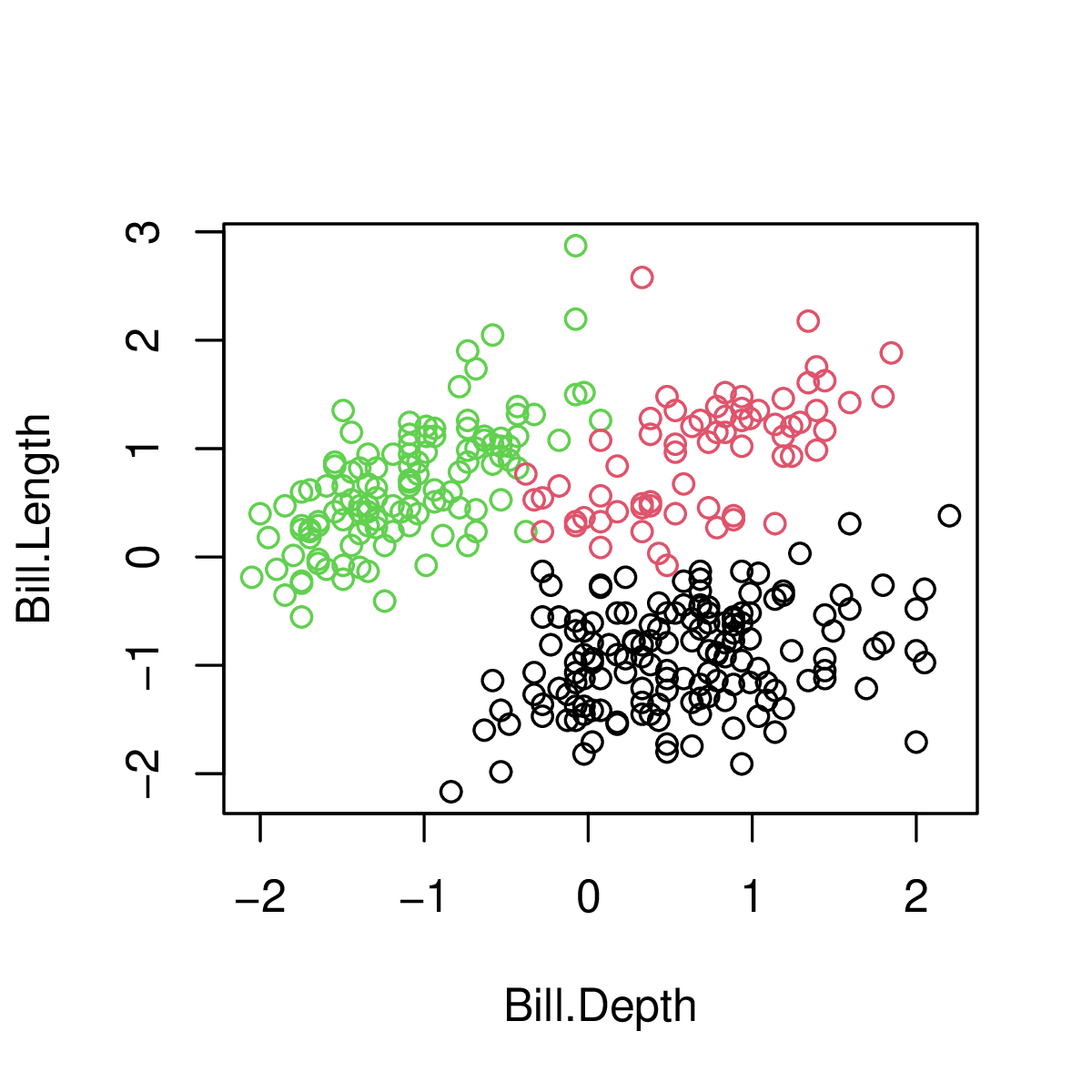}
            \caption{Forest model estimate using $x_i$ as external covariates with $\eta=1$.}
    \end{subfigure} 
        \caption*{Figure S.3: Clustering the penguin data that contain records of bill length and depth. The forest model alone (Panel b) leads to a good estimate (accuracy $94.6\%$). Nevertheless, using external covariates (flipper length and body mass)  gives more accurate estimates  (Panel c: accuracy $95.8\%$,
        Panel d: accuracy $97.3\%$).
        }
        \end{figure}

\subsection{Algorithm for Estimating Multi-view Clustering}

We use $k^{(s)}_i\in \{1,\ldots,\tilde K\}$ to denote the latent assignment $k^{(s)}_i=l$, for $\eta^{(s)}_i=\eta^*_l$. We use the following Gibbs sampling algorithm:
\begin{itemize}
	\item Using $L_{\mathcal T^{(s)}_{1:n,1:n}}$ to denote the Laplacian matrix of the forest graph without auxiliary node 0, we have
$$ z^{(s)} \mid \eta^{(s)}, \mathcal T^{(s)} \sim \text{N}\bigg\{ (L_{\mathcal T^{(s)}_{1:n,1:n}}/\rho + I/\sigma^2_z)^{-1} (\eta^{(s)}/\sigma^2_z), (L_{\mathcal T^{(s)}_{1:n,1:n}}/\rho + I/\sigma^2_z)^{-1} \bigg\}.$$
	\item Sample
\be
\text{Pr}(k^{(s)}_i=l \mid \cdot) \propto 
v_{i,l}
\exp
\big(
 -\frac{1}{2}
 \|z^{(s)}_j-\eta^*_l\|_2^2/\sigma^2_z 
 \big ).
\ee
\item Sample $v_i \sim \text{Dir}( \{ 1/\tilde \kappa + \sum_s 1(k_i^{(s)}=l) \}_{ l=1,\ldots,\tilde \kappa}).$
\item Sample $\mathcal T^{(s)}$ and $\theta^{(s)}$ for all $s$, according to the algorithm in  Section 3.1 of the main text.
\end{itemize}

%We use $k^{(s)}_i\in \{1,\ldots,\tilde K\}$ to denote the latent assignment $k^{(s)}_i=l$, for $z^{(s)}_i=\tilde z^{(l)}_i$. We use the following Gibbs sampling algorithm:
%\begin{itemize}
%	\item Sample
%\be
%(\tilde z^{(l)}_i \mid \cdot) \sim \text{N}
%\bigg\{ 
%\frac
%{\rho^{-1}\sum_{s:k_i^{(s)}=l} \sum_{j:j\neq i} A_{\mathcal{T}^{(s)},i,j}z^{(s)}_j }
%{(1+ \rho^{-1}\sum_{s:k_i^{(s)}=l} \sum_{j:j\neq i} A_{\mathcal{T}^{(s)},i,j} )}
%, 
%\frac{I_d}
%{(1+ \rho^{-1}\sum_{s:k_i^{(s)}=l} \sum_{j:j\neq i} A_{\mathcal{T}^{(s)},i,j} )}
%\bigg\},
%\ee
%and update the values of $z^{(s)}_i$ for those $k_i^{(s)}=l$; repeat for all $i$ and $s$.
%	\item Sample
%\be
%\text{Pr}(k^{(s)}_i=l \mid \cdot) \propto 
%v_{k^{(s)}_1\ldots, k^{(s)}_{i-1}, l,k^{(s)}_{i+1} \ldots k^{(s)}_n}
%\exp
%\bigg\{
% -\frac{1}{2\rho}
% {\sum_{s} \sum_{j:j\neq i} A_{\mathcal{T}^{(s)},i,j} [\tilde z_i^{(l)} -z^{(s)}_j]^2 }
% \bigg\},
%\ee
%and update the value of $z^{(s)}_i$ according to $\tilde z_i^{k^{(s)}_i}$; repeat for all $i$ and $s$.
%\item Sample $v \sim \text{Dir}( \{ \alpha_{i,l} + \sum_s 1(k_i^{(s)}=l) \}_{i=1,\ldots,n, l=1,\ldots,\tilde K}).$
%\item Sample $\mathcal T^{(s)}$ and $\theta^{(s)}$ for all $s$, according to the algorithm in  Section 3.1 of the main text.
%\end{itemize}

\section{Proof of Theorems}
\subsection{Proof of Theorem 1}

\begin{proof}

For ease of notation, in this proof, we use $p=(n+1)$.

{\bf 1. Obtain closed-form of the marginal connecting probability.}

Since $L$ correspond to a connected graph with weight $A_{i,j}=\exp(W_{i,j})$ for $i\neq j$ and $A_{i,i}=0$, hence only has one eigenvalue equal to $0$ and with eigenvector $\vec 1/\sqrt{p}$, therefore we have:
\[
p^{-1}\prod_{i=2}^{n+1}\lambda_{(i)} (L)
= |L +  J/p^{2}|,
\]
where $J=\vec 1 \vec 1^{\rm T}$.
Let $\tilde L= L+ J/p^2$, differentiating $\log |\tilde L|$ with respect to $W_{i,j}$ yields:
\[
M_{i,j}= (\Omega_{i,i}+\Omega_{j,j}-2\Omega_{i,j})A_{i,j},
\]
where $\Omega =  \tilde L^{-1}$, and $M_{i,i}=0$.

{\bf 2. Obtain $M$ as a perturbation form}

Let $L= \sum_{l=1}^p \lambda_{l}\psi_l\psi^{\rm T}_l$ be the eigendecomposition of $L$, and  $N=D^{-1/2}LD^{-1/2}$. Note that,

\be
M_{i,j} = &
(\Omega_{i,i}+\Omega_{j,j}-2\Omega_{i,j})A_{i,j}  \\
= &
(\Omega_{i,i}+\Omega_{j,j}-2\Omega_{i,j})\{-L_{i,j}1(j\neq i)\}\\
\stackrel{(a)}=    & 
\vec b^{\rm T}_{i,j} (L+J/p^2)^{-1} \vec b_{i,j} (-L_{i,j})\\
=    & D_i^{1/2}\vec b^{\rm T}_{i,j} (L+J/p^2)^{-1} \vec b_{i,j}D_j^{1/2} (-N_{i,j})
%\vec b^{\rm T}_{i,j} \{\sum_{l=2}^p \lambda^{-1}_{l}\psi_l\psi^{\rm T}_l +
%p (J/p^2) \}\vec  b_{i,j} (-L_{i,j})\\
%=   & - 
% \{\sum_{l=2}^p \lambda^{-1}_{l}(\psi_{l:i}-\psi_{l:j})^2
% \}   L_{i,j} \\
% =   & -
% \{\sum_{l=2}^p \lambda^{-1}_{l}(\psi_{l:i}-\psi_{l:j})^2
% D_i^{1/2}D_j^{1/2}\}  N_{i,j}\\
\ee
 where (a) is due to $\Omega_{i,i}+\Omega_{j,j}-2\Omega_{i,j}=0$ if $1(j \neq i)=0$, hence $1(k\neq i)$ can be omitted;  $\vec b_{i,j}$  is a binary vector with the $i$th element $1$ and the $k$th
element $-1$, and all other elements $0$.

Let $\alpha_{i,j} := D_i^{1/2}\vec b^{\rm T}_{i,j} (L+J/p^2)^{-1} \vec b_{i,j}D_j^{1/2} $ for $i\neq j$. Since $N_{i,i}=1$, and $x(I-N)$ has the same eigenvectors as $N$ for any scalar $x>0$, we see that $M=-\alpha\circ (I-N)$ is an element-wise perturbation of $x(I-N)$. Therefore, our next task is to show $\alpha$ is close to a simple $xJ$ for some $x>0$.

{\bf 3.  Bound the difference between the $K$ leading eigenvectors.}

Slightly changing the above,
\be
\alpha_{i,j} & = D_i^{1/2}\vec b^{\rm T}_{i,j} (L+J/p^2)^{-1} \vec b_{i,j}D_j^{1/2} \\
& = D_i^{1/2}\vec b^{\rm T}_{i,j}  D^{-1/2}(N+D^{-1/2}JD^{-1/2}/p^2)^{-1} D^{-1/2}\vec b_{i,j}D_j^{1/2}.
\ee

Using $N=I-D^{-1/2}AD^{-1/2}$, we have
\be
(N+D^{-1/2}JD^{-1/2}/p^2)^{-1} &=
\{ I-D^{-1/2} (A-J/p^2) D^{-1/2}\}^{-1}  \\
& \stackrel{(a)}= I+ \sum_{k=1}^{\infty}\{D^{-1/2} (A-J/p^2) D^{-1/2}\}^{k}\\
& = I+ E .
\ee
where (a) uses the Neumann expansion, as  $E$ is not divergent:
\be
E & =D^{1/2} (L+J/p^2)^{-1}D^{1/2} -I  = D^{1/2} (\sum_{l=2}^p \lambda^{-1}_{l}\psi_l\psi^{\rm T}_l + J)D^{1/2} -I,
\ee
as $\lambda_2>0$, $E$ is bounded element-wise for any $D$ with finite-value elements. Further, note that $D_i^{-1/2} (A_{i,j}-1/p^2) D_j^{-1/2}\to 0$ and monotonically decreasing for fixed $A_{i,j}$ and increasing $D_i$ or $D_j$; hence $E$ is always bounded elementwise even as all $D_i\to\infty$. We denote the bound constant by $\max_{i,j}|E_{i,j}|\le \epsilon$.

%Further, we see that $\|D^{-1/2} (A-J/p^2) D^{-1/2}\|_{op}<1$ strictly.

Combining the above,
\be
\alpha_{i,j} & =  D_i^{1/2}D_j^{1/2} \vec b^{\rm T}_{i,j} \left\{
 D^{-1}+D^{-1/2}ED^{-1/2}
 \right\}\vec b_{i,j}
 \\
&= D_i^{1/2}D_j^{1/2} \left\{ (D^{-1}_i+D^{-1}_j) +
(D_i^{-1}E_{i,i}+D_j^{-1}E_{j,j}-2 D_{i}^{-1/2}D_{j}^{-1/2}E_{i,j}) \right\}.
\ee

Now we can bound the difference between $M$ and $x(I-N)$ minimized over $x$:
\be
&\min_x \max_{i,j} |\{(\alpha -x J)\circ (I-N)\}_{i,j} |\\
&=\min_x \max_{i,j} |( \alpha_{i,j} -x) (D_i^{-1/2}D_j^{-1/2}A_{i,j})|\\
 & =  \min_x \max_{i,j} \left | \left \{(D^{-1}_i+D^{-1}_j) - x D_i^{-1/2}D_j^{-1/2}  +
(D_i^{-1}E_{i,i}+D_j^{-1}E_{j,j}-2 D_{i}^{-1/2}D_{j}^{-1/2}E_{i,j})
\right \}A_{i,j}
 \right| \\
 & \le  \min_x \max_{i,j} \left | \left \{(D^{-1}_i+D^{-1}_j) - x D_i^{-1/2}D_j^{-1/2}  +
(D_i^{-1}\epsilon+D_j^{-1} \epsilon+2 D_{i}^{-1/2}D_{j}^{-1/2}\epsilon)
\right \}A_{i,j}
 \right| \\
 & \stackrel{(a)}\le   \max_{i,j} \left \{ (1+\epsilon)(D^{-1/2}_i-D^{-1/2}_j)^2
A_{i,j} \right\},
  \ee
  where (a) chooses  $x= 2(1+\epsilon) +2\epsilon$.

%Using the series form of $E$, $D_{i}^{-1/2}D_{j}^{-1/2}E_{i,j}$ is the $(i,j)$th element (including $i=j$) from matrix:
%\be
% D^{-1/2}ED^{-1/2}
%& = D^{-1} (A-J/p^2) D^{-1}+D^{-1} (A-J/p^2) D^{-1}(A-J/p^2)D^{-1}+\cdots\\
%& = D^{-1}
%\sum_{k=1}^{\infty}\{ (A-J/p^2) D^{-1}\}^{k}
%\ee

Using Theorem 2 from \cite{yu2015useful}, there exists a orthonormal matrix $R\in \mathbb{R}^{K\times K}$, such that,

\env{equation}{
  \|\Psi_{1:K}-\phi_{1:K} R \|_F \le \frac{
  2^{3/2} \min \{ \sqrt{K} \|(\alpha -x J)\circ (I-N)\|_{op}, \ \|(\alpha -x J)\circ (I-N)\|_{F}\}
  }{\xi_{K} -\xi_{K+1}} 
}
for any $x>0$.

Since $|\{(\alpha -x J)\circ (I-N)\}_{i,j}|$ is upper-bounded hence is sub-Gaussian with bound parameter $\sigma_e = \max_{i,j}  \{ (1+\epsilon)(D^{-1/2}_i-D^{-1/2}_j)^2
A_{i,j} \}$.

Using Theorem 1 of Duan, Michailidis and Ding 2020 (arXiv preprint:1910.02471), with probability $1-\delta_t$
\be
\|\Psi_{1:K}-\phi_{1:K} R \|_F \le \frac{
  2^{3/2} (\sqrt{K p} \sigma_e )
  }{\xi_{K} -\xi_{K+1}} t.
\ee
where $\delta_t= \exp[- (t^2/64 -\log(5\sqrt{2})) p]$.
Taking $t=14$, we have $(t^2/64 -\log(5\sqrt{2}))>1.$ Therefore, we have with probability at least $1- \exp(-p)$ \{which is greater than $1-\exp(-n)$\}.
\be
\|\Psi_{1:K}-\phi_{1:K} R \|_F \le \frac{
  40 \sqrt{Kp} \sigma_e 
  }{\xi_{K} -\xi_{K+1}}.
\ee
\end{proof}

\subsection{Proof of Theorem 2 and 3}

Let the conditional probability associated with Gaussian leaf density $f$ be $\t{Pr}\{B(y_1,M_n/2)\mid y_1\}=m_n$, % for all $i=1,\ldots,n$ with some $\delta>0$ and
where $B(y_1,M_n/2)$ stands for an open ball of radius $M_n/2$ around $y_1$. 
If the true number of clusters is $K=K_0$, then $m_n^{n-K-1}$ is the probability that the distances $\{ d_{\ell,n}^0\}_l$ in the minimum spanning tree under null are all below $M$.
Specifically, let $E_n=\{d_{\ell,n}^0\leq M_n/2: 1\leq\ell\leq n-K-1\}$ =.
Then $\t{Pr}(E_n)=m_n^{n-K-1}$.

With $x_i\sim$N$(0, \sigma^{0,n})$, we have
$m_n=\t{Pr}(\frac{\sum_{i=1}^px_i^2}{\sigma^{0,n}}<\frac{M_n^2}{2^2\sigma^{0,n}})=1-\frac{\Gamma\left(p/2,\frac{M^2}{2^3\sigma^{0,n}}\right)}{\Gamma(p/2)}$, where $\Gamma(\cdot,\cdot)$ stand for the upper incomplete gamma function using cumulative distribution function of $\chi^2$ distribution.
Since $p$ belongs to the set of natural numbers, we have $\Gamma(p/2,x)< C_1 x^{p/2}e^{-x}$ except for $p=1.5$ and any $x>0$ with some constant $C_1$ which depends on $p$ \citep{pinelis2020exact}.
However, for large $x$, we have $\Gamma(p/2,x)< C_1 x^{p/2}e^{-x}$ even for $p=1.5$.
We then have $m_n>1-\frac{C_2}{\Gamma(p/2)}(\log n)^{p/2-1} e^{-\tilde m_0\log n}=1-\frac{C_2}{\Gamma(p/2)}\frac{(\log n)^{p/2-1}}{n^{\tilde m_0}}$ where $C_2=C_1(\tilde m_0)^{p/2-1}$. Since $\tilde m_0> (p/2+2)$, we have $m_n^{n-K-1}>1-(n-K-1)\frac{C_2}{\Gamma(p/2)}\frac{(\log n)^{p/2-1}}{n^{\tilde m_0}}\rightarrow 1$ as $n\rightarrow \infty$ as $\frac{\log n}{n}$ goes to $0$.
Hence $\pr (E_n)\rightarrow 1$.

We further have, $\sum_{n\geq K} [1-\{1-\frac{C_2}{\Gamma(p/2)}\frac{(\log n)^{p/2-1}}{n^{\tilde m_0}}\}^{n-K-1}]<\sum_{n} n\frac{C_2}{\Gamma(p/2)}\frac{(\log n)^{p/2-1}}{n^{\tilde m_0}}$.
Thus for $\tilde m_0>2+p/2$, we have $\sum_{n} [1-\{1-\frac{C_2}{\Gamma(p/2)}\frac{(\log n)^{p/2-1}}{n^{\tilde m_0}}\}^{n-K-1}]<\infty$.
Then, by the Borel-Cantelli Lemma, we also have almost sure convergence of this event.

We now show for $y\in E_n$, the ratio of the maximum posterior probability assigned to a “non-true” clustering arrangement to the posterior probability assigned to the “true” clustering arrangement converges to zero.

\noindent \textbf{Proof of Theorem 2}

Note that for any $\sigma$ and a given $R$, the posterior $\Pi(\mathcal{T}_{\textrm{MST},R},\sigma\mid y)$ is maximized at the $\mathcal{T}$, which is a combination of minimum spanning trees constructed within the regions $R_k$'s. Thus, %Thus, in the denominator under the null partition, the posterior distribution is computed at MST. 
%(\AR{Or we may assume the null density will have $\mathcal{T}$=MST of the sample.})

We have $\frac{\Pi(R_0\mid y)}{\Pi(\mathcal{T}_{\textrm{MST}, R^0},\sigma^{0,n}\mid y)}>1$ as $\Pi(R_0\mid y)=\sum_{\mathcal{T}}\Pi(\mathcal{T}, R^0,\sigma^{0,n}\mid y)$. 

\be
    &\frac{\Pi(\mathcal{T}_{\textrm{MST},R},\sigma^{0,n}\mid y)}{\Pi(\mathcal{T}_{\textrm{MST}, R^0},\sigma^{0,n}\mid y)}\\&\leq\left(\frac{\epsilon_2}{\epsilon_1}\right)^{K}\exp\left(-\sum_{\ell=1}^{n-K_0-1}d_{\ell,n}^2/(2\sigma^{0,n})+\sum_{\ell=1}^{n-K-1}(d_{\ell,n}^0)^2/(2\sigma^{0,n})\right),
\ee
%\leo{Shouldn't the above be based on $d_{.}^2$ instead of $d_{.}$?, $d_{.}$ is the Euclidean norm. If that's an error, note that the MST based on $\min\sum_{(i,j)\in T} \t{dist}_{(i,j)}$ is the same as the one based on $\min\sum_{(i,j)\in T} \t{dist}^2_{(i,j)}$, I know how to prove that.}
%\AR{Please add the proof for MST being the same for the two}
where $\sum_{\ell=1}^{n-K-1}d^2_{\ell,n}$ is the total squared norm distance on the minimum spanning tree under the partition regions $R$ excluding the edges with the root node and $\sum_{\ell=1}^{n-K_0-1} (d_{\ell,n}^0)^2$ is the same under $\mathcal{T}_{\textrm{MST}, R^0}$. The above is because based on the Prim's algorithm \citep{prim1957shortest}, the minimum spanning tree is equal to the result of sequential growing a tree starting from one node, each time by adding an edge (along with a node) with the shortest distance between one node in the existing tree and one of the remaining nodes not yet in the tree. Clearly, at each step, the edge choice is unaffected when changing distance from $d$ to $d^2$; therefore, the minimum spanning trees based on the sum of $d^2_{l,n}$ and the sum of $d_{l,n}$ are the same.

%If we have $\sum_{\ell=1}^{n-K-1}d_{\ell,n}>\sum_{\ell=1}^{n-K-1}d_{\ell,n}^0$, then it implies $\frac{\Pi(\mathcal{T}_{R},\sigma^{0,n}\mid y)}{\Pi(\mathcal{T}_{\textrm{MST}, R^0},\sigma^{0,n}\mid y)}\leq C/n$ for some constant $C$ due to assumption (3). 
Since, $\inf_{x\in R_i^0,y\in R_j^0}\|x-y\|_2>M_n$, for all $i\neq j$,  for at least one $\ell$, we must have $d_{\ell,n}>M$. Due to the above result, with probability at least $m_n^n$, we have $\sum_{\ell=1}^{n-K-1}(d_{\ell,n})^2>\sum_{\ell=1}^{n-K-1}(d_{\ell,n}^0)^2 + M_n^2/4$, which implies $\frac{\Pi(\mathcal{T}_{\textrm{MST},R},\sigma^{0,n}\mid y)}{\Pi(\mathcal{T}_{\textrm{MST}, R^0},\sigma^{0,n}\mid y)}< n^{-\tilde m_0}$. And we further have that $m_n^n\rightarrow 1$ as $n\rightarrow 1$.

\noindent \textbf{Proof of Theorem 3}

First, we consider that the alternative partitioning has a lower number of clusters than the null. Let that be $K$, which is less than $K_0$.
Then we have
\be
    &\frac{\Pi(\mathcal{T}_{\textrm{MST},R},\sigma^{0,n}\mid y)}{\Pi(\mathcal{T}_{\textrm{MST}, R^0},\sigma^{0,n}\mid y)}\\&\leq\lambda^{K-K_0}\frac{K_0!}{K!}\left(\frac{\epsilon_2}{\epsilon_1}\right)^{K}\frac{(\sigma^{0,n})^{(K_0-K)/2}}{\epsilon_1^{K_0-K}}\exp\left(-\sum_{\ell=1}^{n-K-1}d_{\ell,n}^2/(2\sigma^{0,n})+\sum_{\ell=1}^{n-K_0-1}(d_{\ell,n}^0)^2/(2\sigma^{0,n})\right),
\ee

We again must have  $\sum_{\ell=1}^{n-K-1}(d_{\ell,n})^2>\sum_{\ell=1}^{n-K_0-1}(d_{\ell,n}^0)^2+M_n^2/4$ with probability $m_n^n\rightarrow 1$ as the alternative partitioning will have edges with length greater than $M_n$.% and then apply Assumption ().

Next, we show the above when the alternative partitioning has a larger number of clusters than the null. Specifically, for $K>K_0$, we replace A3 and vary the conditions on $r(y)$ with $n$.

Then we have
\be
    &\frac{\Pi(\mathcal{T}_{\textrm{MST},R},\sigma^{0,n}\mid y)}{\Pi(\mathcal{T}_{\textrm{MST}, R^0},\sigma^{0,n}\mid y)}\\&\leq\lambda^{K-K_0}\frac{K_0!}{K_2!}\left(\frac{c_2}{c_1}\right)^{K}c_2^{K-K_0}(\sigma^{0,n})^{(K_0-K)/2}\\&\quad\times\exp\left(-(K-K_0)M_n^2/(2\sigma^{0,n})-\sum_{\ell=1}^{n-K-1}d_{\ell,n}^2/(2\sigma^{0,n})+\sum_{\ell=1}^{n-K_0-1}(d_{\ell,n}^0)^2/(2\sigma^{0,n})\right),
\ee
Again, for any $K>K_0$, the above ratio goes to zero as $n\rightarrow \infty$ we have $1/(\sigma^{0,n}n)\rightarrow 0$ with probability at least $m_n^n\rightarrow 1$.

\section{Posterior consistency of the clustering}
Here, we study the clustering consistency of our Bayesian methods when the number of clusters is known. 
\begin{theorem*}
Under some assumptions outlined below, we have ${\Pi(R\neq R_0|y)}\rightarrow 0$ almost surely, unless $R_{i}^0\subseteq R_{\xi(i)}$ for some permutation map $\xi(\cdot)$ when number of clusters in known. 
\end{theorem*}
The total number of possible clusters with $n$ data points and $K$ clusters is $n-1\choose K-1$, which is of order $n^K$.
To show clustering consistency, we require the following assumption,
\begin{itemize}
    \item (S1, Diminishing scale and minimum separation) We let $\sigma^{0,n}=C'(1/n\log^{1+\iota} n$ for some $\iota>0$ and $C'>0$ and $\inf_{x\in R_k^0,y\in R_{k'}^0}\|x-y\|_2>M_n$, for all $k\neq k'$ with some positive constant $M_n>0$ such that $M_n^2/\sigma^{0,n}=8\tilde m_0n\log(n)$  for all $(i,j)$ and is known for some constant $\tilde  m_{0}>p/2+2$. 
 \end{itemize}
In the above assumption, the main requirement is $M_n^2/\sigma^{0,n}=8\tilde m_0n\log(n)$ which is achieved by allowing the scale to decay faster than Assumption A1. Alternatively, one may increase $M_n$ instead of reducing $\sigma^{0,n}$. However, from a practical point of view, one would expect $M_n$ to be a non-increasing function of $n$.
%Under bounded domain assumption, (I think) the squared length of an MST can be upper bounded by $d^p$. 
 % The bounds $\epsilon_1$ and $\epsilon_2$ will depend on

\be
\frac{\Pi(R\mid y)}{\Pi(R_0\mid y)}=\frac{\frac{\Pi(R\mid y)}{\Pi(\mathcal{T}_{\textrm{MST},R},\sigma^{0,n}\mid y)}}{\frac{\Pi(R_0\mid y)}{\Pi(\mathcal{T}_{\textrm{MST}, R^0},\sigma^{0,n}\mid y)}}\frac{\Pi(\mathcal{T}_{\textrm{MST},R},\sigma^{0,n}\mid y)}{\Pi(\mathcal{T}_{\textrm{MST}, R^0},\sigma^{0,n}\mid y)},
\ee

%The cross-cluster distances in $R_0$ are upper-bounded by, $\sqrt{pd}$ and the total distance of an MST within a given cluster in $R_0$ is again upper-bounded $d^p$.
%Now any cluster $R_i$ in $R$ can be written as $\cup_{j=1}^K R_i\cap R_{0,j}$. We thus have that $\Pi(\mathcal{T}_{\textrm{MST},R},\sigma^{0,n}\mid y)\gtrsim \exp[-\{pdK(K-1)+K^2d^p\}/(2\sigma^{0,n})]=\exp[-d_1/(2\sigma^{0,n})]$ as $pdK(K-1)+K^2d^p$ is the maximum possible distance of an MST in any alternative cluster $R\neq R_0$.

%Let $B_n=\{y: -\t{tr}[S(\theta) A_{\mathcal T_{\textrm{MST}}}]>d_n\}=\{y: y^TP_{\mathcal{T}_{\textrm{MST}}}^TD^TDP_{\mathcal{T}_{\textrm{MST}}}^Ty\geq 2d_n(\sigma^{0,n})^2\}$

We have $\frac{\Pi(R_0\mid y)}{\Pi(\mathcal{T}_{\textrm{MST}, R^0},\sigma^{0,n}\mid y)} > 1$ and $\frac{\Pi(R\mid y)}{\Pi(\mathcal{T}_{\textrm{MST}, R},\sigma^{0,n}\mid y)} < n^{n-2}$ (the total number of possible spanning trees with $n$ points) %$\Pi(R\mid y)\leq 1$, 
hence $\frac{\Pi(R\mid y)}{\Pi(R_0\mid y)}\lesssim n^{n-2}\frac{\Pi(\mathcal{T}_{\textrm{MST},R},\sigma^{0,n}\mid y)}{\Pi(\mathcal{T}_{\textrm{MST}, R^0},\sigma^{0,n}\mid y)}$

When the number of clusters is known,
$$
\frac{1-\Pi(R^0\mid y)}{\Pi(R^0\mid y)}=\sum_{R\neq R^0} \frac{\Pi(R\mid y)}{\Pi(R^0\mid y)}\lesssim \exp((n+K-2)\log n)\frac{\Pi(\mathcal{T}_{\textrm{MST},R},\sigma^{0,n}\mid y)}{\Pi(\mathcal{T}_{\textrm{MST}, R^0},\sigma^{0,n}\mid y)}
$$

And applying the steps from our previous section, we have $\frac{1-\Pi(R^0\mid y)}{\Pi(R^0\mid y)}<\exp(-n\log n)$, goes to zero and thus completes the proof.

%When the number of clusters is unknown, the best result we can show is that $\Pi(K=K_0\mid y)$ goes to one. However, it is a very hard problem to address, in general.
%Even for a simple mixture model, it is not yet fully addressed. For a complex model like ours, it is beyond the scope of this paper.

\section{Additional Numerical Experiments}

\subsection{Uncertainty Quantification on Clustering Data from a Mixture Model}
 
 We now present some uncertainty quantification results, for clustering data that are from a mixture model. We experiment with $n=400$ data points in $\mathbb{R}^2$ generated from a two-component mixture distribution:
 \be
 y_i \sim 0.5 \mathcal K(\cdot \mid \mu_1) + 0.5 \mathcal K(\cdot \mid \mu_2),
 \ee
 for $i=1,\ldots,n$, with $\mu_1=(0,0)$ and $\mu_2=(b,b)$ two location parameters. We experiment with two settings, with $\mathcal K$ as (i) independent bivariate Gaussian distribution $\text{N}(\mu_k,I_2)$, (ii) independent bivariate $t$ distribution with $5$ degrees of freedom $t_5(\mu_k)$.
 
 When fitting models, we consider the unknown $K$ scenario, and use the distribution $\Pi(\mathcal T) \propto  \lambda^K$ for the Bayesian forest model, with $\lambda=0.5$. For comparison, we use
 the Dirichlet process Gaussian mixture model (DP-GMM) with a $\text{Gamma}(2,20)$ hyper-prior on the concentration parameter (with prior mean $0.1$). We use the ``dirichletprocess'' package in R \citep{ross2018dirichletprocess} for estimating the posterior distribution from DPMM. Notice that our two choices of $\mathcal K$ above correspond to fitting a Dirichlet process mixture with correctly specified components and one with misspecified components, respectively.
  
 To estimate the posterior, for each model, we ran the MCMC algorithm for 1,000 iterations and discarded the first $500$ iterations. We calculated the posterior co-assignment probability matrix $\text{Pr}(c_i=c_j \mid y)$, and the posterior number of clusters $\text{Pr}(K\mid y)$.
 
          When the data are from the Gaussian mixture (Figure S.5), both the DP-GMM and the forest model lead to satisfactory performances, with the mode of $\text{Pr}(K\mid y)$ equal/close to the ground truth at $K=2$. It is interesting to note that there is a proportion of the posterior sample from the forest model corresponds to $K=1$.
            This is likely due to less parametric assumption imposed on the shape of the clusters, compared to the DP-GMM. Nevertheless, 
            the posterior mode of the forest model correctly falls on $K=2$.

          % Nevertheless, as the cluster centers are close to each other ($b=2$), the DP-GMM seems to suffer from an over-estimation problem, with the mode of $\text{Pr}(K\mid y)$ shifted to 5. A similar over-estimation problem has been previously observed for DP-GMM but under fixed $\alpha=1$ setting \citep{miller2013simple}. One known solution is to change the DP mixture to a finite mixture with a discrete prior on $K$ \citep{miller2018mixture}.
         
         On the other hand, when the data are from the $t_5$ mixture (Figure S.6), we find the DP-GMM always show an over-estimation problem. Such issues are due to the misspecification in the component distribution, and \cite{cai2021finite} have shown that switching to a finite Gaussian mixture with a prior on $K$ does not solve the problem. In comparison, the clustering of the forest model shows much less sensitivity to model specification.
\newpage
\begin{figure}[H]
     \begin{subfigure}[t]{.15\textwidth}
        \includegraphics[height=3cm,width=1\linewidth, trim=.5cm .5cm 0cm .5cm,clip]{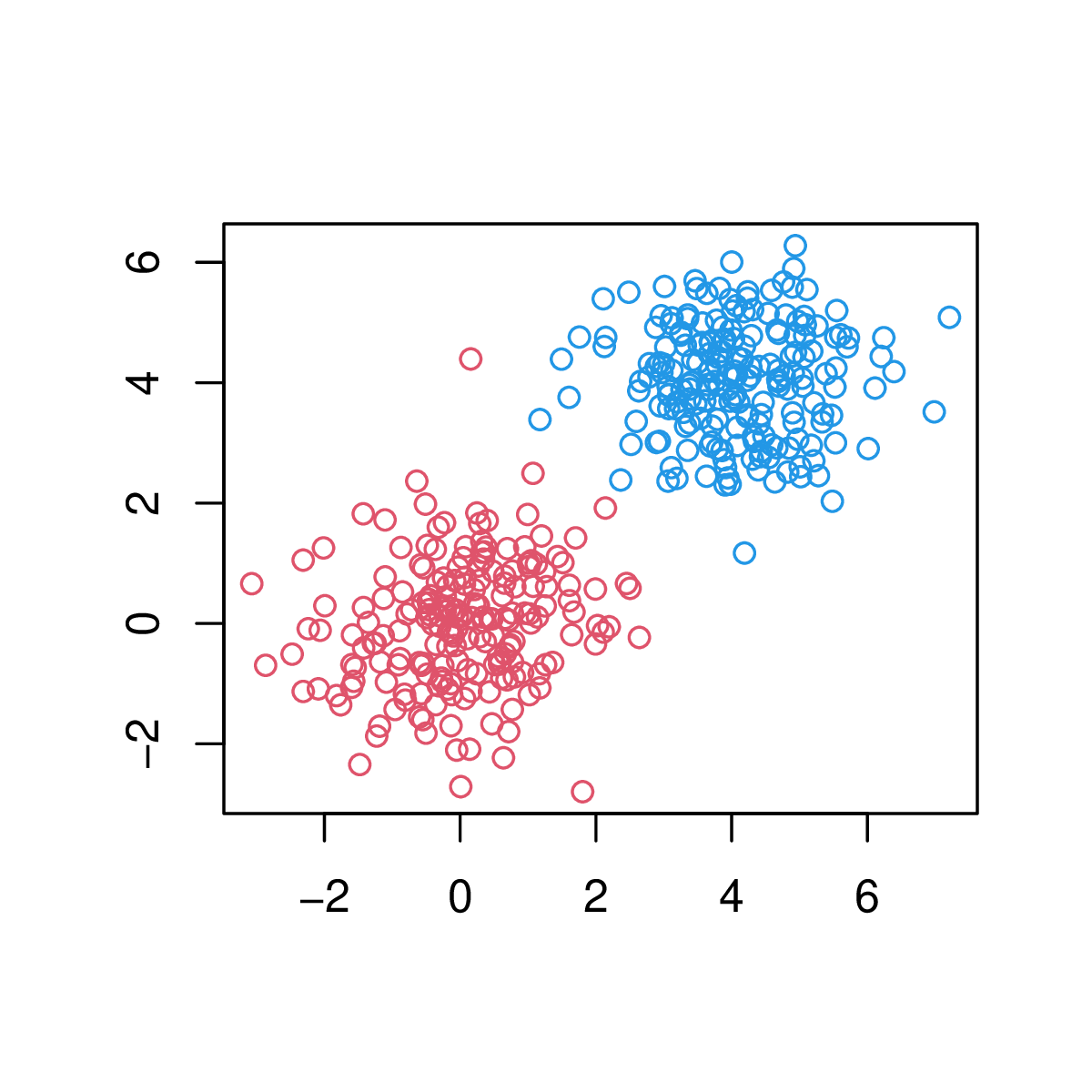}
            \caption{\footnotesize Data from two-component GMM ($b=4$).}
    \end{subfigure} \rulesep
     \begin{subfigure}[t]{.21\textwidth}
        \includegraphics[height=3cm,width=1\linewidth, trim=.5cm .5cm .5cm .5cm,clip]{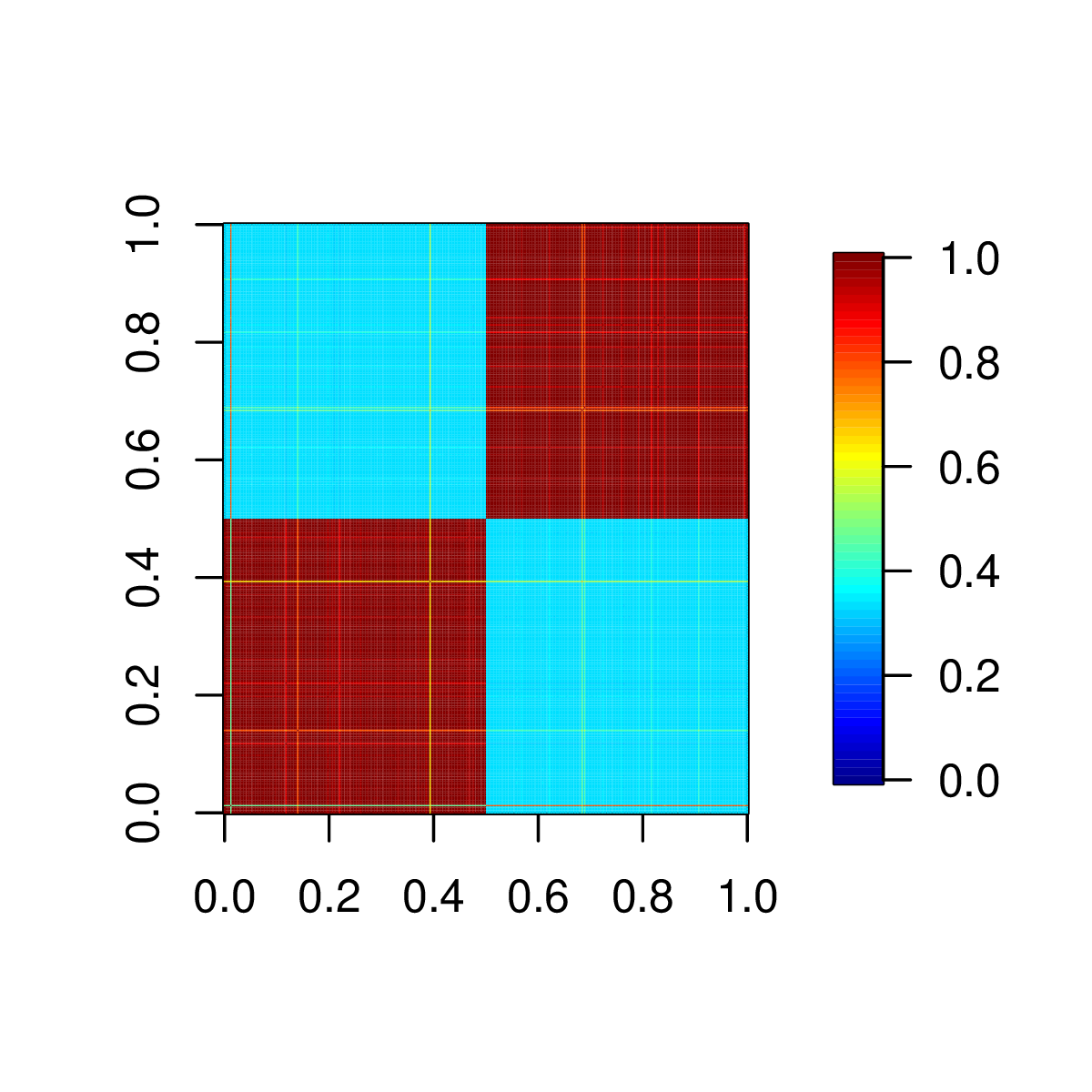} 
            \caption{\footnotesize $\text{Pr}(c_i=c_j \mid y)$ from the forest model.}
    \end{subfigure}\rulesep
    \begin{subfigure}[t]{.19\textwidth}
        \includegraphics[height=3cm,width=1\linewidth, trim=.5cm .5cm .5cm .5cm,clip]{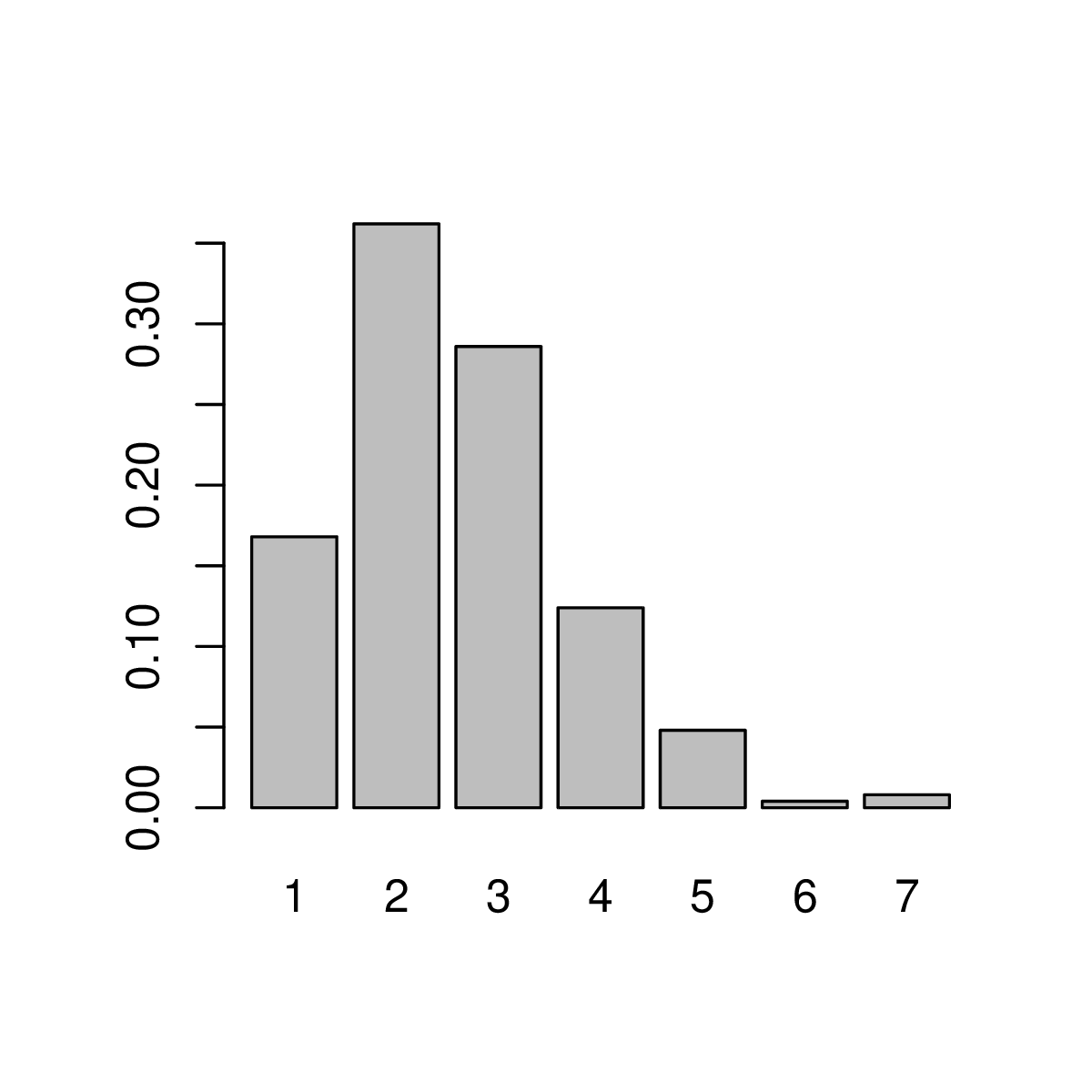}
            \caption{\footnotesize $\text{Pr}(K\mid y)$ from the forest model.}
    \end{subfigure} \rulesep    
   \begin{subfigure}[t]{.21\textwidth}
        \includegraphics[height=3cm,width=1\linewidth, trim=.5cm .5cm .5cm .5cm,clip]{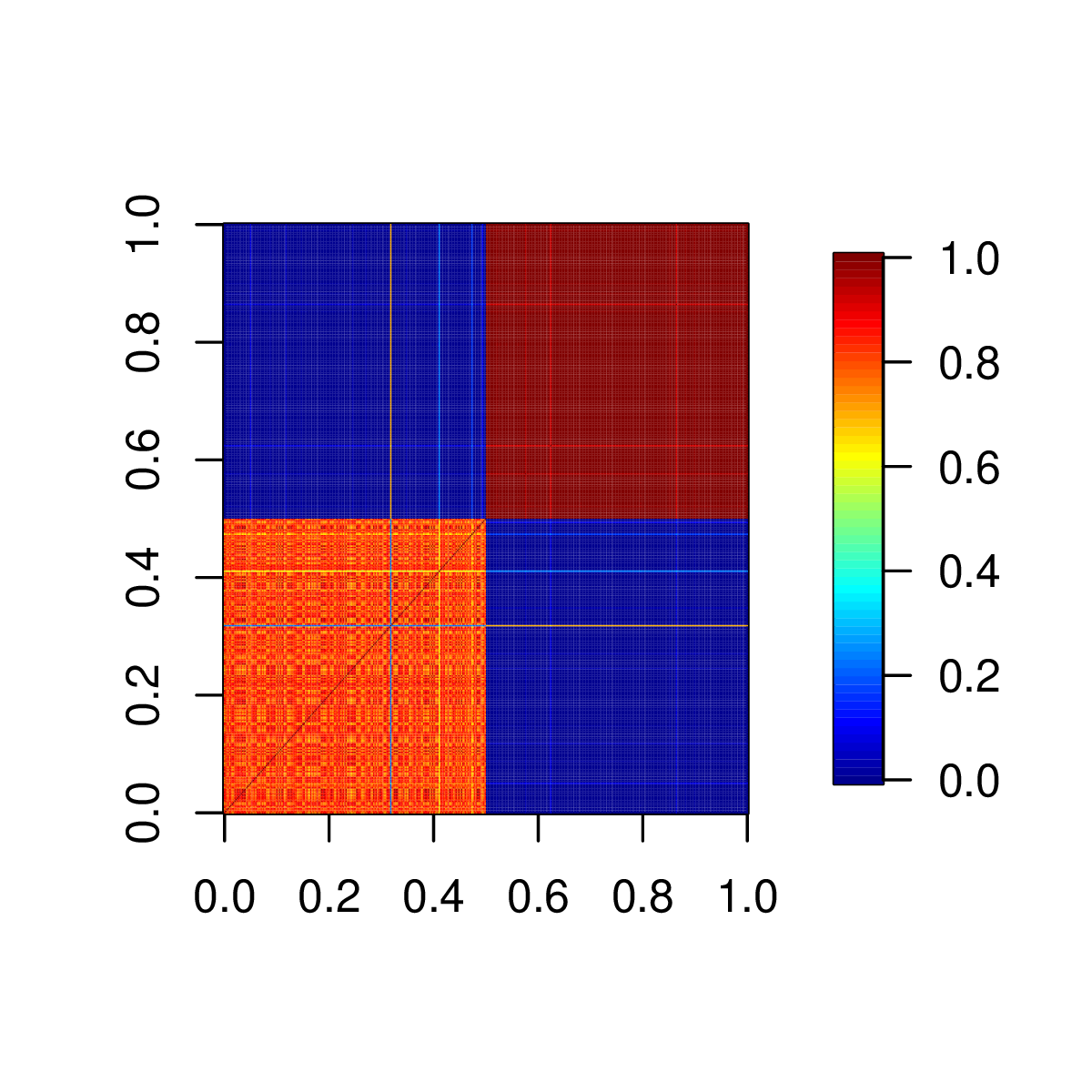}
            \caption{\footnotesize $\text{Pr}(c_i=c_j \mid y)$ from DP-GMM.}
    \end{subfigure} \rulesep
       \begin{subfigure}[t]{.19\textwidth}
        \includegraphics[height=3cm,width=1\linewidth, trim=.5cm .5cm .5cm .5cm,clip]{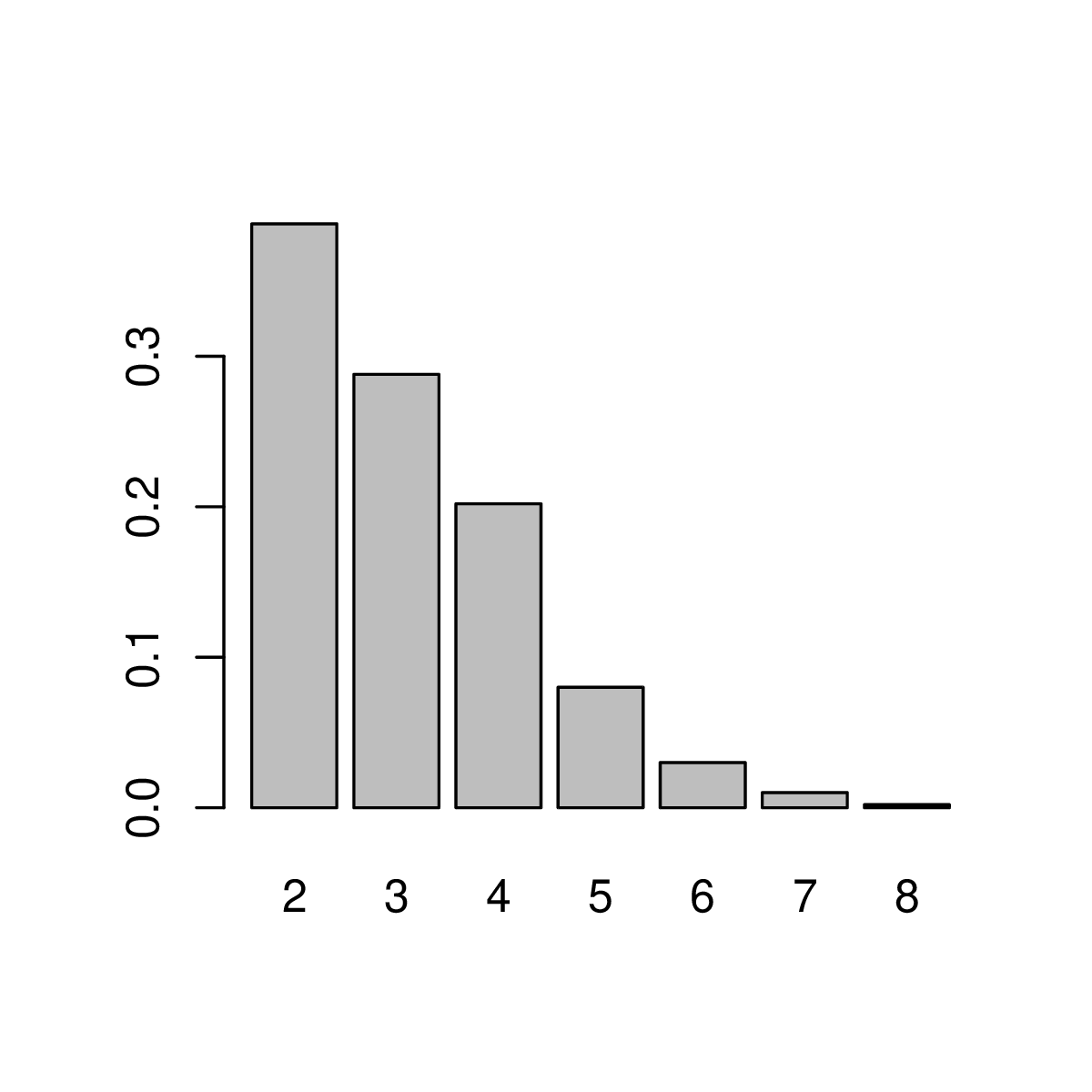}
            \caption{\footnotesize $\text{Pr}(K\mid y)$ from DP-GMM.}
    \end{subfigure} 
     \begin{subfigure}[t]{.15\textwidth}
        \includegraphics[height=3cm,width=1\linewidth, trim=.5cm .5cm 0cm .5cm,clip]{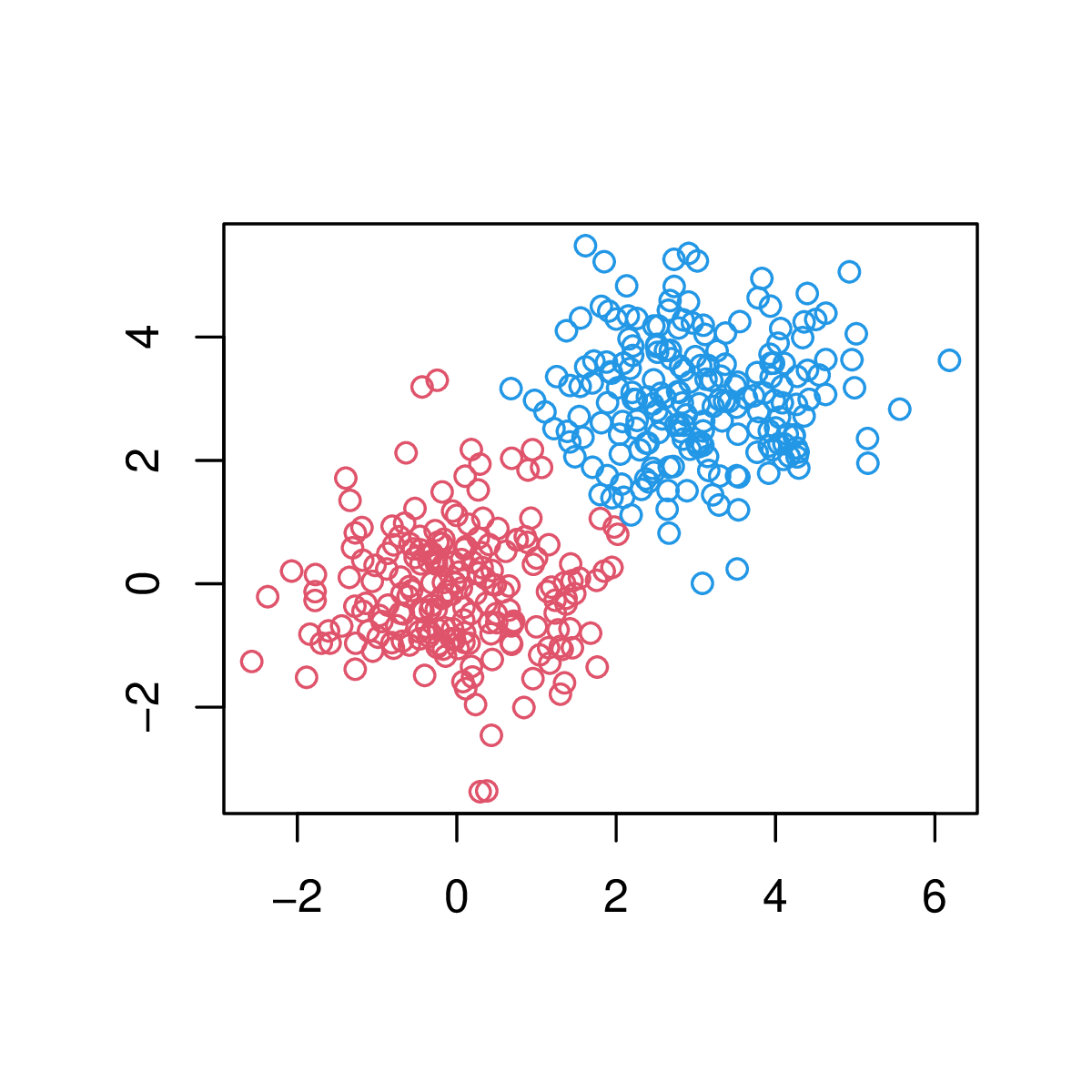}
            \caption{\footnotesize Data from two-component GMM ($b=3$).}
    \end{subfigure} \rulesep
     \begin{subfigure}[t]{.21\textwidth}
        \includegraphics[height=3cm,width=1\linewidth, trim=.5cm .5cm .5cm .5cm,clip]{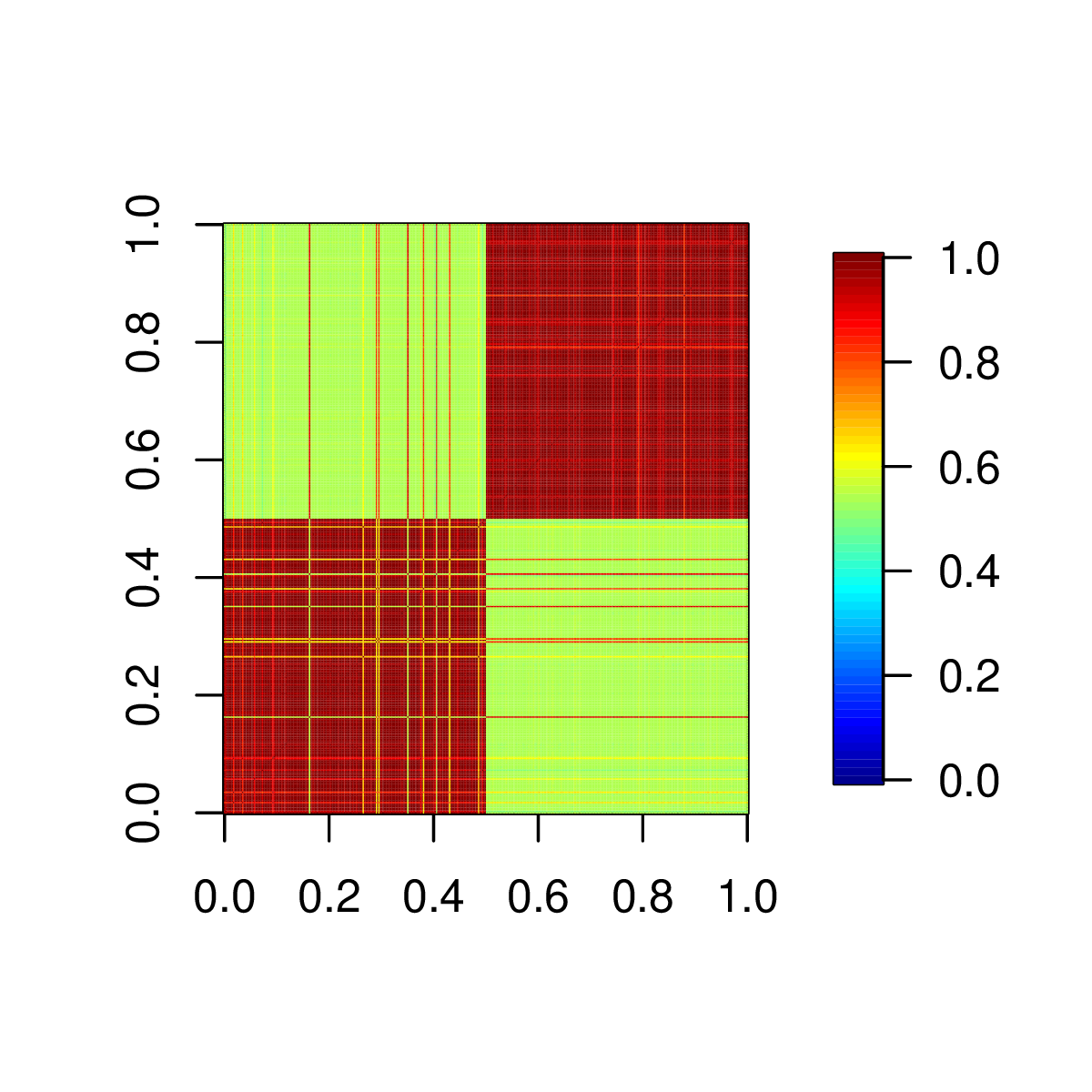}
            \caption{\footnotesize $\text{Pr}(c_i=c_j \mid y)$ from the forest model.}
    \end{subfigure}\rulesep
    \begin{subfigure}[t]{.19\textwidth}
        \includegraphics[height=3cm,width=1\linewidth, trim=.5cm .5cm .5cm .5cm,clip]{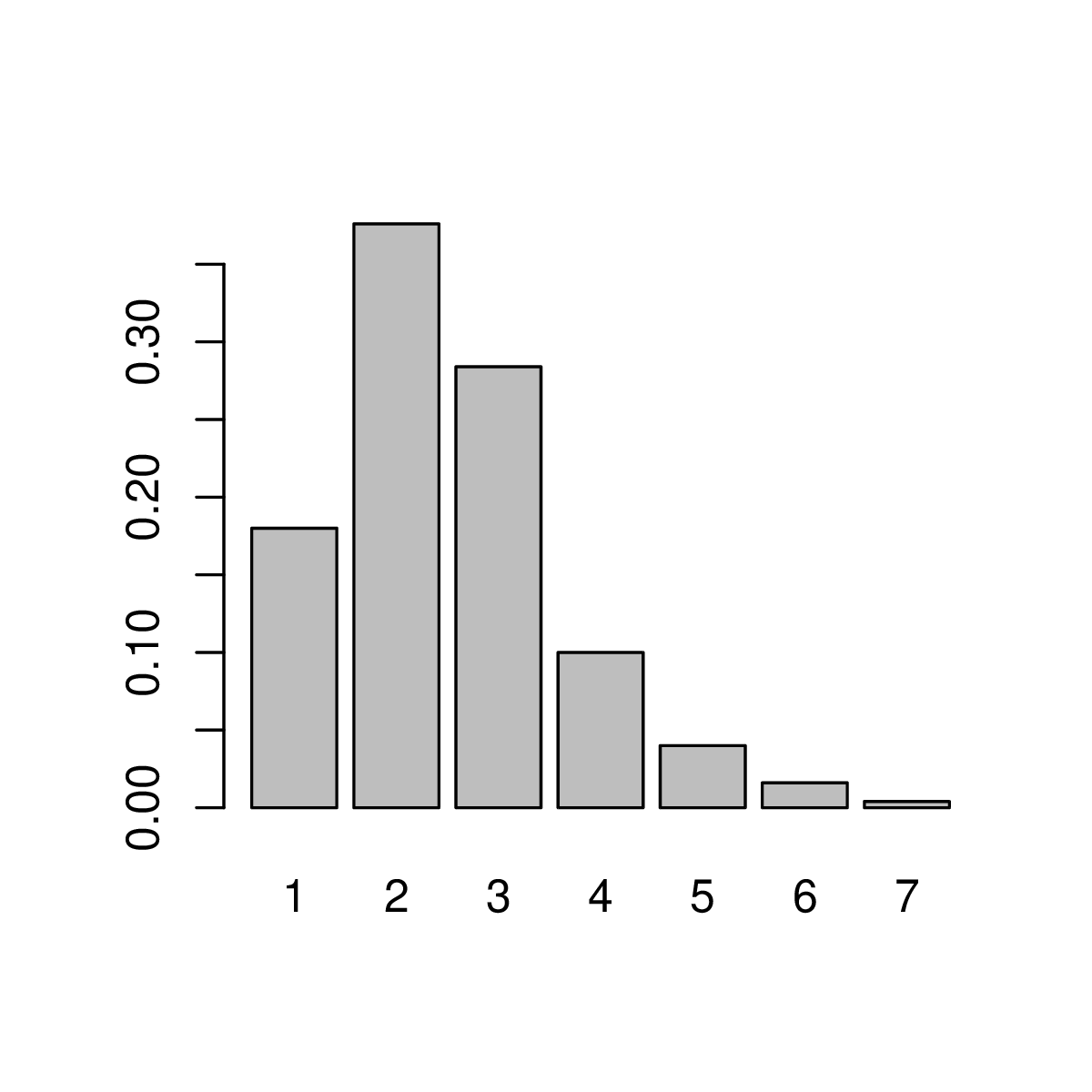}
            \caption{\footnotesize $\text{Pr}(K\mid y)$ from the forest model.}
    \end{subfigure} \rulesep
   \begin{subfigure}[t]{.21\textwidth}
        \includegraphics[height=3cm,width=1\linewidth, trim=.5cm .5cm .5cm .5cm,clip]{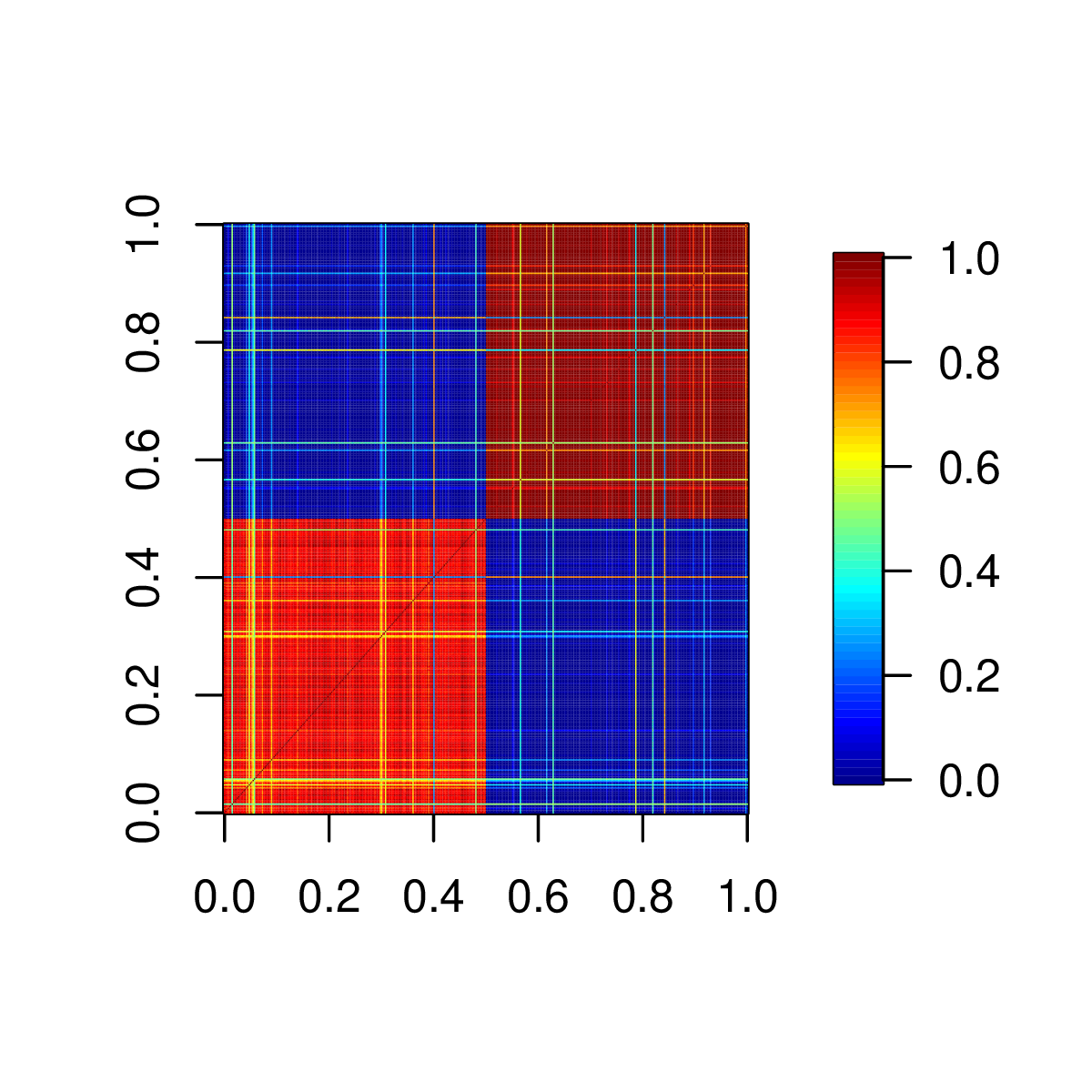}
            \caption{\footnotesize $\text{Pr}(c_i=c_j \mid y)$ from DP-GMM.}
    \end{subfigure} \rulesep
       \begin{subfigure}[t]{.19\textwidth}
        \includegraphics[height=3cm,width=1\linewidth, trim=.5cm .5cm .5cm .5cm,clip]{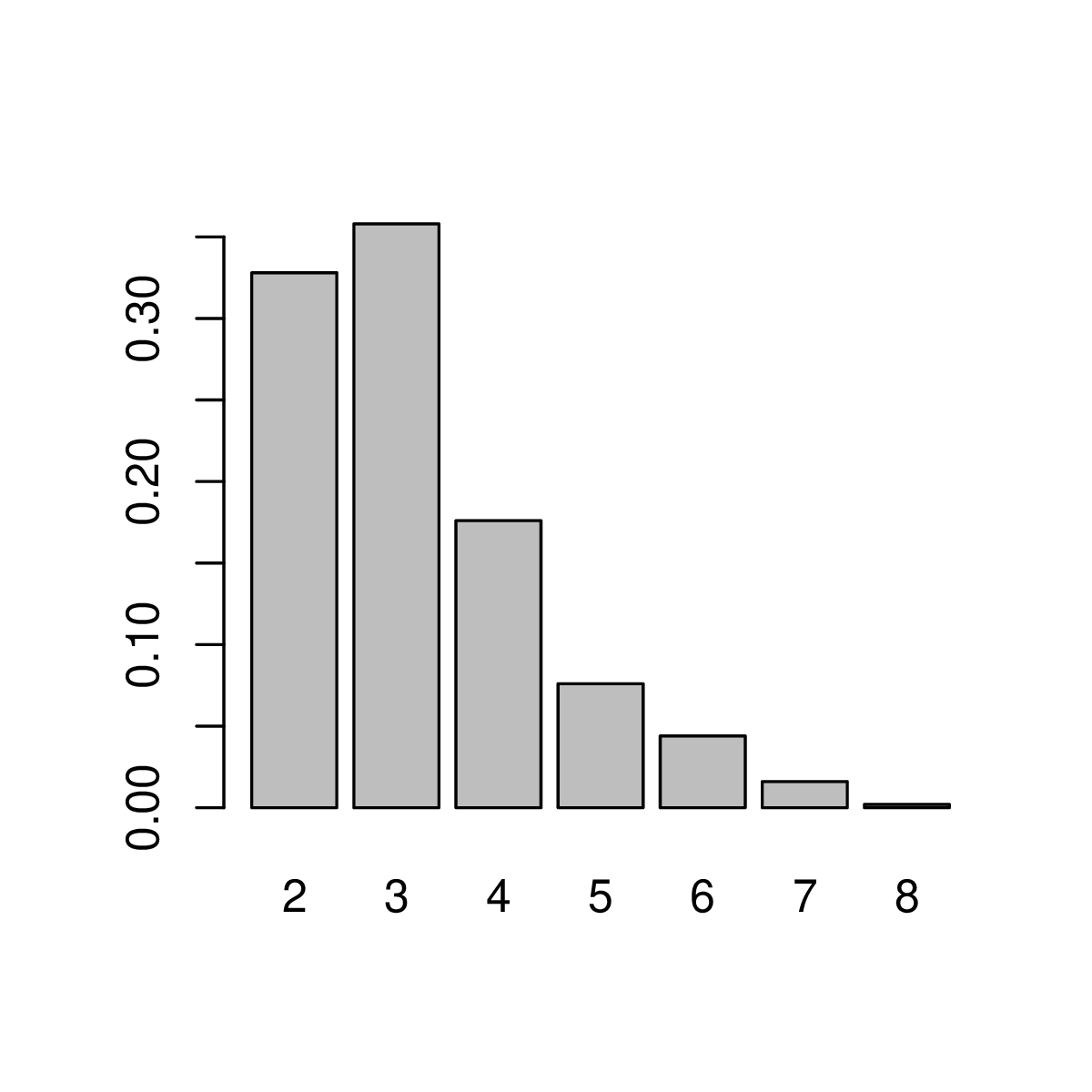}
            \caption{\footnotesize $\text{Pr}(K\mid y)$ from DP-GMM.}
    \end{subfigure} 
     \begin{subfigure}[t]{.15\textwidth}
        \includegraphics[height=3cm,width=1\linewidth, trim=.5cm .5cm 0cm .5cm,clip]{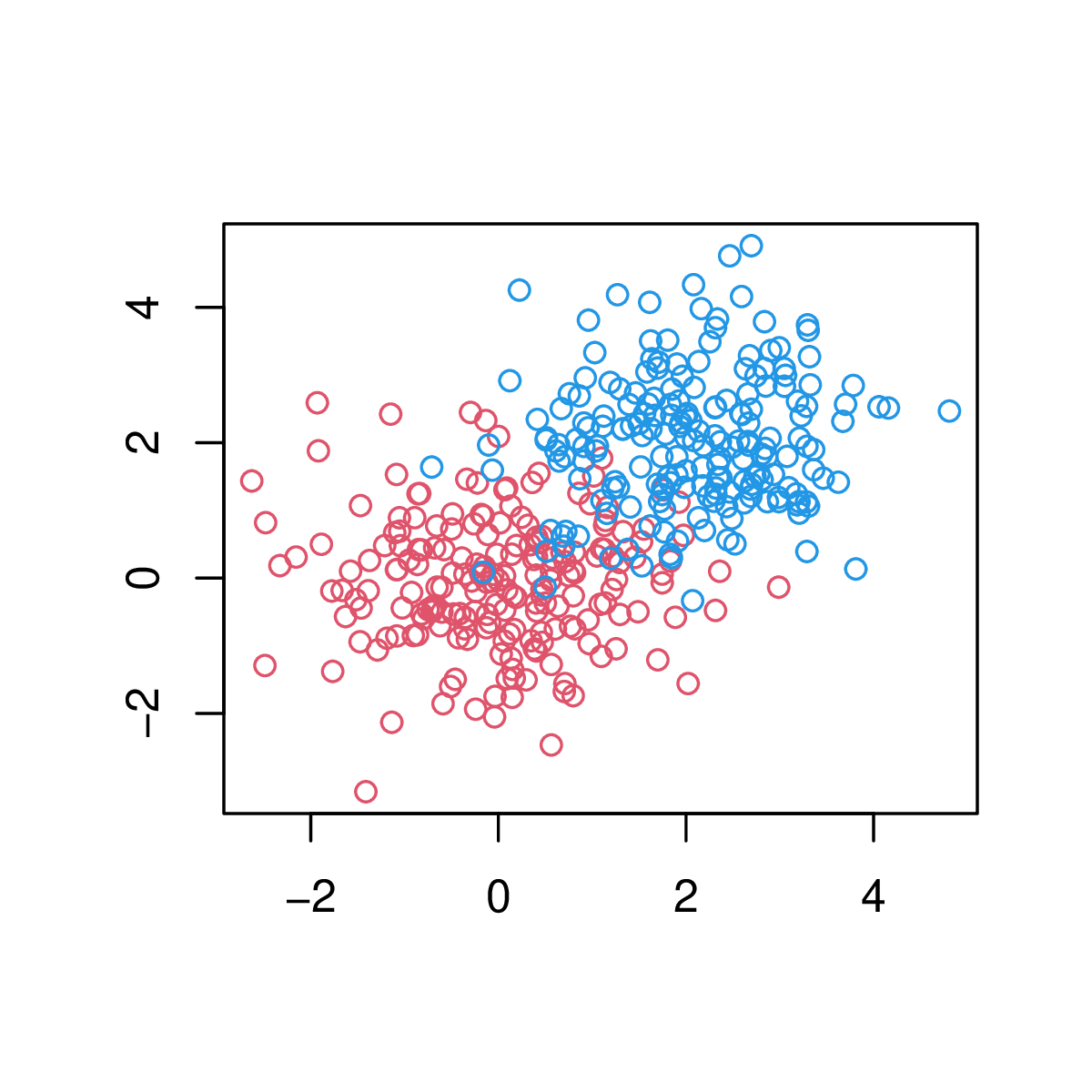}
            \caption{\footnotesize Data from two-component GMM ($b=2$).}
    \end{subfigure} \rulesep
     \begin{subfigure}[t]{.21\textwidth}
        \includegraphics[height=3cm,width=1\linewidth, trim=.5cm .5cm .5cm .5cm,clip]{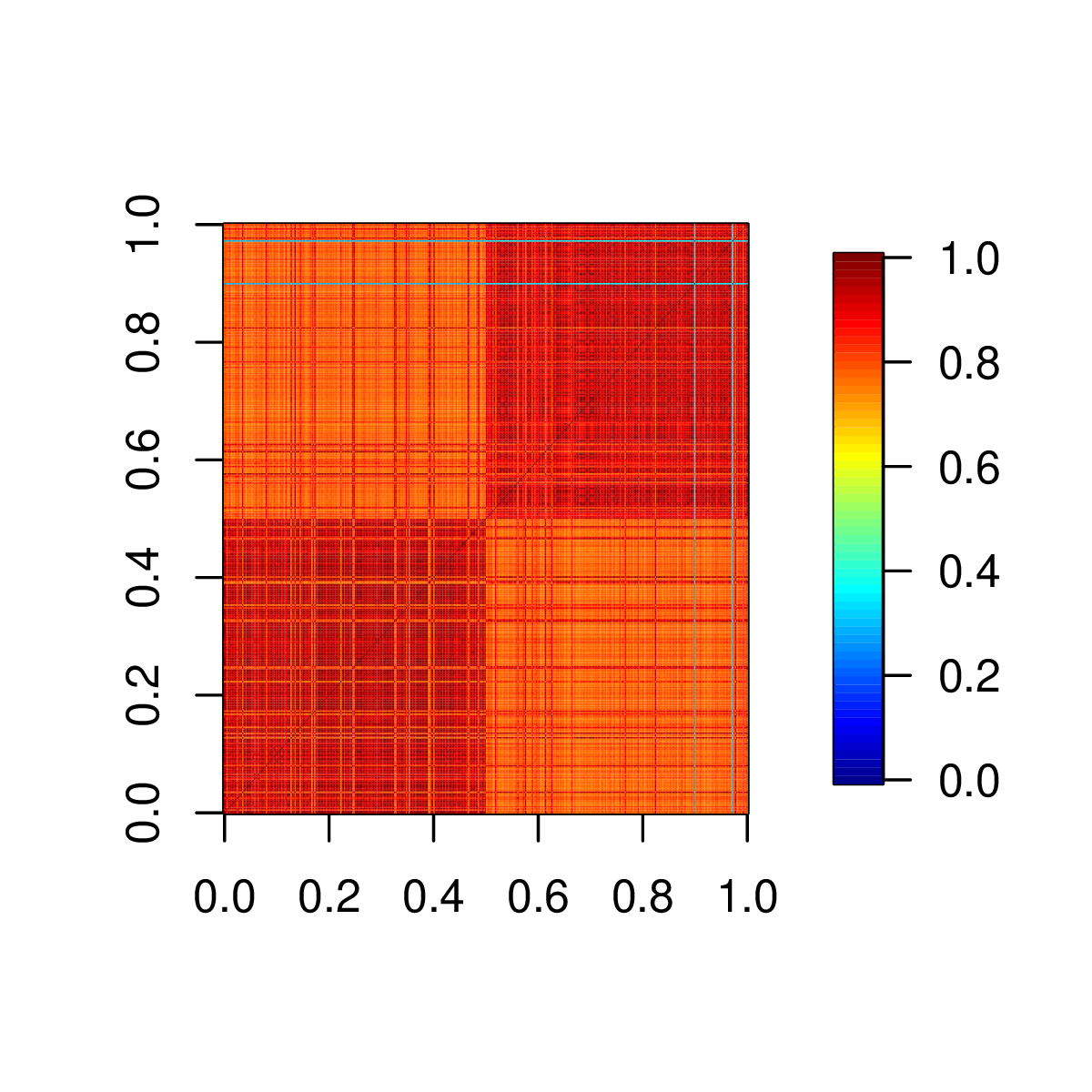}
            \caption{\footnotesize $\text{Pr}(c_i=c_j \mid y)$ from the forest model.}
    \end{subfigure}\rulesep
    \begin{subfigure}[t]{.19\textwidth}
        \includegraphics[height=3cm,width=1\linewidth, trim=.5cm .5cm .5cm .5cm,clip]{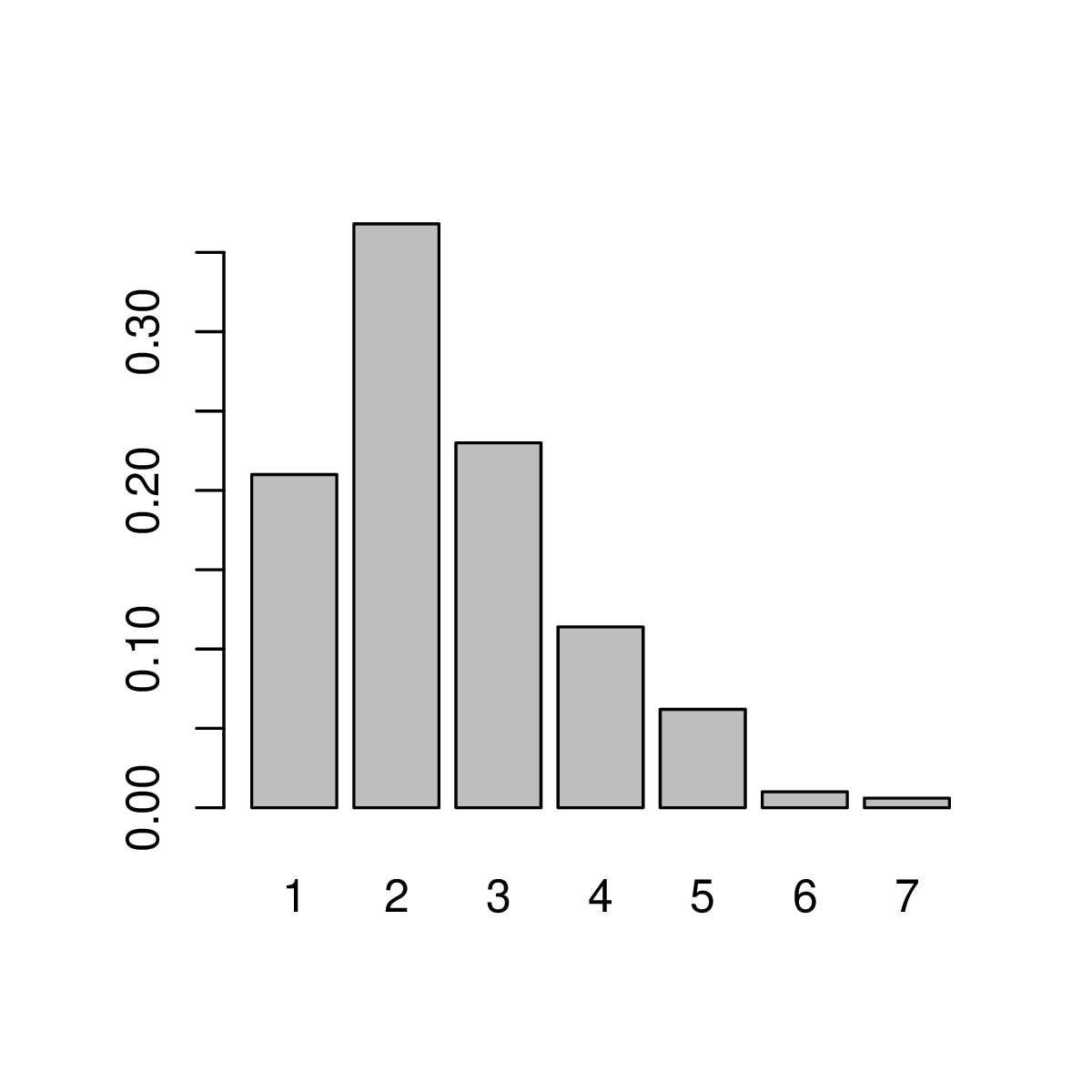}
            \caption{\footnotesize $\text{Pr}(K\mid y)$ from the forest model.}
    \end{subfigure} \rulesep
   \begin{subfigure}[t]{.21\textwidth}
        \includegraphics[height=3cm,width=1\linewidth, trim=.5cm .5cm .5cm .5cm,clip]{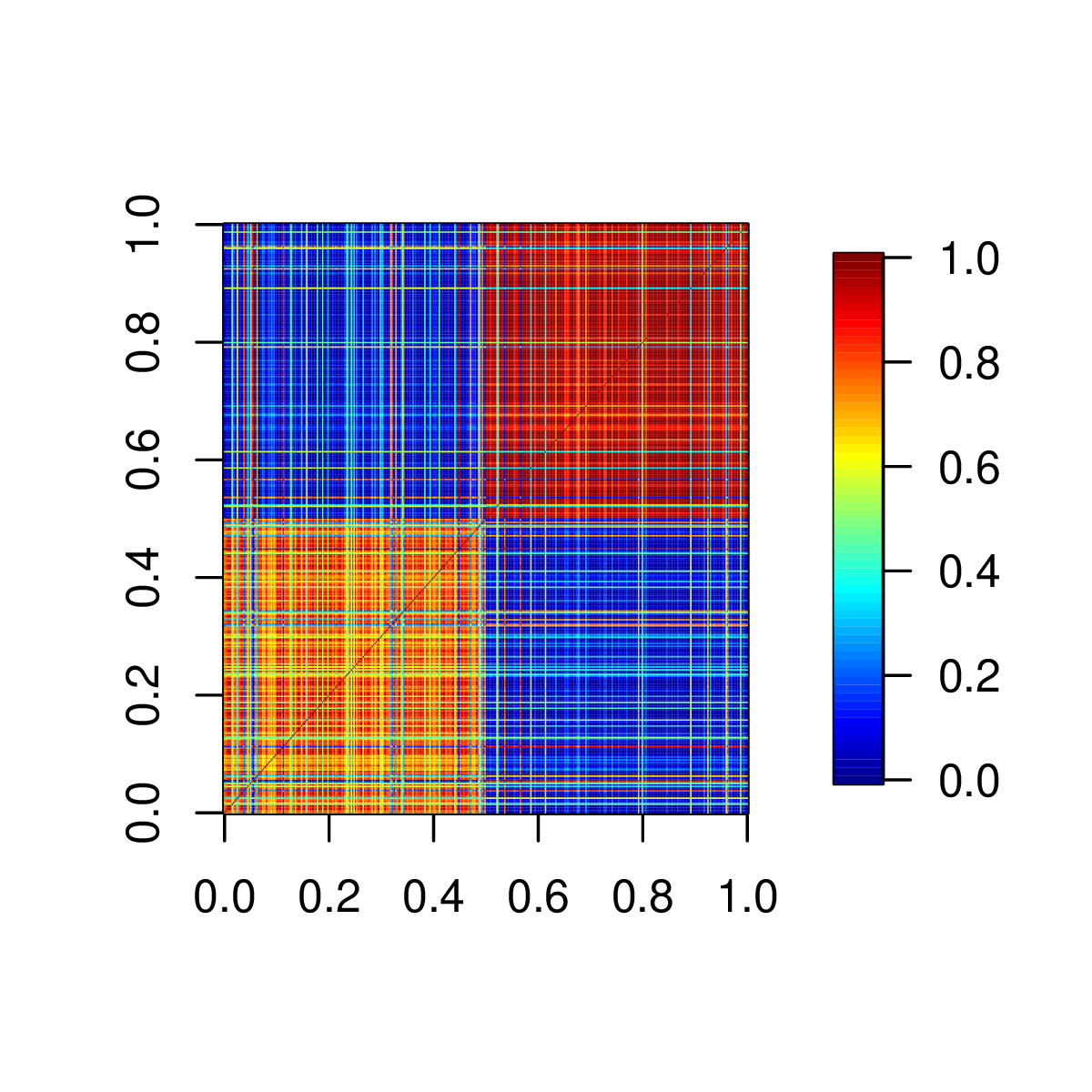}
            \caption{\footnotesize $\text{Pr}(c_i=c_j \mid y)$ from DP-GMM.}
    \end{subfigure} \rulesep
       \begin{subfigure}[t]{.19\textwidth}
        \includegraphics[height=3cm,width=1\linewidth, trim=.5cm .5cm .5cm .5cm,clip]{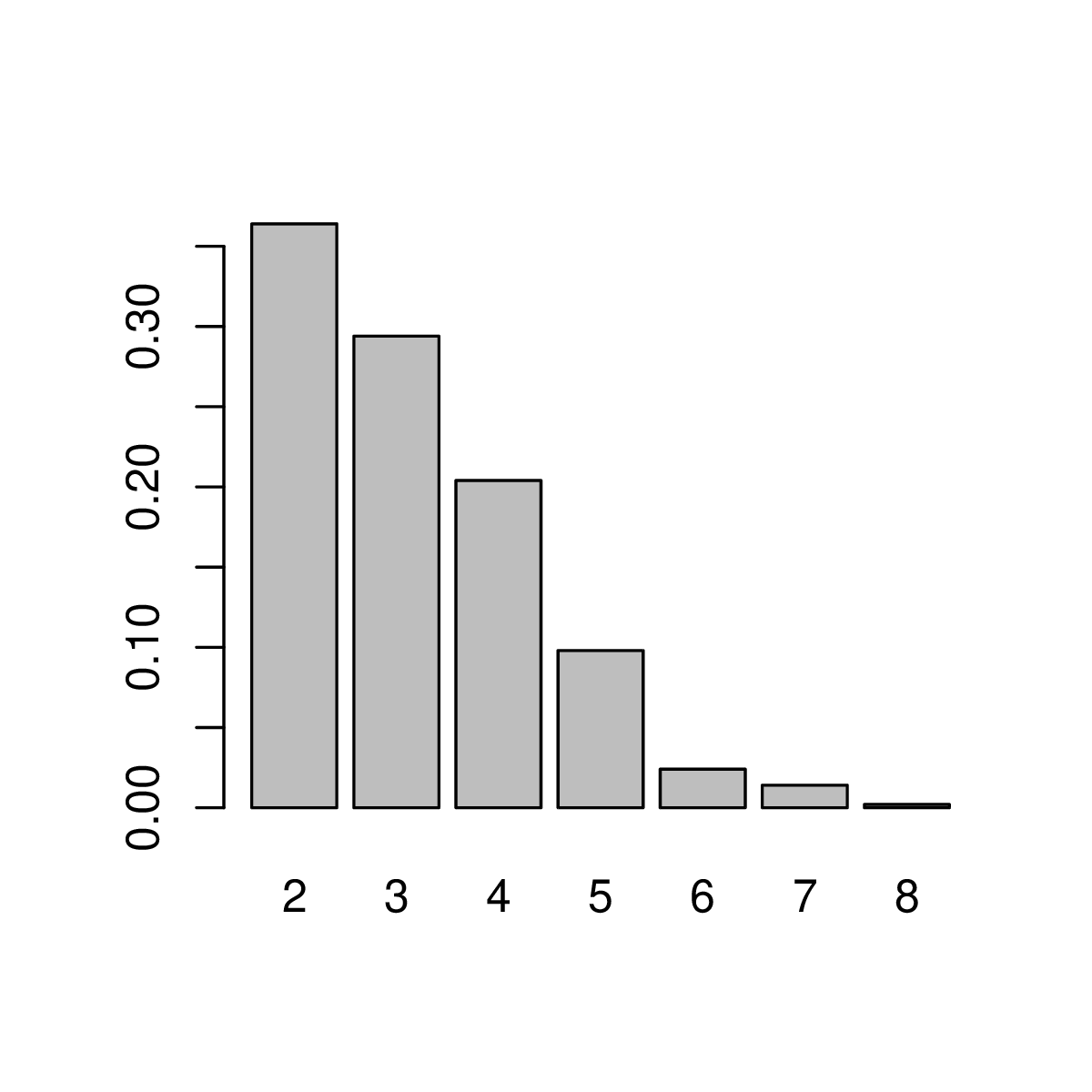}
            \caption{\footnotesize $\text{Pr}(K\mid y)$ from DP-GMM.}
    \end{subfigure} 
            \caption*{Figure S.4: Uncertainty quantification in clustering data generated from a two-component Gaussian mixture model.
        }
        \end{figure}

\begin{figure}[H]
     \begin{subfigure}[t]{.15\textwidth}
        \includegraphics[height=3cm,width=1\linewidth, trim=.5cm .5cm 0cm .5cm,clip]{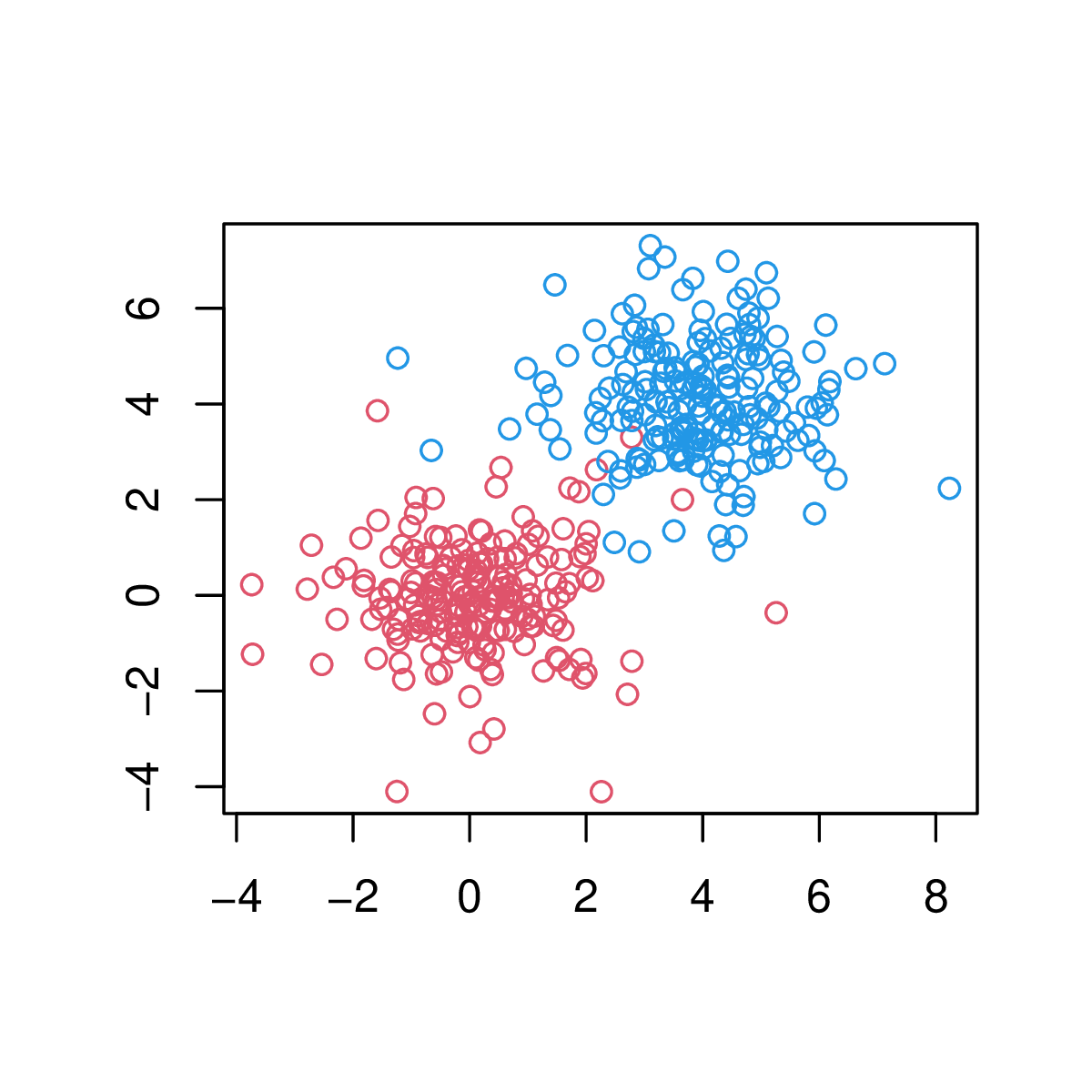}
            \caption{\footnotesize Data from two-component $t_5$-MM ($b=4$).}
    \end{subfigure} \rulesep
     \begin{subfigure}[t]{.21\textwidth}
        \includegraphics[height=3cm,width=1\linewidth, trim=.5cm .5cm .5cm .5cm,clip]{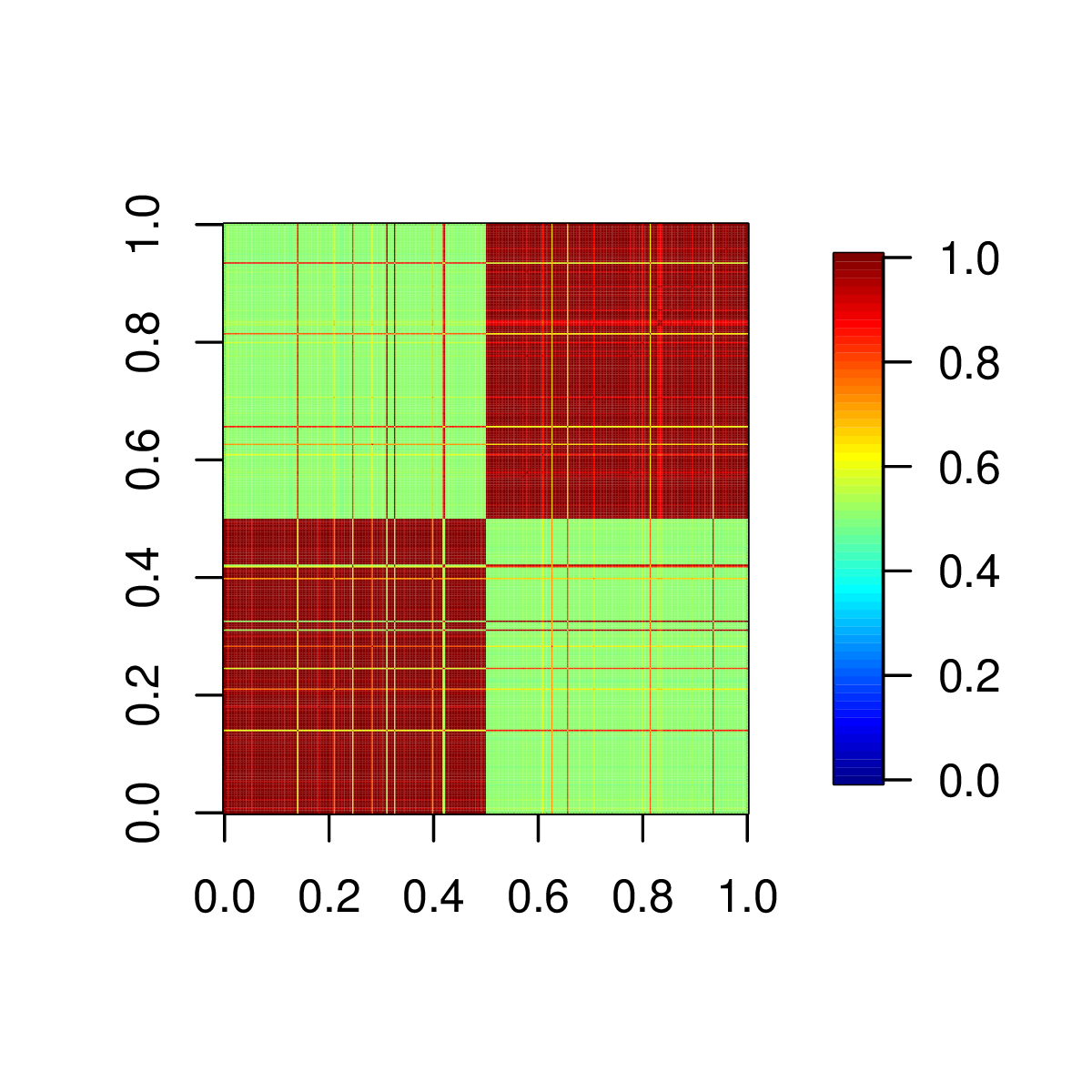}
            \caption{\footnotesize $\text{Pr}(c_i=c_j \mid y)$ from the forest model.}
    \end{subfigure}\rulesep
    \begin{subfigure}[t]{.19\textwidth}
        \includegraphics[height=3cm,width=1\linewidth, trim=.5cm .5cm .5cm .5cm,clip]{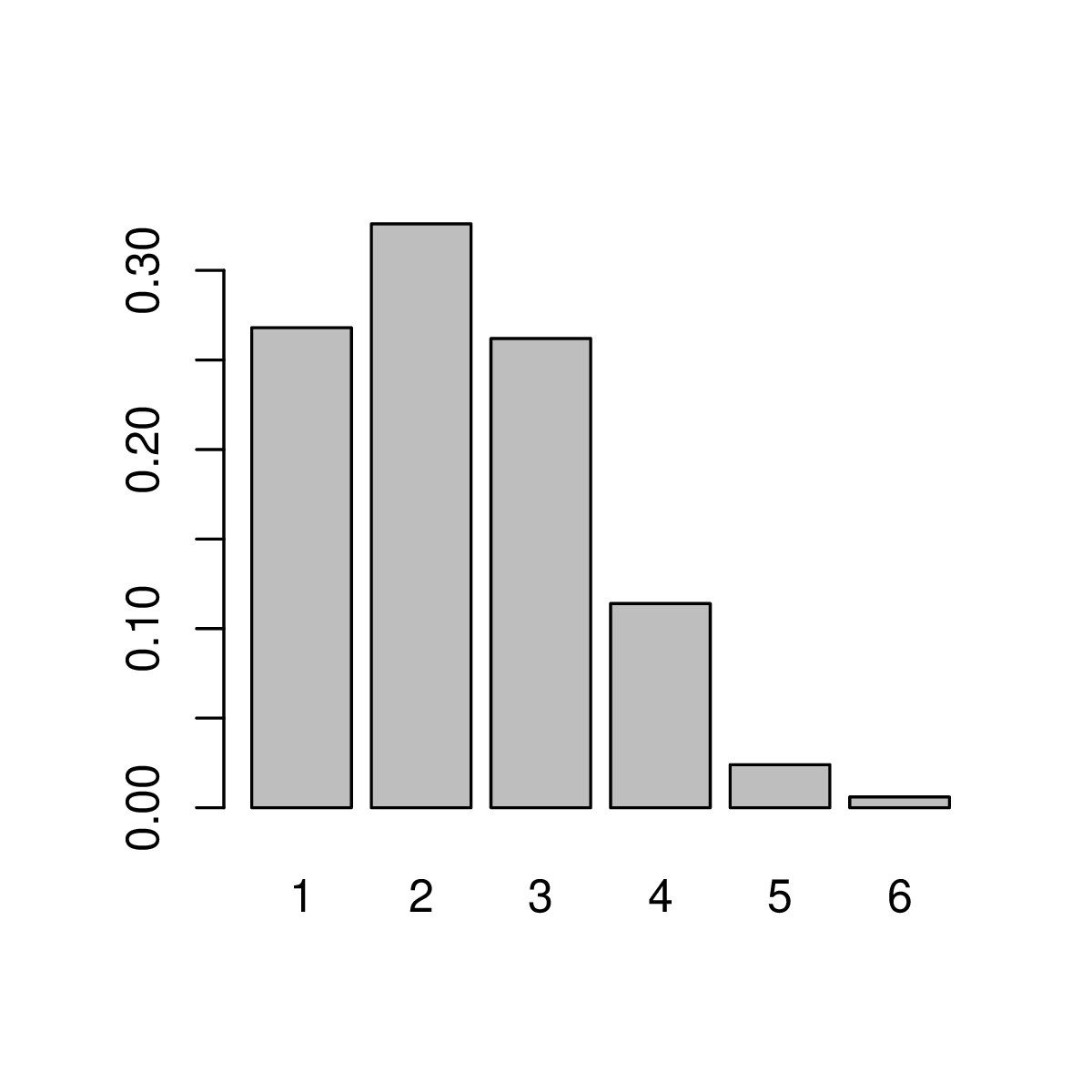}
            \caption{\footnotesize $\text{Pr}(K\mid y)$ from the forest model.}
    \end{subfigure} \rulesep
   \begin{subfigure}[t]{.21\textwidth}
        \includegraphics[height=3cm,width=1\linewidth, trim=.5cm .5cm .5cm .5cm,clip]{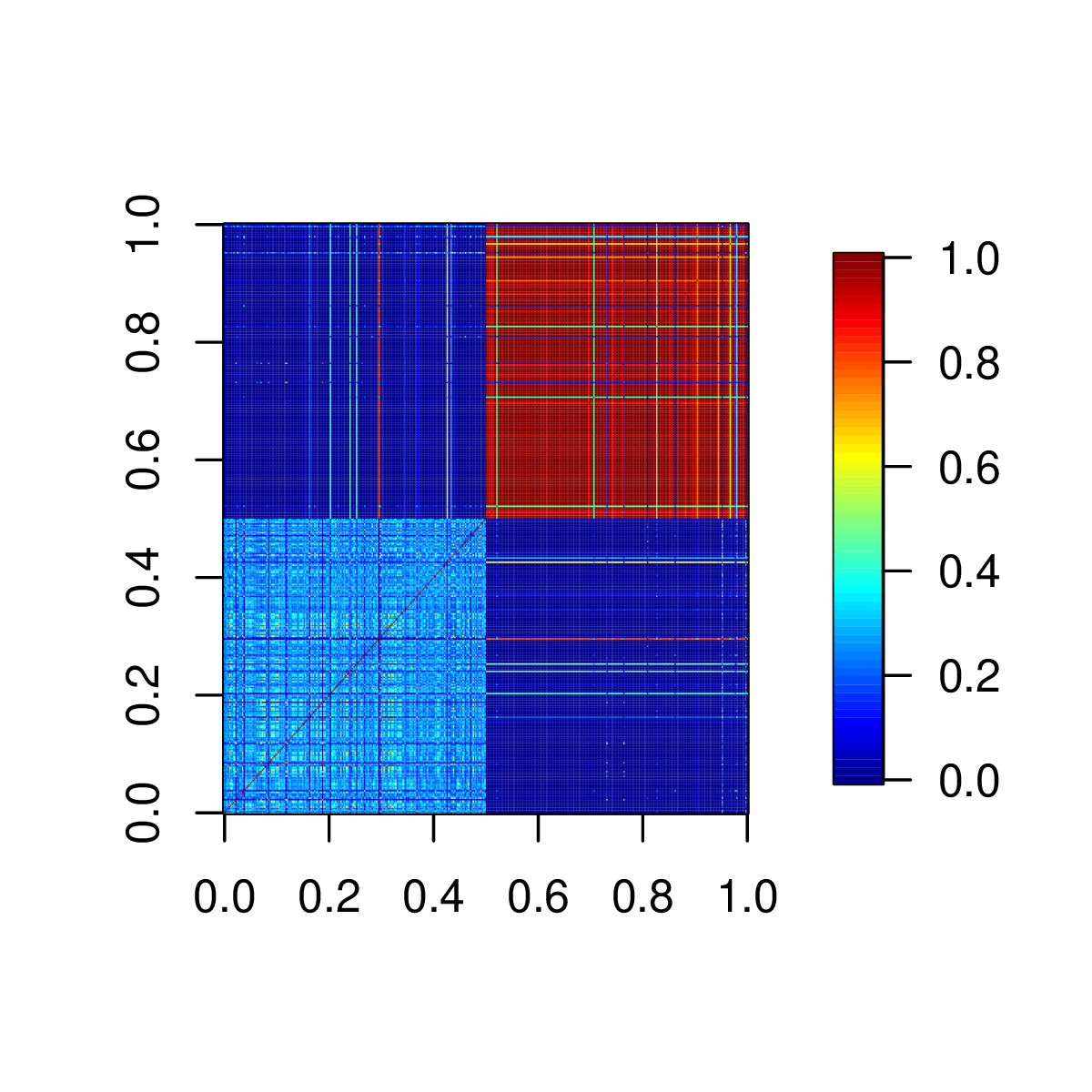}
            \caption{\footnotesize $\text{Pr}(c_i=c_j \mid y)$ from DP-GMM.}
    \end{subfigure} \rulesep
       \begin{subfigure}[t]{.19\textwidth}
        \includegraphics[height=3cm,width=1\linewidth, trim=.5cm .5cm .5cm .5cm,clip]{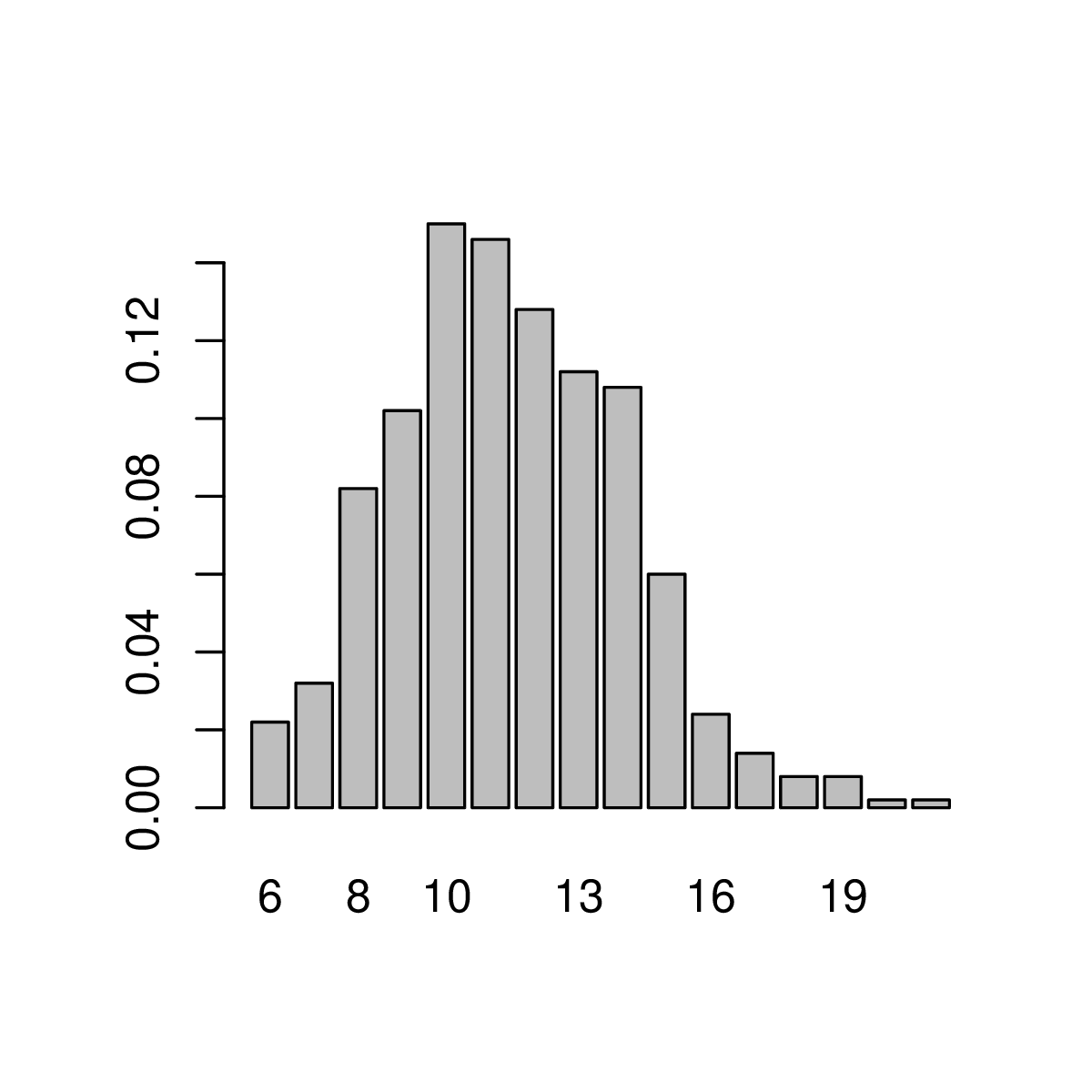}
            \caption{\footnotesize $\text{Pr}(K\mid y)$ from DP-GMM.}
    \end{subfigure} 
     \begin{subfigure}[t]{.15\textwidth}
        \includegraphics[height=3cm,width=1\linewidth, trim=.5cm .5cm 0cm .5cm,clip]{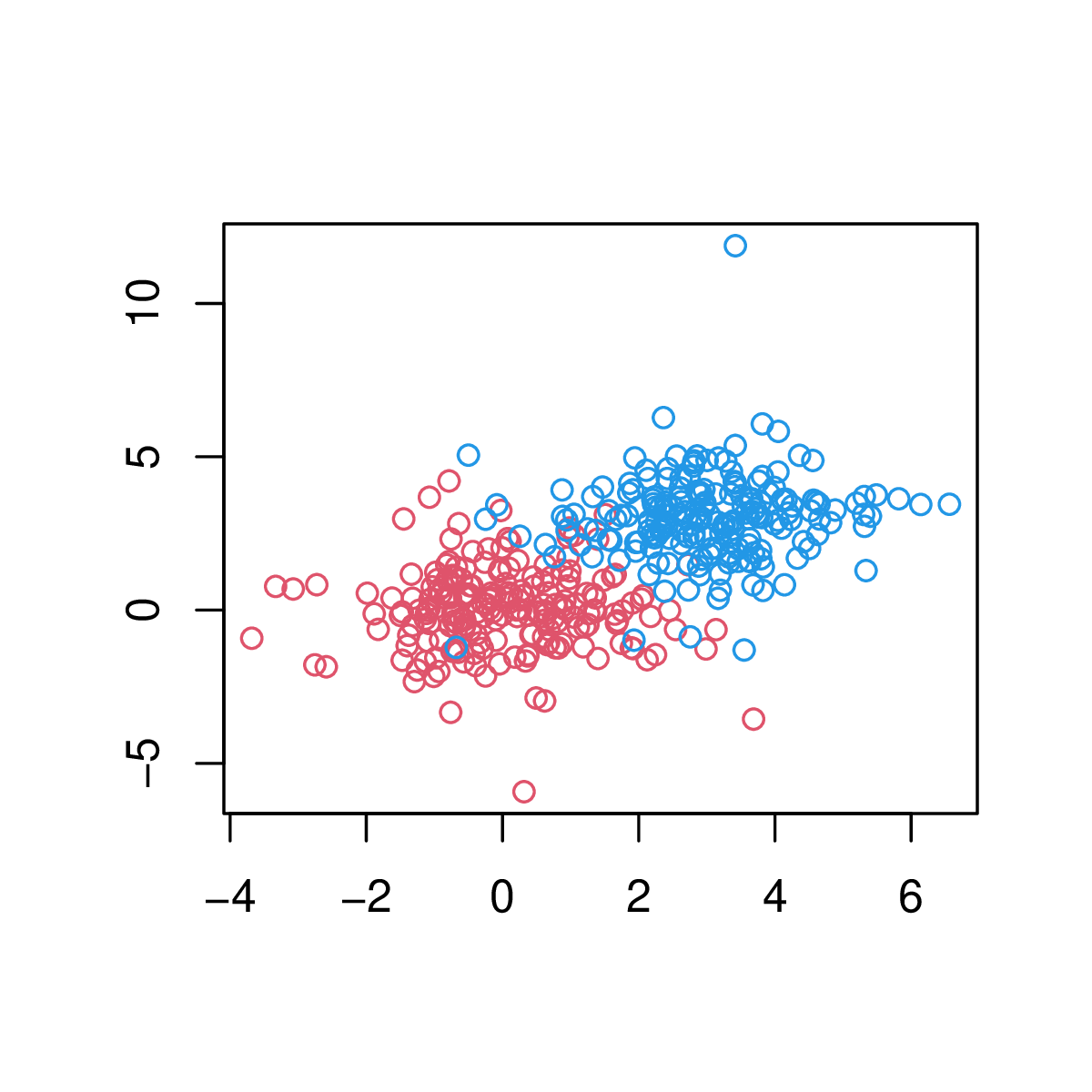}
            \caption{\footnotesize Data from two-component $t_5$-MM ($b=3$).}
    \end{subfigure} \rulesep
     \begin{subfigure}[t]{.21\textwidth}
        \includegraphics[height=3cm,width=1\linewidth, trim=.5cm .5cm .5cm .5cm,clip]{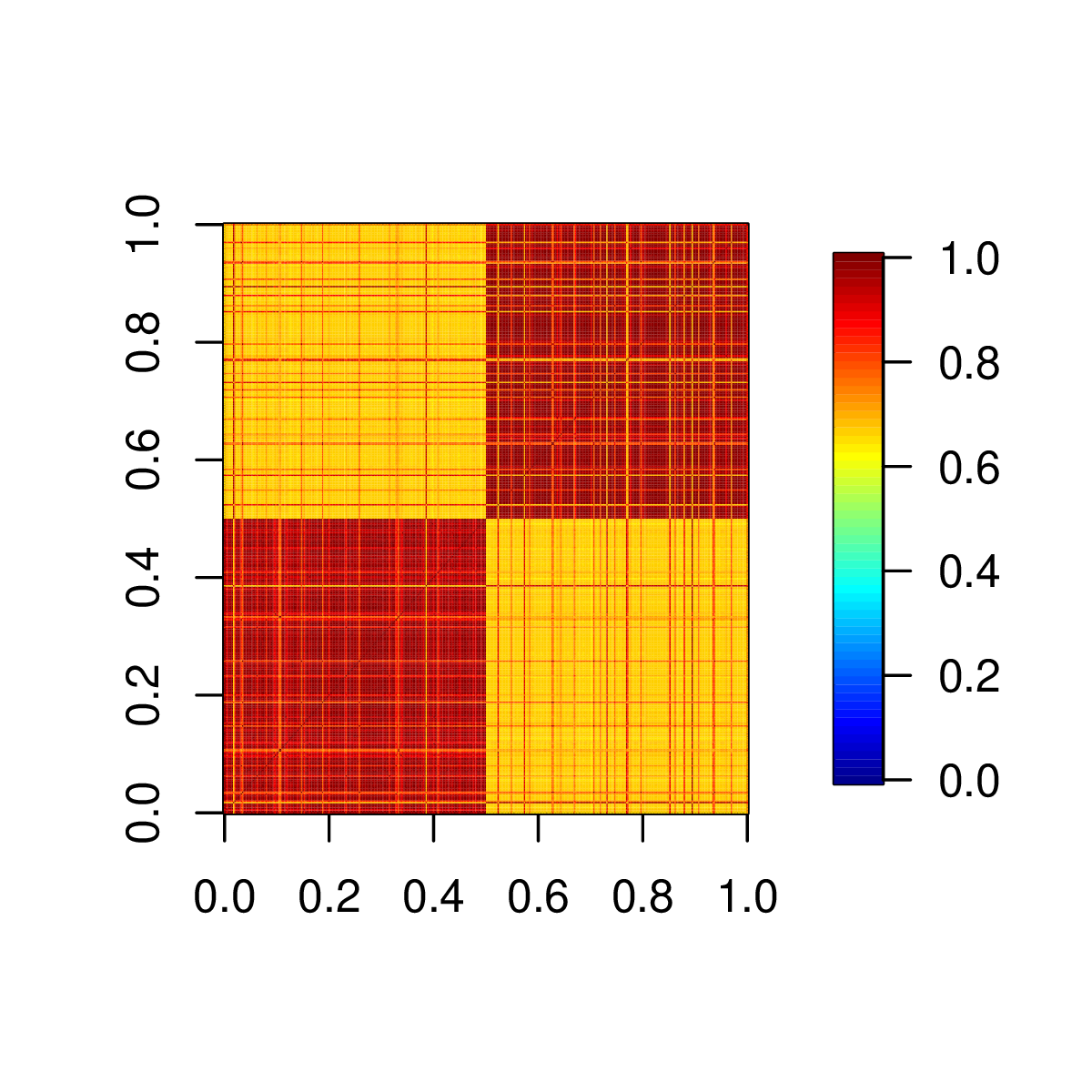}
            \caption{\footnotesize $\text{Pr}(c_i=c_j \mid y)$ from the forest model.}
    \end{subfigure}\rulesep
    \begin{subfigure}[t]{.19\textwidth}
        \includegraphics[height=3cm,width=1\linewidth, trim=.5cm .5cm .5cm .5cm,clip]{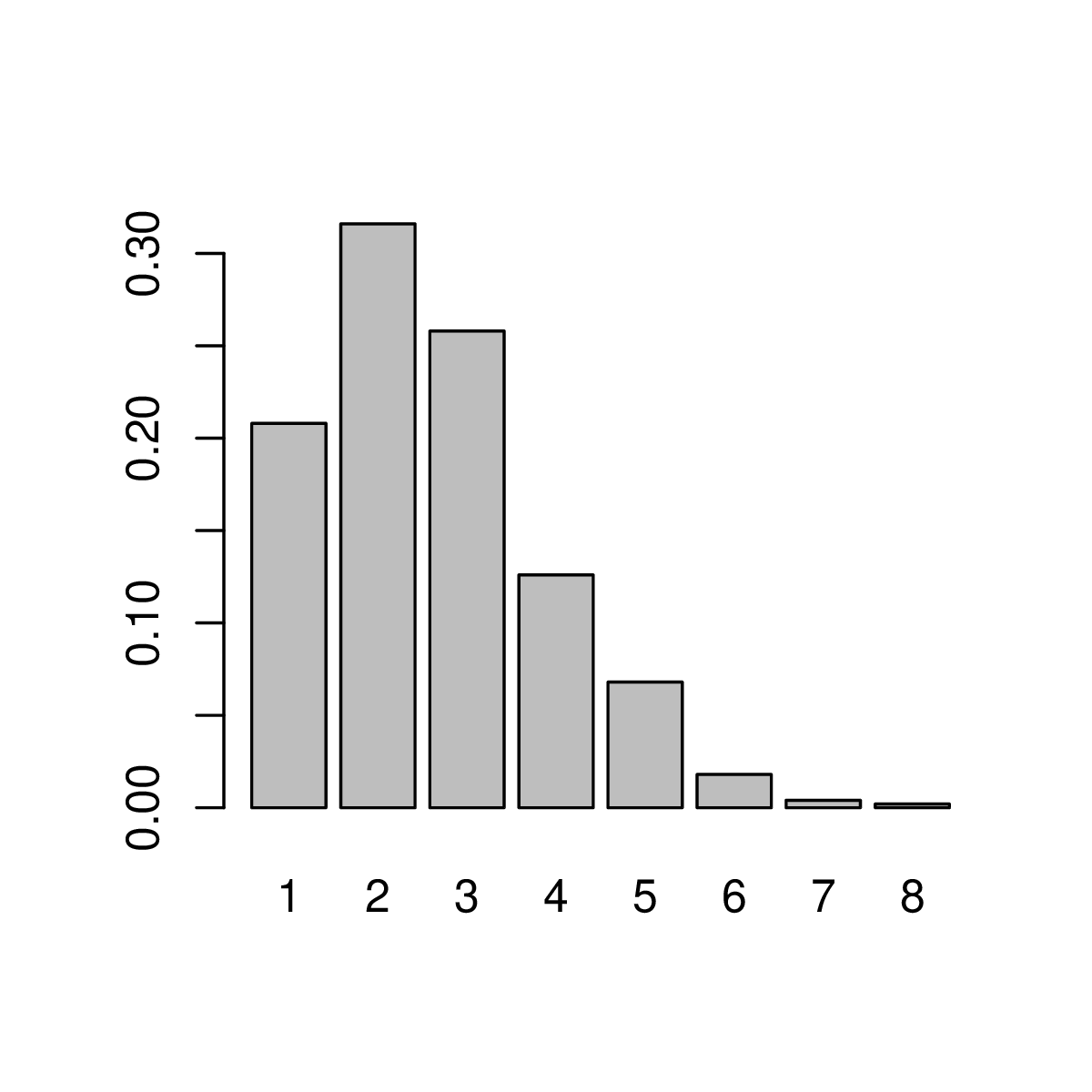}
            \caption{\footnotesize $\text{Pr}(K\mid y)$ from the forest model.}
    \end{subfigure} \rulesep
   \begin{subfigure}[t]{.21\textwidth}
        \includegraphics[height=3cm,width=1\linewidth, trim=.5cm .5cm .5cm .5cm,clip]{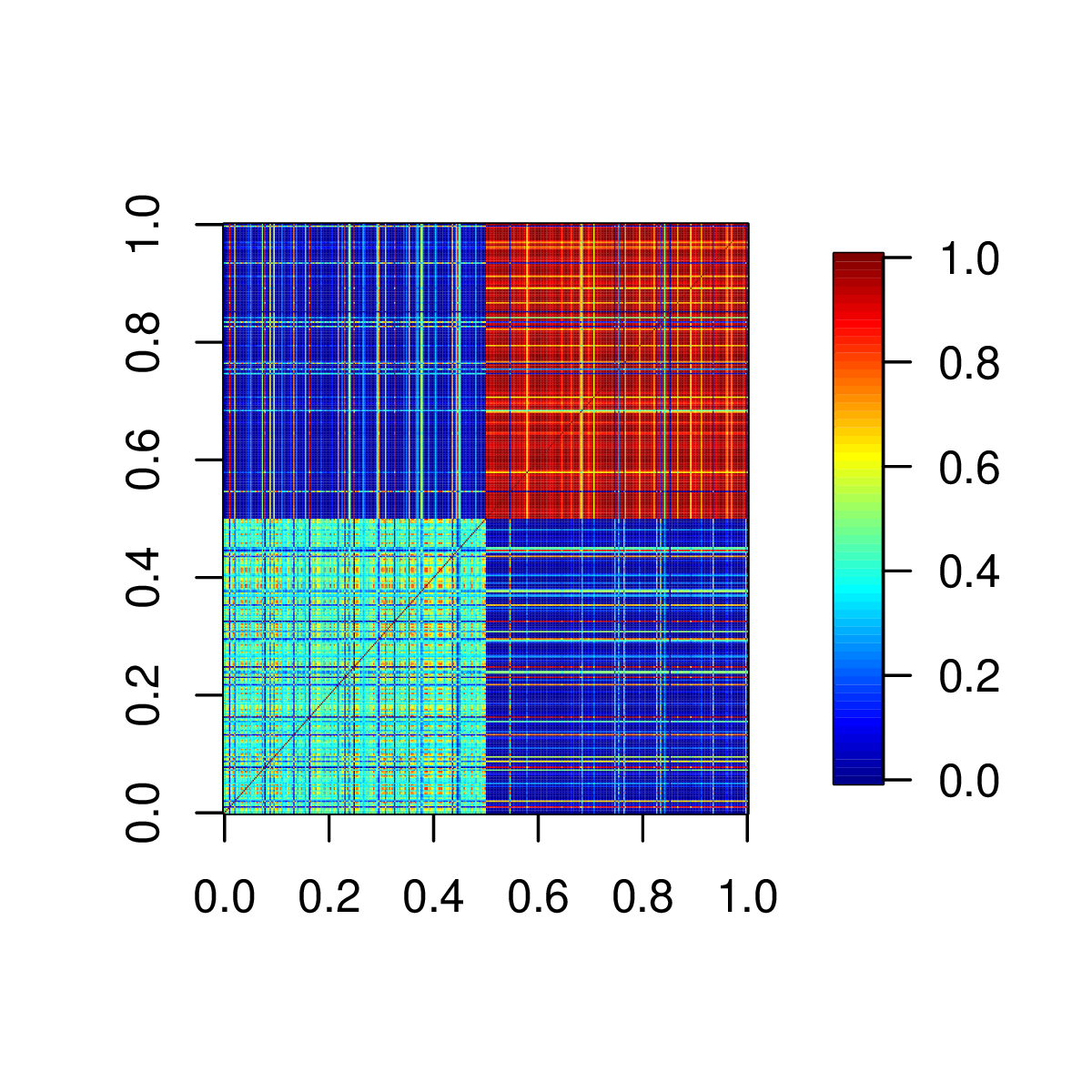}
            \caption{\footnotesize $\text{Pr}(c_i=c_j \mid y)$ from DP-GMM.}
    \end{subfigure} \rulesep
       \begin{subfigure}[t]{.19\textwidth}
        \includegraphics[height=3cm,width=1\linewidth, trim=.5cm .5cm .5cm .5cm,clip]{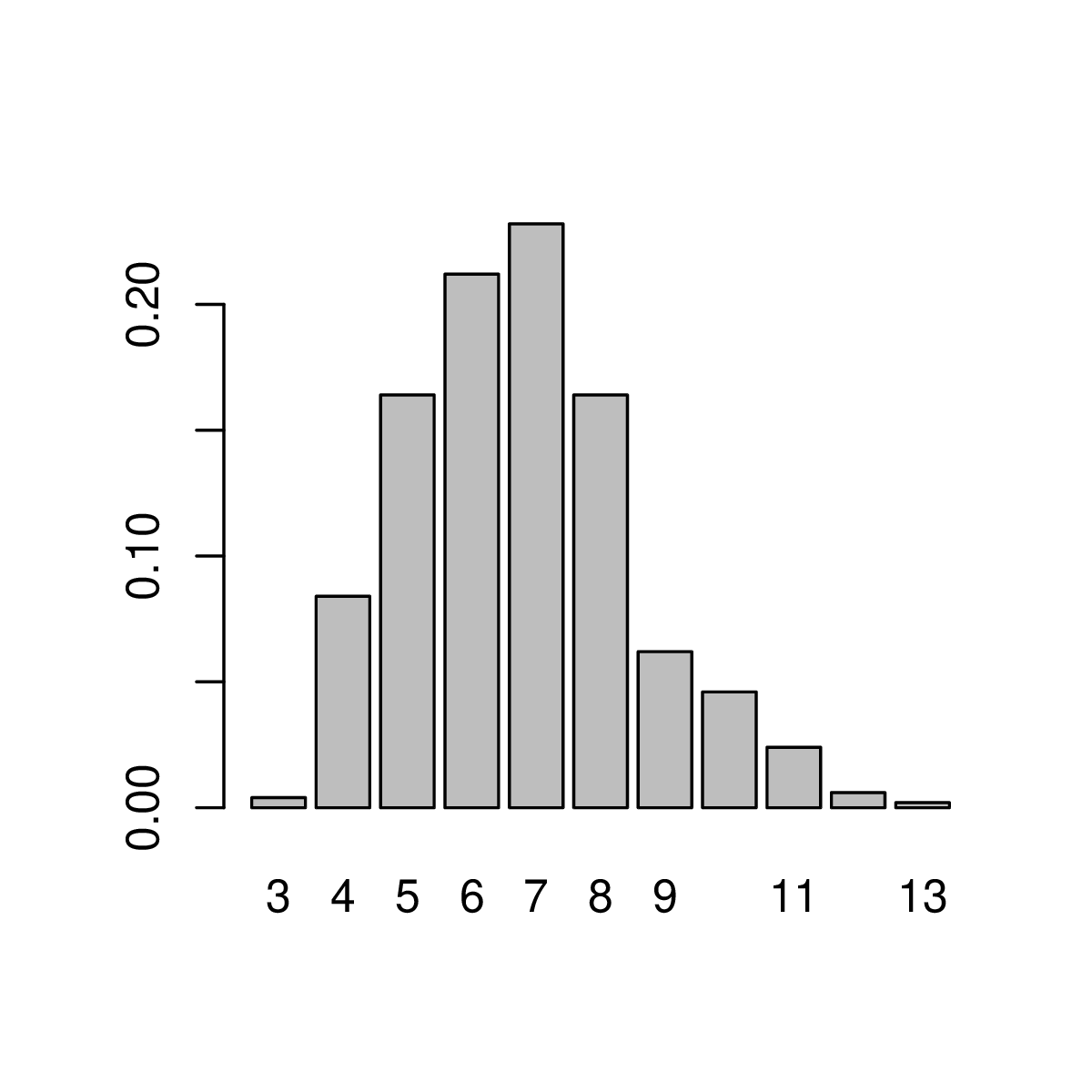}
            \caption{\footnotesize $\text{Pr}(K\mid y)$ from DP-GMM.}
    \end{subfigure} 
     \begin{subfigure}[t]{.15\textwidth}
        \includegraphics[height=3cm,width=1\linewidth, trim=.5cm .5cm 0cm .5cm,clip]{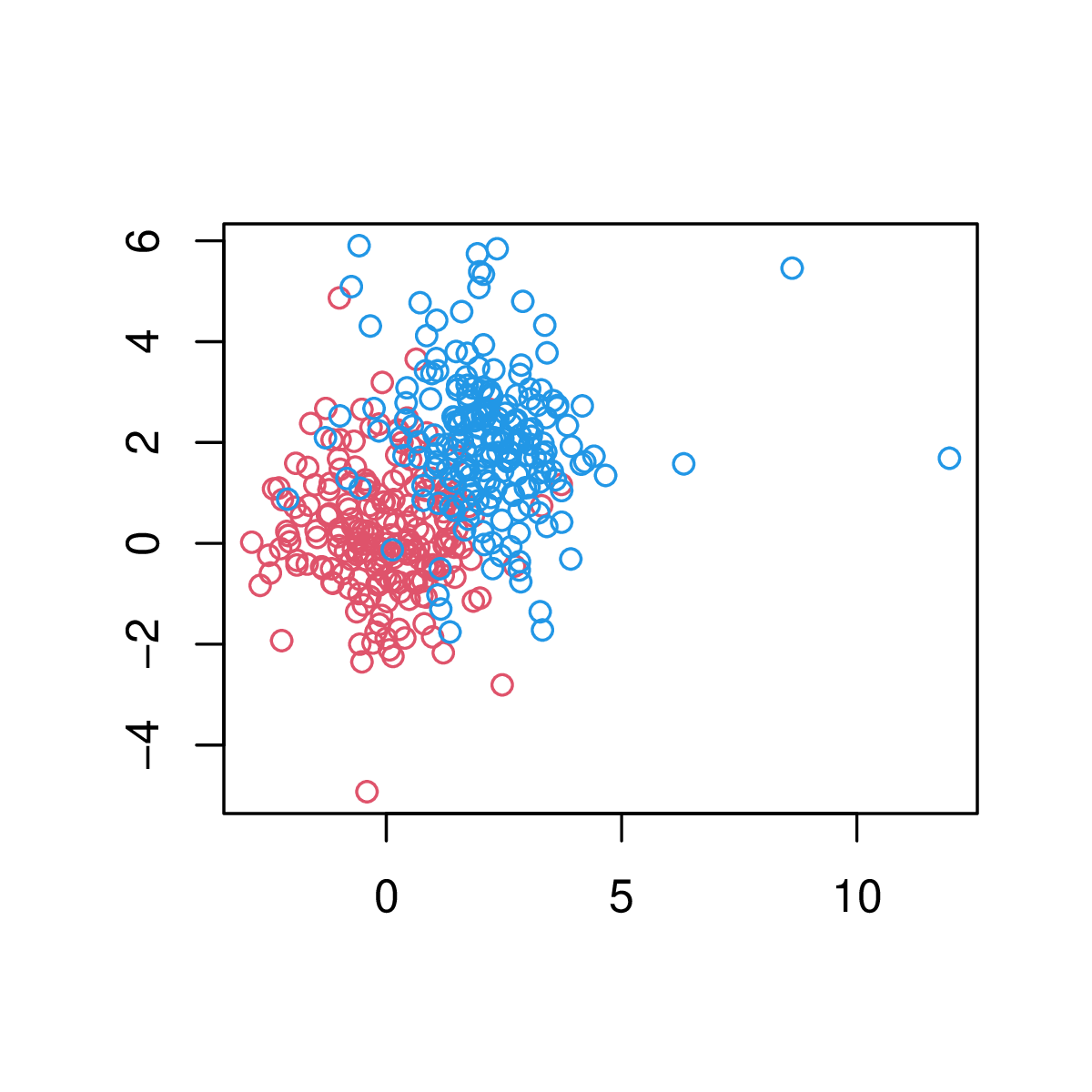}
            \caption{\footnotesize Data from two-component $t_5$-MM ($b=2$).}
    \end{subfigure} \rulesep
     \begin{subfigure}[t]{.21\textwidth}
        \includegraphics[height=3cm,width=1\linewidth, trim=.5cm .5cm .5cm .5cm,clip]{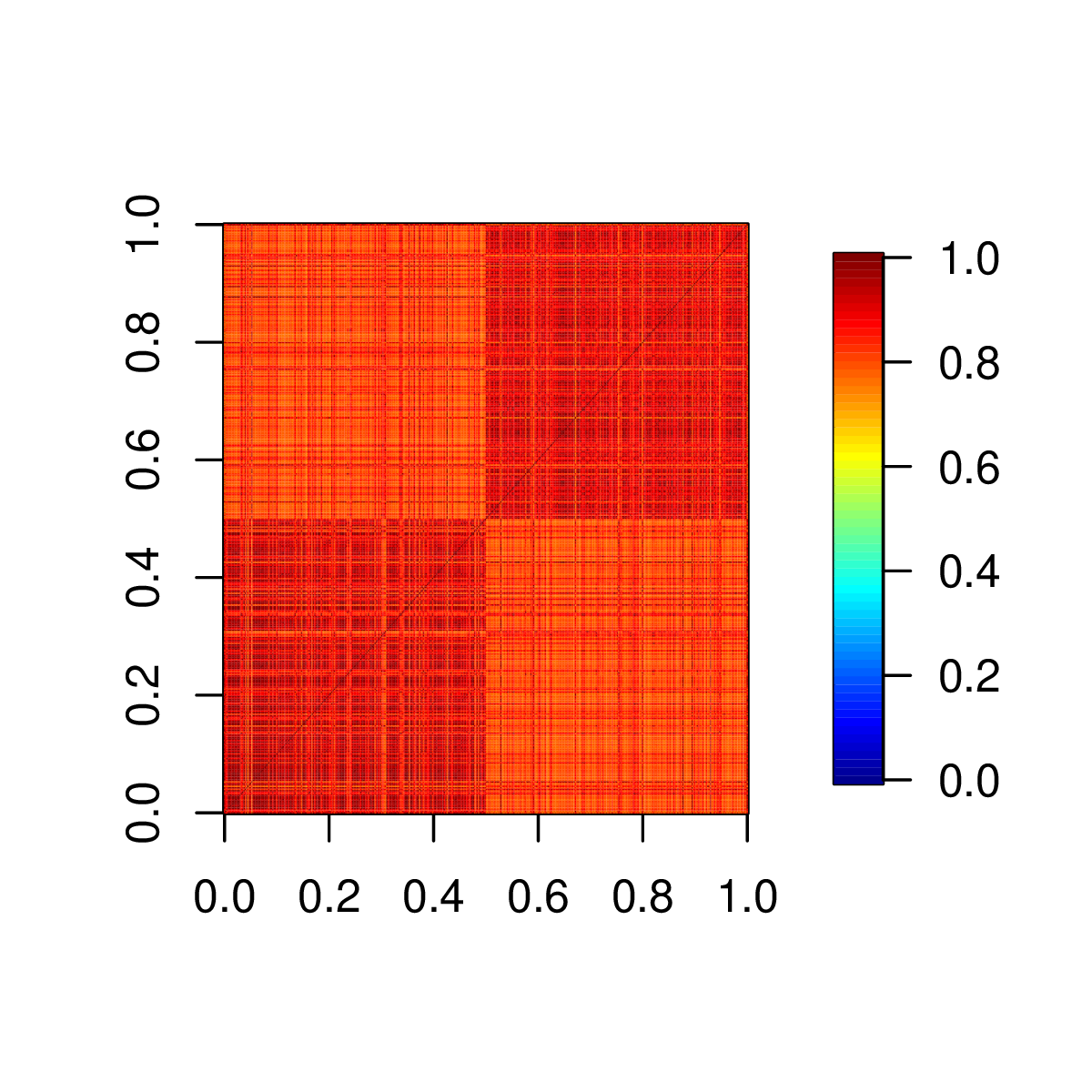}
            \caption{\footnotesize $\text{Pr}(c_i=c_j \mid y)$ from the forest model.}
    \end{subfigure}\rulesep
    \begin{subfigure}[t]{.19\textwidth}
        \includegraphics[height=3cm,width=1\linewidth, trim=.5cm .5cm .5cm .5cm,clip]{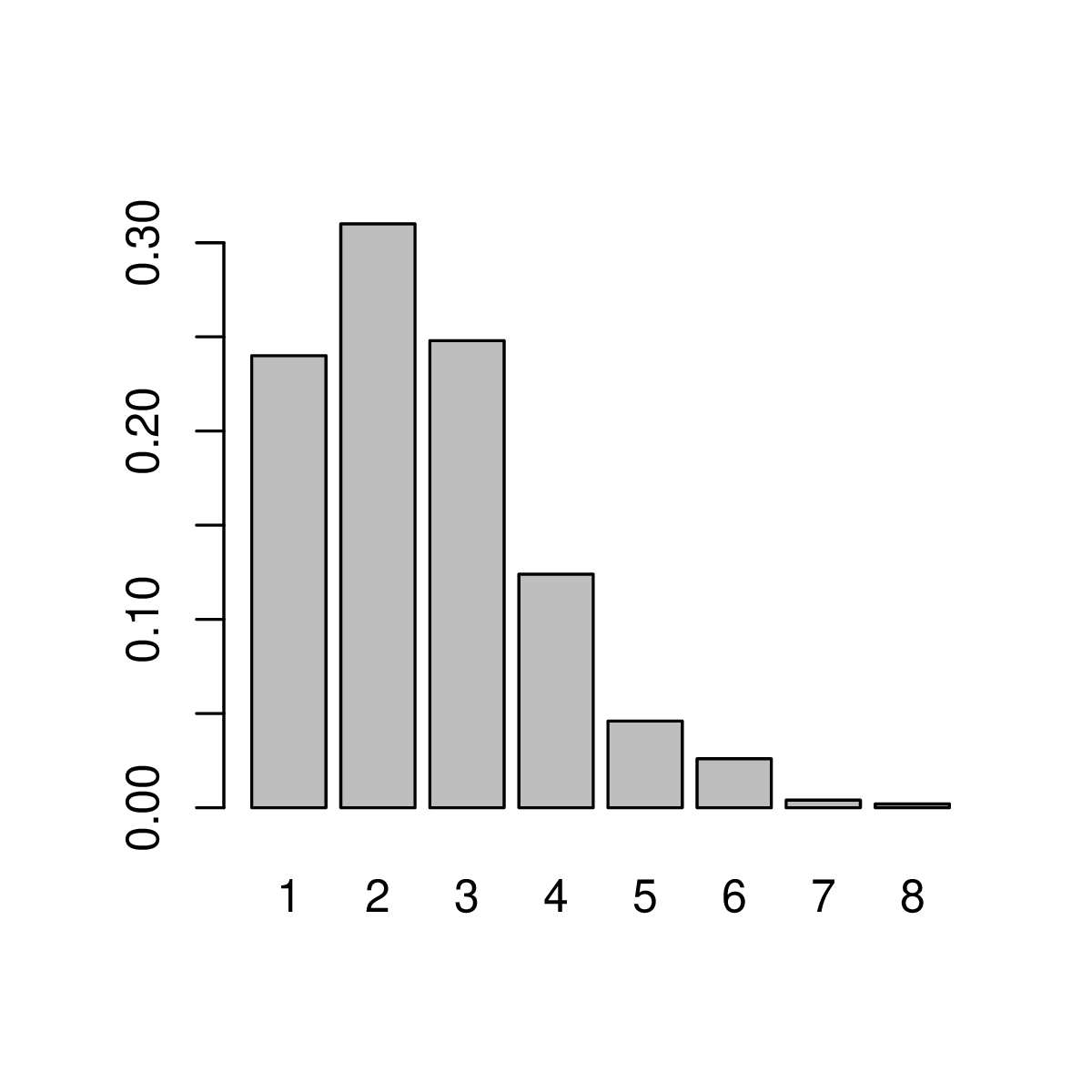}
            \caption{\footnotesize $\text{Pr}(K\mid y)$ from the forest model.}
    \end{subfigure} \rulesep
   \begin{subfigure}[t]{.21\textwidth}
        \includegraphics[height=3cm,width=1\linewidth, trim=.5cm .5cm .5cm .5cm,clip]{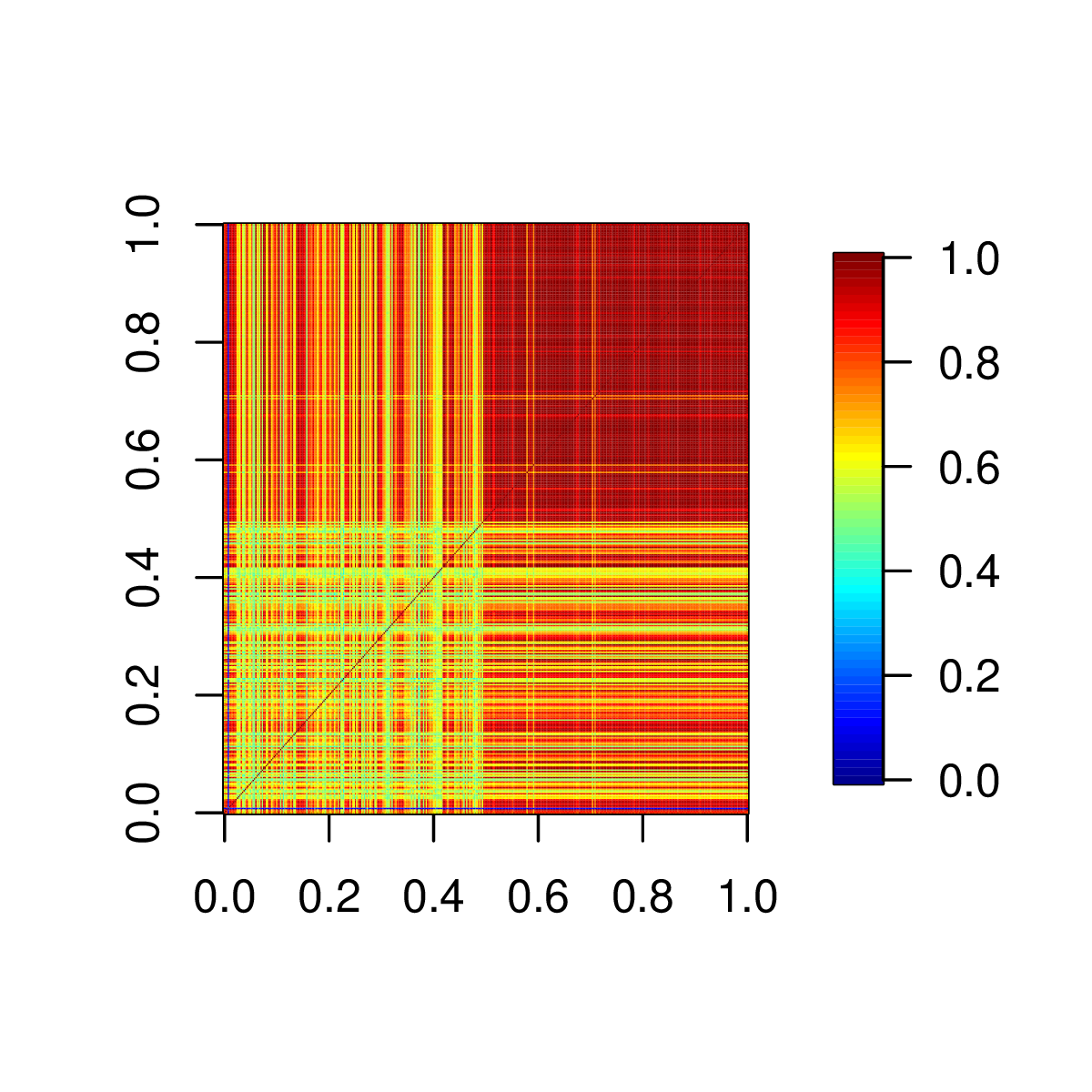}
            \caption{\footnotesize $\text{Pr}(c_i=c_j \mid y)$ from DP-GMM.}
    \end{subfigure} \rulesep
       \begin{subfigure}[t]{.19\textwidth}
        \includegraphics[height=3cm,width=1\linewidth, trim=.5cm .5cm .5cm .5cm,clip]{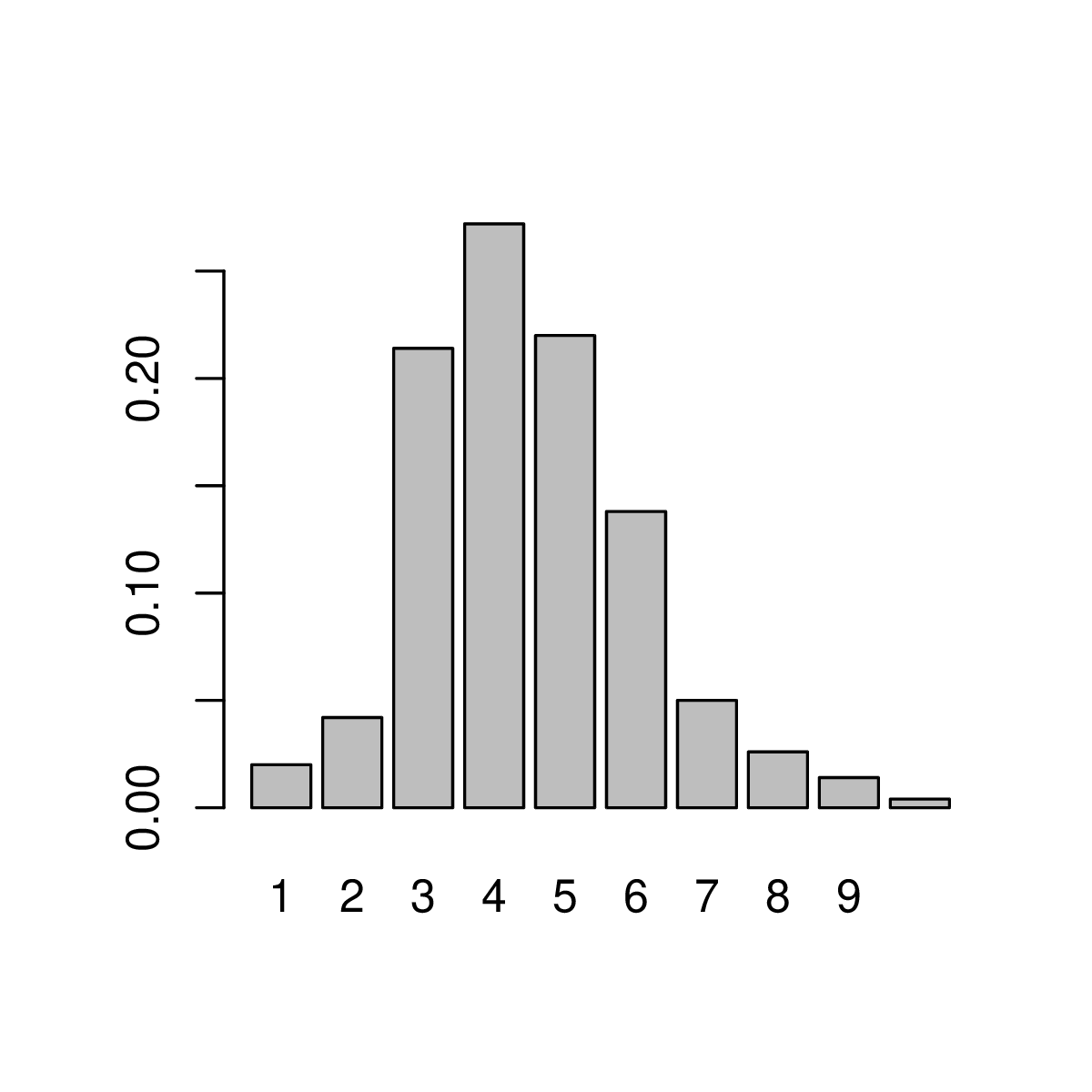}
            \caption{\footnotesize $\text{Pr}(K\mid y)$ from DP-GMM.}
    \end{subfigure} 
            \caption*{Figure S.5: Uncertainty quantification in clustering data generated from a two-component $t_5$ mixture model.
        }
        \end{figure}
        
        \subsection{Additional Experiments on Clustering Near-Manifold Data}
We conduct additional simulations on clustering near-manifold data. The results are shown in Figure S.7.
\begin{figure}[H]
     \begin{subfigure}[t]{.24\textwidth}
        \includegraphics[height=4cm,width=1\linewidth]{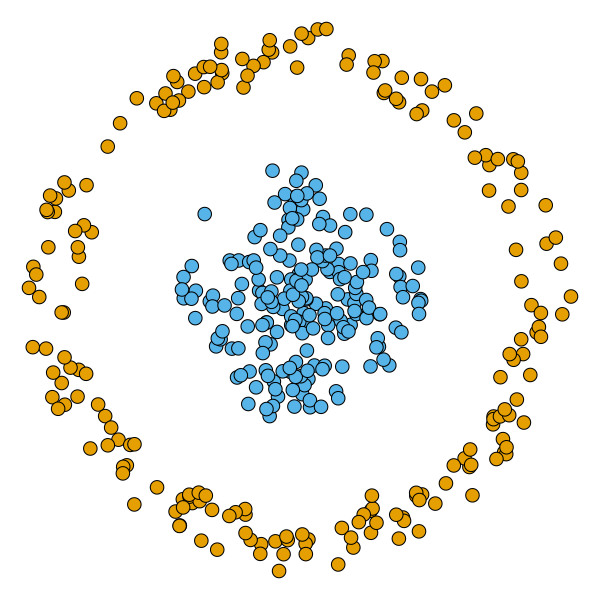}
            \caption{Posterior point estimate.}
    \end{subfigure} 
     \begin{subfigure}[t]{.24\textwidth}
        \includegraphics[width=.8\linewidth]{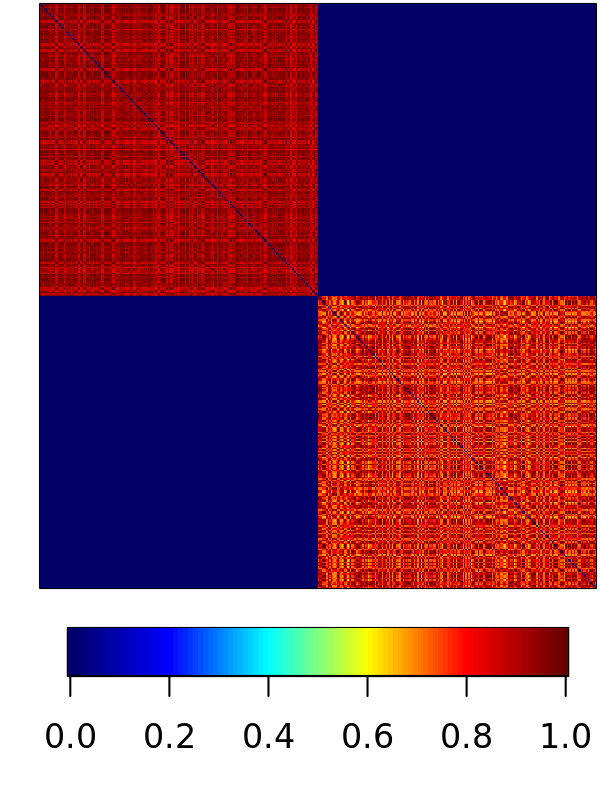}
            \caption{$\text{Pr}(c_i=c_j \mid y)$.}
    \end{subfigure}
    \begin{subfigure}[t]{.24\textwidth}
        \includegraphics[height=4cm,width=1\linewidth]{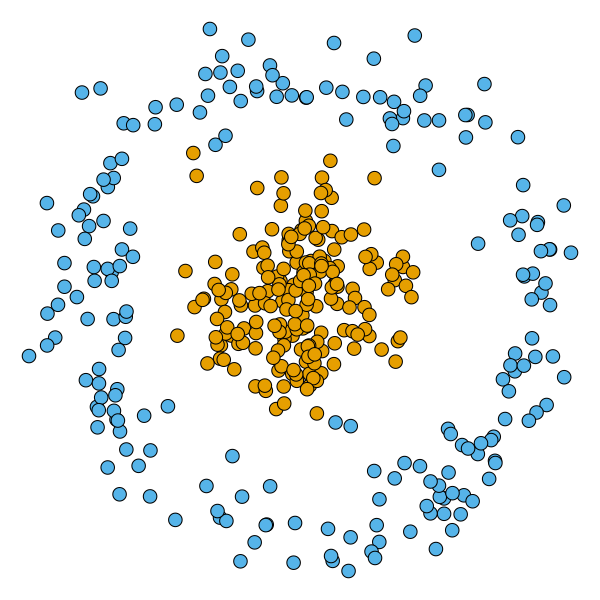}
            \caption{Posterior point estimate.}
    \end{subfigure} 
     \begin{subfigure}[t]{.24\textwidth}
        \includegraphics[width=.8\linewidth]{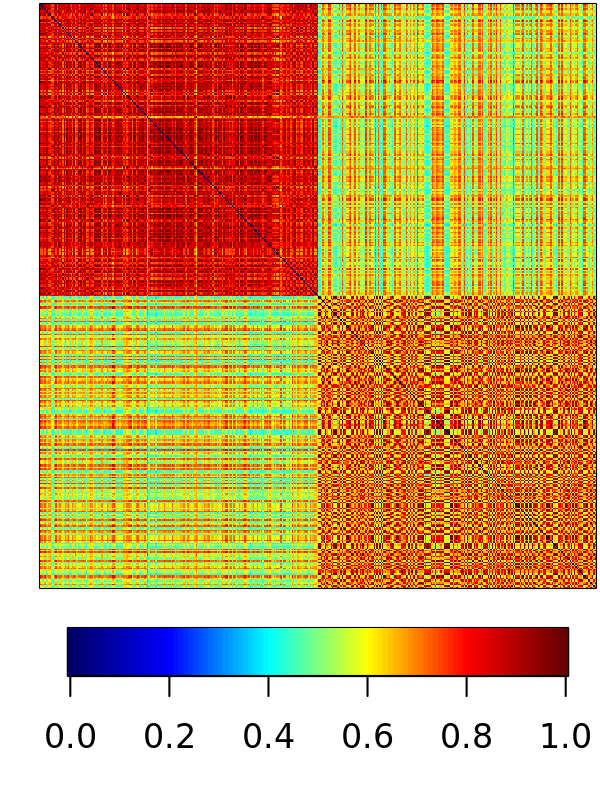}
            \caption{$\text{Pr}(c_i=c_j \mid y)$.}
    \end{subfigure} 
     \begin{subfigure}[t]{.24\textwidth}
        \includegraphics[height=4cm,width=1\linewidth]{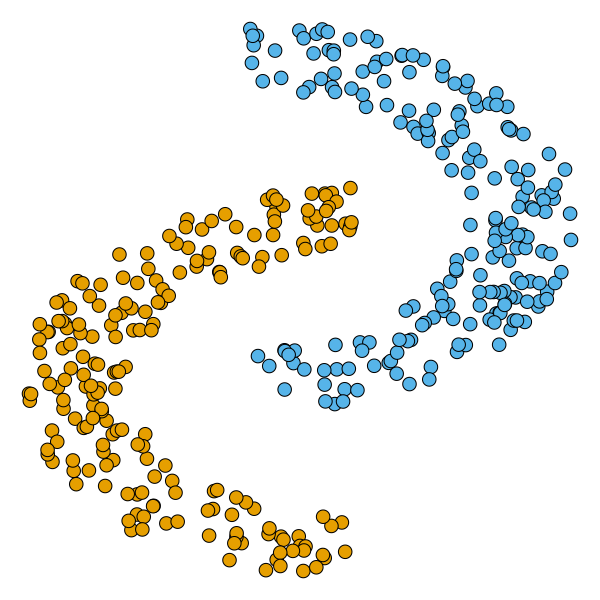}
            \caption{Posterior point estimate.}
    \end{subfigure} 
     \begin{subfigure}[t]{.24\textwidth}
        \includegraphics[width=.8\linewidth]{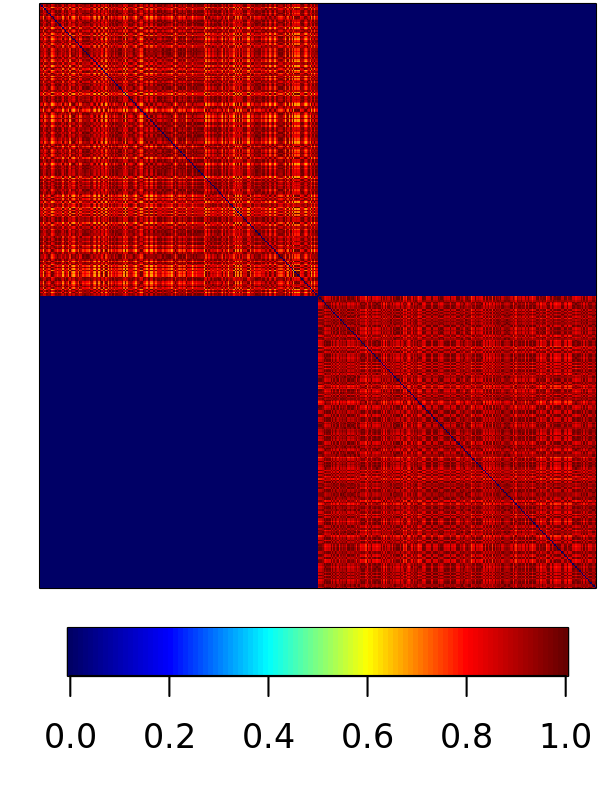}
            \caption{$\text{Pr}(c_i=c_j \mid y)$.}
    \end{subfigure}
    \begin{subfigure}[t]{.24\textwidth}
        \includegraphics[height=4cm,width=1\linewidth]{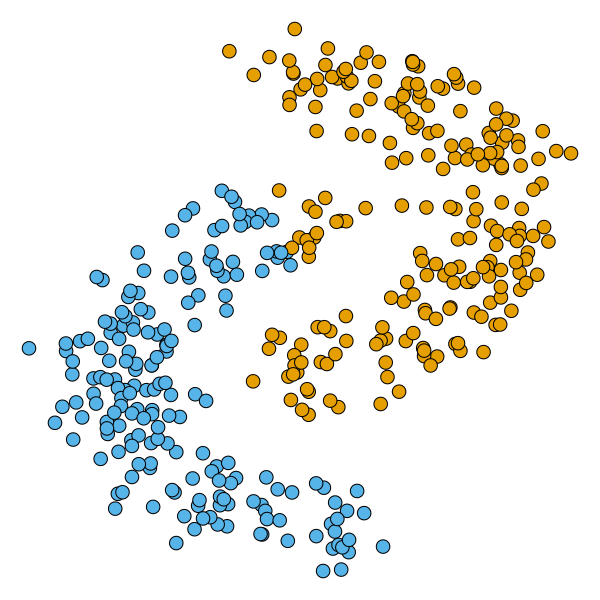}
            \caption{Posterior point estimate.}
    \end{subfigure} 
     \begin{subfigure}[t]{.24\textwidth}
        \includegraphics[width=.8\linewidth]{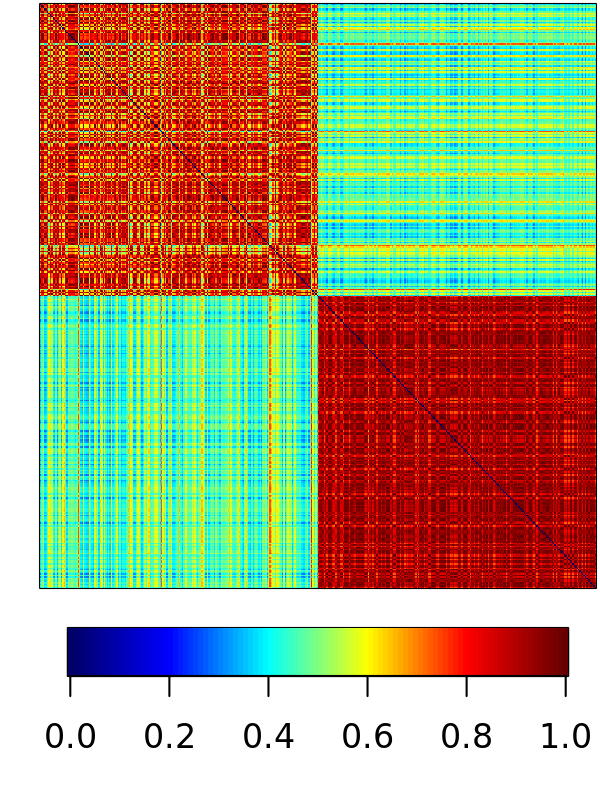}
            \caption{$\text{Pr}(c_i=c_j \mid y)$.}
    \end{subfigure} 
        \caption*{Figure S.6: Clustering data generated near manifolds.
        }
        \end{figure}

\subsection{Additional Simulations on Uncertainty and Clustering Accuracy}

We now compare the uncertainty and clustering accuracy. We consider three possible scenarios as different sources of uncertainties: increasingly imbalanced cluster sizes, an increasing number of clusters, and an increasing number of noisy points between clusters. In addition, we gradually reduce the separation between clusters, so that the uncertainty can increase as well.

When measuring the clustering accuracy of the point estimate, we calculate the adjusted Rand index (ARI), normalized mutual information (NMI), as well as the clustering accuracy rate (the match rate between the point estimate of $\hat c_i$ and each ground-truth label, minimized over all possible label switchings in $\hat c_i$). We run 10 times of experiments under each combination of values, and show the boxplots.

\begin{figure}[H]
         \begin{subfigure}[t]{.3\textwidth}
        \includegraphics[width=1\linewidth]{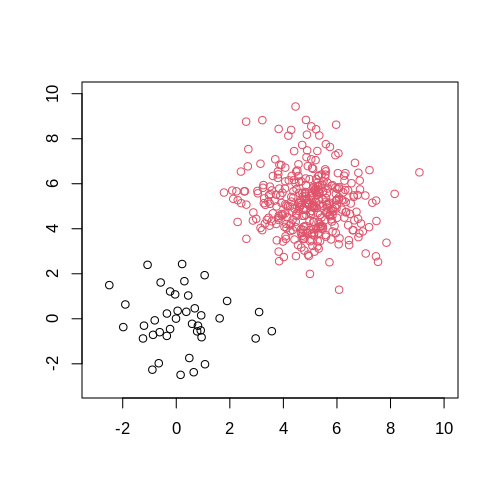}
    \end{subfigure}\;
         \begin{subfigure}[t]{.3\textwidth}
        \includegraphics[width=1\linewidth]{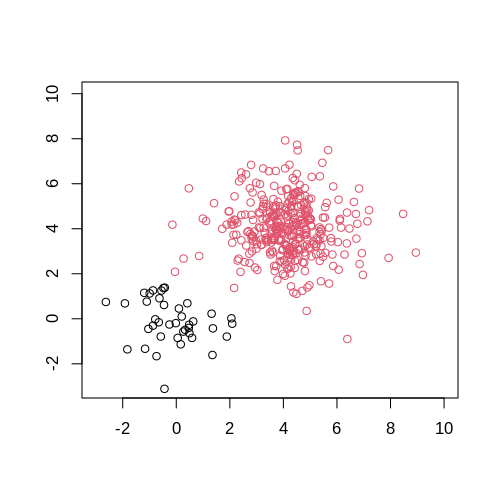}
    \end{subfigure}\;
     \begin{subfigure}[t]{.3\textwidth}
        \includegraphics[width=1\linewidth]{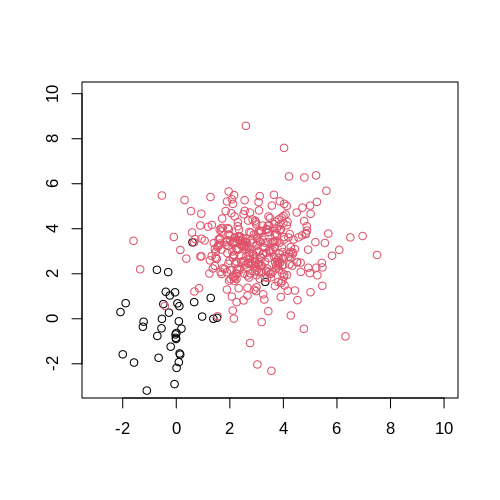}
    \end{subfigure}\\
     \begin{subfigure}[t]{.3\textwidth}
        \includegraphics[width=1\linewidth]{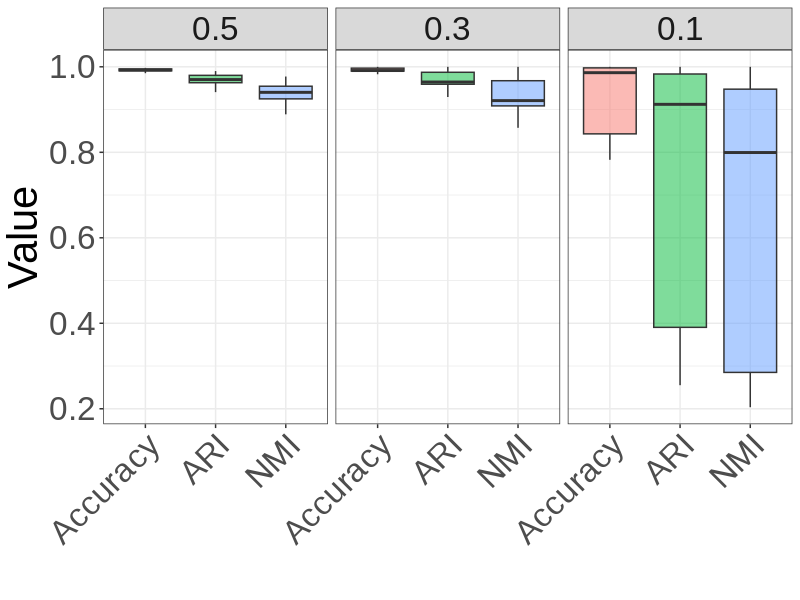}
                    \caption{Two clusters of imbalanced sizes with two means separated by vector $[5,5]$ (above). The experiments are repeated with the proportion of Cluster 1 size taken from $\{0.5, 0.3, 0.1\}$, and the clustering accuracy measures are shown below.}
    \end{subfigure}\;
         \begin{subfigure}[t]{.3\textwidth}
        \includegraphics[width=1\linewidth]{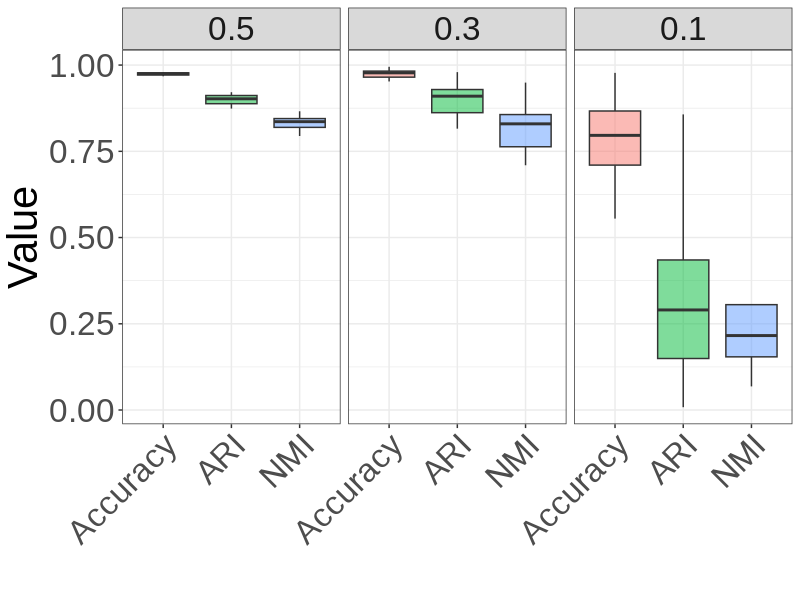}
                    \caption{Two clusters of imbalanced sizes with two means separated by vector $[4,4]$ (above). The experiments are repeated with the proportion of Cluster 1 size taken from $\{0.5, 0.3, 0.1\}$, and the clustering accuracy measures are shown below.}
    \end{subfigure}\;
     \begin{subfigure}[t]{.3\textwidth}
        \includegraphics[width=1\linewidth]{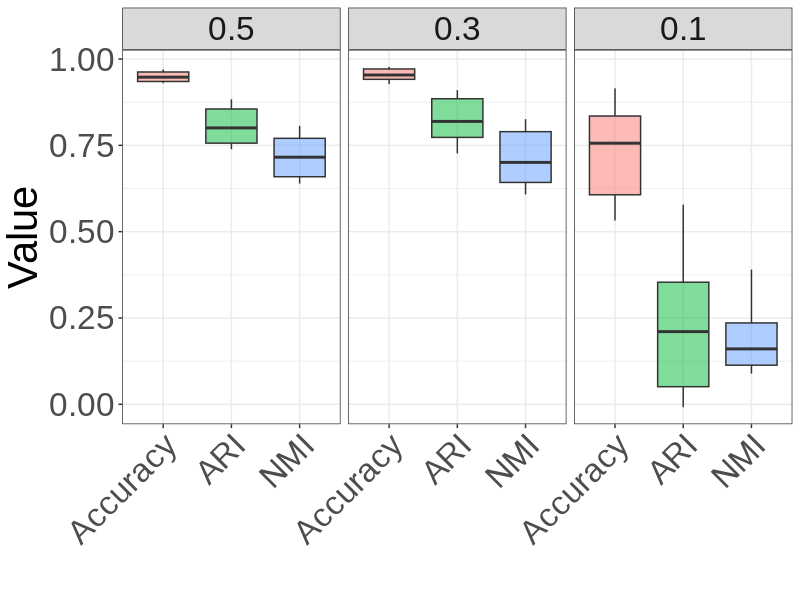}
                    \caption{Two clusters of imbalanced sizes with two means separated by vector $[3,3]$ (above). The experiments are repeated with the proportion of Cluster 1 size taken from $\{0.5, 0.3, 0.1\}$, and the clustering accuracy measures are shown below.}
    \end{subfigure}
        \caption{Clustering accuracy decreases as the cluster sizes become more imbalanced. 
        The adjusted Rand index (ARI), normalized mutual information (NMI), and the clustering accuracy rate (Accuracy, the match rate between $\hat c_i$ and the ground truth, minimized over all possible label switchings in $\hat c_i$) are shown. \label{fig:imbal}
        }
        \end{figure}

For the first scenario, we generate $n=400$ data points from a two-component independent bivariate $t$ distribution with $5$ degrees of freedom,  $y_i \sim  \tilde  w_1  t_5 (\cdot \mid [0,0]) + (1-\tilde  w_1)  t_5(\cdot \mid [\tilde  b, \tilde b])$. We experiment with different values of $\tilde w_1\in \{0.5, 0.3, 0.1\}$ to have different degrees of cluster size imbalance, as well as different values of $\tilde b \in \{5,4,3\}$   to have different degrees of separation between cluster centers. The results are shown in Figure \ref{fig:imbal}.

\begin{figure}[H]
         \begin{subfigure}[t]{.3\textwidth}
        \includegraphics[width=1\linewidth]{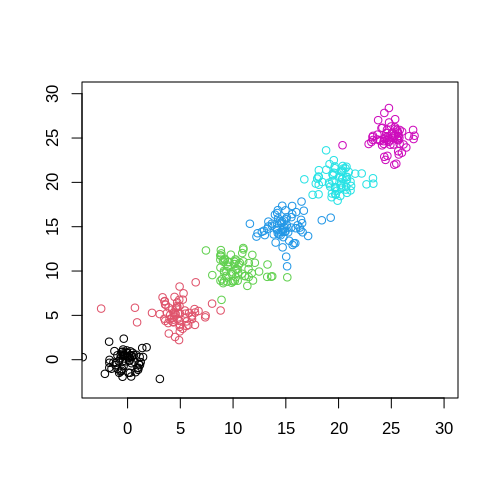}
    \end{subfigure}\;
         \begin{subfigure}[t]{.3\textwidth}
        \includegraphics[width=1\linewidth]{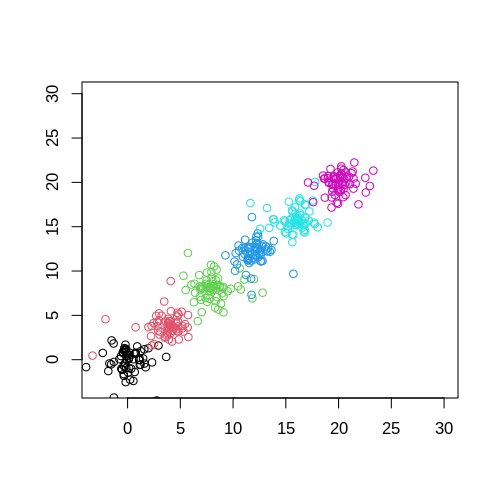}
    \end{subfigure}\;
     \begin{subfigure}[t]{.3\textwidth}
        \includegraphics[width=1\linewidth]{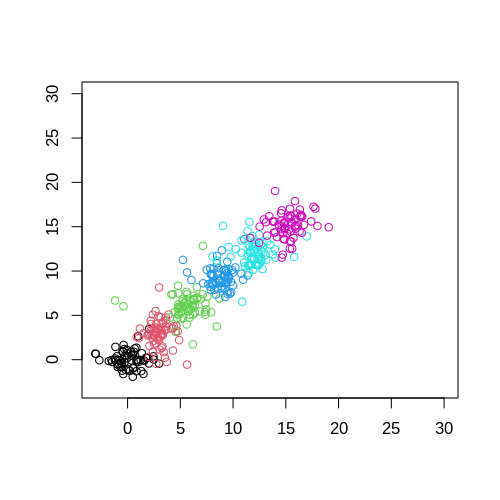}
    \end{subfigure}\\
         \begin{subfigure}[t]{.3\textwidth}
        \includegraphics[width=1\linewidth]{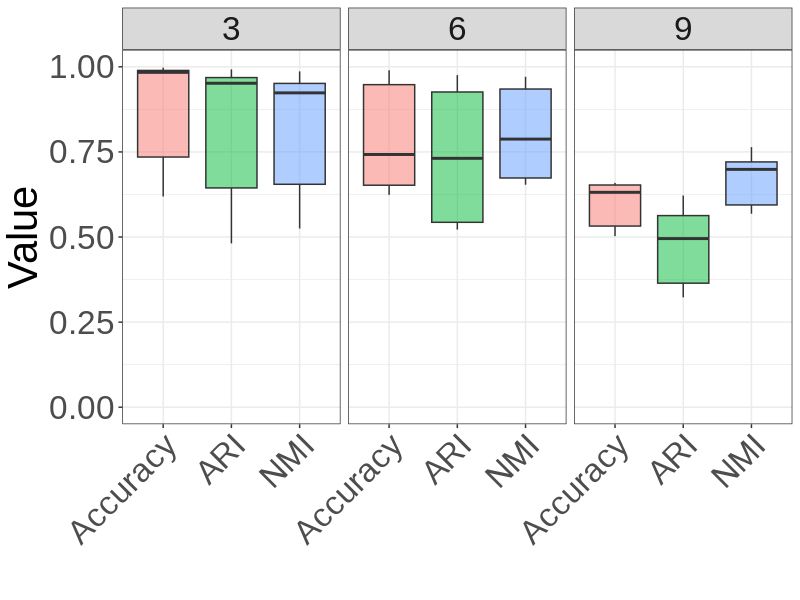}
            \caption{Increasing number of clusters.}
    \end{subfigure}\;
         \begin{subfigure}[t]{.3\textwidth}
        \includegraphics[width=1\linewidth]{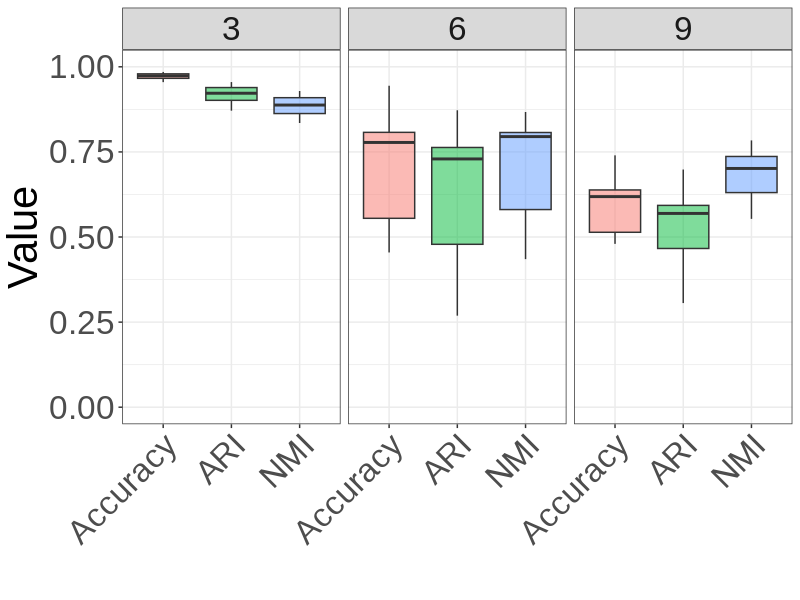}
            \caption{Increasing number of clusters.}
    \end{subfigure}\;
     \begin{subfigure}[t]{.3\textwidth}
        \includegraphics[width=1\linewidth]{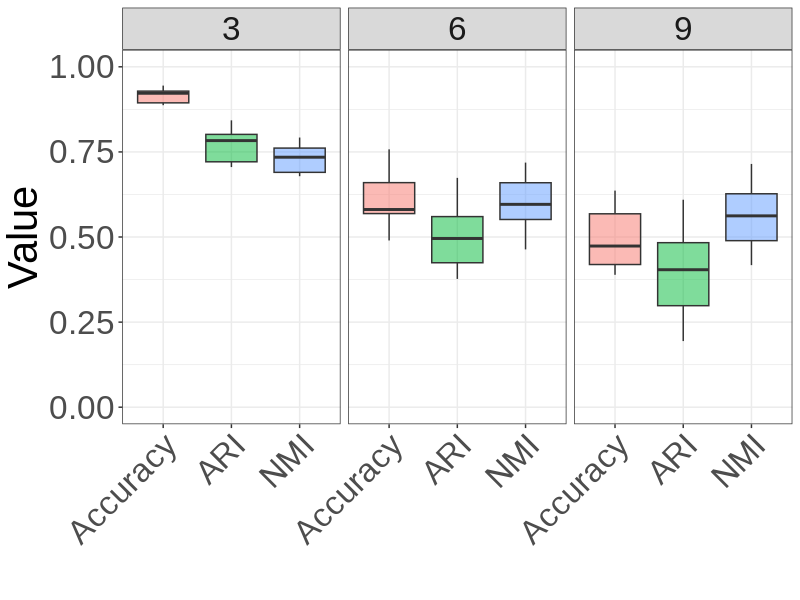}
            \caption{Increasing number of clusters.}
    \end{subfigure}\\
        \caption{Clustering accuracy decreases as the number of clusters increases. 
        The adjusted Rand index (ARI), normalized mutual information (NMI) and the clustering accuracy rate (Accuracy, the match rate between $\hat c_i$ and the ground-truth, minimized over all possible label switchings in $\hat c_i$) are shown. \label{fig:inc_clus_num}
        }
        \end{figure}

For the second scenario, we generate $n=400$ data points from a $\tilde K$-component bivariate $t$ distribution with $5$ degrees of freedom,  $y_i \sim  \tilde  \sum_{k=1}^{\tilde K} (1/\tilde K)  t_5 (\cdot \mid [b_k,b_k])$. We experiment with different values of $\tilde K\in \{3,6,9\}$ to have different numbers of clusters, as well as different values of $b_k=3(k-1)$, $4(k-1)$ or $5(k-1)$ to have different degrees of separation between cluster centers.  The results are shown in Figure \ref{fig:inc_clus_num}.

\begin{figure}[H]
         \begin{subfigure}[t]{.3\textwidth}
        \includegraphics[width=1\linewidth]{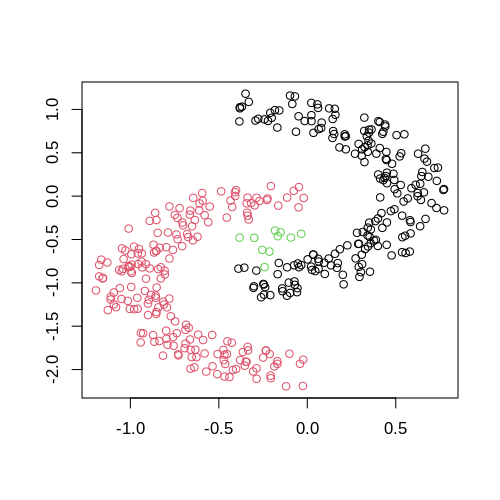}
    \end{subfigure}\;
         \begin{subfigure}[t]{.3\textwidth}
        \includegraphics[width=1\linewidth]{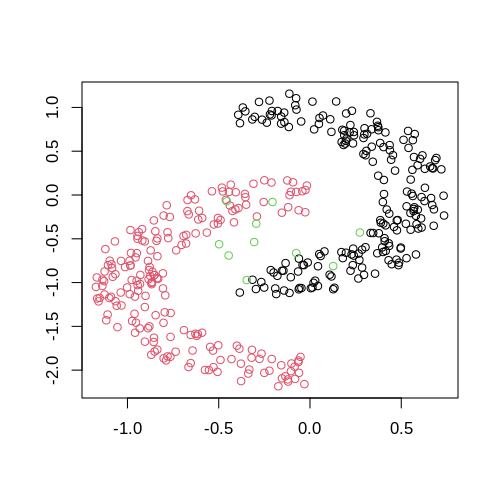}
    \end{subfigure}\;
     \begin{subfigure}[t]{.3\textwidth}
        \includegraphics[width=1\linewidth]{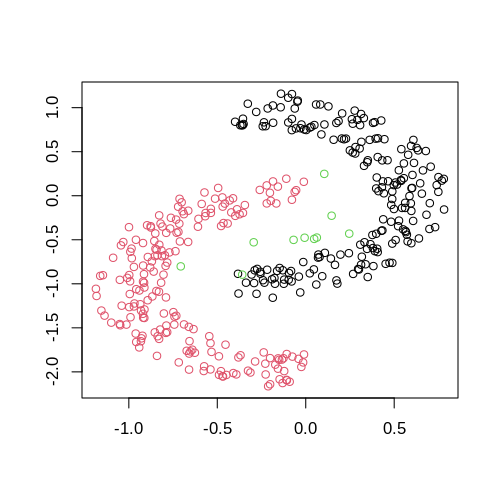}
    \end{subfigure}\\
         \begin{subfigure}[t]{.3\textwidth}
        \includegraphics[width=1\linewidth]{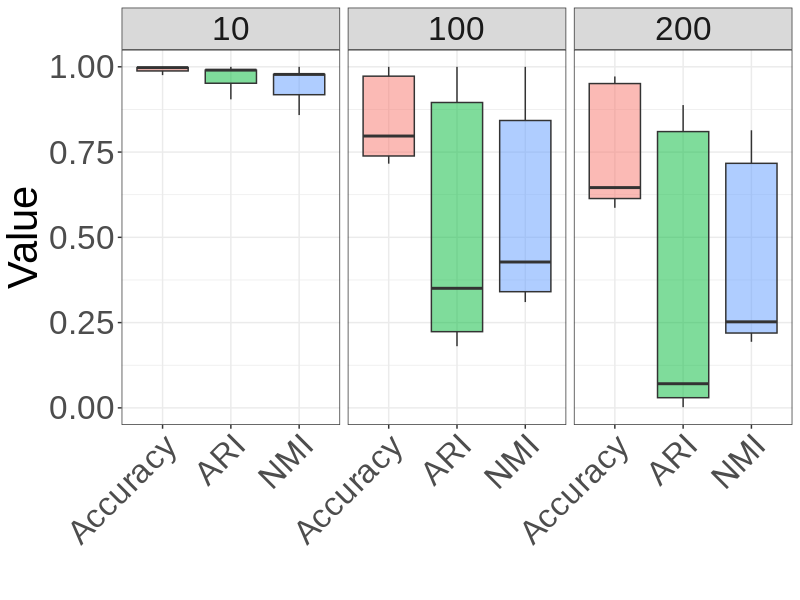}
            \caption{Noisy points (green, with variance $0.1^2$) between clusters (above). The clustering accuracy measures are collected (below) with the number of noisy points  taken from $\{10,100,200\}.$}
    \end{subfigure}\;
         \begin{subfigure}[t]{.3\textwidth}
        \includegraphics[width=1\linewidth]{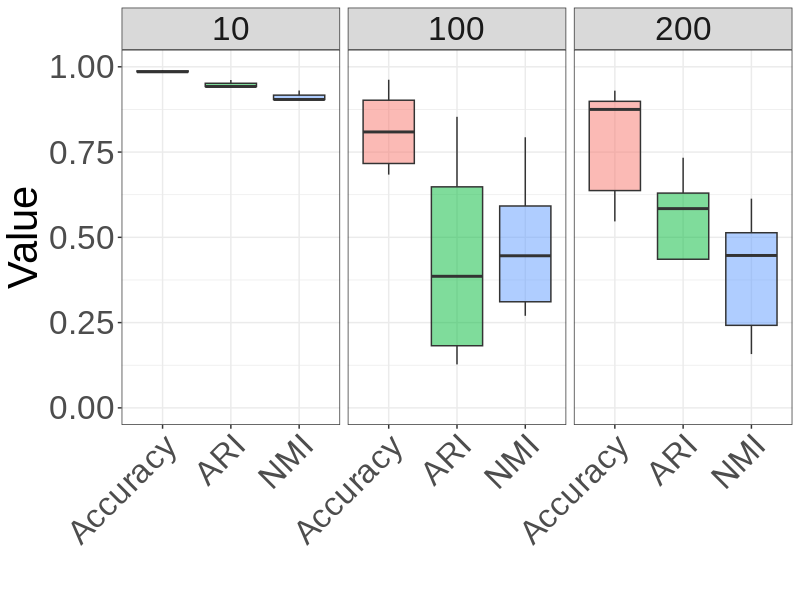}
            \caption{Noisy points (green, with variance $0.2^2$) between clusters (above). The clustering accuracy measures are collected (below) with the number of noisy points  taken from $\{10,100,200\}.$}
    \end{subfigure}\;
     \begin{subfigure}[t]{.3\textwidth}
        \includegraphics[width=1\linewidth]{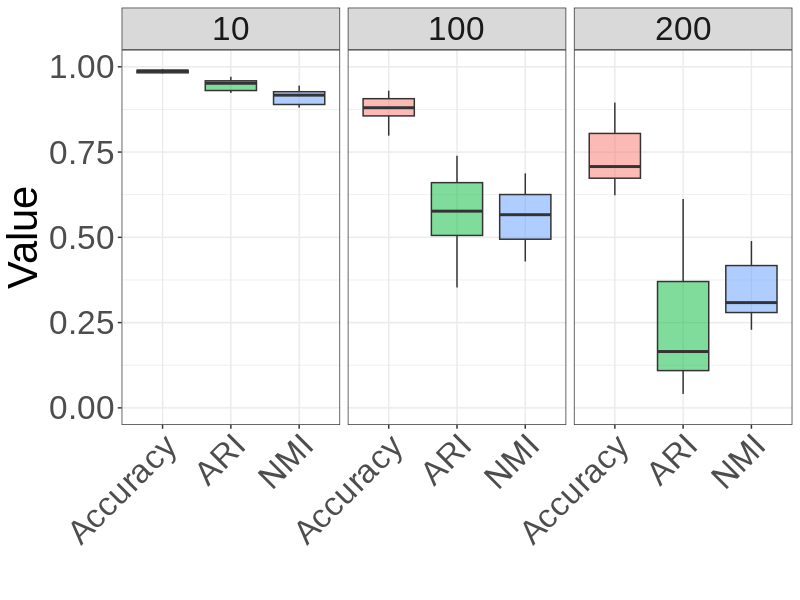}
            \caption{Noisy points (green, with variance $0.3^2$) between clusters (above). The clustering accuracy measures are collected (below) with the number of noisy points  taken from $\{10,100,200\}.$}
    \end{subfigure}\\
        \caption{Clustering accuracy decreases as the number of noisy points between clusters increases. 
        The adjusted Rand index (ARI), normalized mutual information (NMI), and the clustering accuracy rate (Accuracy, the match rate between $\hat c_i$ and the ground truth, minimized over all possible label switchings in $\hat c_i$) are shown. \label{fig:lcg}
        }
        \end{figure}

For the third scenario, we first generate $n=400$ data points near the two moon manifolds that are well separated from one another, then we add $m$ number of points generated from Gaussian distribution with variance $\tilde \gamma^2$, and its center placed between the two manifolds. We experiment with different values of $m\in \{10,100,200\}$, so that the clusters would appear somewhat connected to each other as $m$ increases; we also vary $\tilde \gamma^2\in \{0.1^2,0.2^2,0.3^2\}$ to have different levels of noise.  The results are shown in Figure \ref{fig:lcg}.

\subsection{Diagnostic plots for Markov chain Monte Carlo}

The MCMC algorithm that we describe in the main text shows a fast mixing of Markov chains. To illustrate this, we use the Markov chain collected from the experiment related to Figure S.6(k), and calculate the autocorrelations in (i) the degrees for each node in the forest $D_{i,i}$'s, and (ii) the number of clusters $K$. We plot the results in Figure S.8.
         
\begin{figure}[H]
%     \begin{subfigure}[t]{.3\textwidth}
%        \includegraphics[height=4cm,width=1\linewidth, trim=.5cm 2cm 4cm 8cm,clip]{trace_sigma_tilde}
%            \caption{Traceplot of one scale parameter $\tilde \sigma_1$.}
%    \end{subfigure} \;
     \begin{subfigure}[t]{.45\textwidth}
       \includegraphics[height=5cm,width=1\linewidth, trim=.5cm 2cm 4cm 8cm,clip]{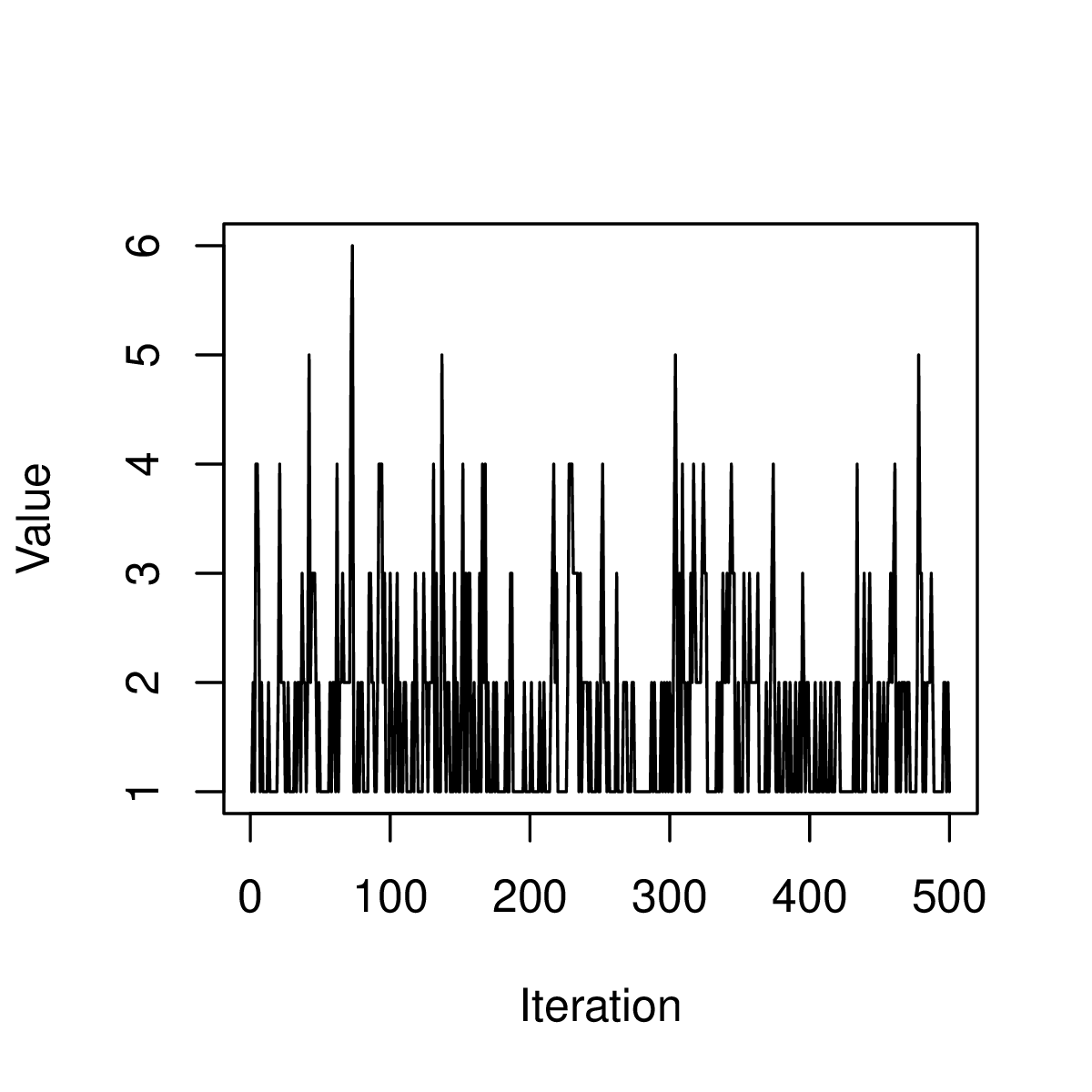}
            \caption{Traceplot of one degree in the forest $D_{1,1}$.}
    \end{subfigure}\quad
    \begin{subfigure}[t]{.45\textwidth}
        \includegraphics[height=5cm,width=1\linewidth, trim=2cm .5cm 4cm 8cm,clip]{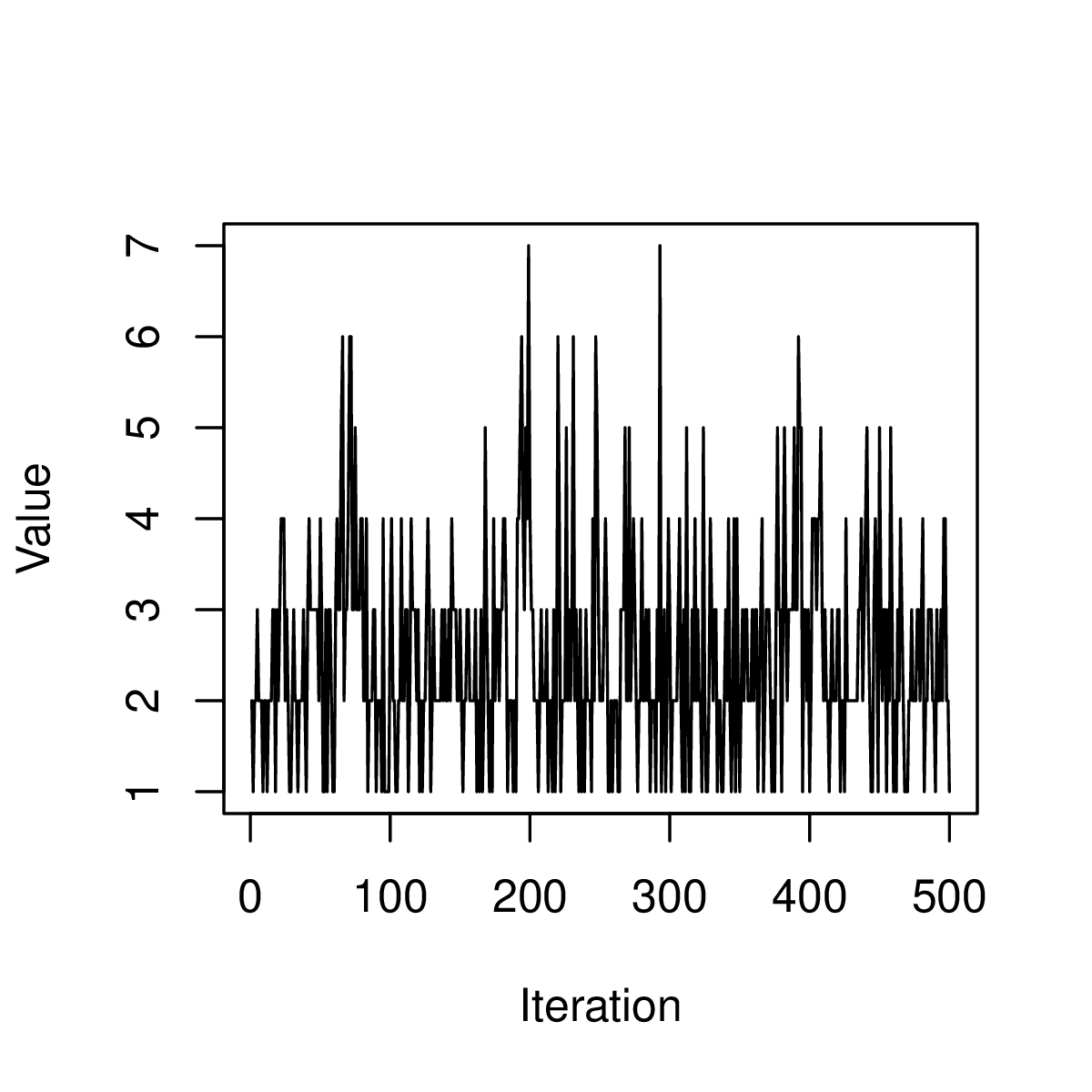}
            \caption{Traceplot of the number of clusters $K$.}
    \end{subfigure}
%         \begin{subfigure}[t]{.3\textwidth}
%        \includegraphics[height=3.5cm,width=1\linewidth, trim=-4.2cm .5cm 0cm .5cm,clip]{acf_sigma_tilde}
%            \caption{Boxplot of the autocorrelations for $\tilde \sigma_i$'s.}
%    \end{subfigure} \;
     \begin{subfigure}[t]{.45\textwidth}
   \includegraphics[height=3.5cm,width=1\linewidth, trim=-6cm .5cm -2cm .5cm,clip]{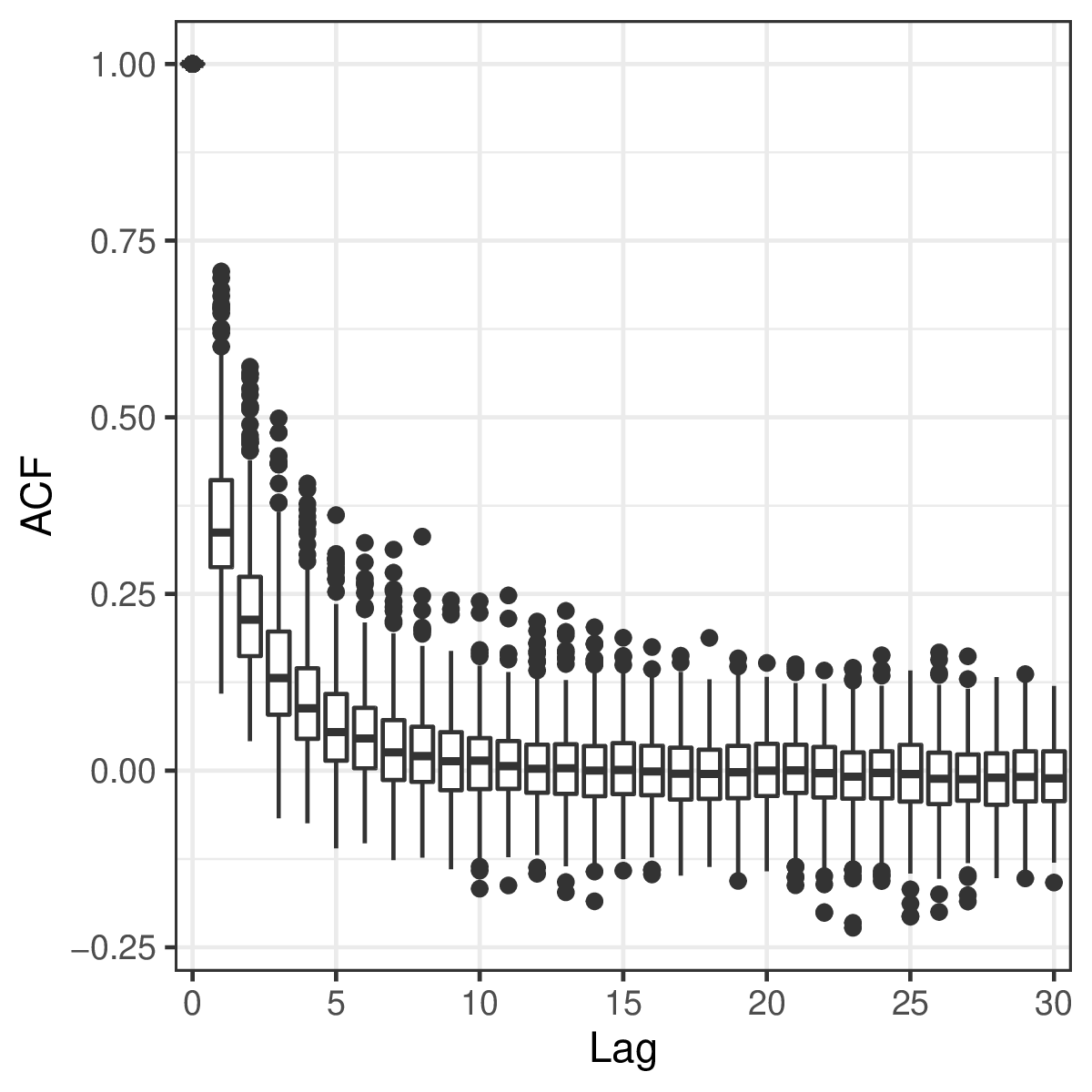}
            \caption{Boxplot of the autocorrelations for $D_{i,i}$'s.}
    \end{subfigure}\quad
    \begin{subfigure}[t]{.45\textwidth}
        \includegraphics[height=4cm,width=1\linewidth, trim=0.6cm 7cm 3cm 5cm,clip]{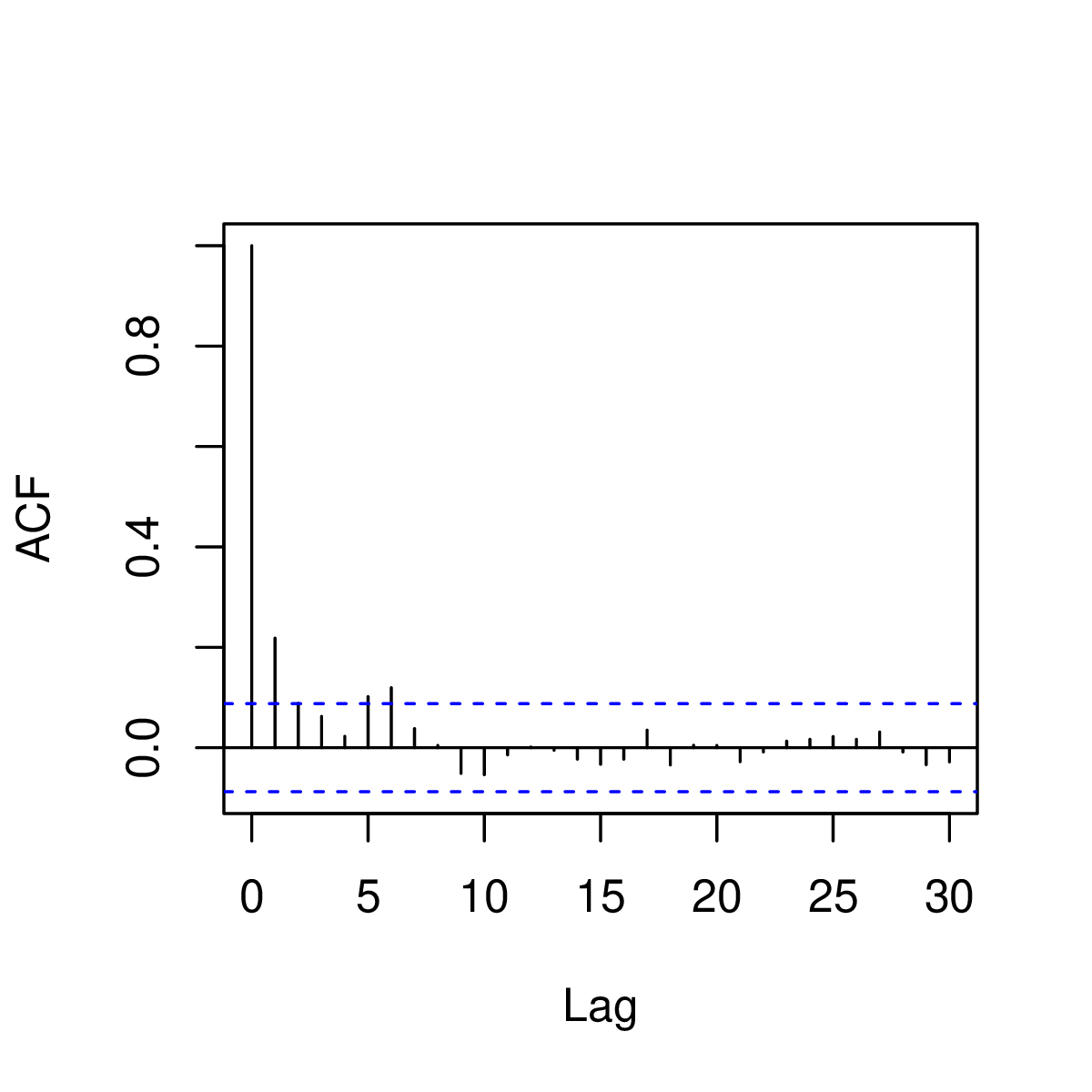}
            \caption{Autocorrelation for the number of clusters $K$.}
    \end{subfigure}
            \caption*{Figure S.10: Traceplots and autocorrelation plots show fast mixing of the MCMC algorithm.
        }
        \end{figure}

 To demonstrate the high efficiency of updating $\mathcal T$ in a block via the random-walk covering algorithm  \citep{broder1989generating,aldous1990random,mosbah1999non}, we plot the sampled $\mathcal T$ at three contiguous iterations (after burn-ins) in Figure S.9. The forest shows a rapid change from iteration to iteration --- indeed, the proportion of edge changes (the number of edges $\{(i,j): (i,j)\in \mathcal T_{[t]}, (i,j)\not \in \mathcal T_{[t+1]}\}$ divided by the total number of edges) is around 50\% at each iteration.

        \begin{figure}[H]
     \begin{subfigure}[t]{.23\textwidth}
        \includegraphics[height=4cm,width=1\linewidth, trim=1cm 1cm 1cm 1cm,clip]{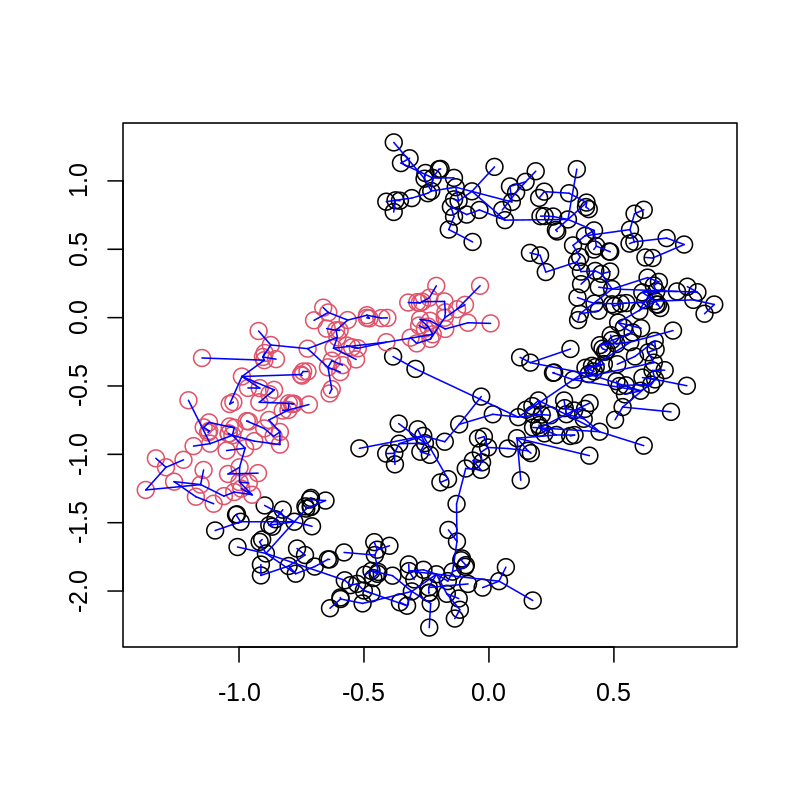}
            \caption{\small Sampled $\mathcal T$ at iteration 1 after burn-ins.}
    \end{subfigure} \;
     \begin{subfigure}[t]{.23\textwidth}
        \includegraphics[height=4cm,width=1\linewidth, trim=1cm 1cm 1cm 1cm,clip]{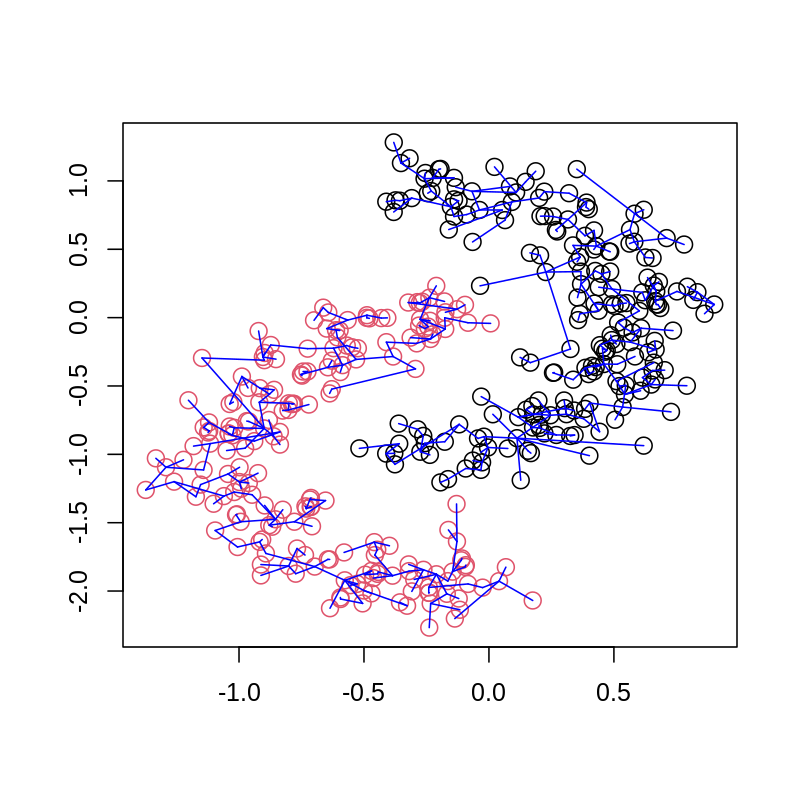}
            \caption{\small Sampled $\mathcal T$ at iteration 2 after burn-ins.}
    \end{subfigure}\;
     \begin{subfigure}[t]{.23\textwidth}
        \includegraphics[height=4cm,width=1\linewidth, trim=1cm 1cm 1cm 1cm,clip]{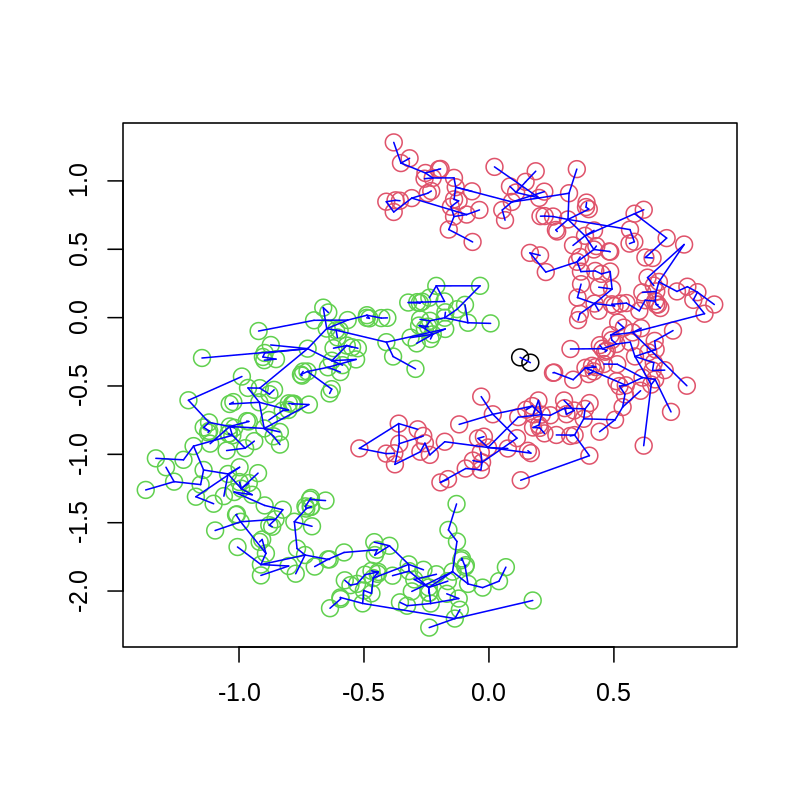}
            \caption{\small Sampled $\mathcal T$ at iteration 3 after burn-ins.}
    \end{subfigure}
     \begin{subfigure}[t]{.23\textwidth}
     \begin{overpic}[height=4cm,width=1\linewidth, trim=0cm 0.5cm 1cm 0.3cm]
			{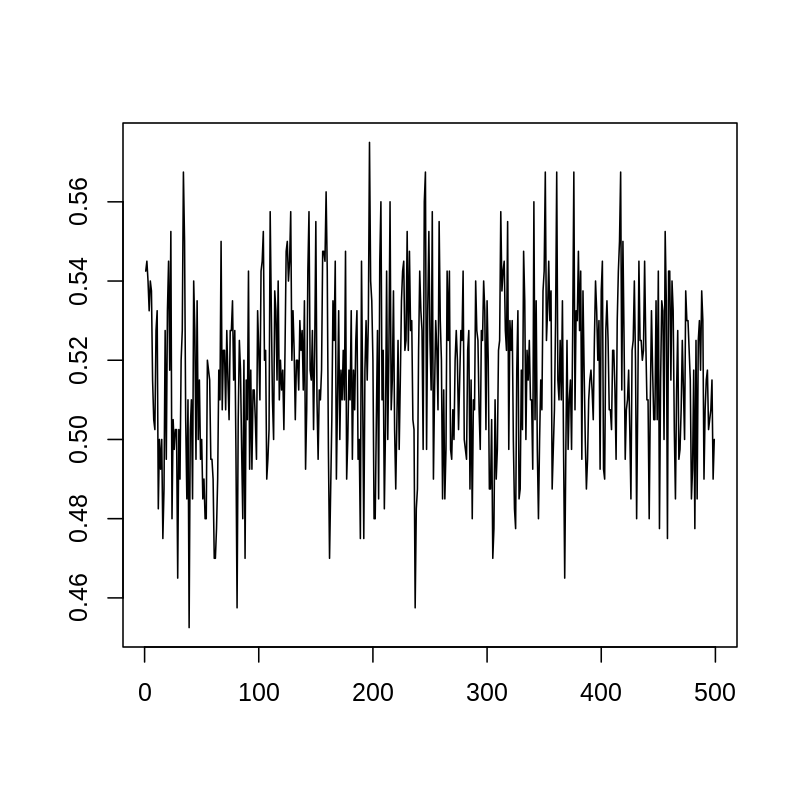}
			\put(40,0){\footnotesize Iteration}
			\end{overpic}
            \caption{\footnotesize Proportion of edge changed from $\mathcal T_{[t]}$ to $\mathcal T_{[t+1]}$.}
    \end{subfigure} \;
     \begin{subfigure}[t]{.23\textwidth}
        \includegraphics[height=4cm,width=1\linewidth, trim=1cm 1cm 1cm 1cm,clip]{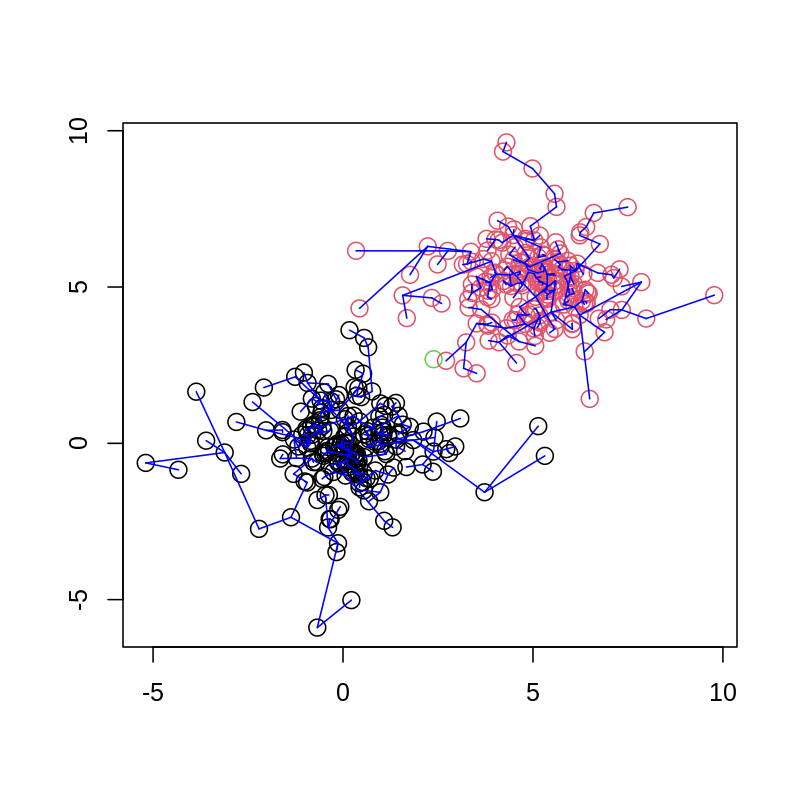}
            \caption{\small Sampled $\mathcal T$ at iteration 1 after burn-ins.}
    \end{subfigure} \;
     \begin{subfigure}[t]{.23\textwidth}
        \includegraphics[height=4cm,width=1\linewidth, trim=1cm 1cm 1cm 1cm,clip]{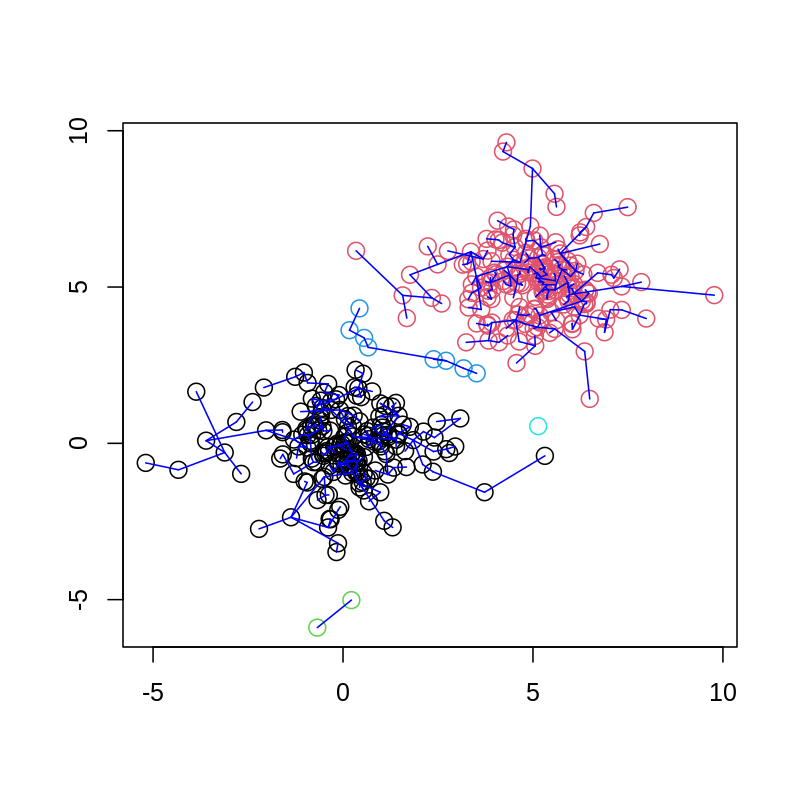}
            \caption{\small Sampled $\mathcal T$ at iteration 2 after burn-ins.}
    \end{subfigure}\;
     \begin{subfigure}[t]{.23\textwidth}
        \includegraphics[height=4cm,width=1\linewidth, trim=1cm 1cm 1cm 1cm,clip]{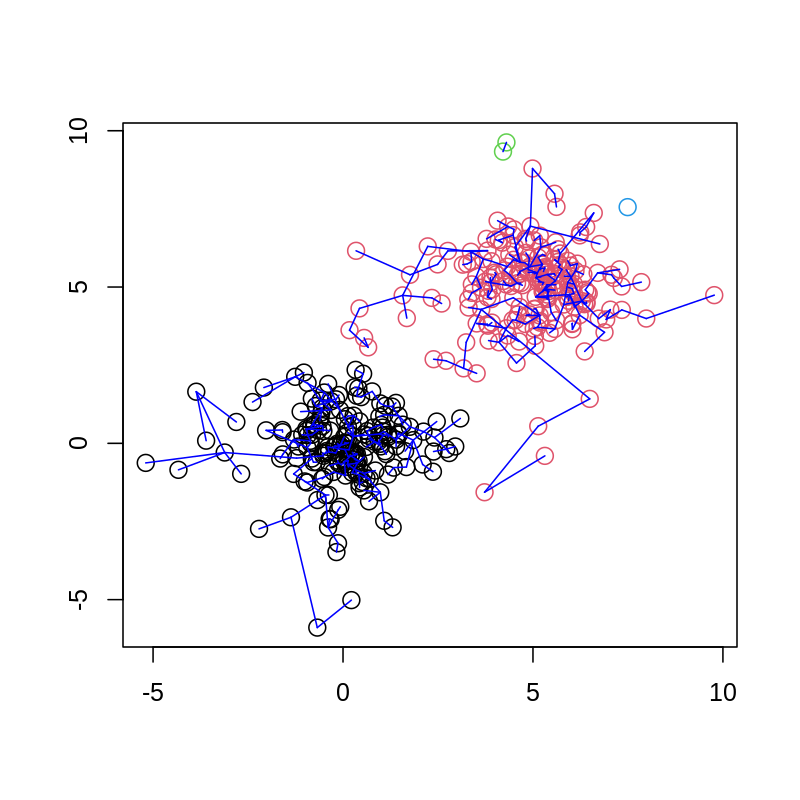}
            \caption{\small Sampled $\mathcal T$ at iteration 3 after burn-ins.}
    \end{subfigure}\;
     \begin{subfigure}[t]{.23\textwidth}
     \begin{overpic}[height=4cm,width=1\linewidth, trim=0cm 0.5cm 1cm 0.3cm]
			{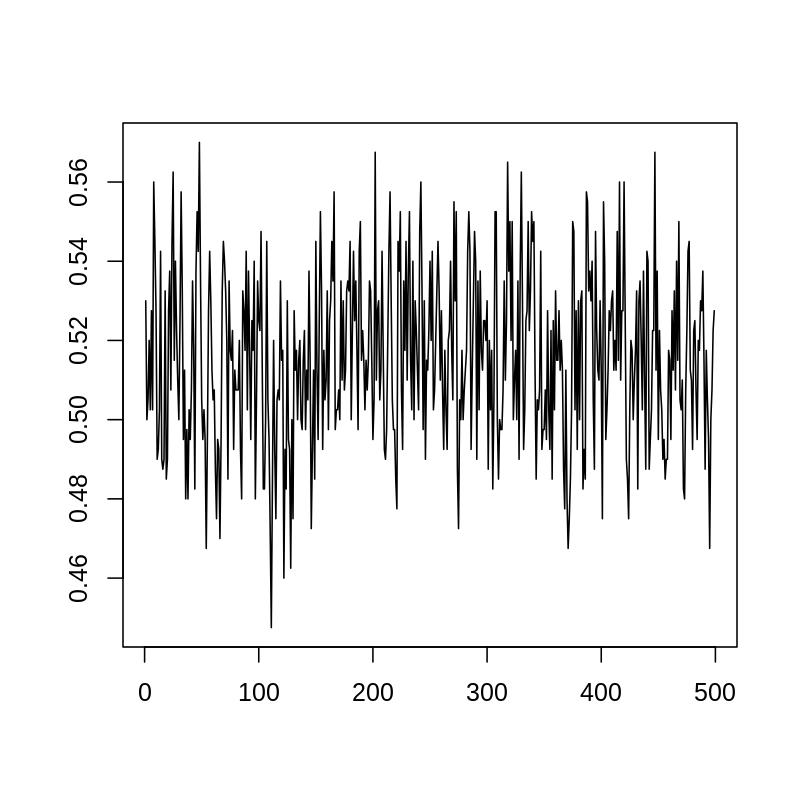}
			\put(40,0){\footnotesize Iteration}
			\end{overpic}
            \caption{\footnotesize Proportion of edge changed from $\mathcal T_{[t]}$ to $\mathcal T_{[t+1]}$.}
    \end{subfigure} \;
            \caption*{Figure S.11: The forest $\mathcal T$ changes rapidly from one iteration to another: Panels a-c plot the forests (blue) at three contiguous iterations, and Panel d shows the proportion of edge changes in each iteration, as measured by the number of edges $\{(i,j): (i,j)\in \mathcal T_{[t]}, (i,j)\not \in \mathcal T_{[t+1]}\}$ divided by the total number of edges.  Panels e-h show the results from another experiment. In both cases, the forest $\mathcal T$ has about 50\% of edges changed from one iteration to the next.
        }
        \end{figure}

Further, we assess the convergence by randomly initializing $(\mathcal T_{[0]},\theta_{[0]})$ at 5 different start points, and run 5 separate Markov chains. Specifically, for the elements $\tilde\sigma_i$ and $\gamma$ in $\theta_{[0]}$, we initialize them randomly from $\text{Inverse-Gamma}(0.5,0.5)$, then we draw $\mathcal T_{[0]}\sim\Pi(\mathcal T\mid \theta_{[0]}, y)$. Figure S.10 shows two randomly initialized $\mathcal T$'s. The traceplots of the parameters show the convergence of 5 Markov chains, and we calculate the Gelman--Rubin statistics (potential scale reduction factor, \cite{gelman1992inference}) and find all of them smaller than $1.1$, which indicates convergence.

\begin{figure}[H]
 \begin{subfigure}[t]{.23\textwidth}\centering
     \begin{overpic}[height=4cm,width=4cm, trim=0cm 0.5cm 1cm 0.3cm]
			{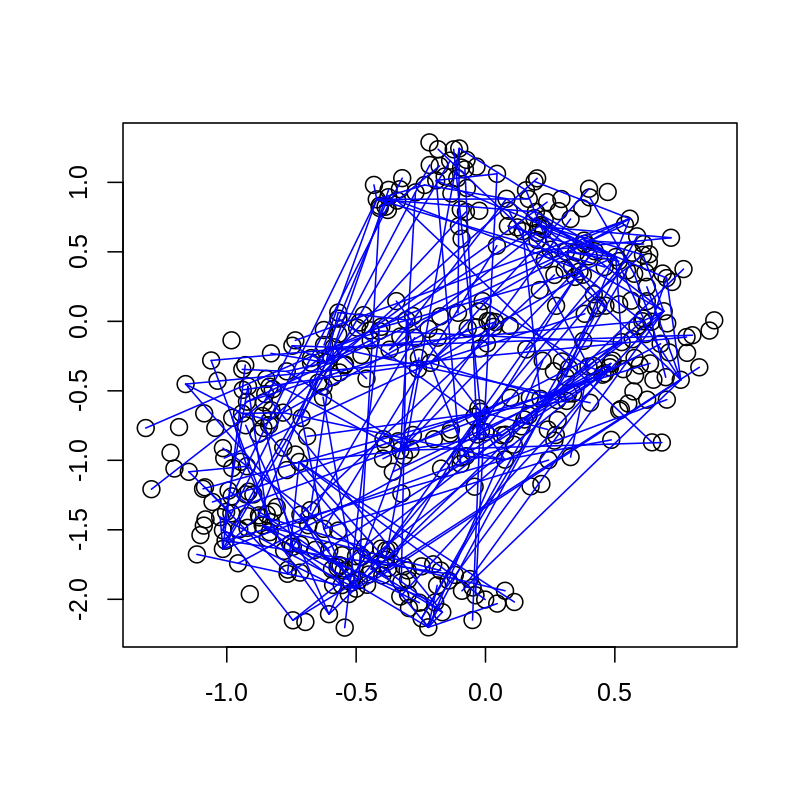}
			\end{overpic}
            \caption{\footnotesize Randomly initialized $\mathcal T$ in Chain 1.}
    \end{subfigure} \;
     \begin{subfigure}[t]{.23\textwidth}\centering
     \begin{overpic}[height=4cm,width=4cm, trim=0cm 0.5cm 1cm 0.3cm]
			{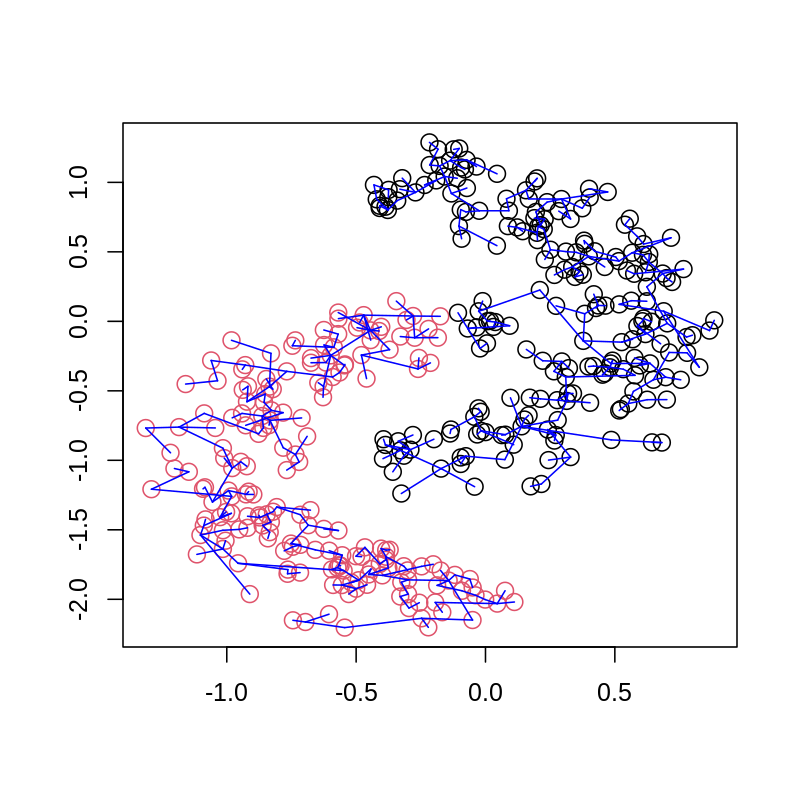}
			\end{overpic}
            \caption{\footnotesize One sampled $\mathcal T$ in Chain 1 after burn-ins.}
    \end{subfigure} \;
     \begin{subfigure}[t]{.23\textwidth}\centering
     \begin{overpic}[height=4cm,width=4cm, trim=0cm 0.5cm 1cm 0.3cm]
			{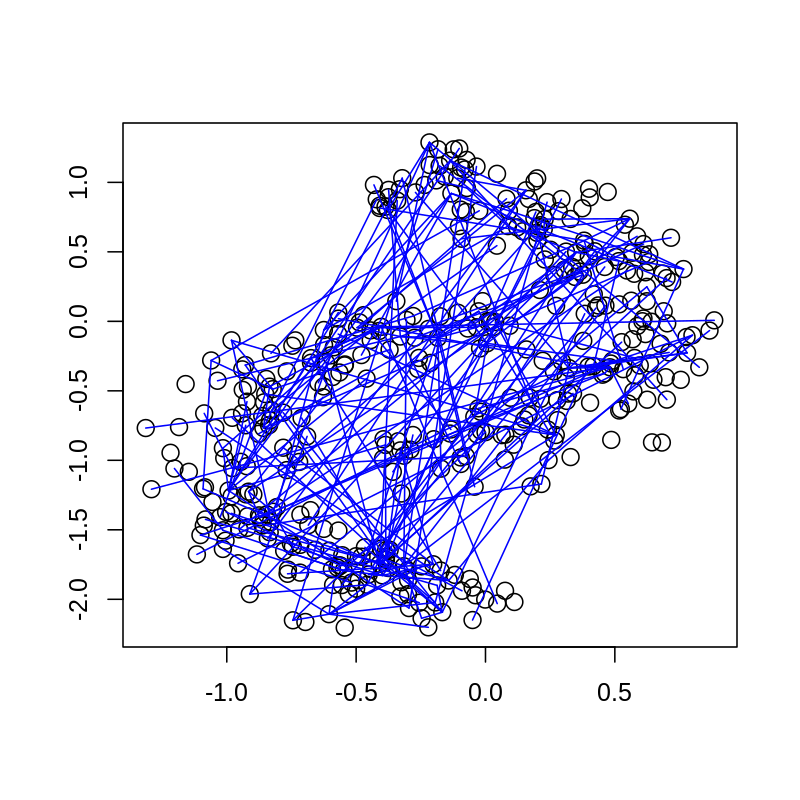}
			\end{overpic}
            \caption{\footnotesize Randomly initialized $\mathcal T$ in Chain 2.}
    \end{subfigure} \;    
     \begin{subfigure}[t]{.23\textwidth}\centering
     \begin{overpic}[height=4cm,width=4cm, trim=0cm 0.5cm 1cm 0.3cm]
			{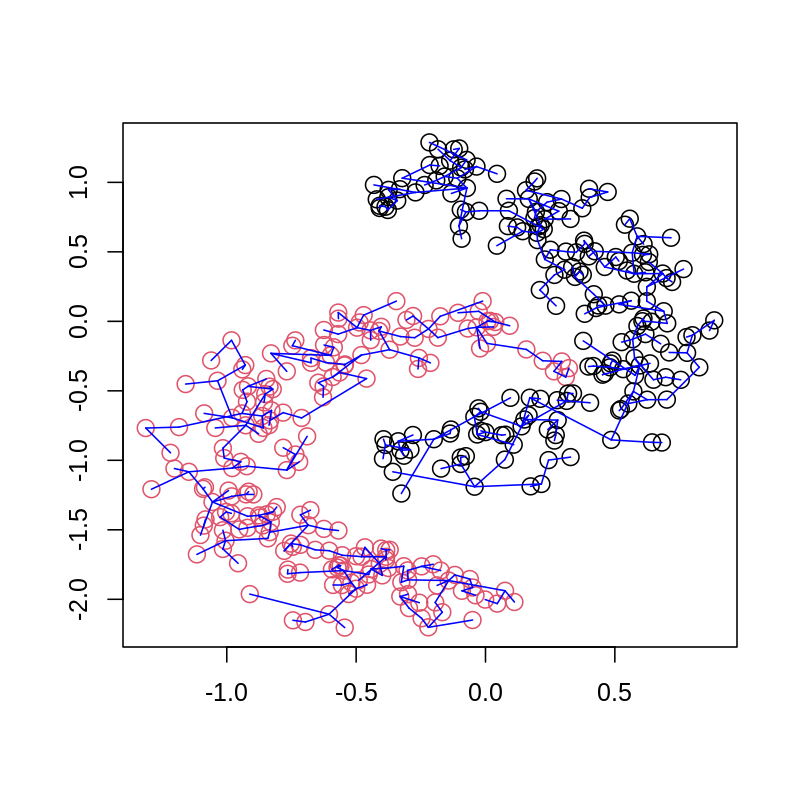}
			\end{overpic}
            \caption{\footnotesize One sampled $\mathcal T$ in Chain 2 after burn-ins.}
    \end{subfigure} \;
 \begin{subfigure}[t]{.3\textwidth}
     \begin{overpic}[height=4cm,width=1\linewidth, trim=0cm 0.5cm 1cm 0.3cm]
			{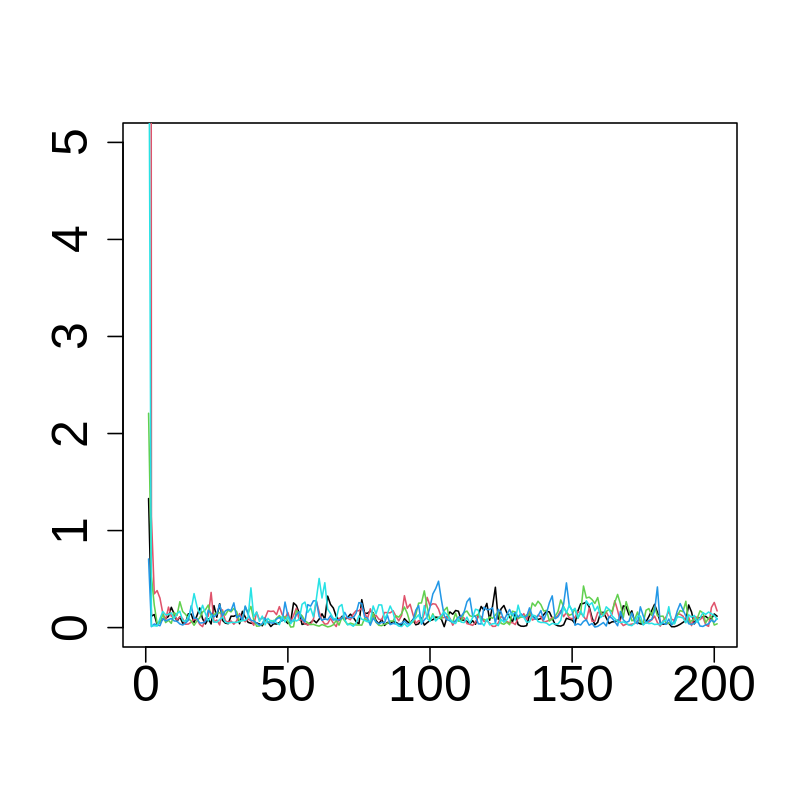}
			\put(40,0){\footnotesize Iteration}
			\end{overpic}
            \caption{\footnotesize Traceplot for the parameter $\tilde\sigma_1$ from 5 chains.}
    \end{subfigure} \;
     \begin{subfigure}[t]{.3\textwidth}
     \begin{overpic}[height=4cm,width=1\linewidth, trim=0cm 0.5cm 1cm 0.3cm]
			{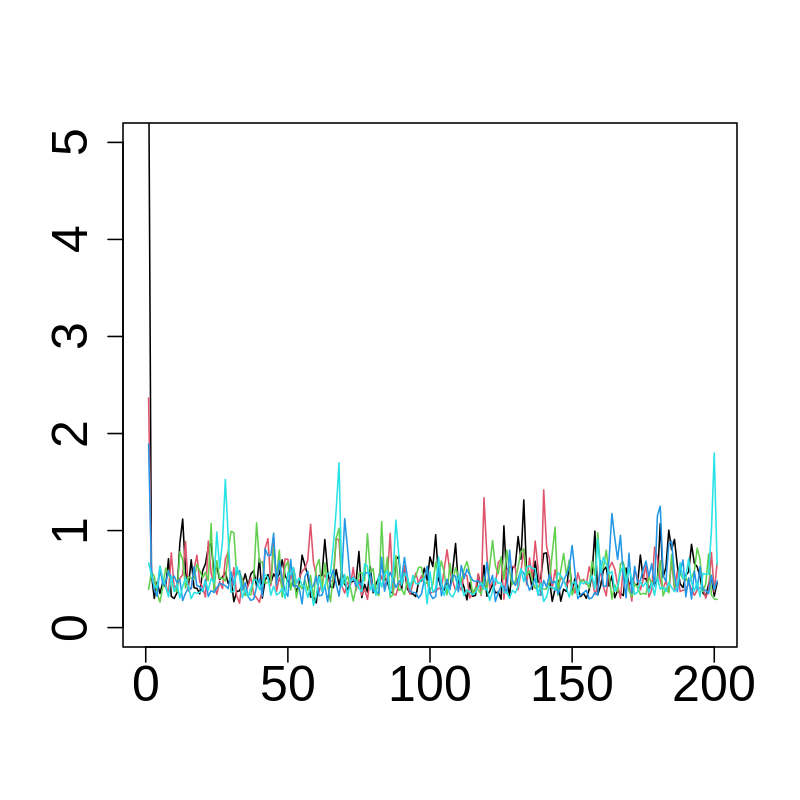}
			\put(40,0){\footnotesize Iteration}
			\end{overpic}
            \caption{\footnotesize Traceplot for the parameter  $\gamma$  from 5 chains.}
    \end{subfigure} \;    
     \begin{subfigure}[t]{.32\textwidth}
     \centering
     \begin{overpic}[height=4cm, trim=0cm 0.5cm 1cm 0.3cm]
			{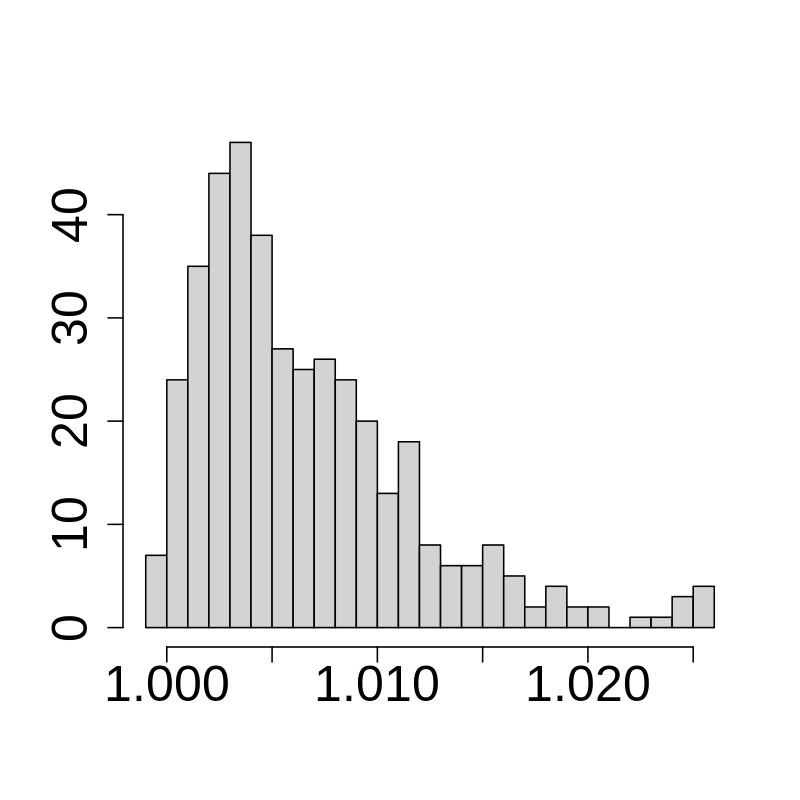}
			\put(40,0){\footnotesize Iteration}
			\end{overpic}
            \caption{\footnotesize Histogram of the Gelman--Rubin statistics for  all $\tilde\sigma_i$'s and $\gamma$.}
    \end{subfigure} 
            \caption*{Figure S.12: The convergence of five randomly initialized Markov chains.
        }
        \end{figure}

\subsection{Additional Details on the Multi-view Clustering in the Alzheimer's Disease Study}

\begin{figure}[H]
 \begin{subfigure}[t]{.45\textwidth}\centering
     \begin{overpic}[height=4cm,width=4cm, trim=0cm 0.5cm 1cm 0.3cm]
			{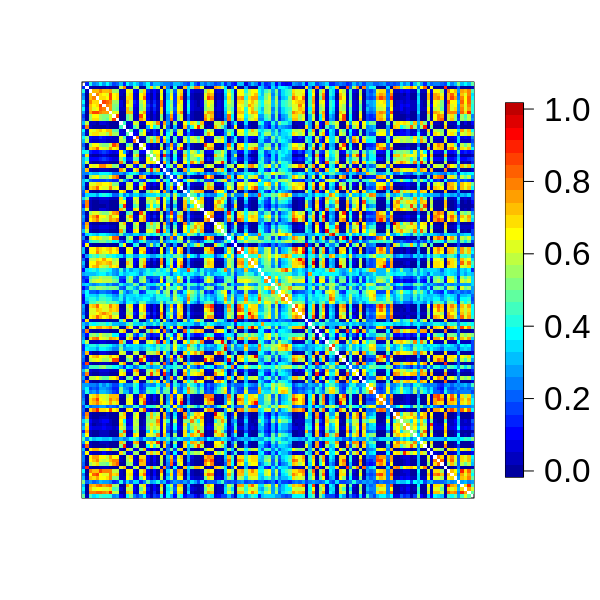}
			\end{overpic}
            \caption{\footnotesize The average $\text{Pr}(c_i=c_j \mid y)$ for the ROIs in the healthy group.}
    \end{subfigure} \;
     \begin{subfigure}[t]{.45\textwidth}\centering
     \begin{overpic}[height=4cm,width=4cm, trim=0cm 0.5cm 1cm 0.3cm]
			{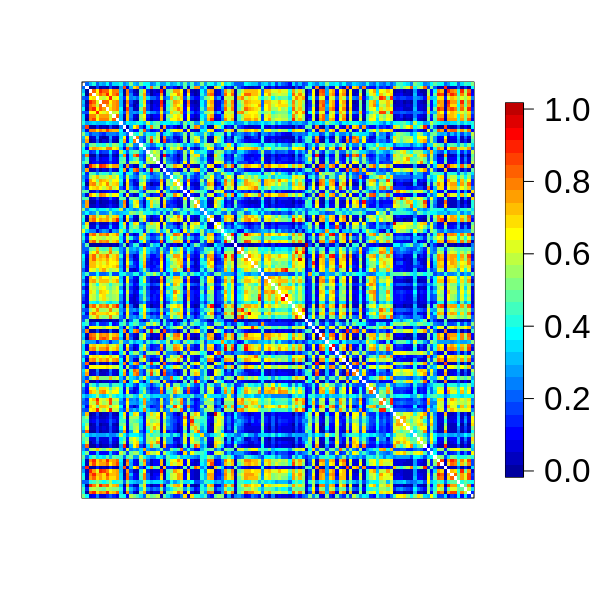}
			\end{overpic}
            \caption{\footnotesize The average $\text{Pr}(c_i=c_j \mid y)$ for the ROIs in the diseased group.}
    \end{subfigure} \;
     \begin{subfigure}[t]{.45\textwidth}\centering
     \begin{overpic}[height=4cm,width=4cm, trim=0cm 0.5cm 1cm 0.3cm]
			{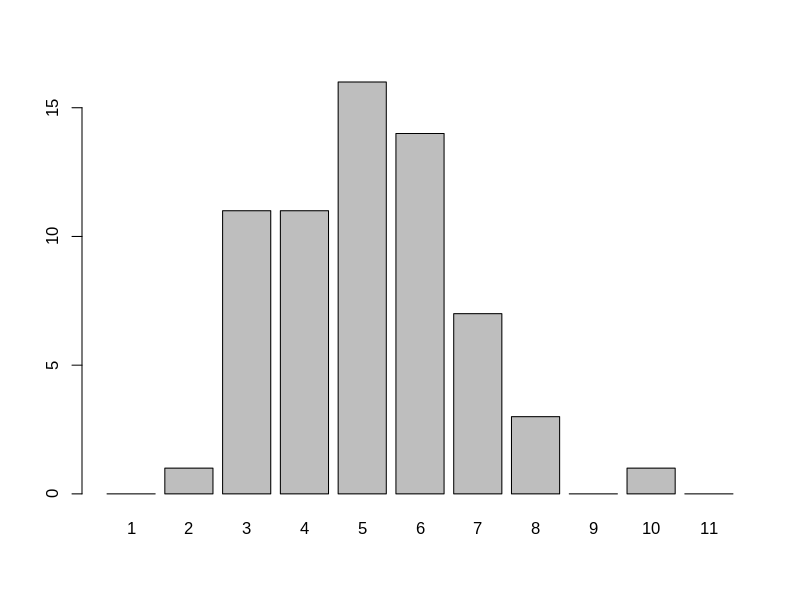}
			\end{overpic}
            \caption{\footnotesize Frequency plot of the number of clusters of ROIs for each subject in the healthy group.}
    \end{subfigure} \;
     \begin{subfigure}[t]{.45\textwidth}\centering
     \begin{overpic}[height=4cm,width=4cm, trim=0cm 0.5cm 1cm 0.3cm]
			{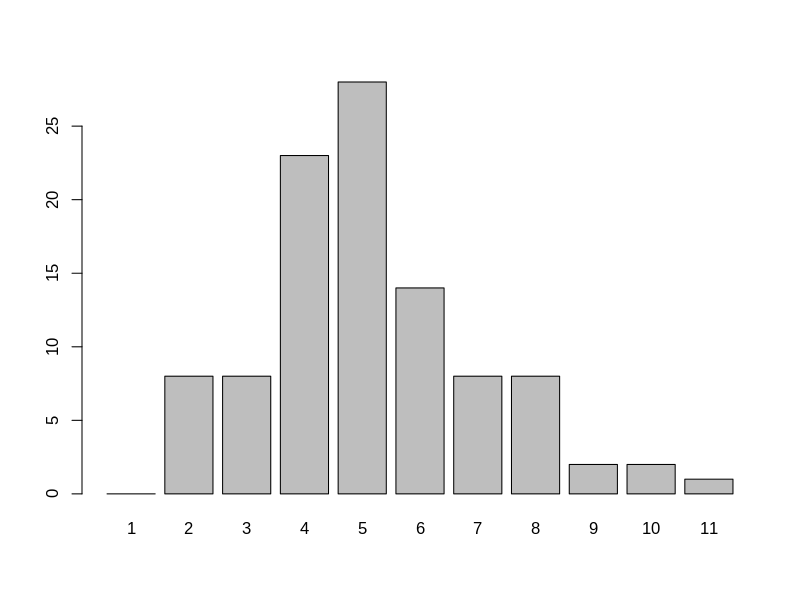}
			\end{overpic}
            \caption{\footnotesize Frequency plot of the number of clusters for each subject in the diseased group.}
    \end{subfigure}
            \caption*{Figure S.13: Clustering estimates for the healthy and diseased group.
        }
        \end{figure}

\subsection{Comparison with Minimum Spanning Tree-based Cut}\label{sec:compare_vs_mst_cut}
Since our Bayesian forest model uses spanning trees, it is natural to compare with the clustering algorithm based on cutting the minimum spanning tree (MST). To formalize, the minimum spanning tree-based cut (MST-Cut) algorithm first finds the MST:
$\hat T = {\arg\min}_{T\in \mathbb{T}_n}\sum_{(i,j)\in T} \|y_i-y_j\|,
$
where $\mathbb{T}_n$ denotes all spanning trees that connect $n$ nodes, with $\|y_i-y_j\|$ as some distance between the two points. Then, one removes the longest $(K-1)$ edges (with  length defined as $\|y_i-y_j\|$) to create $K$ clusters.
This algorithm is shown to be equivalent to the single-linkage clustering algorithm \citep{Hartigan1981}.

As we could imagine, such MST-Cut algorithms work well when clusters are well separated. In that case, those clusters will more likely be connected by the longest few edges. However, such algorithms will suffer sensitivity issues, when any one or more of the following happens: 1) when clusters are close to each other; 2) when a few isolated points are lying between two clusters; 3) when one or more clusters are from a heavy-tailed distribution, with a few points away from the bulk of the cluster. As a result, the longest edges in the MST may not be ideal for partitioning data.

\begin{figure}[H]
     \begin{subfigure}[t]{.3\textwidth}
        \includegraphics[width=1\linewidth, trim=20cm 10cm 15cm 10cm,clip]{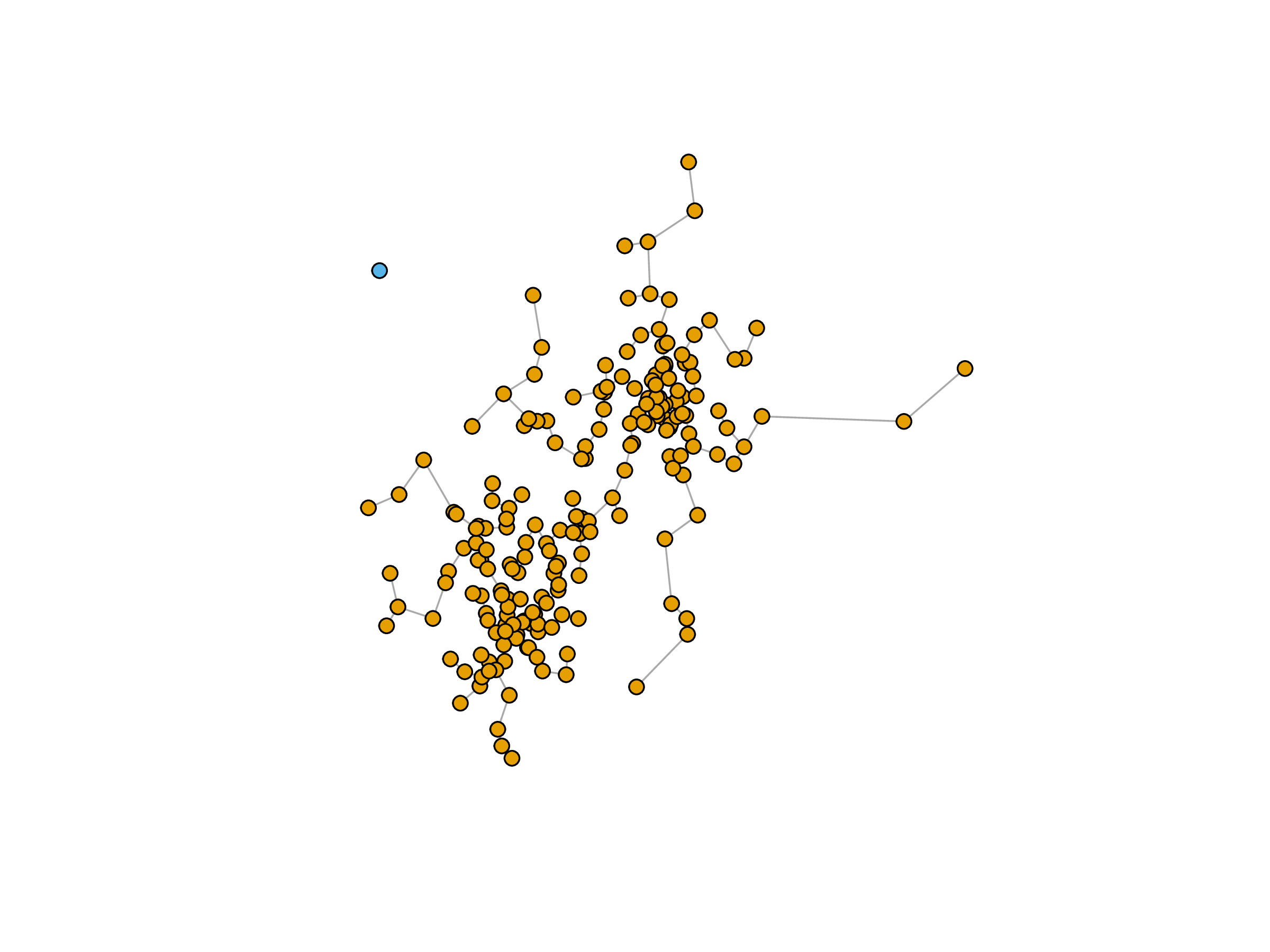}
            \caption{Partitioning the data by cutting the longest edge in the minimum spanning tree.}
    \end{subfigure}\;
         \begin{subfigure}[t]{.3\textwidth}
        \includegraphics[width=1\linewidth, trim=20cm 10cm 15cm 10cm,clip]{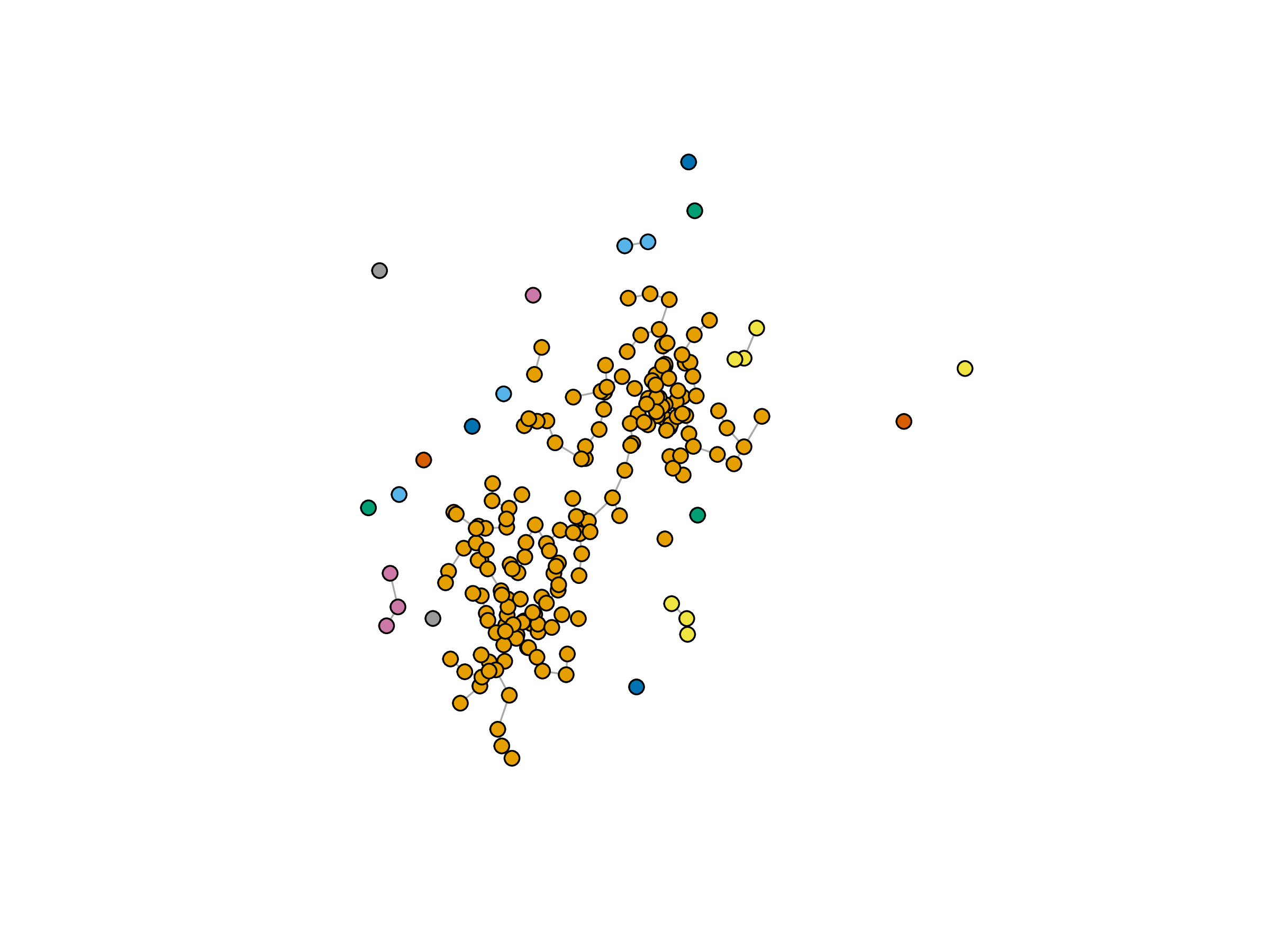}
            \caption{Partitioning the data by cutting the top $10\%$ longest edges in the minimum spanning tree.}
    \end{subfigure}\;
     \begin{subfigure}[t]{.3\textwidth}
        \includegraphics[width=1\linewidth, trim=20cm 10cm 15cm 10cm,clip]{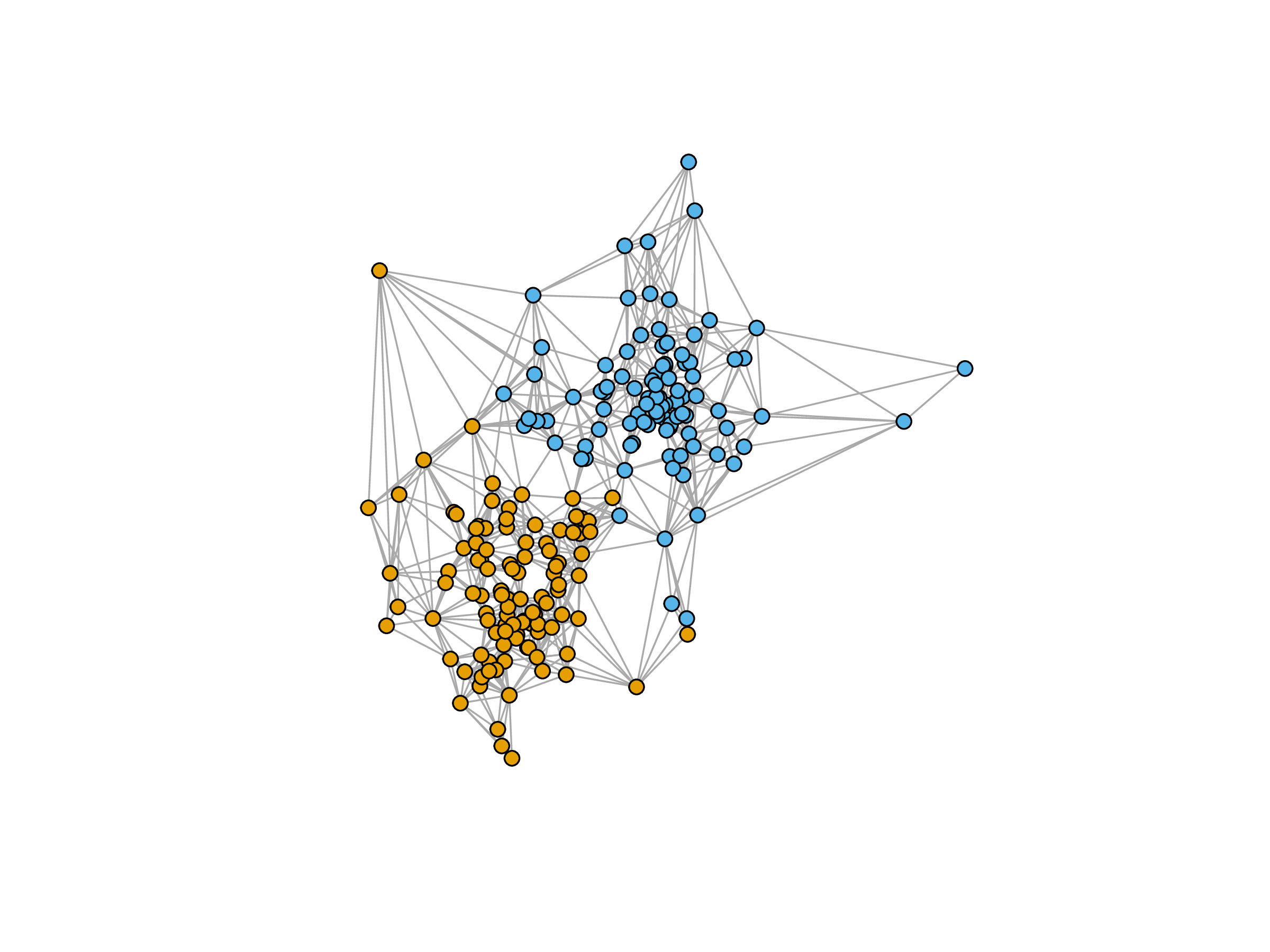}
            \caption{Partitioning the data using the Bayesian spanning forest model.}
    \end{subfigure}
        \caption{
            Comparing point estimates from the minimum spanning tree-based cut (MST-Cut) algorithms and the Bayesian spanning forest model. \label{fig:sensitivity_mst_cut}
        }
        \end{figure}

To illustrate this problem, we use a simulation with data from a two-component $t$-distribution in $\mathbb{R}^2$ with $3$ degrees of freedom. One component has the mean $(0,0)$ and the other has $(4,4)$, and both have the scale parameter equal to $I_2$. And we generate $n=200$ data points. As shown in Figure \ref{fig:sensitivity_mst_cut}(a), due to the heavy tail and closeness of the two clusters, cutting the longest edge in the MST (using Euclidean distances) yields a trivial and sub-optimal partition. Further, cutting the top $10\%$ longest edges still does not produce the desired result (Panel b).

Fundamentally, the reason is that relying on the minimum spanning tree (that is, one tree) leads to an underestimation of the graph uncertainty. Different from the MST-Cut algorithms, the Bayesian forest model effectively uses the marginal distribution  \eqref{eq:marginal_lik} incorporating the multiplicity of those likely trees (with edges shown in Panel c). As the result, it leads to better performance than the MST-Cut algorithm.

\subsection{Empirical Evidence for the Fast Convergence of Eigenvectors}

\begin{figure}[H]
        \centering
  % \begin{subfigure}[t]{.48\textwidth}
        \includegraphics[width=.6\linewidth]{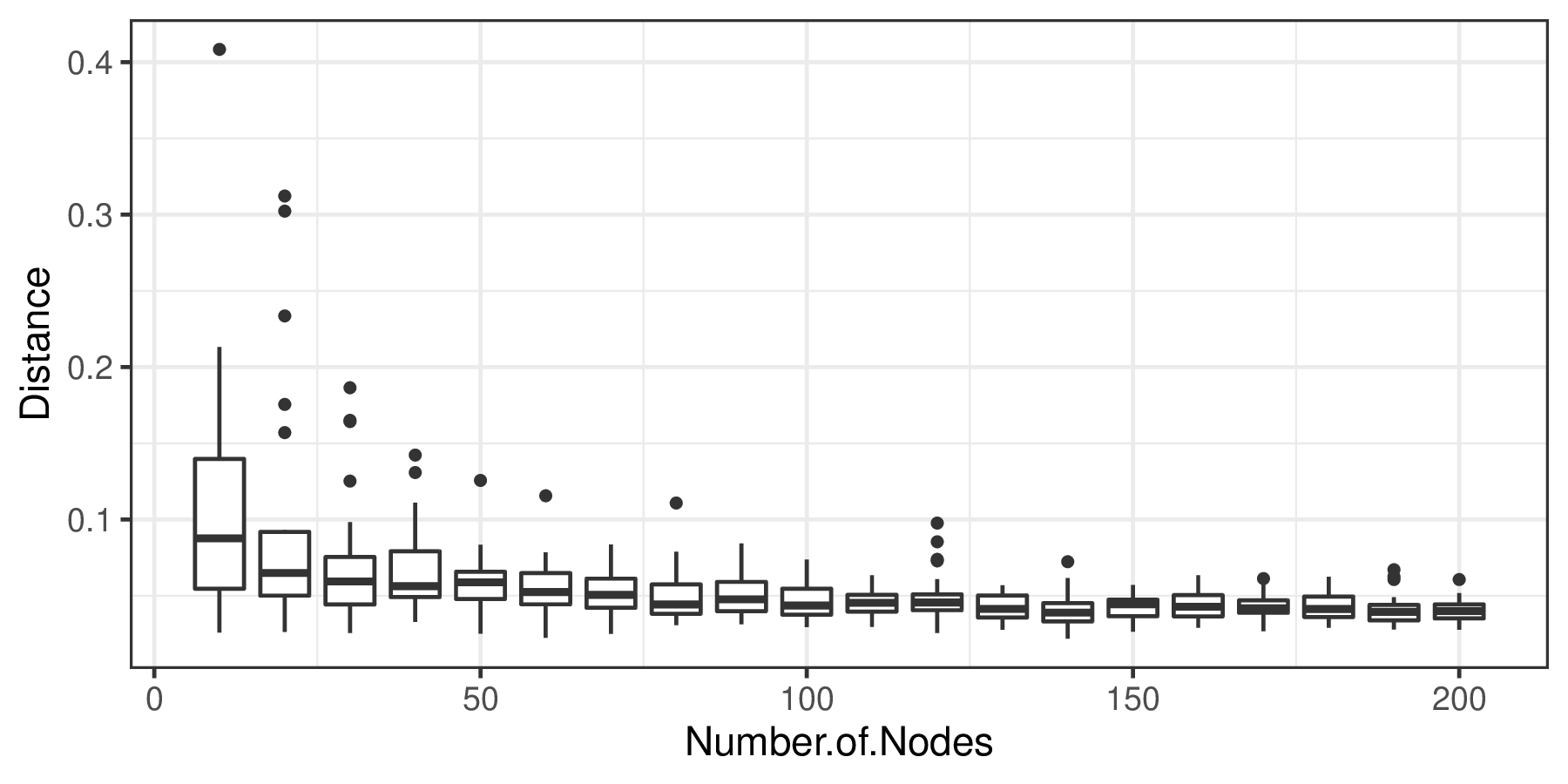}
               \caption*{Figure S.15: The difference between eigenvectors converges to zero rapidly as $n$ increases. \label{fig:ev_compare}}
    % \end{subfigure}
        % \caption{Numeric experiments showing that the top few eigenvectors of a marginal connecting probability matrix $M$ and the ones of normalized Laplacian become almost indistinguishable starting from $n=50$. 
%        }
        \end{figure}

We now use simulations to illustrate the closeness between the $K$ leading eigenvectors of the marginal connecting probability matrix $M$ and the ones of the normalized Laplacian $N$. It is important to clarify that such closeness does not depend on how the data are generated. Therefore, for simplicity, we generate $y_i$ from a simple three-component Gaussian mixture in $\mathbb{R}^2$ with means in $(0,0),(2,2),(4,4)$ and all variances equal to $I_2$, then we fit our forest model, and estimate $\sigma_{i,j}$'s using posterior mean. Based on the posterior mean of $\sigma_{i,j}$, we compute $M$ and $N$, and then compute distances between their leading eigenvectors $\min_{R: RR' =I_K}\|\Psi_{1:K}-\Phi_{1:K}R\|$. We conduct such experiments under different sample sizes $n$ ranging from $10$ to $200$; for each $n$, we repeat experiments for $30$ times. As our theory requires a spectral gap $\xi_{K} -\xi_{K+1}$ not too close to zero, we choose to compare the top $K=5$ eigenvectors. As shown in the boxplot of Figure S.4, the distance between two sets of eigenvectors quickly drops to near zero, for $n\ge 50$.

\section{Alternative Model for the Scale Parameters}

As an alternative to specifying a prior on the scale parameter $\tilde \sigma_i$ in the leaf density, the heuristic of setting $\tilde\sigma_{i}$ to a low order statistic of  $\{\|y_i-y_j\|_2\}_{j=1}^n$ is shown to enjoy a good empirical performance in spectral clustering  \citep{zelnik2005self}. In this section, we extend this heuristic to a formal model-based solution.

To start, we first relate the small distances to the $\tilde k$-nearest neighbor density estimator. \cite{loftsgaarden1965nonparametric} show that for $y_1, \cdots, y_{n}$ iid from a distribution with probability density $f$, with a growing $\tilde k\to \infty$, $\tilde k/n\to 0$ as $n$ increase, if $f$ is continuous at $y_i$, the $\tilde k$-nearest neighbor density estimator $f_n(y_i)=\tilde k /\left[ n V_{\tilde k}(y_i)\right]$ is consistent for estimating $f(y_i)$, where $V_{\tilde k}(y_i)$ is the volume of the ball centered at $y_i$ and with radius equal to the distance to the $\tilde k$-th nearest neighbor, denoted by $d^{(\tilde k)}_{i}$ from now on.

Although we no longer consider data as iid, $d^{(\tilde k)}_{i}$ is still informative on how dense the data points are near $y_i$. Therefore, to bring information from $d^{(\tilde k)}_{i}$ into the spanning forest model-based clustering, we consider a generative model that simultaneously depends on a spanning forest (with $K$ component trees) and a $\tilde k$-nearest neighbor graph $\tilde G_{nn}$. We can use a likelihood
\be
\mathcal L(y; \tilde G_{nn}, \mathcal T, \theta) \propto &\bigg\{ \prod_{i=1}^n  \frac{ (\tilde\sigma_i)^{\alpha_\sigma}}{\Gamma(\alpha_\sigma)}[\frac{d^{(\tilde k)}_i}{\sqrt{p}}]^{-\alpha_\sigma -1}\exp \bigg[ - \frac{\tilde\sigma_i}{d^{(\tilde k)}_i /\sqrt{p}} \bigg]\bigg\} \\
& \cdot \prod_{k=1}^K\bigg\{ r(y_{k^*}; \theta)\prod_{(i,j)\in  T_k}{(2\pi\sigma_{i,j})^{-p/2}} \exp \left ( - \frac {\|y_i-y_j\|_2^2}{2\tilde\sigma_{i}\tilde \sigma_j} \right) \bigg\}.
\ee
And one can verify that each term on the right-hand side is integrable in $y_i$, even if  $d^{(\tilde k)}_{i}=\|y_i-y_j\|_2$ happened (that is, when $(i,j)\in T_k$ and $j$ happened to be the $\tilde k$-nearest neighbor of $i$); therefore, the right-hand side forms a proper likelihood. We choose the inverse-gamma distribution for each $d^{(\tilde k)}/\sqrt{p}$, as it leads to an equivalent Gamma$(\alpha_\sigma+1,d^{(\tilde k)}_i/\sqrt{p})$ distribution for $\tilde\sigma_i$ that produces a shrinkage effect on $\tilde \sigma_i$ \citep{brown2010normalgamma}, and it enjoys closed-form Gibbs sampling update via the generalized inverse gaussian distribution. We test the above model using $\tilde k=  \lceil n^{1/10} \rceil$, and $\alpha_\sigma=1$ on all the examples presented in the article, and the results are quite similar to the ones shown in the main text. We provide the implementation in the R source code.

\section{Software Source Code}

The software source code for this paper can be found under \href{https://github.com/royarkaprava/Bayesian_forest_clustering/tree/code_for_the_paper}{\color{blue}this link} on github.

\end{document}